\documentclass[aps,nofootinbib,preprintnumbers,eqsecnum,superscriptaddress]{revtex4-1}
\usepackage[
      colorlinks=true,
      linkcolor=blue,
      urlcolor=blue,
      filecolor=black,
      citecolor=red,
      pdfstartview=FitV,
      pdftitle={},
        pdfauthor={Nabil Iqbal, Hong Liu, Mark Mezei},
        pdfsubject={},
        pdfkeywords={},
        pdfpagemode=None,
        bookmarksopen=true
      ]{hyperref}

\usepackage[normalem]{ulem}
\usepackage{amsmath}
\usepackage{enumerate}
\usepackage{amsfonts}
\usepackage{yfonts}

\usepackage{subfigure}
\usepackage{psfrag}

\usepackage{epsfig}
\usepackage[latin1]{inputenc}
\usepackage{float}
\usepackage{graphicx}
\usepackage{cancel}
\usepackage{mathrsfs}
\usepackage{amssymb}
\usepackage{amsfonts}
\usepackage{amsmath}
\usepackage{slashed}

\usepackage{amssymb}
\usepackage{latexsym}

\newcommand \tr {\mbox{{\bf Tr}}}

\usepackage{graphicx}
\usepackage{bm}

\def\({\left(}
\def\){\right)}
\def\[{\left[}
\def\]{\right]}
\def\<{\langle}
\def\>{\rangle}
\def\tr{\mathop{\rm tr}}

\newcommand\half{{\ensuremath{\frac{1}{2}}}}
\newcommand\p{\ensuremath{\partial}}

\newcommand\field[1]{{\ensuremath{\mathbb{{#1}}}}}

\newcommand\vev[1]{{\ensuremath{\left\langle{#1}\right\rangle}}}

\newcommand{\RR}{\field{R}}

\newcommand{\be}{\begin{equation}}
\newcommand{\ee}{\end{equation}}
\newcommand{\bea}{\begin{eqnarray}}
\newcommand{\eea}{\end{eqnarray}}
\newcommand{\bwt}{\begin{widetext}}
\newcommand{\ewt}{\end{widetext}}

\newcommand{\bi}{\begin{itemize}}
\newcommand{\ei}{\end{itemize}}
\newcommand{\ben}{\begin{enumerate}}
\newcommand{\een}{\end{enumerate}}
\newcommand{\bca}{\begin{cases}}
\newcommand{\eca}{\end{cases}}
\newcommand{\bln}{\begin{align}}
\newcommand{\eln}{\end{align}}
\newcommand{\bst}{\begin{split}}
\newcommand{\est}{\end{split}}

\newcommand\al{{\alpha}}
\newcommand\ep{\epsilon}
\newcommand\sig{\sigma}
\newcommand\Sig{\Sigma}
\newcommand\lam{\lambda}
\newcommand\Lam{\Lambda}
\newcommand\om{\omega}
\newcommand\Om{\Omega}

\newcommand\ga{{\ensuremath{{\gamma}}}}
\newcommand\Ga{{\ensuremath{{\Gamma}}}}
\newcommand\de{{\ensuremath{{\delta}}}}
\newcommand\De{{\ensuremath{{\Delta}}}}

\newcommand\ze{{\zeta}}

\newcommand\Th{{\Theta}}
\def\th{{\theta}}

\newcommand\ov{\over}
\newcommand\ha{{\half}}

\def\le{\left}
\def\ri{\right}

\newcommand\sA{{\ensuremath{{\mathcal A}}}}
\newcommand\sB{{\ensuremath{{\mathcal B}}}}

\newcommand\sD{{\ensuremath{{\mathcal D}}}}

\newcommand\sG{{\ensuremath{{\mathcal G}}}}

\newcommand\sL{{\ensuremath{{\mathcal L}}}}

\newcommand\sN{{\ensuremath{{\mathcal N}}}}
\newcommand\sO{{\ensuremath{{\mathcal O}}}}
\newcommand\sR{{\ensuremath{{\mathcal R}}}}

\newcommand\bpsi{{\bar \psi}}

\renewcommand{\Im}{\textrm{Im}\,}

\newcommand{\vk}{{\vec k}}
\newcommand{\ka}{{\kappa}}

\newcommand{\ta}{\tilde a}

\newcommand\ut{{\underline{t}}}
\newcommand\ur{{\underline{r}}}

\newcommand{\chis}{{\chi_*}}

\newcommand{\slql}{{semi-local quantum liquid}}
\newcommand{\Slql}{{SLQL}}

\newcommand\psinorm{\boldsymbol{\psi}}

\newcommand\Psinon{\mathfrak{Y}}

\long\def\nicefootnote[#1]#2{\begingroup%
\def\thefootnote{\arabic{footnote}}\footnote{#2}\endgroup}

\begin{document}

% UNCOMMENT THESE LINES FOR WS FORMAT

%\chapter{Lectures on holographic non-Fermi liquids and quantum phase transitions}

%\author[N. Iqbal, H. Liu, and M. Mezei]{Nabil Iqbal, Hong Liu, and Mark Mezei}
%\address{Center for Theoretical Physics, Massachusetts Institute of Technology, Cambridge, MA 02139}

% comment out from here to 
\title {Lectures on holographic non-Fermi liquids and quantum phase transitions}

\preprint{MIT-CTP 4282}

\author{Nabil Iqbal}
\affiliation{Kavli Institute for Theoretical Physics, University of California,
Santa Barbara, CA 93106 }
\author{ Hong Liu}
\affiliation{Center for Theoretical Physics,
Massachusetts
Institute of Technology,
Cambridge, MA 02139 }
\author{M\'ark Mezei}
\affiliation{Center for Theoretical Physics, Massachusetts Institute of Technology,
Cambridge, MA 02139 }
%.....HERE for WS format

\begin{abstract}

In these lecture notes we review some recent attempts at searching for non-Fermi liquids and novel quantum phase transitions in holographic systems using gauge/gravity duality. 
We do this by studying the simplest finite density system arising from the duality, obtained by turning on a nonzero chemical potential for a $U(1)$ global symmetry of a CFT, and described on the gravity side by a charged black hole. We address the following questions of such a finite density system:
\ben

\item Does the system have a Fermi surface?  \\ 
What are the properties of low energy 
excitations near the Fermi surface?

\item Does the system have an instability to condensation of scalar operators?  \\
What is the critical behavior near the corresponding quantum critical point?

\een
We find interesting parallels with those of high $T_c$ cuprates and heavy electron systems. Playing a crucial role in our discussion is a universal intermediate-energy phase, called a ``semi-local quantum liquid,'' which underlies the non-Fermi liquid and novel quantum critical behavior of a system. It also provides a novel mechanism for the emergence of lower energy states such as a Fermi liquid or a superconductor.

\vskip 2cm

{\it Lectures by HL at TASI 2010, Boulder, June 2010; KITPC workshop/school on ``AdS/CM duality and other approaches,'' Beijing, November, 2011; School on ``Holographic View of Condensed Matter Physics''
at International Institute of Physics, the Federal University of Rio Grande do Norte,  Natal, Brazil, July 2011.}

\end{abstract}

\today

% COMMENT OUT NEXT TWO LINES FOR WS FORMAT
\maketitle
\tableofcontents
% /end commetning out. 

%\newpage

\section{Introduction and motivation}

Understanding phases of matter for which there is no quasiparticle description 
presents some of the most challenging problems in physics. Examples include the quark-gluon plasma (QGP) created in heavy ion collisions at RHIC during the past decade and now at LHC and ultracold atomic systems at unitarity.  In condensed matter physics prominent examples include the quantum spin liquid phase of a magnetic insulator, the  ``strange metals'' occurring in the normal state of the high temperature cuprate superconductors, and heavy electron systems near a quantum phase transition. There has been an accumulation of examples, but so far no satisfactory theoretical framework exists to describe them. Finding solvable examples of quantum phases with no quasiparticles should provide a valuable guide for the search for such a framework. 

During the last decade, developments in string theory have revealed surprising and profound
connections between gravity and many-body systems.  The so-called gauge/gravity duality (or AdS/CFT, or holographic duality), relates  a classical gravity theory in a weakly curved $(d + 1)$-dimensional anti-de Sitter (AdS$_{d+1}$) spacetime to a strongly-coupled $d$-dimensional quantum
field theory living on its boundary~\cite{Maldacena:1997re,Witten:1998qj,Gubser:1998bc}. 
This suggests that complicated questions in strongly interacting many-body physics can be mapped to simple problems in classical gravity, i.e. to {\it geometry}, potentially giving rise to a new paradigm for approaching many-body dynamics. 
In particular, black holes appear to play a universal role in characterizing quantum phases without quasiparticles, giving important insight into dynamical behavior of such systems. 

Black holes are known for their simplicity and universality; assuming rotational symmetry, the geometry of a black hole is fully specified by its mass and certain conserved charges such as electric charge, independent of other details of a gravity system (such as matter content). Moreover, for a given mass, there is a maximal allowed charge. A quark-gluon plasma has almost equal number of ``quarks'' and ``anti-quarks'' and the temperature is the main characteristic energy scale. Such a state is 
 described on the gravity side by a black hole of almost zero charge, i.e. a Schwarzschild black hole. 
In contrast, systems which are of interest in condensed matter physics, like strange metals, are finite density systems, with temperature much smaller than the chemical potential (a strange metal has only electrons but no positrons!). On the gravity side these finite density systems are described by black holes having close to the maximum charges. Thus, in some sense black holes provide a unified description of such seemingly distant systems like a QGP and a strange metal. 
 
Using gravity dual to explore dynamical properties of holographic QGPs has provided important insights into the understanding of the QGP at RHIC (see e.g. \cite{Gubser:2009md,CasalderreySolana:2011us} and references therein). The goal of these notes is to bring these holographic tools to bear on strongly correlated systems at {\it finite density}. In particular, we review recent progress in finding new universal classes of non-Fermi liquids and novel quantum phase transitions 
using this approach.% In the rest of this introduction we will first outline some challenges from condensed matter physics.

\subsection{Challenge: non-Fermi liquids} \label{fermiliquid}

One of the cornerstones of condensed matter physics is Landau's Fermi liquid theory, 
which underlies our understanding of almost all metals, semi-conductors, superconductors and superfluids. Since 1980's there has been an accumulation of metallic materials whose thermodynamic and transport properties differ significantly from those predicted by Fermi 
liquid theory.  These so-called non-Fermi liquids, which include the strange metal phase of cuprate superconductors and heavy fermion materials near a quantum phase transition, present important theoretical challenges. Here we first highlight some salient aspects of the Fermi liquid theory and then point out where it breaks down.

%The theory is based on the following two simple postulates regarding an interacting {\it fermionic} 
%system:

%\begin{enumerate}
%\item{
%The ground state of the system is characterized by a Fermi surface in momentum space with all states filled for $k<k_F$.
%}
%\item{
%Low energy excitations are described by quasi-particles and quasi-holes, which have the same charge as fundamental fermions and they satisfy Fermi statistics. %They are different from the free theory excitations in the following aspects:
%}
%\een
%These postulates were motivated from  considering 
%Let us start with a brief summary of
Let us start by recalling the story for a non-interacting Fermi gas (e.g. a gas of electrons in a box) for which the many-particle states can be obtained by simply filing single-particle energy eigenstates following the Pauli exclusion principle. The ground state is then given by filling 
all the (single-particle) states inside a sphere\nicefootnote[1]{In these lectures we will only consider rotationally invariant theories, hence we will not encounter Fermi surfaces with shapes other than the sphere.} in momentum space with radius $k_F$ 
determined by the density of fermions and with 
all states outside the sphere empty. 
The locus of points in momentum space at the boundary of this sphere, $k \equiv |\vec{k}|=k_F$, is called the Fermi surface. 
The low-energy excitations of the system are given by either 
filling a state slightly outside the Fermi surface or removing a fermion from a filled state slightly inside the Fermi surface, and are called a {\it particle} and {\it hole} respectively. These excitations are gapless and have a linear dispersion (for $k-k_F \ll k_F$):
\be \label{disp}
\ep(k) = E_0(k)- \mu ={k_F \ov m}(k-k_F) \equiv v_F (k-k_F) 
\ee
where $E_0(k) = {k^2 \ov 2m}$ denotes the free particle energy and we are working in the grand 
canonical ensemble,  in which the energy of a excitation is measured from the chemical potential 
$\mu = E_F= {k_F^2 \ov 2m}$.
 Particles and holes\nicefootnote[1]{ Note that while all excitations have positive energies, it is convenient to represent a hole using negative energy, i.e. in terms of a filled particle state. 
} are distinguished by the sign of $k - k_F$.   At a field theoretical level, these excitations manifest themselves as poles in the complex frequency plane of the retarded Green's function $G_R (\om,k)$ for the electron operator in momentum space
\be \label{freegr}
G_R(\om,\vec{k})={1\ov \om -\ep(k)+ i 0_+}
\ee
where $\om$ is again measured with respect to the chemical potential $\mu$.
 Note that the retarded function is relevant as it describes the causal response of the system if we ``add'' an electron to the system. Fourier transforming~\eqref{freegr} back in time we see that the propagator describes the propagation of a particle of energy $\ep(k)$:
\be \label{proh}
G_R(t,\vec{k})\sim  \th (t) e^{-i\ep(k)t}  \
\ee
which is of course as it should be for a free theory.

The situation becomes  complicated once we switch on interactions between fermions, since 
the notion of single-particle states no longer makes sense. While one may intuitively expect that 
the qualitative picture for a non-interacting gas should still apply for sufficiently weak interactions, 
it is {\it a priori}  not clear what should happen at finite or strong couplings.  
Landau's Fermi liquid theory postulates that the above qualitative picture for a non-interacting gas in fact persists for {\it generic} interacting fermionic systems. In particular, it assumes that

\begin{enumerate}
\item{
The ground state of an interacting fermionic system 
is characterized by a Fermi surface in momentum space at $k=k_F$. 
}
\item{
Despite (possibly strong) interactions among fundamental fermions, the {\it low energy} excitations near the Fermi surface nevertheless behave like weakly interacting 
particles and holes, which are called quasi-particles and quasi-holes. 
They have the same charge as fundamental fermions and satisfy Fermi statistics. The dispersion of a quasi-particle (similarly for a quasi-hole) resembles~\eqref{disp} in free theory 
\be
\ep(k) =v_F(k-k_F)+\dots \qquad v_F={k_F \ov m_*} \ ,
\ee
where $m_*$ can be considered as the effective mass of the quasi-particle and is in general different from the original fermion mass $m$, from renormalization by many-body interactions.
}
\een
Given the existence of a Fermi surface, the second postulate above is {\it self-consistent}.  When turning on interactions, a particle (or hole) can now decay into another particle 
plus a number of particle-hole pairs. 
But it is not difficult to check that  the exclusion principle constrains the phase space around a Fermi surface so much, that given {\it generic local interactions} %\footnote{Note that the long range interactions between electrons are screened at a finite density and interactions among between them can be considered as short-ranged. } 
among particles and holes, the decay (or scattering) rate of
a particle (or hole) obeys 
\be \label{fermide}
\Gamma\sim {\ep^2 \ov \mu} \ll \ep \ 
\ee
where $\ep$ is the energy of a particle (or hole). Thus, despite potentially strong interactions, particle or hole excitations near the Fermi surface have a long lifetime and an approximate particle picture still applies. 
(Below for simplicity we will refer to quasi-particles and quasi-holes collectively as quasi-particles.)
Thus for a Fermi liquid, equation~\eqref{proh} should be modified to  
\be \label{proh1}
G_R(t,\vec{k})\sim e^{-i\ep(k)t - {\Ga \ov 2} t} \ , \qquad \Ga \sim \ep^2 (k) \ 
\ee
which implies that near the Fermi surface the retarded function for the electron operator should have the form 
\be 
G_R(\om,\vec{k})={Z\ov \om - v_F (k-k_F) + \Sig (\om, k) }+\dots \ , \qquad 
 \label{eq:qpprop1}
\ee
with the self-energy $\Sig (\om, k)$ 
\be \label{selffer}
\Sig = {i \Ga \ov 2} \sim i \om^2 \ . 
\ee
The residue $Z \leq 1 $ of the pole, which is called the quasiparticle weight,
can be interpreted as the overlap between the (approximate) one-quasiparticle state with the state generated by acting the electron operator on the vacuum.  

The concept of quasi-particles is extremely powerful and makes it possible to develop a general low energy theory  -- {\it Fermi liquid theory} -- independently of the precise microscopic details of a system. With some phenomenological input, the theory can then be used to predict essentially all the low energy 
behavior of the system. 
For example, the theory predicts that the specific heat is linear in temperature (see e.g.~\cite{coleman})
\be \label{fermispe}
C_e=\gamma T+\dots \qquad \gamma \sim m_*
\ee
and that the low temperature resistivity increases with temperature quadratically
\be \label{fermires}
\rho_e=\rho_0+AT^2+\dots \ .
\ee
The theory has been tremendously successful in explaining {\it almost} all metallic states in nature.

It is important to emphasize that Fermi liquid theory does not require interactions among the fundamental constituents to be weak. 
For example for ${^3 \text{He}}$, one finds that $m_*=2.8m_{He}$,
indicating that interactions among ${^3 \text{He}}$ atoms are clearly not weak. There also exist 
so-called heavy electron compounds for which the effective mass for electron quasi-particles can 
be as large as $10^2- 10^3 $ times of the electron mass.
% resulting from hybridizing with localized magnetic moments in those compounds.

 It is rather remarkable that  weakly interacting quasiparticles can emerge as the low energy collective excitations of a  strongly interacting many-body system, with only certain parameters~(such as the effective mass) renormalized compared to the fundamental constituents of the system. 
The self-consistency of Fermi liquid theory can also be understood from an effective field theory perspective using the renormalization group~\cite{Benfatto:1990zz, Polchinski:1992ed, Shankar:1993pf}. Assuming the existence of quasi-particles, one can then try to write down the most general  {\it local} effective field theory for them.
One finds that due to kinematical constraints from the Fermi surface all interactions are {\it irrelevant} at low energies except for the forward scatterings\nicefootnote[1]{
The forward scatterings give rise to interactions among quasi-particles via their densities, which are incorporated in the Fermi liquid theory. Note that such interactions do not give rise to widths.} and BCS-type pairing instabilities leading to a superconductor. 
Note that while the renormalization group analysis shows that the Fermi liquid theory is a stable fixed point (up to superconducting instabilities), it does {\it not} tell us whether or why a specific microscopic theory will flow to this fixed point.

\begin{figure}[h]
\begin{center}
\includegraphics[scale=0.4]{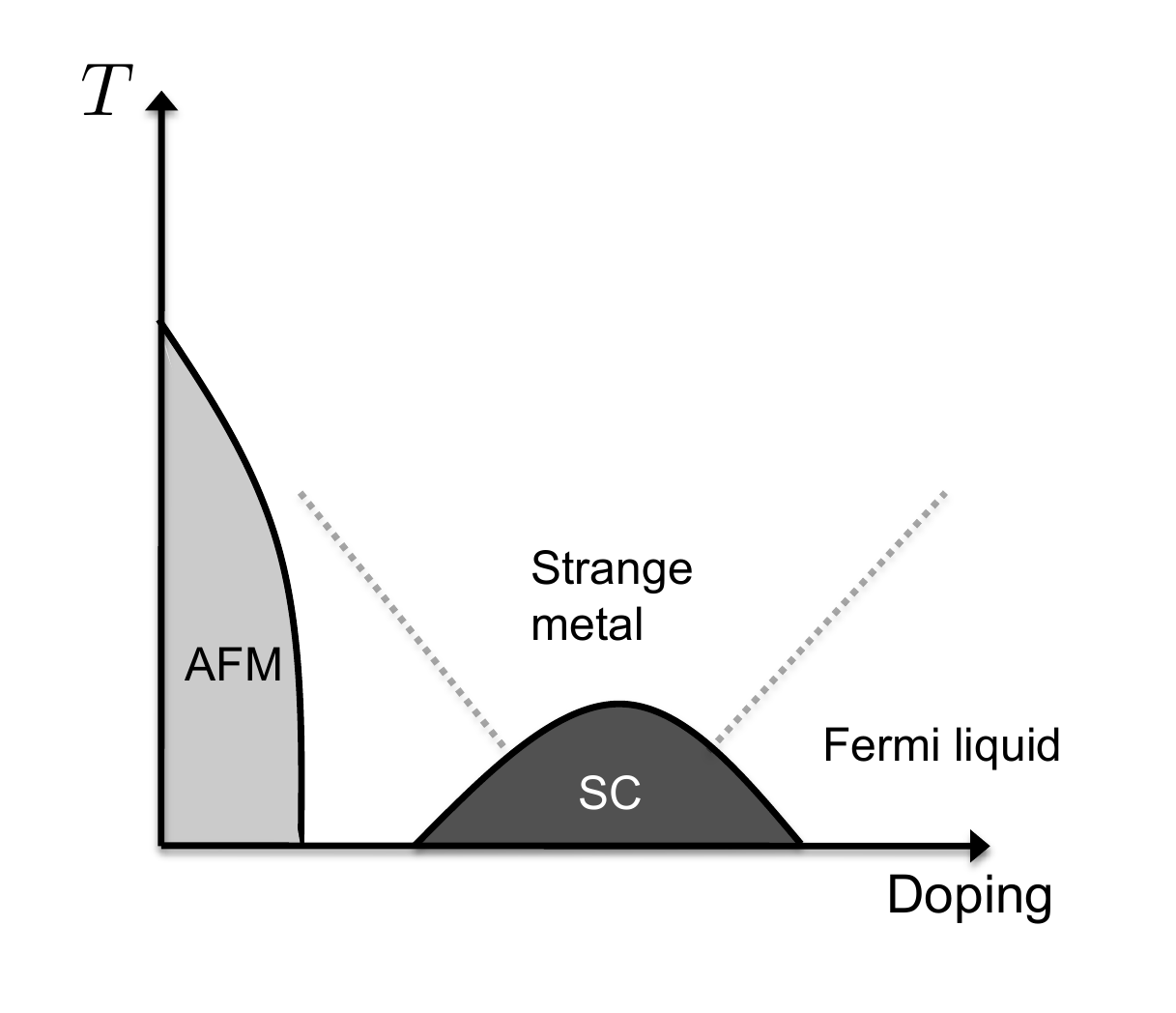} \qquad \includegraphics[scale=0.4]{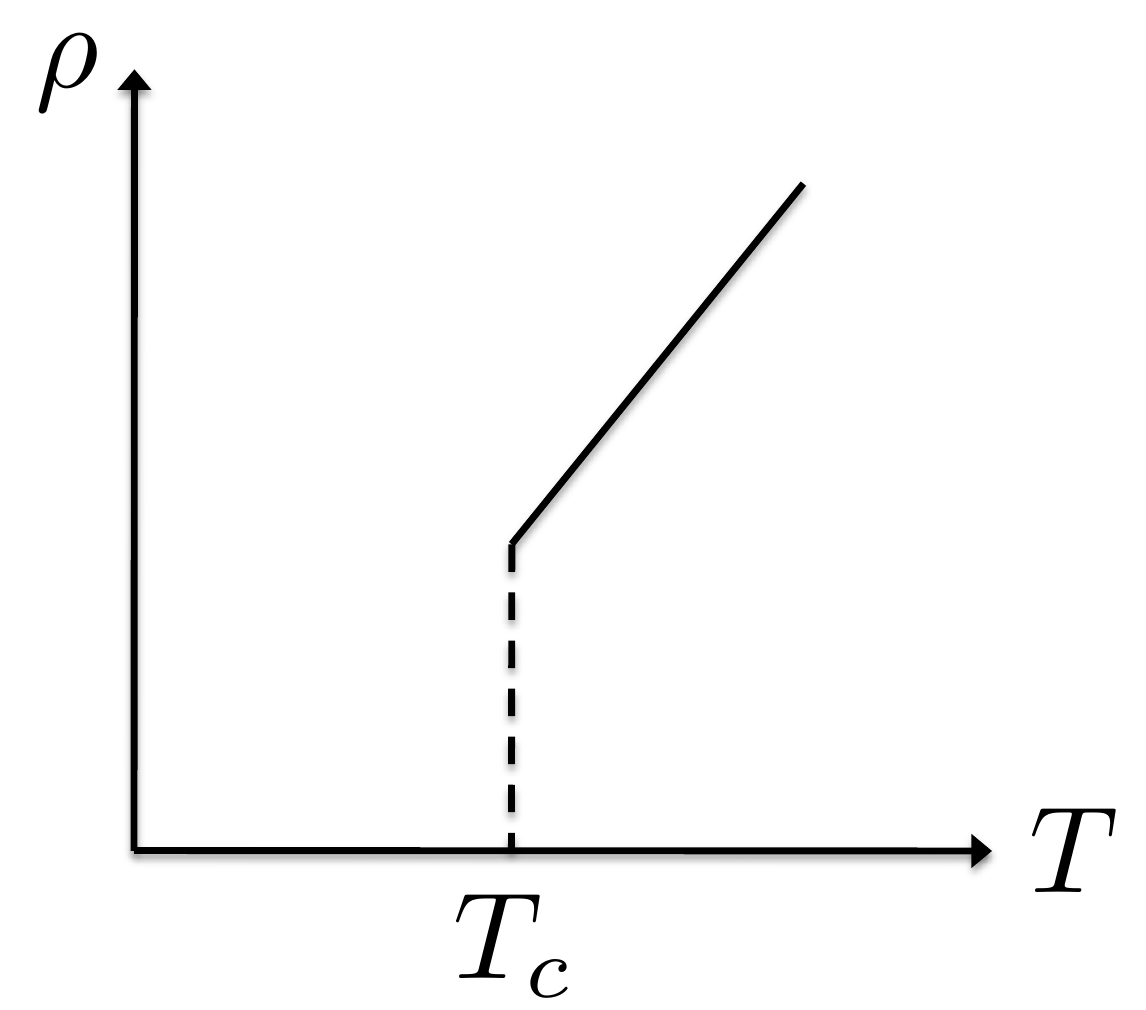}
\end{center}
\caption{Left: a cartoon picture of the phase diagram of cuprate superconductors; Right: a cartoon picture of linear temperature dependence of the resistivity in the strange metal phase.} 
\label{fig:cuprates}
\end{figure}

Even though the Fermi liquid theory has been tremendously successful, nature always has interesting surprises for us. Since the 1980's, there has been an accumulation of metallic materials whose 
thermodynamic and transport properties differ significantly from those predicted by Fermi liquid theory (see e.g.~\cite{stewart}). One prominent class of examples of these so-called non-Fermi liquids is the strange metal phase of the cuprate superconductors, which refers to the  
metallic state above the superconducting transition temperature $T_c$ near optimal doping (see Fig.~\ref{fig:cuprates}). 
The strange metal phase exhibits thermodynamic and
transport behavior significantly different from those of an ordinary metal. A particularly striking property of the strange metal phase
is that the electrical resistivity increases linearly with temperature, in contrast to
the quadratic temperature dependence~\eqref{fermires} of an ordinary metal. This remarkably simple
behavior is very robust, existing over a wide range of temperatures (sometimes until
a material melts), and universal, appearing in all cuprate superconductors. Nonetheless,
it has resisted a satisfactory explanation for more than twenty years. 

The anomalous behavior of a strange metal implies that the Fermi liquid theory breaks down. The immediate question is whether one or both of the postulates stated earlier break down. Fortunately, this question can be answered with the help of photoemission experiments which 
can probe directly a Fermi surface and its low energy excitations.  In an Angular Resolved Photoemission Spectropy (ARPES)
experiment, incident photons knock out electrons from the sample and the intensity $I(\om,\vec{k})$ of the electron beam is proportional to $A(\om,\vec{k})f(\om,\vec{k})$, where $f(\om,\vec{k})$ is the Fermi-Dirac distribution and $A(\om,\vec{k})$
is the electron spectral function defined by
\be
A(\om,\vec{k}) = {1 \ov \pi} \Im G_R (\om, k) \ .
\ee
Near the Fermi surface, the spectral function for a Fermi liquid~\eqref{eq:qpprop1} is given by a Lorentzian peak centered at $\ep (k) = v_F (k-k_F)$ with a width proportional to $\ep^2$. In particular, as  the Fermi surface is approached 
\be \label{delZ}
A(\om, k) \quad \to \quad Z \delta (\om - v_F (k-k_F)), \qquad  (k \to k_F)  \ .
\ee

For a non-Fermi liquid, it is {\it a priori} not clear what to expect of $A(\om, k)$. Nevertheless, as a working definition, a Fermi surface $k=k_F$ can be defined as the surface in momentum space where there are gapless excitations, which in turn should result in non-analytic behavior of $A(\om,\vec{k})$ at $k=k_F$ and $\om =0$. 
Thus a Fermi surface and low energy excitations around it can be diagnosed by examining the non-analytic behavior of $A(\om, k)$ obtained from ARPES experiments. 

For high $T_c$ cuprates in the strange metal region,  ARPES experiments indicate that a Fermi surface still exists, but excitations exhibit a much broader peak than that for a Fermi liquid. The experimental results can be fit well to the following 
expression~\cite{abrahams2000}, postulated as ``Marginal Fermi liquid'' (MFL) in~\cite{Varma89},
\be \label{marsp}
G_R (\om, k) = {h \ov  \om - v_F (k-k_F)  +  \Sig (\om,k)}
\ee
with the self-energy $\Sig (\om,k)$ given by
\be  \label{selM}
 \Sig (\om) \approx  c \om \log \om +  d \om, \qquad c \;\; {\rm real}, \quad d \;\; {\rm complex} \ .
 \ee
 We see from~\eqref{marsp} that the system appears to possess gapless excitations of dispersion relation  $\om = v_F (k-k_F) $. However, the decay rate $\Ga$ of such excitations, which is given by 
the imaginary part of $\Sig$, is now proportional to $\om$ in contrast to $\om^2$~\eqref{selffer} 
for a Fermi liquid.  The decay rate, which is comparable to $\om$, is so large, that an excitation will already have decayed before it can propagate far enough (i.e. one wavelength) to show its particle-like properties.  As a result,  such an excitation can no longer be treated as a quasiparticle. Also note that the residue for the pole in the complex plane scales like $Z \sim1/ \log(k-k_F)$ as the Fermi surface is approached and thus the quasiparticle weight vanishes logarithmically. Thus at the Fermi surface the overlap of an excitation with original electrons vanishes. 
Mathematically, the singularity of $A(\om, k)$ at $k=k_F$ and $\om =0$ is much softer than that~\eqref{delZ} for a Fermi liquid.
 
%For heavy fermion materials, while ARPES experiments are not yet available, measurements from Hall effects also indicate that Fermi surface still exists in the region of the phase diagram which exhibits non-Fermi liquid behavior. 

%We thus conclude that for 
Thus the strange metal phase of cuprates  has a Fermi surface (i.e. there still exist a surface in momentum space which has gapless excitations), but the quasiparticle picture breaks down. In other words, we are dealing with a {\it Fermi surface without quasiparticles}. From the perspective of renormalization group, it must be that the system is flowing to a nontrivial fixed point distinct from that for a Fermi liquid. 
 
Known field theoretical examples of non-Fermi liquids which also exhibit the phenomenon of a Fermi surface without long-lived quasiparticles include Luttinger liquids in $1+1$ dimensions, and a free fermion gas coupled to some gapless bosonic excitations, which can be either a transverse gauge field or certain order parameter fluctuations near a quantum critical point (see e.g.~\cite{Holstein:1973zz,sungsikgaugefield,Metlitski:2010pd,Mross:2010rd}). In the latter case gapless bosonic excitations generate {\it nonlocal} interactions among fermionic excitations near a Fermi surface and destroy the quasiparticles. 
Neither of these, however, is able to explain the behavior of a strange metal phase. The former is specific to $1+1$ dimensional kinematics. In the latter class of examples, the influence of gapless bosons is mostly along the forward direction, and is not enough, for example, to account for the linear temperature dependence of the resistivity. Even for this class of examples, there still exists significant technical challenges in understanding the precise nature of the IR fixed point~\cite{sungsikgaugefield,Metlitski:2010pd,Mross:2010rd}.

\subsection{Challenge: novel quantum phase transitions} \label{sec:nqpt}

We now switch gears, turning to another challenge faced by contemporary condensed matter theory. Consider a continuous phase transition; these are generally be driven by fluctuations that can be both quantum and thermal. 
Thermal phase transitions are by now very well developed, 
based on the so-called Landau-Ginsburg-Wilson paradigm, which consists of 
the following postulates: 

\bi

\item Different orders are characterized by different symmetries. 
    Phase  transitions between orders is a consequence of symmetry breaking. 

\item The symmetry breaking is characterized by an order parameter and critical behavior for the 
phase transition is solely captured by low energy dynamics of the order parameter. For example, the free energy can be obtained from 
\be \label{pathph}
Z = \int D \phi (\vec x) \, e^{-S [\phi]}
\ee
where $S[\phi]$ is the effective action for the order parameter $\phi$.\nicefootnote[1]{Note that one can ignore the dependence of $\phi$ on the time direction as such dependence only give rise to gapped excitations due to compactness of the Euclidean time direction at a finite temperature.} 

\ei

Continuous {\it quantum} phase transitions, which happen at zero temperature from tuning non-thermal control parameters, have also been traditionally formulated within the Landau paradigm~\cite{Hertz.76,Sachdev_book,mucio,Sondhi,Vojta,Natphys.08,Gegenwart.08,Lohneysen.07}. Now the Euclidean time direction is uncompact and the quantum critical fluctuations of the order parameter live both in space and in time. As the critical point is approached in addition to a divergent correlation length $\xi$, one also finds a divergent correlation time $\xi_\tau \sim \xi^z$, with $z$ the so-called the dynamical exponent. Thus 
the order parameter now lives effectively in
$d+z$ dimensions, where $d$ is the number of spatial dimensions and 
$z$ is the effective time dimension.  Again one expects that the critical behavior be captured by a path integral like that in~\eqref{pathph} with now $\phi$ depending on both space and time. 

This paradigm, however, 
has been challenged on several fronts in recent times. Firstly, there are quantum phases of matter whose order cannot be captured by symmetries alone, with examples including topologically ordered states such as the quantum Hall phases~\cite{xiaogang0,xiaogang}. Phase transitions 
among or out of such phases cannot be described in terms of order parameter fluctuations.  
Secondly, during the last decade, heavy fermion compounds have provided a variety of fascinating examples of quantum phase transitions (see e.g.~\cite{coleman06,Gegenwart.08,heavy2} for recent reviews), which appear not to be explainable using the standard Landau paradigm. In these examples, while the phase transition can still be characterized as symmetry breaking, new modes, which are inherently quantum and are beyond order-parameter fluctuations, appear to emerge as part of the quantum critical excitations.\nicefootnote[1]{In some examples, the phase transition can involve a jump in the size or shape of the Fermi surfaces.} These experimental examples also inspired new theoretical findings of quantum critical points lying outside the Landau paradigm, including ``local quantum critical points''~\cite{Si.01} and ``deconfined quantum critical points''~\cite{Senthil.04}.

Quantum criticality is hence considerably richer and more delicate than its thermal classical counterpart. In particular, quantum critical behavior has been seen to be associated with some of the most interesting phenomena in condensed matter physics including non-Fermi liquids, novel superconductivity etc. 
In turn, new methods are needed to search for, study, and characterize strongly coupled quantum critical systems.

 For a quantum phase transition which does have an order parameter, an important set of observables to characterize the dynamical nature of the quantum phase transition
are susceptibilities of the order parameter. Suppose the order parameter is given by the expectation value of some bosonic operator $\sO$, then the corresponding susceptibility $\chi (\om, \vec k)$ is given by the retarded function  for $\sO$, which captures the linear responses of the system to an infinitesimal source\nicefootnote[1]{For example if $\sO$ is the magnetization of the system, then the corresponding source is a magnetic field.} conjugate to $\sO$.

In a stable phase in which $\sO$ is uncondensed, turning on an infinitesimal source will result in an 
expectation value for $\sO$ which is proportional to the source with the proportional constant given by the  susceptibility. However, if the system has an instability to the condensation of $\sO$, turning on an infinitesimal source will lead to modes exponentially growing with time. Such growing modes are reflected in the presence of singularities of  $\chi (\om, \vk)$ in the {\it upper} complex $\om$-plane. Thus the dynamical susceptibility provides an important diagnostic of potential instabilities of a system. Similarly, at the onset of an instability (i.e. a critical point, both thermal and quantum), the static susceptibility typically diverges, reflecting that the 
tendency of the system to develop an expectation value of $\sO$ even in the absence of an external source. The divergence is characterized by a critical exponent $\ga$ (see Appendix~\ref{app:critexp} for a review of definitions of other critical exponents)
\be
\chi(k=0,\om=0) \sim \le| g-g_c  \ri|^{-\gamma} 
\ee
where $g$ is the tuning parameter (which is temperature for a thermal transition) with $g_c$ the critical point.

\subsection{``Local'' quantum critical behavior}  \label{sec:lqc}
  
A particularly interesting example which illustrates the nontrivial nature of quantum phase transitions 
for heavy fermions is provided by $CeCu_{6-x}Au_x$  with the doping parameter $x$ as the tuning parameter~\cite{schroeder:00}. There is a quantum critical point at $x_c =0.1$ above which the system is described by an antiferromagnetic phase. One finds near the quantum critical point the dynamical  susceptibility behaves as~\cite{schroeder:00} 
\be \label{cechi}
\chi (\om, \vk) = {Z \ov (\vec k-\vec Q)^2 + x_c - x + T^{\al} f \le({\om \ov T}\ri)}
\ee 
where $\vec Q$ is the ordering vector for the antiferromagnetic order. Equation~\eqref{cechi} exhibits an
interesting contrast between momentum and frequency dependence: 

\ben 

\item  The spatial part of the susceptibility exhibits ordinary mean-field behavior.

\item  The self-energy depends only on frequency and  exhibits nontrivial $\om/T$  
scaling (the exponent $\al = 0.75$).  

\een
The above behavior has been called ``local quantum critical behavior''~\cite{coleman99,Si.01},  and cannot be explained using the standard Landau paradigm.  

Interestingly, local quantum critical behavior is also present in the retarded response function for the Marginal Fermi Liquid~\eqref{marsp}. The self-energy~\eqref{selM} and its finite temperature generalization depend only on $\om$ and exhibits $\om/T$ scaling~\cite{Varma89}. In contrast, the spatial momentum dependence is the same as that for a free fermion Fermi surface. Note that in~\cite{Varma89}, the fermionic response function~\eqref{marsp} can be obtained if we couple the electrons to spin and charge fluctuations whose spectral function satisfies  
\be
  \label{eqn:fluc}
  \Im \chi(\om) \sim
\left\{ \begin{array}{ll}
  \chi_0  { \omega \over T} & {\rm for} \quad {|\om| \ll  T} \\
\chi_0 {\rm sign}\, \om & {\rm for} \quad {T \ll |\om| \ll \om_c}
\end{array} \right.
\ee
where $\chi_0$ is a constant and $\om_c$ is a UV cutoff. Note that this expression is $k$-independent and again exhibits local quantum critical behavior.  

The local quantum critical behavior described above naturally arises in theoretical models of Dynamical Mean Field Theory (DMFT)~\cite{DMFT},  based on a large spatial dimension mean field approximation.  In DMFT one maps a lattice many body problem into an impurity problem and  the self-energy arises from that of the impurity. 
Thus by construction the self-energy is  momentum independent and its scaling behavior can be modeled from the critical behavior of the impurity. Note that since an impurity typically has a nonzero entropy, in general such an approximation leads to a finite entropy density, indicating that it describes intermediate-energy rather than ground state physics. So it is an interesting question whether the observed 
local quantum critical behavior in heavy fermion materials and cuprates could be due to intermediate energy effects.\nicefootnote[1]{See also an argument which points to this direction at the end of Sec.~\ref{sec:inter}. We have more to say on this issue in the conclusion section.} In the course of this review we will actually be led to very similar locally quantum critical behavior arising from holography. 

\subsection{Scope of this review}

Holography is the surprising statement that certain quantum field theories can be exactly equivalent to theories of quantum gravity that live in one extra spatial dimension. 
This equivalence is obviously conceptually startling, having far-reaching implications for our understanding of quantum gravity. Remarkably, it is also practically useful; 
which stems from the fact that it is a strong/weak duality: generally when the field theoretical side is strongly correlated, the gravitational description is weakly coupled and tractable. This suggests that complicated questions in strongly interacting many-body physics can be mapped to simple problems in classical gravity, i.e. to {\it geometry}. 

%During the last few years an increasing amount of research has gone into using the duality to extract dynamical insights into strongly coupled quantum many-body systems.   

In the most well-studied examples of the duality, gravity (and associated matter fields) propagating on a weakly curved anti-de Sitter spacetime in $d+1$ dimensions (AdS$_{d+1}$) is mapped to a strongly coupled conformally invariant quantum field theory (CFT$_d$) that lives at its boundary (with $d$ dimensions). Many examples are now known in various spacetime dimensions, including $\sN=4$ Super-Yang-Mills theory in $d=4$, and ABJM theory in $d=3$~\cite{Bagger:2007vi,Gustavsson:2007vu,Aharony:2008ug}. These theories essentially consist of elementary bosons and fermions interacting with
non-Abelian gauge fields. At a heuristic level one may visualize such a theory as 
the continuum limit of a lattice system where the number of degrees of freedom at each lattice site is of order $O(N^2)$, with $N$ the rank of the gauge group. 
The classical gravity approximation in the bulk corresponds to the strong coupling regime and the large $N$ limit.  In addition to these theories, there also exist vastly many asymptotically-AdS vacua of string theory, each of which is believed to give rise to an example of the correspondence,
though an explicit description of the dual field theory is not known for most vacua.

The goal of these notes is to bring holographic tools to bear on the problem of strongly-interacting field theory states at {\it finite density}, with an attempt to make contact with the physical problems alluded to in the above. 
We do this by studying the simplest finite density system arising from the duality, obtained by turning on a nonzero chemical potential for a $U(1)$ global symmetry of a CFT$_d$, which is described on the gravity side by a charged black hole. We ask the following questions of such a finite density system:
\ben

\item Does the system have a Fermi surface?  What are physical properties of low energy excitations near the Fermi surface?

\item Does the system have instabilities to condensation of scalar operators?  What is the quantum critical behavior near such a critical point?

\een
In general for a  strongly coupled system, these questions are difficult to answer. Fortunately here they can be answered with the help of the gravity dual. In addition to thermodynamical quantities, the main dynamical observables that we will use to probe the system are retarded Green's function of gauge invariant operators, which could be a scalar or a spinor,
\be \label{retF}
G_R(t,\vec{x}) \equiv i \th(t) \tr\le(\rho[\sO(t,\vec{x}), \sO(0)]_\pm\ri),
\ee
where $\rho$ is the density matrix characterizing the ensemble and the notation $[,]_{\pm}$ refers to a commutator for scalars and an anticommutator for spinors. The retarded Green's function carries all of the real-time information about the response of the system and its imaginary part is proportional to the spectral function (or spectral weight) $A(\om,k) = \frac{1}{\pi} \Im G_R(\om, k)$, which captures the density of states which can be accessed by acting $\sO$ on the ensemble. 
As discussed in Sec.~\ref{fermiliquid} and~Sec.~\ref{sec:nqpt}, the retarded function for a fermionic operator can be used to probe the existence of a Fermi surface and the nature of low energy excitations around the Fermi surface, while that for a bosonic order parameter provides a convenient diagnostic for possible instabilities, and encodes 
important dynamical information for quantum phase transitions.  For real-life systems the spectral function of various operators can be directly measured; for example that for the electron operator by photo-emission experiments and that for spin polarizations by neutron scatterings. 
   
We will see that, even in this simplest context, holographic systems offer us plenty of opportunities for studying exotic phenomena associated with Fermi surfaces without long-lived quasiparticles and 
novel quantum phase transitions,  some of which share intriguing parallels with cuprates and heavy fermion systems discussed earlier. Gravity provides us with fascinating ways to characterize them. Our discussion will
be mainly based on~\cite{Lee09,Liu09,Cubrovic09,Faulkner09,Faulkner:2010zz,Faulkner:2011tm,Faulkner:2010tq,Iqbal:2011in,Iqbal:2010eh,Faulkner:2010gj,sdwnew}.\nicefootnote[1]{The application of holographic methods to condensed matter physics was initiated in~\cite{Herzog:2007ij} and other reviews include~\cite{quantcritbh,Hartnoll:2009sz,herzogr,Horowitz:2010gk,soojong,Sachdev:2010ch,Hartnoll:2011fn,Sachdev:2011wg}.}
In particular, we will find a  {\it universal intermediate-energy phase}, called ``semi-local quantum liquid'' in~\cite{Iqbal:2011in}, which does not have a quasiparticle description, and is characterized by: (1) a finite spatial correlation length; (2) an infinite correlation time and associated nontrivial scaling behavior in the time direction; (3)  a nonzero entropy density. In a semi-local quantum liquid, the self-energy for an operator depends on spatial momentum\nicefootnote[1]{We will only consider rotationally symmetric systems in this review. $k$ refers to the magnitude of momentum.} $k$  only through the ratio $k/\mu$, where $\mu$ is the chemical potential. Thus in the kinematic regime where momentum variation is much smaller than the chemical potential (e.g. near a Fermi surface),  the momentum dependence can be neglected and one finds the ``local quantum critical behavior'' described in Sec.~\ref{sec:lqc}. For example, we will find Fermi surfaces  which exhibit the MFL behavior~\eqref{marsp} and quantum phase transitions whose dynamical susceptibility has the behavior~\eqref{cechi} or~\eqref{eqn:fluc}.

For a holographic system at a nonzero chemical potential, this unstable semi-local quantum liquid phase sets in at an energy scale of order of the chemical potential, and then orders into other phases at lower energies, including  superconductors, antiferromagnetic-type states, and Fermi liquids.  While the precise nature of the lower energy state depends on the specific dynamics of the individual system, the semi-local quantum liquid arises universally from these lower energy phases through deconfinement (or in condensed matter language, fractionalization).

In this review we will be mainly interested in searching for new physical scenarios and dynamical mechanisms which can arise for a class of holographic systems, rather than detailed phase structure or dynamical effects for a specific theory. 
%concerned with extracting universal features for general theories with a gravity dual, rather than any specific theory. % with a gravity dual,  
Thus we will take the so-called ``bottom-up'' approach, i.e.
 we will just consider a certain type of operator spectrum without referring to a specific theory. The results  should apply to any theory which contains the given spectrum. 
This complements the so-called ``top-down'' approach which starts with specific string theory embeddings and works out the precise phase structure for these theories. See~\cite{Gauntlett:2011mf,Belliard:2011qq,Bah:2010cu,Bah:2010yt,Gauntlett:2009dn,Gauntlett:2009bh,Gubser:2009qm,Aprile:2011uq,Donos:2011ut} for some examples of the top-down approach. 

% (see e.g.~\cite{Aprile:2011uq,Donos:2011ut}). 

The plan of the paper is as follows. In next section we discuss various aspects of gauge/gravity duality for a finite density system. In particular, we introduce the notation of an intermediate-energy semi-local quantum liquid phase and work out its physical properties. 
In Sec.~\ref{sec:nfl} we turn to study of holographic (non)-Fermi liquids and in Sec.~\ref{sec:qpt} possible scalar instabilities and associated quantum phase transitions. We conclude in Sec.~\ref{sec:conc} with a summary of main points and a discussion of possible implications  for real-life systems.

\section{Gauge/gravity duality at a finite density}

In this section we discuss various aspects of gauge/gravity duality at a finite density, in preparation for our discussion of non-Fermi liquids and quantum phase transitions in subsequent sections.

\subsection{Some aspects of AdS/CFT} \label{sec:ads}

In this subsection we highlight certain aspects of gauge/gravity duality that will be needed for our discussion. More detailed expositions on AdS/CFT are available: for example, the TASI lectures of Polchinski \cite{Polchinski:2010hw}. Other recent reviews include \cite{CasalderreySolana:2011us,Hartnoll:2009sz, McGreevy:2009xe}. % focus on the applications of holography to many-body physics. 
The reader who is already familiar with the basic ideas of holography is encouraged to skip this subsection, and the reader who only wants to compute things can probably skip to the results \eqref{vevdef} and \eqref{GRdef}.

The gravitational action in AdS  takes the form
\be
S={1\ov 2 \kappa^2} \int  d^{d+1} x \ \sqrt{-g}\le(\sR +{d (d-1) \ov R^2}\ri)\ \label{basicac}
\ee
where $R$ is the AdS radius, and Newton's 
constant  
\be \label{kasc}
{2 \ka^2 \ov R^{d-1}} \sim {1 \ov N^2}
\ee
 is small in the large $N$ limit of the boundary theory.
The simplest solution satisfying the field equations arising from \eqref{basicac} is pure anti-de Sitter space, which takes the form 
\be
ds^2 = R^2\le(\frac{dz^2 + \eta_{\mu\nu} dx^{\mu} dx^{\nu}}{z^2}\ri) = {R^2 \ov z^2} (dz^2 - dt^2 + d \vec x^2)
\label{pureAdS}
\ee
Here the $x^{\mu} = (t, \vec x)$ run over the coordinates of the field theory, and $z$ is the extra ``holographic'' coordinate in the gravitational description, with the AdS boundary lying at $z=0$. 
As the simplest and most symmetric solution on the gravitational side this geometry represents the {\it vacuum} of the dual field theory. Indeed, the vacuum of the CFT should be invariant under the conformal group in $d$-dimensions, which is precisely the same as the isometry group of AdS$_{d+1}$. We draw special attention to the following isometry:
\be
z \to \lambda z \qquad x^{\mu} \to \lambda x^{\mu} \label{scalisom}
\ee
which represents  {\it scaling}. 
%This is the isometry representing {\it scaling}; we see that as we scale the energy we must scale the holographic coordinate $z$ as well. This is indicative of the fact that the holographic coordinate represents the ``energy scale'' at which we consider the theory, with the UV at $z \to 0$ and the IR at $z \to \infty$. Thus the bulk {\it geometrizes the RG flow} of the field theory: we will make this more precise as we go along.

One of the most important aspects of the duality is that the bulk spacetime {\it geometrizes the renormalization group (RG) flow} of the field theory, which can be seen as follows. 
Due to the warp factor ${R^2 \ov z^2}$ in front of the Minkowski metric in~\eqref{pureAdS}, local proper energy and length scales along field theory directions in AdS are related to those in the dual field theory by a $z$-dependent rescaling. More explicitly, consider a physical process with {\it proper} energy $E_{loc}$ at some value of $z$ in AdS. The bulk energy $E_{loc}$ is measured in units of the local proper time $d \tau = {R \ov z} dt$ and when viewed in terms of boundary time $t$, it corresponds to an excitation of boundary theory energy $E$ given by
\be 
E = {R \ov z} E_{loc} \ . % \sim {1 \ov z} \ 
\ee 
%where in the second equality 
%Thus corresponding boundary theory energy scale depends on where the objects sits in the bulk. 
 We thus see that physical processes in the bulk with identical proper energies but occurring at different radial positions correspond to different field theory processes with energies that scale as $1/z$. In other words, given that typical local bulk energies are given by the curvature scale, i.e. $E_{loc} \sim {1 \ov R}$, a field theory process with a characteristic energy $E$ is associated with a bulk process at $z \sim1/E$~\cite{Maldacena:1997re,Susskind:1998dq,Peet:1998wn}. This relation %between the radial direction $z$ and boundary energy 
 implies that the $z$ can be identified as renormalization group scale of the boundary theory. %i.e. the bulk {\it geometrizes the RG flow} of the field theory!
 In particular, the high-energy (UV) limit $E \to \infty$ corresponds to $z \to 0$, i.e. to the near-boundary region, while the low-energy (IR) limit $E \to  0$ corresponds to $z \to \infty$, i.e. deep in the interior. In a conformal theory, there exist excitations of arbitrarily low energies, which in the bulk is reflected in the fact that the geometry~\eqref{pureAdS} extends all the way to $z \to \infty$. 
If instead we consider a boundary theory with a mass gap $m$, the corresponding bulk geometry will then end smoothly at a finite value $z_0 \sim 1/m$. Similarly, at a finite temperature $T$, which provides an effective IR cutoff, the bulk spacetime will be cut off by an event horizon at a finite $z_0 \sim 1/T$. 

Another important aspect of the duality is the field/operator mapping, {\it i.e.} to each (conformally primary) operator one can associate a field in the bulk. % with mass $m$. 
For illustration, we will consider a scalar operator $\sO$ in the boundary theory. This is {\it dual} to some
massive bulk scalar field $\phi$ with mass $m$, whose action can, for example, be written as 

\be
S_{\phi} = -\frac{1}{2}\int d^{d+1} x \, \sqrt{-g} \le((\nabla\phi)^2 + m^2 \phi^2 + 
\dots \ri) \ . 
\ee 
By studying the equations of motion arising from the action above one can show that near the boundary a solution to the scalar wave equation has the expansion
\be
\phi(z \to 0,x^{\mu}) \sim A(x) z^{\Delta_-} + B(x)z^{\Delta_+} \label{nearBdyScal}
\ee
where the two exponents satisfy
\be
\Delta_{\pm} = \frac{d}{2} \pm \nu_U \qquad \nu_U = \sqrt{\frac{d^2}{4} + m^2 R^2}. \label{UVdim}
\ee

Then one can show that:
%This mapping contains the following elements, 

\bi

\item $\Delta_+$ is the conformal dimension of the dual operator $\sO$.

\item $A(x)$, which is the coefficient of the more dominant term in~\eqref{nearBdyScal} (in the $z \to 0$ limit)  can be identified as the source for $\sO$, i.e. a nonzero $A(x)$ corresponds to adding to the boundary theory Lagrangian a source term 
\be 
\de S_{\rm boundary} = \int d^d x \, A(x) \sO (x) \ . \label{singtrace}
\ee

\item $B(x)$, which is the coefficient of the subdominant term in~\eqref{nearBdyScal} (in the $z \to 0$ limit), can be identified with the expectation value of the operator $\sO(x)$, i.e.
\be
\langle \sO(x) \rangle = 2\nu_U B(x). \label{vevdef}
\ee
For example, if one finds a regular solution where $A = 0$ but $B \neq 0$, this implies that the operator $\sO$ has spontaneously developed an expectation value,  in the absence of a source. 
%Note that by studying the transformation of $B$ under the scaling isometry \eqref{scalisom} we can confirm the claim that the conformal dimension of $\sO$ is $\Delta_+$.

\item Given~\eqref{singtrace} and~\eqref{vevdef}, one immediately sees that the linear response  function (in momentum space) for $\sO$ should be given by 
\be
G_R(\om, k) = 2\nu_U \frac{B(\om, k)}{A(\om, k)} \  \label{GRdef}
\ee
where $B(\om, k)$ and $A(\om, k)$ are the Fourier transform along the boundary directions of the corresponding quantity in~\eqref{nearBdyScal}.
%proportional to the ratio $\frac{B}{A}$. 
In order to determine the ratio in~\eqref{GRdef}, one needs to provide an additional boundary condition at the interior of the spacetime, typically a regularity condition. In the situation that a horizon develops in the interior, the boundary condition for computing a {\it retarded} correlator is that the solution for $\phi$ should be {\it infalling} at the horizon~\cite{Son:2002sd}. 

\item  For $\nu_U \in (0,1)$, both terms in~\eqref{nearBdyScal} are normalizable and 
there are two ways to quantize $\phi$ by imposing Dirichlet or Neumann conditions at the AdS boundary,  which are often called standard and alternative quantizations respectively. This means that depending on the boundary conditions the exact same bulk gravity action can be dual to two {\it different} CFTs. Our discussion above refers to the standard quantization. In the alternative 
quantization the roles of $A(x)$ and $B(x)$ in~\eqref{nearBdyScal} are reversed, i.e. $B(x)$  corresponds to the source and $A(x)$ to the expectation value. Also the conformal dimension 
dimension for $\sO$ is now given by $\De_-$ which lies between $({d \ov 2}-1, {d \ov 2})$; note that the lower limit (corresponding to $\nu_U \to 1$) is now approaching that of a free massless scalar in $d$ spacetime dimension. In the alternative quantization, the double trace operator $\sO^2$ is relevant, upon turning on which the system flows to the standard quantization in the IR. Thus this construction provides us with a relevant operator that can be tuned; this will play an important role in our discussion of quantum phase transitions below. See Appendix~\ref{app:doub} for more discussion. 

\ei

At a practical level, the above (and their generalizations to spinor fields) are all that we will need to perform the computations in the rest of these lectures. Note that while the factors of $2\nu_U$ in \eqref{vevdef} and \eqref{GRdef} can be important, we will generally omit them in our discussion below for 
notational simplicity. Before moving on we summarize our understanding so far:

\begin{center}
\begin{tabular}{ c c c } 
Boundary CFT$_d$ & & Bulk AdS$_{d+1}$\\
\hline 
Complicated many-body dynamics & $\qquad\leftrightsquigarrow\qquad$ & Physics encoded in geometry  \\ 
RG flow of the system & $\qquad\leftrightsquigarrow\qquad$ & Bulk ``holographic'' coordinate $z$\\
Bulk scalar field $\phi$ & $\qquad\leftrightsquigarrow\qquad$ & CFT scalar operator $\sO$ \\
$N\to \infty$  and strong coupling & $\qquad\leftrightsquigarrow\qquad$ & classical GR\\ 
\end{tabular}
\end{center}
The last entry in the table deserves further explanation; generally the specific field theories we are considering are gauge theories with gauge group $SU(N)$. It is a well-known fact~\cite{SCole} that such theories are in some sense classical (but not trivial!) in the large-$N$ limit, by which we mean that correlators factorize and that the physics should be determined by some classical equations of motion, which are not those of the naive $\hbar \to 0$ classical limit. AdS/CFT now tells us that the relevant classical equations of motion are actually those of higher-dimensional Einstein gravity. One must always keep in mind the important caveat when comparing to more-conventional physical systems that the observables computed using classical AdS/CFT are generally exact only in the $N \to \infty$ limit. Indeed there exist interesting effects that are visible only at finite $N$; these are dual to quantum effects in the bulk, and we will discuss some of them in later sections (see e.g. Sec.~\ref{sec:transport}
for perturbative corrections and Sec.~\ref{sec:HFL} for non-perturbative effects).

\subsection{Finite density states and AdS$_2$}

Our primary interest is to study systems at a finite density (and temperature), the simplest way of realizing this in AdS/CFT is to consider a conformally invariant theory with a $U(1)$ global symmetry and turn 
on a nonzero chemical potential $\mu$ for the $U(1)$. 
%More specifically, we would like to study field theory states carrying a charge under a $U(1)$ current that we will call
The corresponding conserved current $J^{\mu}(x)$ is dual to a bulk gauge field $A_M(z,x)$, and  the global $U(1)$ symmetry in the field theory is represented by a $U(1)$ {\it gauge} symmetry in the gravitational description. 

%To create a nonzero charge density $\rho$ we should 
Turning on a finite chemical potential $\mu$ for the current corresponds to perturbing the boundary theory by the operator
\be
\delta S_{\rm boundary} = \mu\int d^d x \; J^t . \label{chempot} 
\ee
Being conserved, $J^t$ has scaling dimension $d-1$ and is thus a relevant operator. As a result,  under~\eqref{chempot}, the system should flow nontrivially in the IR. Note that since~\eqref{chempot} also breaks Lorentz symmetry, one expects the flow to be {\it different} in the time and spatial directions. For a strongly interacting many-body system, understanding the IR physics under the relevant deformation~\eqref{chempot} is very difficult, and in general is not possible using conventional techniques. Fortunately, for a theory with a gravity dual, we can readily extract answers from the gravity side, to which we now turn. 
%For definiteness, for the remainder of this paper we will now specialize to $d = 3$. 

 Analogously to the scalar case discussed in the previous subsection, the boundary value of the gauge field $A_{\mu}$ is equal to the value of the field theory source for $J^{\mu}$: thus the perturbation \eqref{chempot} means that we should study a classical gravity solution where the gauge field $A_t(z)$ at the boundary takes some nonzero value $A_t(z \to 0) = \mu$. We will focus on the minimal sector on the gravity side which is~\eqref{basicac} but now with the addition of the Maxwell term for the gauge field,
\be
S={1\ov 2 \kappa^2} \int  d^{d+1} x \ \sqrt{-g}\le[\sR +{d (d-1)\ov R^2}+{R^2\ov g_F^2}F_{MN}F^{MN}\ri]\ . \label{maxact}
\ee
$g_F$ is a bulk gauge coupling; the factors of $R$ in its definition combine with the dimensional Newton's constant $\kappa^2$ out front to make it dimensionless. The solution that satisfies the relevant boundary condition is the Reissner-Nordstrom charged black hole\nicefootnote[1]{Here we say ``black hole'' even though the horizon is planar with topology $\mathbb{R}^{d-1}$; it would perhaps be more accurate to call this solution ``black brane.''}\cite{Romans:1991nq,Chamblin:1999tk}, whose metric and background gauge field are given by
\be \label{RNmetric}
ds^2 = {R^2 \ov z^2} (-fdt^2 + d \vec x^2) + {R^2 \ov z^2}{dz^2 \ov f}
\ee
with
 \be \label{bhga2}
 f = 1 + { Q^2 z^{2d-2}} - {M  z^d}, \qquad A_t = \mu \le(1- {z^{d-2} \ov  z_0^{d-2}}\ri) \ ,
 \ee
where $Q, M, z_0$ are constants. 

In a specific string realization of such a gravity dual, there are many other matters fields in addition to those in~\eqref{maxact} and in general it is not warranted that~\eqref{RNmetric} will lift to a solution of the full string theory.\footnote{This depends on whether the Maxwell-Einstein action~\eqref{maxact} is a consistent truncation of the full theory. For example, if the Maxwell field is coupled to a scalar field $\phi$ as $e^{\phi} F^2$, then~\eqref{RNmetric} will no longer be a solution as $\phi$ will also be excited~\cite{Gubser:2009qt,Goldstein:2009cv}.} Nevertheless, there are many examples (including certain $U(1)$ subgroups of $\sN=4$ SYM and ABJM theories, and an infinite family of theories discussed in~\cite{Denef:2009tp}) in which~\eqref{RNmetric} does survive the lift to the full gravity description. Within these examples, the charged black hole~\eqref{RNmetric} thus provides a ``universal'' %\footnote{Here we use the term ``universal'' rather loosely. We say that a class of systems possess certain universality if they share properties which do not immediately follow from their symmetries and 
  geometric description of many different systems at a finite chemical potential, independent of specific microscopic details. 
Different systems will have different matter field spectrum on the gravity side, but the underlying geometry is the same. 

Now we proceed to describe properties of~\eqref{RNmetric}. Compare it to \eqref{pureAdS}: the nontrivial function $f(z)$ indicates that the physics is changing with scale. 
The horizon lies at $z=z_0$ where 
\be \label{hord}
f(z_0) =0, \qquad \to \qquad M = z_0^{-d} + {Q^2 z^{d-2}_0} \ .
\ee
$Q$ can be expressed in terms of the chemical potential $\mu$ as 
 \be \label{chem}
Q = \sqrt{2 (d-2) \ov d-1} {\mu  \ov g_F z_0^{d-2}}     \ .
 \ee
The temperature of the system can be identified with the Hawking temperature of the black hole, which 
can be written as 
\be \label{rre1}
T = {d  \ov 4 \pi z_0} \le(1 - {d-2  \ov d} Q^2 z_0^{2d-2}\ri) \ .
\ee
The solution~\eqref{RNmetric}--\eqref{bhga2} has two independent parameters $\mu$ and $z_0$, with $\mu$ setting the unit of scale and the solution is characterized by the dimensionless number $\mu z_0$ (which can in turn be expressed in terms of the dimensionless ratio $T/\mu$).\nicefootnote[1]{ 
Note that since the system is scale invariant in the vacuum, at finite density the system is characterized by the dimensionless ratio $\frac{T}{\mu}$.}

The field theory quantities dual to this black hole are given by the following:
\bea
&&\text{Charge density:}\qquad  \rho = \sqrt{2 (d-1) (d-2)} { R^{d-1}  \ov \kappa^2}{Q \ov  g_F} \\ \label{chargeD1}
&&\text{Entropy density:}\qquad   s=2 \pi  {R^{d-1} \ov \kappa^2} {1 \ov z_0^{d-1}}\\
\label{entropy1}
&&\text{Energy density:}\qquad  \ep={d-1 \ov 2} {R^{d-1}  \ov  \kappa^2} M \ .
\label{ener1}
\eea
From~\eqref{kasc}, all these quantities are of order $O(N^2)$. 
It can be readily checked from the above equations that the first law of thermodynamics is satisfied
 \be
 d \ep = T ds + \mu d \rho \ .
 \ee
Up to some constant factors $Q$ is essentially the charge density of the boundary theory.
It is convenient to parameterize it as
 \be
 Q \equiv \sqrt{d \ov d-2} \, {1 \ov z_*^{d-1}} \ 
 \ee
 by introducing a length scale $z_*$ (which is fixed by boundary charge density).  In terms of $z_*$, various quantities can be written as 
 \be \label{newpa}
 \rho ={R^{d-1} \ov \kappa^2}  {1 \ov z_*^{d-1}} {1 \ov e_d}, \qquad % \\
\mu  =  {d (d-1) \ov d-2} {1 \ov z_*} \le({z_0 \ov  z_*} \ri)^{d-2} e_d,
%\label{mudefJ}\\
 \qquad T = {d  \ov 4
\pi z_0} \le(1 - {z_0^{2d-2} \ov  z_*^{2d-2}} \ri) 
 \ee
where we have introduced
 \be \label{defed}
 e_d \equiv {g_F \ov \sqrt{2d (d-1)}}  \ .
 \ee
Note that $z_* \geq z_0$ as otherwise the metric will have a naked singularity (rather than a black hole).

From~\eqref{newpa}, at zero temperature,  we have $z_0 = z_*$ and the black hole metric can be written as 
\be \label{zeroQ}
f = 1+ {d \ov d-2} {z^{2d-2} \ov z_*^{2d-2}} - {2 (d-1) \ov d-2} {z^d \ov z_*^d}, \qquad  A_t = \mu \le(1 - {z^{d-2} \ov z_*^{d-2} } \ri)
\ee
where $\mu$ and $z_*$ are related by
\be \label{pp1}
\mu_* \equiv {1 \ov z_*}  
 = 2 (d-2) e_d {\mu  \ov g_F^2}  \ .
\ee
For convenience we introduce the appropriately rescaled $\mu_*$, which will be used often below as it avoids having various numerical factors and $ g_F$ flying around. The corresponding 
field theory quantities can now be written as 
\be \label{entropy}
  \rho %=2  (d-2) {R^{d-1} \ov \ka^2} {\mu \ov g_F^2 z_*^{d-2}} 
= {R^{d-1} \ov \ka^2} {1 \ov e_d} {1 \ov  z_*^{d-1}}, \qquad  %\\ \label{chargedens}
   s={R^{d-1} \ov  \kappa^2} {2 \pi \ov z_*^{d-1}}, \qquad %\\
\qquad  \ep={R^{d-1} \ov  \kappa^2} {(d-1)^2 \ov d-2} {1 \ov z_*^d} \ .
%\label{ener}
\ee
Note that the entropy density is nonzero even at $T=0$. This finite ground state entropy density appears to violate the Third Law of Thermodynamics; we will discuss the implications of this in Sec.~\ref{sec:inter}. 

For small $z \ll z_*$ we have $f \sim 1$, and the metric resembles that of pure AdS \eqref{pureAdS}; this is telling us that physics at energy scales much larger than the chemical potential $\mu$ is simply that of the conformal vacuum, as expected. Let us first consider $T=0$. At $z \sim z_*$ the geometry is very different: $f$ has a double zero at the horizon $z=z_*$, with  
 \be
 f (z) \approx d (d-1) {(z_* -z)^2 \ov z_*^2} + \dots, \quad (z \to z_* ) \ .
 \ee 
This implies that the horizon is actually an infinite proper distance away, as one can readily check from \eqref{RNmetric}. In fact, the near-horizon geometry factorizes into AdS$_2 \times \mathbb{R}^{d-1}$:
\be \label{ads2M}
ds^2 = \frac{R_2^2}{\zeta^2}(-dt^2 + d\zeta^2) +\mu_*^2 R^2 d\vec{x}^2, \qquad A = {e_d \ov  \zeta} dt.
\ee
Here we have defined a new radial coordinate $\zeta$ and $R_2$ is the curvature radius of AdS$_2$,
\be
\zeta \equiv  {z_*^2 \ov d (d-1) (z_*-z)}, \qquad R_2 \equiv {R \ov \sqrt{d (d-1)}}  \ .
\label{defZ}
\ee
The metric~\eqref{ads2M} applies to the region ${z_*-z \ov z_*} \ll 1$ which translates into $\mu \zeta \gg 1$. From~\eqref{pp1} and~\eqref{defZ}, a relation which will useful later is that 
\be \label{pp2}
e_d = {g_F R_2 \ov \sqrt{2} R} \ .
\ee
See Fig.~\ref{fig:RN} for a cartoon picture of the geometry of an extremal charged black hole.

At finite temperature, in the regime ${z_* - z_0 \ov z_*} \ll 1$, i.e. $T/\mu \ll 1$, the near horizon metric is 
obtained from replacing the AdS$_2$ factor in~\eqref{ads2M} by an AdS$_2$ black hole \cite{Faulkner09},
 \be \label{ads2T}
 ds^2 =  {R^2_2 \ov \ze^2} \le( - \le(1- {\ze^2 \ov \ze_0^2} \ri)  dt^2 +
 {d \ze^2 \ov 1- {\ze^2 \ov \ze_0^2}} \ri)
 + \mu_*^2 R^2 d \vec x^2
 \ee
 with
\be
A_t =  {e_d  \ov \ze}  \le(1-{\ze \ov \ze_0} \ri), \qquad  \ze_0 \equiv  {z_*^2 \ov d (d-1) (z_*-z_0)}
\ee
and the temperature becomes 
 \be
 T ={1 \ov 2 \pi \ze_0} \ .
 \ee
At a temperature comparable to $\mu$,  the  AdS$_2$ structure will be lost.  
% small temperatures there is an alternative scaling limit in which we take $T \to 0$, $\frac{\om}{T} = \mbox{const}$; this results in the AdS$_2$ identified above being replaced by  
 
\begin{figure}[h]
\begin{center}
\includegraphics[scale=0.6]{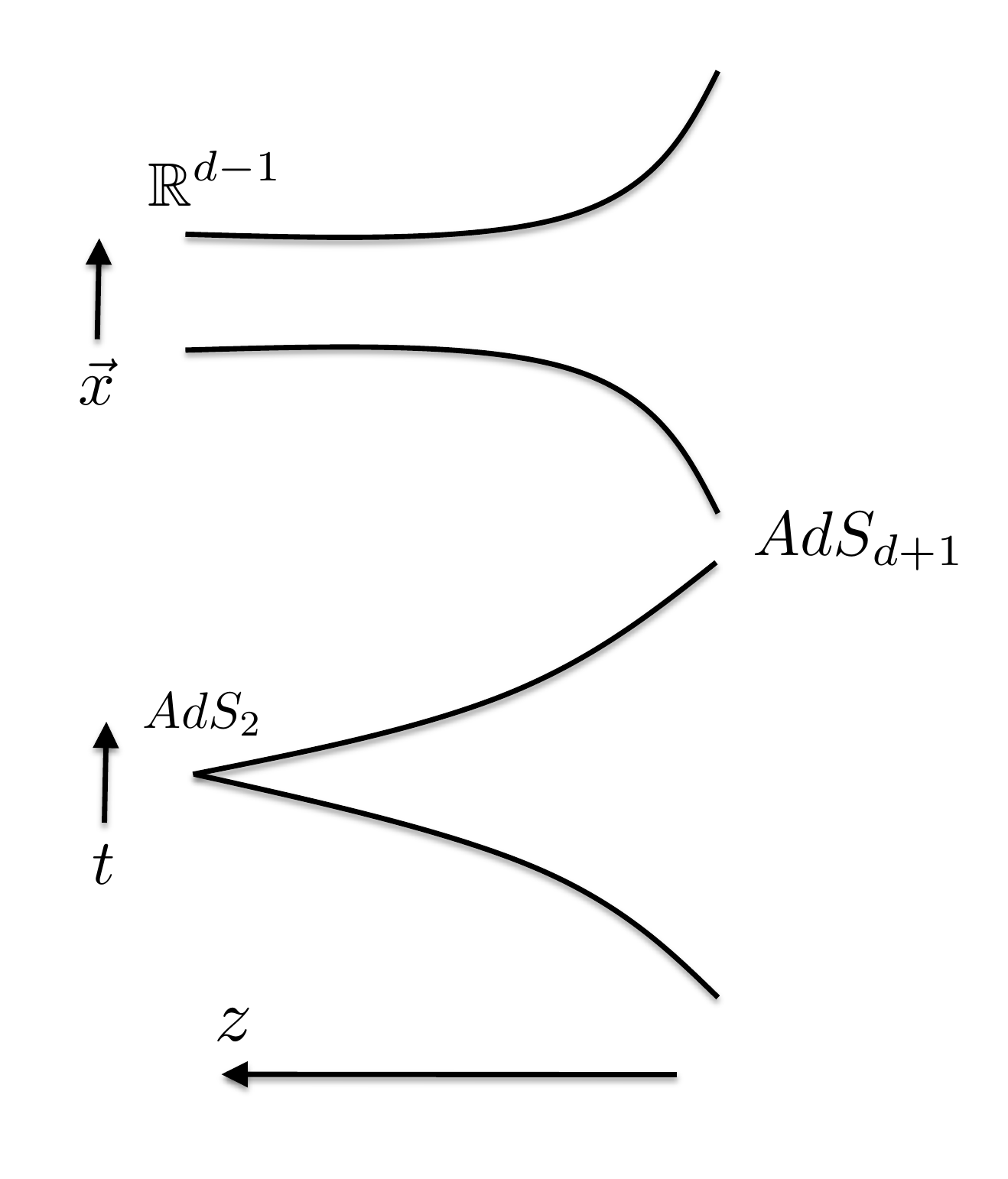}
\end{center}
\caption{A cartoon picture of the geometry of an extremal charged black hole. The horizon lies at an infinite proper distance away and in the near horizon region the warp for the spatial part approaches a constant, while that for the time direction shrinks to zero at the horizon.} 
\label{fig:RN}
\end{figure}

As discussed in the last subsection, the radial direction $z$ can be identified as the RG scale of the boundary theory. The emergence of  the near-horizon  AdS$_2$ region thus holds the key to understanding the low energy physics of the boundary theory at finite density. %~\cite{Faulkner09}.  
The warp factor in~\eqref{ads2M} and the fact that the horizon lies an infinite proper distance away imply that the system should again possess gapless degrees of freedom. In particular,  %We now note that
the metric~\eqref{ads2M} has a scaling isometry 
\be \label{newscal}
t \to \lam t, \qquad \zeta \to \lam \zeta , \qquad \vec x \to \vec x 
\ee
under which only the time coordinate scales (compare this to the relevant scaling symmetry for Lorentz-invariant AdS$_{d+1}$ \eqref{scalisom} in which the spatial and time coordinates scale in the same way.) Note that the AdS$_2$ region is in fact invariant under a full $SL(2,\mathbb{R})$ symmetry group, and likely a Virasoro symmetry. 

Thus gravity tells us that at low energies the boundary system flows to a fixed point which is
dual to AdS$_2 \times  \mathbb{R}^{d-1}$~\cite{Faulkner09}. See Fig.~\ref{fig:schematic} and~\ref{fig:flow} for illustration. This fixed point has previously been called the ``IR CFT$_1$'' to indicate that there is conformal symmetry only in one dimension (the time direction). However, this hides the fact that the dependence on the spatial directions can have important consequences, and thus we believe that a more descriptive name for this fixed point is a {\it semi-local quantum liquid} (or \Slql\;for short)~\cite{Iqbal:2011in}, for reasons that we will elaborate on in the next subsection. We will use this term in what follows. 

In addition to translational and rotational symmetries of the boundary spatial directions, the \Slql\ is thus characterized by conformal symmetries of a $(0+1)$-dimensional conformal quantum mechanics 
including the scaling symmetry in the time direction. 
It is crucial to note that this conformal invariance has nothing to do with the CFT$_d$ in the ultraviolet, whose conformal invariance was completely broken by the chemical potential. This new conformal symmetry is {\it emergent}, and has to do with the collective motion of the large number of charged excitations sustaining the background charge density.  

Note that the \Slql\ is specified by a single scale $\mu_*$. While $\mu_*$ corresponds to the chemical potential for the full UV theory, near the IR fixed point \Slql, $\mu_*$ simply defines a unit for spatial momenta.

 \begin{figure}[h]
\begin{center}
\includegraphics[scale=0.5]{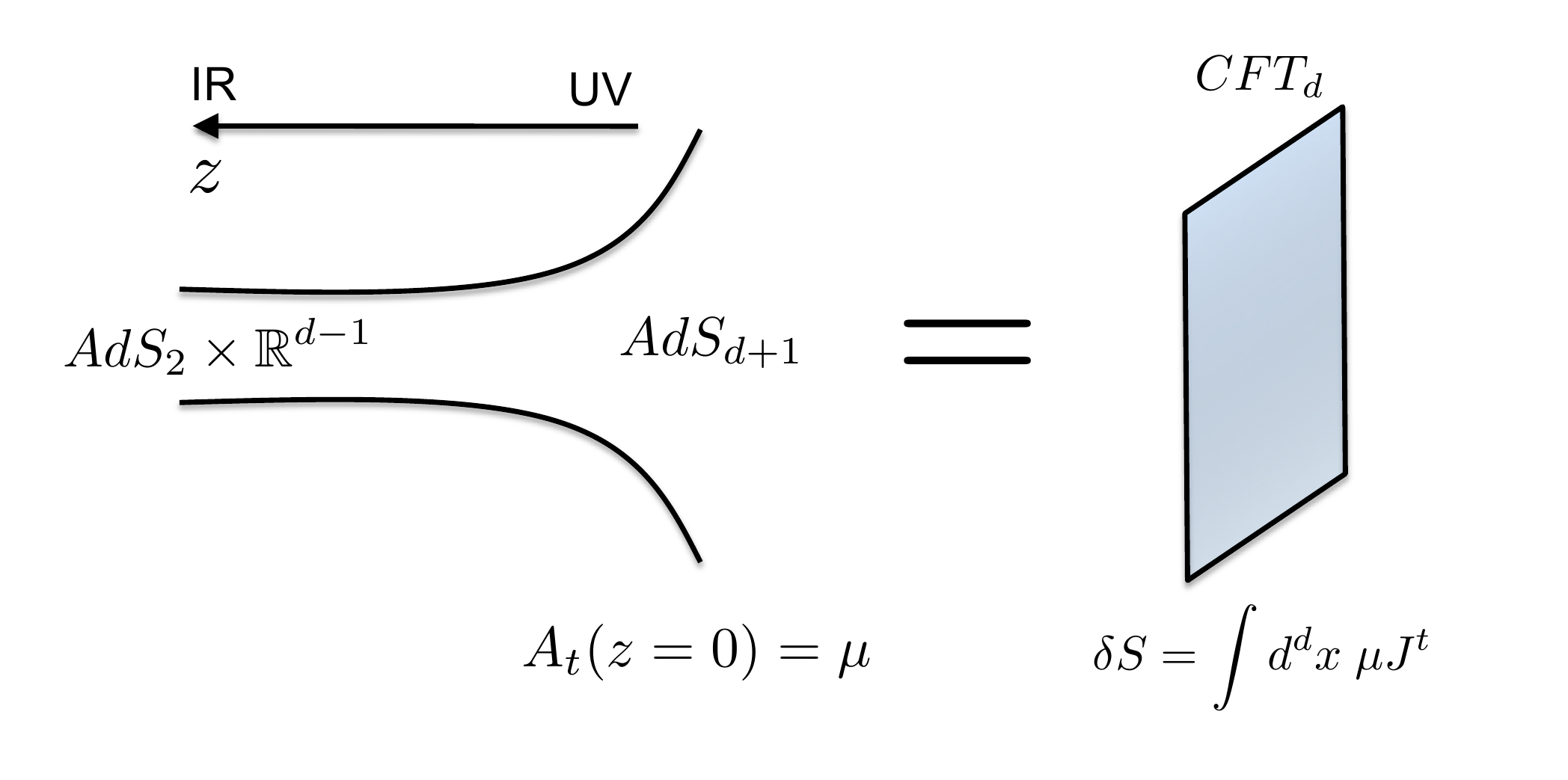}
\end{center}
\caption{Charged black hole with nonzero gauge field is dual to field theory state with nonzero chemical potential; in the infrared there is an emergent conformal symmetry corresponding to the AdS$_2 \times \mathbb{R}^{d-1}$ part of the geometry.} 
\label{fig:schematic}
\end{figure}

 \begin{figure}[h]
\begin{center}
\includegraphics[scale=0.4]{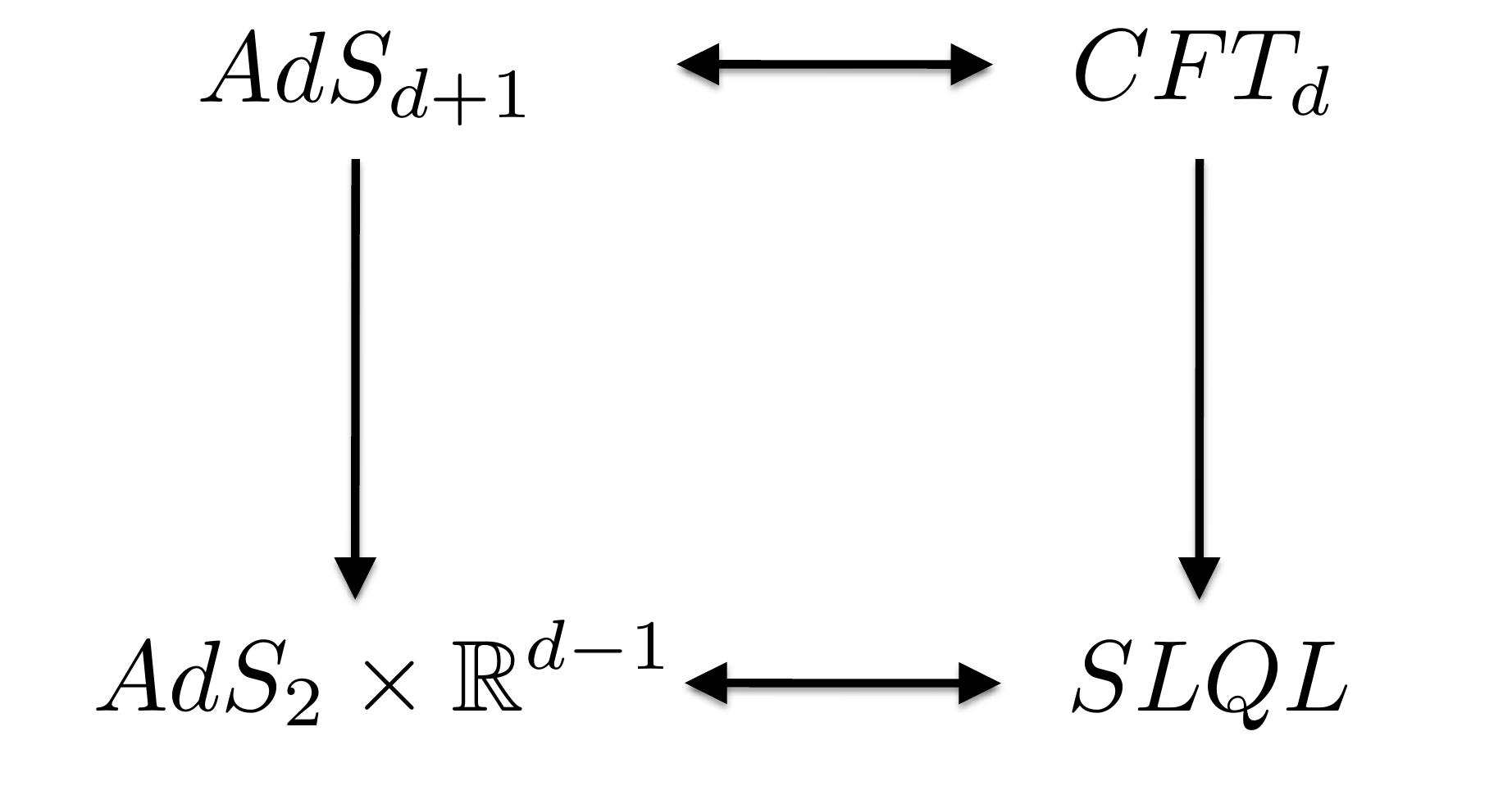}
\end{center}
\caption{At a finite chemical potential, a CFT$_d$ flows in the IR to \Slql. On the gravity side this is realized geometrically via the flow of the AdS$_{d+1}$ near the boundary to AdS$_2 \times \mathbb{R}^{d-1}$ near the horizon.
} 
\label{fig:flow}
\end{figure}

\subsection{Semi-local quantum liquids}  \label{sec:slql}

Consider an operator $\sO (t, \vec x)$ in the boundary theory, which is dual to some bulk field $\phi (t,\vec x, z)$ with mass $m$ and charge $q$.  
At the UV, i.e. CFT$_d$, the dimension of $\sO$ is given in terms of bulk quantities by \eqref{UVdim}. At the IR fixed point, $\sO$ should match to some scaling operator $\Phi$ (under scaling symmetry~\eqref{newscal}) in the  \Slql\ dual to the bulk field $\phi$ 
in AdS$_2 \times \RR^{d-1}$.  

Now in AdS$_2 \times \RR^{d-1}$,  the Fourier transform\nicefootnote[1]{To simplify notations, we will always use the same symbol to denote the Fourier transform of a field, distinguishing them only by their arguments.} $\phi (t, \vk,z)$ of $\phi (t, \vec x, z)$ along the spatial directions has an effective AdS$_2$ mass depending on spatial momentum $k = |\vk|$ (as in a Kaluza-Klein reduction), which will result in a $k$-dependent scaling dimension for boundary operators. It is then convenient to label operators by their spatial momentum, with $\phi (t, \vec k, z)$  in AdS$_2 \times \RR^{d-1}$ dual to some operator $\Phi_\vk (t)$ in the \Slql.
The Fourier transform $\sO_{\vec k}(t)$ of $\sO (t, \vec x)$ should then match to $\Phi_\vk (t)$ at the IR fixed point. Note that the distinction between $\sO_\vk$ and $\Phi_\vk$ is important; $\sO$ is an operator in the full theory, while $\Phi$ only exists at the IR fixed point.

The scaling dimension $\de_k$ of $\Phi_\vk (t) $ and its correlation functions in \Slql\ can be obtained from solving the relevant wave equations for $\phi$ in the AdS$_2 \times \RR^{d-1}$ geometry~\eqref{ads2M}~\cite{Faulkner09,Iqbal:2011in}. We present the results here, leaving their derivation to Appendix~\ref{app:ads2}. One finds that %~(see discussion around equation~\eqref{nuej} in Appendix~\ref{app:master} for a derivation),
  \be
\delta_k = \frac{1}{2} + \nu_k   \  \label{nudef}
\ee
with
\be \label{opep}
 \nu_k  =\sqrt{{m^2 R^2_2}- q_*^2 + {1 \ov 4}  +  {k^2 R_2^2 \ov  \mu_*^2 R^2} } =
 {1 \ov \sqrt{d (d-1)} } \sqrt{m^2 R^2 -  {g_F^2 q^2 \ov 2} + {d (d-1) \ov 4} + {k^2  \ov  \mu_*^2 } }
 \ee
 and
 \be
  k = |\vec k|, \quad q_* = {q e_d} \ ,
 \ee
 where in the second equality of~\eqref{opep} we have used~\eqref{pp2}.  The retarded function of $\Phi_\vk$ in the \Slql\nicefootnote[1]{We stress that this is not the full two-point function of the operator $\sO$ in the UV theory, but rather just the correlation function at the IR fixed point; the relation between these two is given in \eqref{roep1} and is discussed in~Sec.~\ref{lowgreen}.} can be written as 
 \be \label{iiRc}
\sG_k(\omega) = c (\nu_k) (-i \om)^{2 \nu_k}
\ee 
where
\be
c (\nu_k) =  2^{2 \nu_{k}} \frac{\Ga(-2\nu_k)}{\Ga(2 \nu_k)} {\Ga(\ha + \nu_k- iq_*) \ov \Ga (\ha -\nu_k - iq_*)}
 \ . \label{expform}
\ee
%This can be written as $\sG_2(\om) = c(k) \om^{2\nu_k}$ where $c(k)$ has an imaginary part. 
%See Appendix~\ref{app:ads2} for a derivation of~\eqref{iiRc}.

Let us now examine what~\eqref{opep}--\eqref{iiRc} tells us regarding the physical properties of a \Slql.
Firstly, the dimension~\eqref{opep} depends on the momentum
momentum $k$: as a result operators with larger $k$ become less important in the IR.  Also $\nu_k$ decreases with $q$, i.e. an operator with larger $q$ will have more significant IR fluctuations (given the same mass $m$). From the bulk point of view, this $q$-dependence in operator dimension is a consequence of the fact that 
in AdS$_2$, the electric field following from~\eqref{ads2M} blows up as we approach the AdS$_2$ boundary.\nicefootnote[1]{In other words, the electric field is non-renormalizable and specifies the corresponding boundary theory.} 

Secondly, the spectral weight, which is defined by the imaginary part of the retarded function~\eqref{iiRc}, scales with $\om$ as a power for any momentum $k$, which indicates the presence of low energy excitations for all momenta\nicefootnote[1]{Exercise for students: compare this result to more ``ordinary'' gapless systems, like a free massless boson or a Lorentz-invariant CFT} (although with a larger scaling dimension the weight will be more suppressed at larger momenta).
 
Thirdly, in coordinate space, the system has an infinite correlation time, but a finite correlation length in the spatial directions.  This is intuitively clear from the presence of (and lack of) scaling symmetry in the time (spatial) directions. It can be seen more explicitly by examining the Euclidean correlation function $G_E (\tau = i t , \vec x )$  in coordinate space by Fourier transforming~\eqref{iiRc}. For this purpose, we note that $\nu_k$ can be rewritten as 
\be \label{defnk}
\nu_k = {1\ov \sqrt{d (d-1)}\mu_*} \sqrt{k^2 + {1 \ov \xi^2}}, 
\ee
where
\be \label{defnk1}
 \xi  \equiv {1 \ov \mu_*} {1 \ov \sqrt {{m^2 R^2}- {g_F^2 q^2 \ov 2} + {d(d-1) \ov 4}}} 
=  {1 \ov \sqrt{d (d-1)} \nu_{k=0} \mu_*} \ .
\ee
$\nu_k$ has a branch point at  $k= i \xi^{-1}$, which under the Fourier transform leads to exponential decay in spatial directions at large distances with a correlation length given by $\xi$ (see Appendix~\ref{app:fourier}).  Explicit calculation yields that $G_E (\tau , \vec x )$ has two distinct regimes: 

\ben 
\item  For $x \equiv |\vec x|  \ll \xi$, 
\be \label{timde}
G_E (\tau, x) \sim {1 \ov \tau^{2 \de_{k=0}}} \ .
\ee

\item   For $x \gg \xi$, the correlation function decays at least 
exponentially as 
\be \label{spaco}
G_E (\tau, x) \sim e^{- {x \ov \xi}} \ .
\ee 

\een
Fig.~\ref{fig:local} provides a heuristic visualization of the behavior: the system roughly separates into independent clusters of size of order $\xi$, with the dynamics of each cluster controlled by 
a conformal quantum mechanics, with a power law correlation (i.e. infinite relaxation time)~\eqref{timde}. Domains separated by distances larger than $\xi$ are uncorrelated with one another. 
Given that the system has a nonzero entropy density, each cluster has a nonzero entropy that counts the number of degrees of freedom inside the cluster. 

 \begin{figure}[h]
\begin{center}
\includegraphics[scale=0.45]{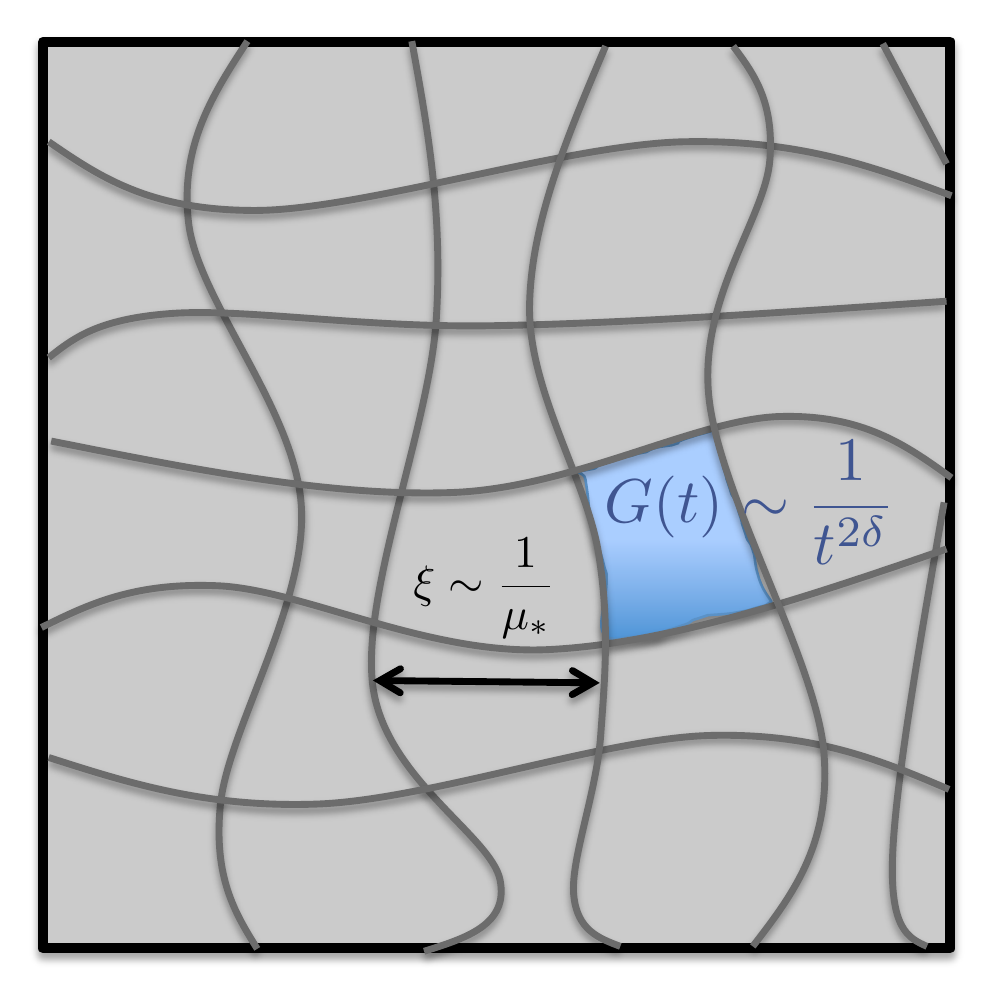}
\end{center}
\caption{A cartoon picture: the system separates into domains of size $\xi \sim {1 \ov \mu}$. 
Within each domain a conformal quantum mechanics governs dynamics in the time direction with a power law correlation (i.e. infinite relaxation time).} 
\label{fig:local}
\end{figure}

While~\eqref{timde} and~\eqref{spaco} can be found by doing Fourier transforms explicitly, they can also be seen geometrically using a geodesic approximation to calculate field-theoretical correlation functions using the Euclidean analytic continuation of~\eqref{ads2M},
\be \label{euec}
ds^2 = \frac{R_2^2}{\zeta^2}(d\tau^2 + d\zeta^2) + R^2 \mu_*^2\;d \vec x^2 \ .
\ee
Consider a geodesic that starts at $\zeta \to 0$, moves into AdS$_2$ at finite $\zeta$, eventually turns around, and returns to the boundary but at a spatial separation $x$ and a Euclidean temporal separation of $ \tau$. In the geodesic approximation the boundary theory Green's function in coordinate space $G_E(x,\tau) \sim e^{- m L (\tau, x)}$ where $L (\tau, x)$ is the proper distance along the geodesic. Since the metric~\eqref{euec} is just a direct product, we can simply find the distance moved in each factor and add them using Pythagoras. The distance moved in the $\mathbb{R}^{d-1}$ factor is $R \mu_* x$. A standard calculation shows that the distance moved in the AdS$_2$ factor is $2 R_2 \log\le(\frac{\tau}{\ep}\ri)$ where $\ep$ is an IR cutoff of the AdS$_2$.\nicefootnote[1]{The easiest way to understand this result is to note that its exponential must reproduce the conformal result, $\tau^{-2 m R_2}$.} 
Note that since the AdS$_2$ is embedded in the full spacetime~\eqref{RNmetric}, the cutoff 
$\ep$ should be provided by the scale ${1 \ov \mu}$. From IR/UV connection, an IR cutoff $\ep$ of AdS$_2$ translates into a short-distance cutoff in the \Slql, thus we will restrict to $\tau > \ep$. Combining these results we find
\bea
G_E (\tau, x) &\sim & \exp\le(-m  \sqrt{4 R_2^2 \log^2 \le(\frac{\tau}{\ep}\ri) +  \mu_*^2 R^2 x^2}\ri)  \cr
& \sim &  \exp \le(-\sqrt{4 \de^2  \log^2 \le(\frac{\tau}{\ep}\ri)  + {x^2 \ov \xi^2}} \ri)
 \label{geodcorr}
\eea
with $\de = mR_2$ and $\xi^{-1} = m R \mu_* $. The geodesic approximation applies to $m R \gg 1$ and in this regime,~\eqref{timde} and~\eqref{spaco} are indeed recovered from~\eqref{geodcorr}. 

The fact that at low energies different points on the $\mathbb{R}^{d-1}$ can be thought of as being in different disconnected domains with size $\xi \sim {1 \ov \mu}$ can also be seen geometrically 
as follows. Consider two spacetime points on a hypersurface of given $\zeta \to \infty$. To see whether observers at those locations can communicate with each other we look at time-like geodesics 
in~\eqref{ads2M} which connect the two points. Simple calculations  show that there is a maximal separation in $\mathbb{R}^{d-1}$
directions for two points to communicate with each other, given by
\be \label{maxd}
\Delta x_{\rm max} = \pi R_2  {1 \ov \mu_* R}  = \frac{\pi}{\mu_* \sqrt{d (d-1)}} \ .
\ee
The first factor $\pi R_2$ in the first equality above is the time for a timelike geodesic (see Fig.~\ref{fig:geo}) to approach the boundary and come back and the second factor ${1 \ov \mu_* R}$ is the effective velocity in $\mathbb{R}^{d-1}$ (see~\eqref{ads2M}). Equation~\eqref{maxd} is consistent with~\eqref{defnk} up to a prefactor.

 \begin{figure}[h]
\begin{center}
\includegraphics[scale=0.6]{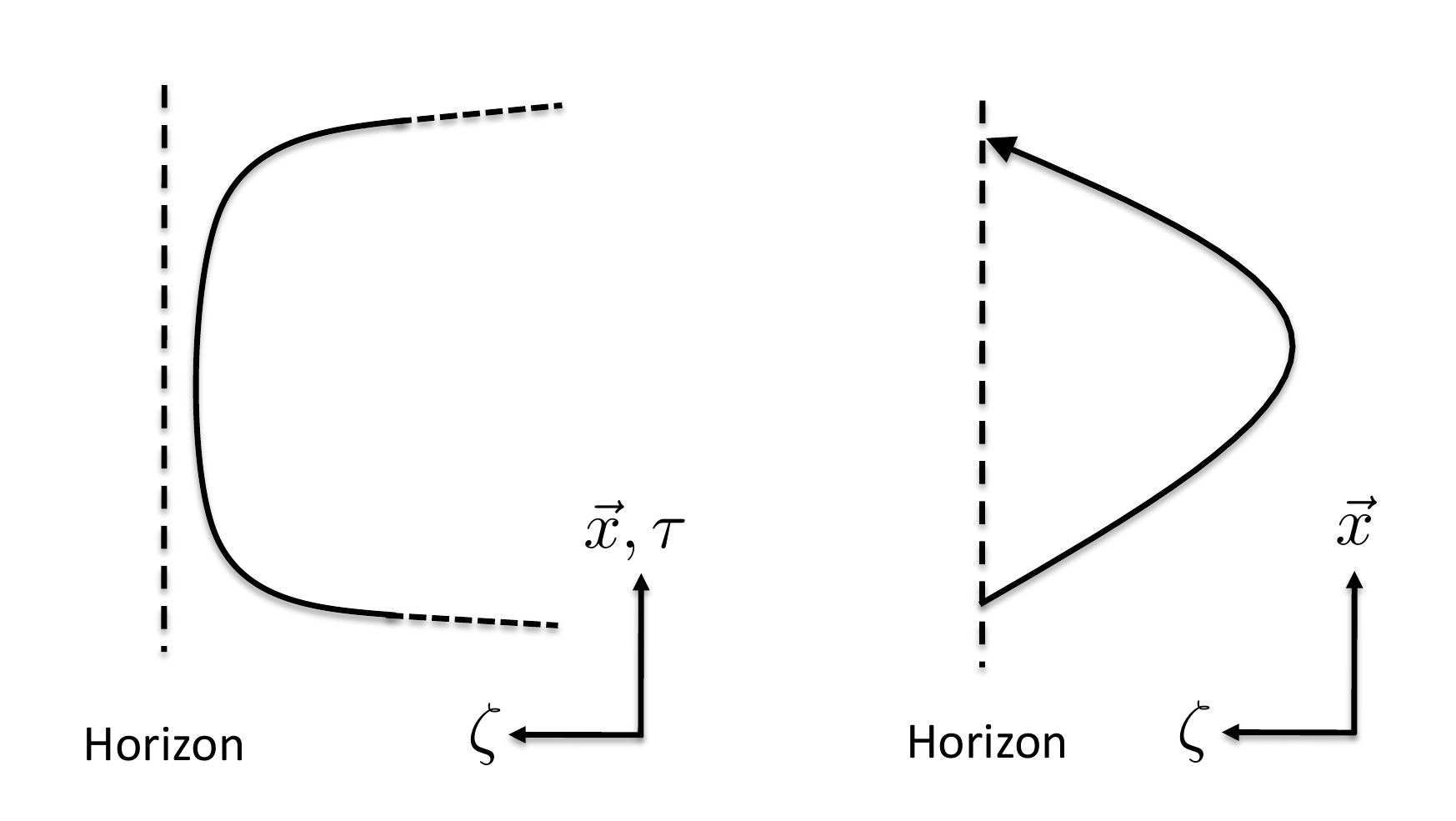}
\end{center}
\caption{Geodesics in the near horizon geometry. In both cases the horizon is the dotted line at $\zeta \to \infty$, and the boundary is to the right towards $\zeta \to 0$. The left plot gives the geodesic for calculating the
Euclidean correlation function~\eqref{geodcorr}; the right plot gives a timelike geodesic connecting two points on 
a constant-$\zeta$ hypersurface.  The maximal boundary time it takes for a timelike geodesic to 
approach the boundary and come back is $\pi R_2$. }  
\label{fig:geo}
\end{figure}

To summarize, equation~\eqref{iiRc}  describes a disordered state in which the space factorizes into independent domains of correlation length $\xi$. Within each domain one has scale invariance along the time direction.   
This behavior is reminiscent of
 various theoretical models based on a large spatial dimension mean field approximation~\cite{DMFT}, such as the gapless quantum liquids of~\cite{SY,zhuetal,bgg} and the ``local quantum critical point'' of~\cite{Si.01}\nicefootnote[1]{Some suggestions for a connection have been made in~\cite{Sachdev:2010um,yamamoto,Subir2,Subir3}.}. %and the local quantum critical behavior discussed earlier in Sec.~\ref{sec:lqc}. 
 We should emphasize that the scaling behavior within each cluster here describes, however, not the behavior of a single site, but the collective behavior of a large number of sites (if one considers our systems as a continuum limit of a lattice) over a size of order $\xi$. It is clear that while there is nontrivial scaling only  in the time direction, the scaling dimension~\eqref{opep} and correlation functions~\eqref{iiRc} depend nontrivially on $k$. As discussed earlier around~\eqref{defnk} it is precisely this dependence that gives the spatial correlation length of the system. Also, while at a generic point in parameter space the dependence of $\nu_k$ and $\sG_k$ on $k$ is analytic and only through $k/\mu$, as we will see in Sec.~\ref{sec:qpt}, near certain quantum critical points, the dependence becomes nonanalytic at $k=0$ and is important for understanding the behavior around the critical point. 
It is also important to emphasize that, despite the scaling behavior in~\eqref{iiRc}  we are describing a {\it phase}, not a critical point.  We thus call it a {\it \slql}, or a \Slql\ for short.

We now consider some generalizations of~\eqref{opep} and~\eqref{iiRc}.  At a finite temperature the background geometry is given by an AdS$_2$ black hole~\eqref{ads2T} and equation~\eqref{iiRc} generalizes to~\cite{Faulkner09, Faulkner:2011tm} 
\be  \label{finiteTch1}
\sG_k^{(T)}(\om) = T^{2\nu_k}  g_b \le(\nu_k,{\om \ov 2 \pi T} \ri)
\ee
where $g_b$ is a scaling function given by
\be \label{univgF}
g_b \le(\nu_k, y \ri) = (4 \pi)^{2\nu_k} \frac{\Ga(-2\nu_k)}{\Ga(2 \nu_k)} {\Ga(\ha + \nu_k- iq_*) \ov \Ga (\ha -\nu_k - iq_*)}  \frac{\Ga\le(\frac{1}{2} + \nu_k - iy + i q_* \ri)}{\Ga\le(\frac{1}{2} - \nu_k - iy  + i q_* \ri) } \ .
\ee

For a spinor field one finds that, by studying the behavior of the Dirac equation in the AdS$_2$ region, the dimension $\delta_k$ of a spinor operator in \Slql\ is given by
\be\label{ferdim}
\delta_k = \ha + \nu_k \qquad \nu_k = \sqrt{m^2 R_2^2 - q_*^2 + \frac{k^2 R_2^2}{R^2 \mu_*^2} }
=
 {1 \ov \sqrt{d (d-1)} } \sqrt{m^2 R^2 -  {g_F^2 q^2 \ov 2}  + {k^2  \ov  \mu_*^2 } } \ .
\ee
Note this is slightly different from the corresponding result for the charged scalar \eqref{opep}. Similarly, the retarded Green's function is given by\nicefootnote[1]{The following expressions are for half of the components. The expressions for other components are obtained by taking $k \to -k$. See~\cite{Faulkner09, Faulkner:2011tm} for details.} 
\be \label{iiFc}
\sG_k (\om,k) = c(\nu_k) (-i \om)^{2 \nu_k}
\ee
with $c (\nu_k)$ now given by
\be \label{ffcc}
c (\nu_k) = 2^{2\nu_k} \frac{\Ga(-2\nu_k)}{\Ga(2\nu_k)} \frac{\Ga(1 + \nu_k - iq_*)}{\Ga(1 - \nu_k - iq_*)}\frac{\le(m - \frac{i k R}{\mu_*}\ri)R_2 - iq_* - \nu_k}{\le(m - \frac{i k R}{\mu_*}\ri)R_2 - iq_* + \nu_k} \ .
\ee
Similarly, at finite temperature, the spinor AdS$_2$ Green's function equation~\eqref{iiFc} generalizes to 
\be  \label{FerFT}
\sG_k^{(T)}(\om) = T^{2\nu_k}  g_f \le(\nu_k,{\om \ov 2 \pi T} \ri)
\ee
with $g_f$  given by
\be \label{univgFF}
g_f \le(\nu_k, y \ri) = (4 \pi)^{2\nu_k} \frac{\le(m - \frac{i k R}{\mu_*}\ri)R_2 - iq_* - \nu_k}{\le(m - \frac{i k R}{\mu_*}\ri)R_2 - iq_* + \nu_k} 
\frac{\Ga(-2\nu_k)}{\Ga(2 \nu_k)} {\Ga(1 + \nu_k- iq_*) \ov \Ga (1 -\nu_k - iq_*)}  \frac{\Ga\le(\frac{1}{2} + \nu_k - iy + i q_* \ri)}{\Ga\le(\frac{1}{2} - \nu_k - iy  + i q_* \ri) } \ .
\ee

\subsection{Low energy behavior of retarded Green functions} \label{lowgreen}

Having understood the behavior of the retarded function  at the IR fixed point \Slql, we now turn to the low energy behavior  of the retarded function for $\sO$ in the full theory, which we will denote as $G_R (\om, k)$ to distinguish it from $\sG_k (\om)$ at the IR fixed point. 
We will consider a scalar operator at $T=0$ and  mention the generalization to a finite temperature and a spinor at the end.  
We are interested in the low frequency regime $\om \ll \mu$. Note that since we are working in the grand canonical ensemble the frequency $\om$ in our discussion of this paper always refers to energy measured from the chemical potential.\nicefootnote[1]{More explicitly, for an operator of charge $q$, the effective chemical potential is then $\mu q$ and $\om$ is the energy measured from $\mu q$. See also discussion below equation~\eqref{eomM} in Appendix~\ref{app:master}.}

$G_R (\om, k)$ can be computed explicitly by solving the equation of motion for the corresponding bulk scalar field in the full charged black hole geometry~\eqref{RNmetric}
and following the procedure outlined in Sec.~\ref{sec:ads}.  Basically our task is to understand how the expansion coefficients $A$, $B$ in \eqref{GRdef} behave at low frequencies.  
We present only the final result here, 
\be \label{roep1}
G_R (\om, \vk) = \mu_*^{2\nu_U} {b_+(\om, k)+ b_- (\om, k) \sG_k (\om) \mu_*^{-2 \nu_k} \ov
 a_+ (\om, k) + a_- (\om, k) \sG_k (\om) \mu_*^{-2 \nu_k}} + \dots \ ,
\ee
relegating the details to Appendix \ref{app:master}.
The various objects appearing in this formula deserve explanation. $\sG_k(\om)$ is the retarded Green's function in the \Slql\ discussed in Sec.~\ref{sec:slql} and is proportional to $\om^{2\nu_k}$ with $\nu_k$ defined in \eqref{opep}. $a_{\pm}$ and $b_{\pm}$ are quantities that arise from the solving the equation of motion in the region of the black hole geometry not too close to the horizon (called the UV region in Appendix~\ref{app:master}); the key fact here is that they are analytic in $\om$ and have a smooth $\om \to 0$ limit. See Appendix~\ref{app:master} for further discussion of their properties. 
Since $a_{\pm}, b_{\pm}$ in \eqref{roep1} are analytic in $\om$, we can expand them as follows
\be \label{omex}
a_{+}(\om, k) = a_{+}^{(0)}(k) + a_+^{(1)}(k) \om + a_+^{(2)}(k) \om^2 + ...
\ee
and keep only the leading order term after which~\eqref{roep1} becomes 
\be \label{roep12}
G_R (\om, \vk) = \mu_*^{2\nu_U} {b_+^{(0)}(k)+ b_-^{(0)}(k) \sG_k (\om) \mu_*^{-2 \nu_k} \ov
 a_+^{(0)}(k) + a_-^{(0)}(k) \sG_k (\om) \mu_*^{-2 \nu_k}} + \dots \ . 
\ee

Equation~\eqref{roep1} contains two sets of data:  ``universal'' data from $\sG_k (\om)$ that depends only on the IR fixed point (or the near-horizon AdS$_2 \times \RR^{d-1}$ region of the black hole geometry), and non-universal data $a_\pm, b_\pm$ which depends on the rest of the black hole geometry and can be thought of as encoding the  the effects of the UV degrees of freedom on the low energy physics.
This is consistent with general expectations from a Wilsonian understanding of renormalization group flow,
and we now show that the form of~\eqref{roep1} follows simply from general considerations of a suitable
low energy effective action. 
 
We expect that the low energy effective action of the system can be written as
\be \label{fulef}
S_{eff} = S_{\Slql} + S_{UV} 
\ee
where $S_{\Slql}$ is the action for the IR fixed point, for which we do not have an explicit Lagrangian description, but as discussed in Sec.~\ref{sec:slql} whose observables can be computable from gravity in AdS$_2 \times \RR^{d-1}$.  $S_{UV}$ arises from integrating out higher energy degrees of freedom, and can be expanded in terms of scaling operators in $S_{\Slql}$. 
 The part relevant for $\sO$ can be written as\nicefootnote[1]{Note the expressions below are written somewhat heuristically to exhibit the main structure. For example, $J_\vk^2$ should be interpreted as $J_{\vk} J_{-\vk}$ and similarly with $\Phi_\vk^2$. }
 \be \label{eff1}
S_{UV}=  
{1 \over 2} \int \Lam (k, \om) J_{\vec k}^2  + \int \eta(k,\om) \Phi_{\vec k} J_{-\vec k}  - {1 \over 2} \int  \, \xi (k,\om)
\Phi_{\vec k}^2 + \dots
\ee
where we have written the action in momentum space since the dimension of \Slql\ operator $\Phi_{\vec k} (t)$ is momentum-dependent, and the integral signs should be understood as 
$\int = \int d\om \int {d \vec k }$. We have introduced a source $J_{\vec k}$ for $\sO_{\vec k} $ and $\dots$ denotes higher powers of $\Phi_{\vec k}$ and $J$. Note again the distinction between $\sO$ and $\Phi$; $\Phi$ is the operator at the IR fixed point to which $\sO$ matches. Since we are only interested in two-point functions it is enough to keep $S_{UV}$ to quadratic order in $\Phi$ and $J$. For simplicity we have assumed to the quadratic order there is no mixing of $\sO$ with other operators.\nicefootnote[1]{This corresponds to that the equation for the dual bulk field $\phi$ does not mix with other fields at linear order.}  The discussion can be easily generalized to the situation with mixing. The functions $\Lam (k,\om), \eta(k,\om)$ and $\xi (k,\om)$ parametrize the ``UV data'' coming from integrating out higher energy modes. In particular they 
should be {\it real} and have an analytic expansion in $\om$ with a nonzero $\om \to 0$ limit, i.e. when written in terms of $t$, the effective action~\eqref{eff1} should have a well defined derivative expansion. 

The (Euclidean) two-point function for $\sO_\vk$ can be obtained by differentiating the Euclidean partition function of $S_{eff}$ with respect to $J_\vk$. It contains two parts. The first part is simply $\Lam (\om, k)$  from the first term in~\eqref{eff1}, which comes purely from the UV and is independent of any IR fields.  The second part
comes from the correlation function $\Phi_\vk$ in the \Slql\ (up to a factor of $\eta$). To the order written explicitly in~\eqref{eff1}, the full action~\eqref{fulef} describes the \Slql\ deformed by a double trace operator with coupling $\xi$. 
The two-point function of $\Phi$ in this deformed theory can be obtained from the general formula~\eqref{Eudou}  in  Appendix~\ref{app:doub}. Combining the two parts and analytically continuing to Lorentizian signature, we find that the retarded function $G_R$ for the full theory can be written as 
\be \label{g1}
G_R (\om, k) = \Lam(k,\om)  + {\eta^2 (k, \om) \ov \sG_k^{-1} (\om) + \xi (k,\om)} \ .
\ee
Equation~\eqref{g1} has exactly the same form as the gravity result~\eqref{roep1} and in fact comparing them we can identify
\be  \label{coffG}
\Lam = \mu_*^{2 \nu_U} {b_+ \ov  a_+} , \qquad \xi  = \mu_*^{-2 \nu_k} { a_- \ov  a_+}, \qquad \eta^2  = {W \ov   a_+^2} \mu_*^{2 \nu_U- 2 \nu_k}, \qquad 
W \equiv a_+ b_- - a_- b_+  \ .
\ee
The low energy effective action~\eqref{fulef},~\eqref{eff1} with identifications~\eqref{coffG} can also be derived directly following the formalism of holographic Wilsonian RG introduced recently in~\cite{Faulkner:2010jy,Heemskerk:2010hk}. Note that while our main goal here is to derive the retarded function~\eqref{roep1}, the effective action~\eqref{fulef}--\eqref{eff1} clearly 
has much wider applications and we will see some of them in later sections. 

Since here the spatial directions do not scale, the last term 
in equation~\eqref{eff1} is an irrelevant perturbation from \Slql\  since $\Phi_\vk$ has dimension $\ha + \nu_k > \ha$. This is what we should expect; \Slql\ is the IR fixed point and the last term in~\eqref{eff1} is simply the lowest irrelevant operator taking it to the UV. Indeed from~\eqref{roep12} (or equivalently~\eqref{g1}) we find that as $\om \to 0$, 
\be \label{genK}
\Im G_R(\om, k)= \mu_*^{2\nu_U }{\nu_k \ov \nu_U }{1 \ov (a_+^{(0)} (k))^2} \Im \sG_k (\om) + ... \sim \om^{2\nu_k} + ... \
\ee
where we have used~\eqref{pp3} and that $a_\pm, b_\pm$ are real. 
As expected, up to a non-universal prefactor (which can be understood as a wave-function renormalization between the UV and IR operators), at low energies the spectral function is given
by that of the IR fixed point. In particular, as already emphasized in Sec.~\ref{sec:slql}~(one paragraph before that of~\eqref{defnk}), the system has gapless degrees of freedom at any spatial momentum. 

It turns out there are circumstances in which~\eqref{genK} does not apply, and the physics is in fact much richer. This will be the subject of the next subsection.

We conclude this subsection with some generalizations:

\ben

\item  At a finite temperature $T \ll \mu$, equation~\eqref{roep1} can be generalized by replacing 
$\sG_k (\om)$ with its finite-temperature generalization $\sG_k^{(T)} (\om)$ of~\eqref{finiteTch1}. 
In addition, there will also be analytic finite-$T$ corrections to $a_{\pm}, b_{\pm}$, which can be expanded perturbatively in $T$. 

\item  An analogous calculation can be done for a Dirac spinor field in the bulk. Some of the details are different  but at the end of the day the formula \eqref{roep1} still applies to each of the {\it eigenvalues} of the Green's function, which is now a matrix in spinor space. We discuss this briefly in Sec.~\ref{sec:setupF}; see ~\cite{Faulkner09,Iqbal:2009fd} for details.
%ee Appendix~\ref{app:spinors} for details. 

\een

\subsection{Possible low energy physics: Fermi surfaces and scalar instabilities} \label{sec:bosvsferm}

As discussed in Sec.~\ref{fermiliquid} and~Sec.~\ref{sec:nqpt}, the retarded function for a fermionic operator can be used to probe the existence of a Fermi surface and the nature of low energy excitations around the Fermi surface, while that for a bosonic order parameter provides a convenient diagnostic for possible instabilities, and encodes important dynamical information for quantum phase transitions. 
If~\eqref{genK} were the full story, then that would be somewhat boring, as there will be no Fermi surface or instability (thus no quantum phase transition). However, equation~\eqref{genK} breaks down when one of the following occurs and interesting new physics emerges,

\ben

\item  $a_+^{(0)} (k)$ could {\it vanish} at some momentum; \label{it1}

\item  $ \nu_k$ which characterizes the scaling dimension of an operator in the SLQL becomes imaginary. Note in this case $a_\pm, b_\pm$ become complex. See~\eqref{UVsolnS} and~\eqref{comab} in Appendix~\ref{app:master}. 
\label{it2} 

\een
From the RG point of view, item~\ref{it1} implies that the effective action~\eqref{eff1}  breaks down as~\eqref{coffG} implies that the coefficients of~\eqref{eff1} become singular in the $\om \to 0$ limit at the momentum 
where $a_+^{(0)} (k)$ vanishes. 
When $\nu_k$ is pure imaginary (say equal to $- i \lam_k$), the dimension of $\Phi_\vk$ in the \Slql\ is now complex $\ha - i \lam_k$; a complex conformal dimension is somewhat peculiar and seems likely to indicate that we are no longer using the correct effective theory. Thus in either situation we can no longer conclude that the theory is described by \Slql\ alone in the IR.

In the rest of this subsection we discuss possible physical interpretations of the above two possibilities assuming they 
 do occur, leaving more in-depth study (including the parameter range over which they could occur) to Sec.~\ref{sec:nfl} and~\ref{sec:qpt}.

\subsubsection{Inhomogeneous black hole hair: Fermi surface v.s. scalar instability} \label{sec:hair}

 We now examine the first possibility, i.e.  $a_+^{(0)} (k)$ could {\it vanish} at some momentum.  Recall from equation~\eqref{12as} in Appendix~\ref{app:master} that $a_{+}^{(0)}$ is the leading expansion coefficient near the AdS$_{d+1}$ boundary of the UV region solution $\eta_+^{(0)}$~\eqref{IRdefUVsoln}. By construction $\eta_+^{(0)}$ is {\it normalizable} at the horizon. %At $k_F$, given $a_{+}^{(0)} (k_F) = 0$, 
 When $a_+^{(0)}$ vanishes $\eta_+^{(0)}$ becomes normalizable also at the AdS$_{d+1}$ boundary and thus  is now a genuine normalizable mode in the black hole geometry. One might call it {\it black hole hair.} Clearly, the condition that $a_+^{(0)} (k) =0$ is not generic; it requires fine-tuning in one of the parameters entering the differential equation and so can only have solutions at most at some discrete values of momentum $k$. When this happens at a nonzero momentum, one finds {\it inhomogeneous} hair. The discussion also applies identically to the spinor case.

Now suppose that $a_{+}^{(0)}$ indeed vanishes at some value of $k = k_F \neq 0$: i.e. $a_{+}^{(0)}(k_F) = 0$.
Let us consider what happens to~\eqref{roep1} near $k_F$. First at $\om =0$, we find simply a pole 
\be \label{grpole}
G_R(\om = 0, k) \approx \mu_*^{2 \nu_U} \frac{b_{+}^{(0)}(k_F)}{\p_k a_{+}^{(0)}(k_F) (k-k_F) } + \dots \ . %\qquad k_\perp = k - k_F
\ee
For a spinor operator, this is a zero-frequency singularity at a shell in $k$-space, and so is precisely the signature of a Fermi surface! Now further turning on a small frequency $\om$ near $k \sim k_F$ and keeping the first few terms in the relevant expansion of \eqref{roep1}, we find
\be
G_R(\om, k) \approx \mu_*^{2 \nu_U} \frac{b_+^{(0)}(k_F)}{\p_k a_+^{(0)}(k_F) (k-k_F) + \om a^{(1)}_+(k_F) + a_-^{(0)}(k_F) \sG_{k_F} (\om)\mu_*^{-2\nu_{k_F}}} + \dots \ 
\ee
which can be further written as  
\be
%\boxed{
G_R(\om, k) \approx \mu_*^{2 \nu_U} \frac{h_1}{k-k_F - \frac{1}{v_F}\om - \Sigma(\om)} + \dots
 \label{FSexp1}
\ee
where 
 \be
 v_F \equiv -\frac{\p_k a_+^{(0)} (k_F)}{a_+^{(1)} ((k_F)}, \qquad
 h_1 \equiv \frac{b_+^{(0)}(k_F)}{\p_k a_+^{(0)}(k_F)}
 \ee
 and
  \be \label{selfM1}
\Sigma(\om) =  h {\sG_{k_F}(\om) \ov \mu_*^{2\nu_{k_F}}}= h c(\nu_{k_F}) \le({\om \ov \mu_*}\ri)^{2 \nu_{k_F}}, \qquad
h \equiv - \frac{a_-^{(0)} (k_F)}{\p_k a_+^{(0)}(k_F)} \ .
\ee
Equation~\eqref{FSexp1}, which has a similar (but more general) form as the Fermi liquid~\eqref{eq:qpprop1}, is indeed  consistent with the behavior of low energy response near a Fermi surface.  $\Sig (\om)$ can be interpreted as the self-energy for fermionic excitations near the Fermi surface and is controlled by the retarded function $\sG_{k_F}$ of the IR fixed point (i.e \Slql) evaluated at $k=k_F$. Recall now that for a Landau Fermi liquid~\eqref{selffer} we have $\Sigma_{FL}(\om) \sim \om^2$, following from basic kinematics of interactions about the Fermi surface. Here, however, we have 
\be
\Sigma \sim \om^{2\nu_{k_F}}\  \label{width}
\ee 
with a nontrivial scaling exponent $\nu_{k_F}$. Thus we find a {\it non-Fermi liquid}; a system with a Fermi surface but whose excitations appear to {\it not} be Landau quasiparticles.  
The dispersion and decay width of small excitations near the Fermi surface can be extracted from the poles of~\eqref{FSexp1}. 
The nature of such excitations turns out to be rather different depending on the value of $\nu_{k_F}$, which we will discuss in detail in Sec.~\ref{sec:nfl}, along with when such a $k_F$ could exist. 

%the dispersion of the pole for the retarded function gives 
%the dispersion of small  
%which for~\eqref{FSexp1} correspond to the solution to the equation 
%\be
%k - k_F - \frac{1}{v_F}\om_c - \Sigma(\om_c) = 0 \label{dispeqn1} %\qquad \om_c(k) = \om_*(k) - i \Ga(k)  
%\ee
%in the complex $\om$-plane. 

The above discussion works well for a spinor operator, but how do we interpret~\eqref{grpole} and~\eqref{FSexp1} for a scalar operator? This turns out to be related to an important question for the spinor case: does the pole of~\eqref{FSexp1} always lie in the lower complex $\om$-plane as is required from analyticity? Note that at $k= k_F$, the pole  lies at the origin of the complex $\om$-plane. 
 If we move slightly away from $k = k_F$ the pole will move off the origin and into the complex $\om$ plane. What will the motion of the pole be? Based on the following property of 
 the scalar and spinor retarded functions  ($\om >0$)
  \bea
 \label{s2}
&&{\rm scalars:} \quad  \Im G_R (-\om) < 0, \quad 
\Im G_R (\om) > 0  \quad \\
 &&{\rm spinors:} \quad  \Im G_R (-\om) > 0, \quad \Im G_R (\om) > 0 \  ,
 \label{f2}
 \eea
one can show that~\cite{Faulkner09}

\ben

\item  For a spinor the pole always lies in the lower-half complex $\om$-plane.

\item For a scalar, the pole crosses from the upper half plane to the lower half plane 
as $k$ is increased from below to above $k_F$. Thus for a scalar there is a pole for $k < k_F$ in the upper half plane.

\een
% {\bf The argument is very simple and is given in Appendix~\ref{app:poles}.}
We note that~\eqref{s2} and~\eqref{f2} simply 
follow from the definition~\eqref{retF} that the retarded function for a fermionic (bosonic) operator is defined to be an anticommutator (commutator). 
Although in the bulk we are dealing with classical equations of scalars and spinors and have not imposed any statistics, the self-consistency of AdS/CFT implies that the classical equations for bulk scalars and spinors should encode the {\it quantum statistics} of the boundary theory.
 
A pole for $k < k_F$ in the upper half $\om$-plane for a scalar gives rise to exponentially growing modes    (recall the discussion of Sec.~\ref{sec:nqpt}).
Thus instead of finding a Fermi surface, black hole scalar hair at some $k_F \neq 0$ 
indicates an instability to condensing the scalar operator.  
  In Sec.~\ref{sec:qpt} we will see that it is possible to vary parameters of the system
so that $k_F$ can be smoothly tuned to zero, after which the instability disappears. The value of the external parameter at which $k_F =0$ thus corresponds to a quantum critical point, whose critical behavior will be the focus of Sec.~\ref{sec:qpt}. 
 
\subsubsection{Low energy effective theory}  \label{sec:semi}
 
Now let us turn to the low effective theory~\eqref{fulef}--\eqref{coffG} to examine what happens when $a_+^{(0)}$ vanishes.
 From equation~\eqref{coffG} and~\eqref{omex}, when $a_+^{(0)}=0$, all the coefficients in the $S_{UV}$ diverge as $\om \to 0$, i.e. the effective action becomes non-local. 
 This means we must have integrated out some gapless modes, which of course should be  the small excitations around $k_F$ which we have just identified from the poles of~\eqref{FSexp1}. 
 In order to have a well-defined local effective action, we need to put such gapless modes back to the low energy effective action.  We will introduce a new field $\Psi$ in the low energy theory and ``un-integrate''~\eqref{eff1}. Clearly there is no unique way of doing this\nicefootnote[1]{We can for example make a field redefinition in $\psi$ as 
$\psi \to Z_1 \Psi + Z_2 \Phi$.} 
 and the simplest choice is
 \be \label{eff3}
S_{UV} =  - {1 \over 2} \int \ka (\om, k) \Phi^2_\vk +\int \, {\lambda (k, \om) } \Phi_\vk \, \Psi_\vk 
- {1 \over 2} \int \, \Psi_{-\vk} \, K (\om,k) \, \Psi_{\vk} + 
  \int \Psi J
\ee
where 
\be \label{newpa1}
K  = \Lam^{-1} = {a_+ \ov  b_+} , \qquad \lam  = {\eta  K} = {\sqrt{W} \ov b_+}, \qquad 
 \ka = \xi +  {\lam^2 \ov K} = {b_- \ov b_+}  \ .
\ee
Now all coefficients in~\eqref{eff3} have local expansion in $\om$. This discussion applies to both a spinor or a scalar operator.

Near $k_F$, $K$ has the expansion
\be \label{freef}
K = {1 \ov h_1} \le(\om - v_F (k-k_F) + \dots \ri)  \ 
\ee
and thus for a spinor operator, the third term in~\eqref{eff3} can simply be interpreted as the action for a free fermion near its Fermi surface at $k_F$. The full low energy effective action~\eqref{fulef} can now be written as~\cite{Faulkner09,Faulkner:2010tq}
\be \label{hybid}
S_{eff} = \tilde S_{\Slql} [\Phi] + \int \, {\lambda (k, \om) } \Phi_\vk \Psi_\vk  + S_{\rm free \; fermion} [\Psi]
+ \int \Psi J
\ee
where 
\be \label{freefer}
S_{\rm free \; fermion} [\Psi] = - {1 \over 2 h_1} \int dt d \vk \, \Psi_{-\vk}  \le(i \p_t- v_F (k-k_F) + \dots \ri) \Psi_{\vk} 
\ee
is the action for a free fermion near a Fermi surface and 
\be \label{newsl}
\tilde S_{\Slql} = S_{\Slql} - {1 \over 2} \int \ka (\om, k) \Phi^2_\vk
\ee 
is the action for $\Phi$ which is \Slql\ deformed by a double-trace term. The low energy action~\eqref{hybid} describes a free fermion field $\Psi$ {\it hybridized} with a strongly coupled  fermionic operator $\Phi$ whose dynamics is in turn controlled by \Slql. \eqref{hybid} also offers a simple field theoretical interpretation for the expression~\eqref{FSexp1}: the self-energy $\Sig (\om)$ simply arises from the mixing of $\Psi$ with $\Phi$ and is thus given by the retarded function of $\Phi$, see Fig.~\ref{fig:hybr}. As we will see in Sec.~\ref{sec:nfl}, the action~\eqref{hybid} also provides an efficient way to understand the behavior of small excitations near the Fermi surface. 

\begin{figure}[h]
\begin{center}
\includegraphics[scale=0.5]{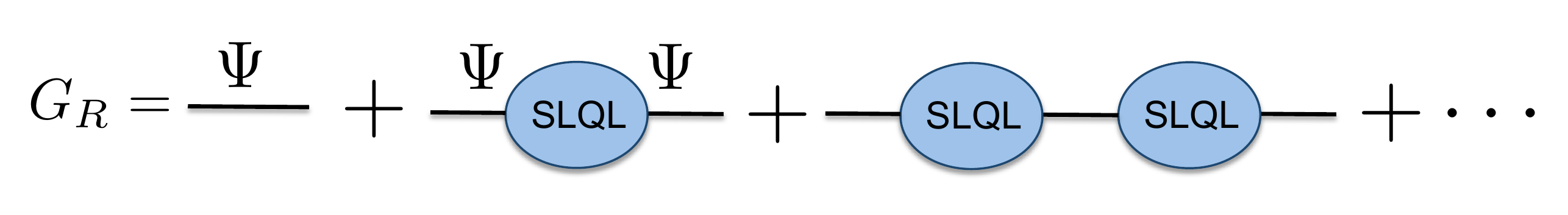}
\end{center}
\caption{The action~\eqref{hybid} gives a simple field theoretical interpretation of~\eqref{FSexp1} through hybridization between a free and a strongly interacting sectors. The propagator for the free fermion is given by $K$~\eqref{freef}.}
\label{fig:hybr}
\end{figure}

The above discussion also applies to the scalar case. For $k_F \neq 0$, one is, however, dealing with an 
unstable state and its physical meaning is not so clear. But for $k_F =0$, i.e. near a quantum critical point mentioned earlier, then $\Psi$ can be interpreted as the order parameter and the analogue of~\eqref{hybid} provides an effective action for the order parameter.  In particular, the free-fermion part in~\eqref{hybid} becomes the standard Landau-Ginsburg 
action. We will discuss this further in Sec.~\ref{sec:qpt}.

\subsubsection{Pair production by a black hole: complex IR dimensions} \label{sec:osc} %bulk Fermi surface and Bose-Einstein condensate

Let us now consider what happens when the conformal dimension $\ha + \nu_k$ of an operator in an SLQL becomes complex. Recall that for a scalar the conformal dimension  is given by~\eqref{opep}, which we copy here for convenience,
\be
 \nu_k  ={1 \ov \sqrt{d (d-1)} } \sqrt{m^2 R^2 -  {g_F^2 q^2 \ov 2} + {d (d-1) \ov 4} + {k^2  \ov  \mu_*^2 } }  \ee
 with a similar expression~\eqref{ferdim} for a spinor. We thus find that for 
\be  \label{Bcon1}
k_o^2 \equiv \mu_*^2 \le({g_F^2 q^2 \ov 2} - m^2 R^2 - {d (d-1) \ov 4}\ri) > 0
\ee
 $\nu_k$ becomes {\it imaginary} for $k^2 < k_o^2$ with
 \be \label{oepp1}
  \nu_k =- i \lam_k, \qquad \lam_k = {1 \ov \sqrt{d (d-1)} \mu_*}  \sqrt{k_o^2 - k^2} \ .
 \ee
Note on the gravity side, equation~\eqref{Bcon1} corresponds to violating the Breitenlohner-Freedman bound~\cite{Breitenlohner:1982jf} of AdS$_2$~\cite{Gubser:2005ih,Hartnoll:2008kx,Gubser:2008pf,Denef:2009tp}.

For a spinor~\eqref{Bcon1} is replaced by 
\be \label{Fcon1}
 k_o^2 \equiv \mu_*^2  \le({g_F^2 q^2 \ov 2} - m^2 R^2 \ri)  > 0 \ .
 \ee
Basically we see that the background electric field acts through the charge as an effective ``negative mass square'', making it possible to have a negative total mass square and
  resulting in an imaginary conformal dimension. 
 
 Using~\eqref{oepp1} we now find from \eqref{roep12} the leading small frequency behavior
\be
G_R(\om, k) = \mu_*^{2\nu_U}\frac{b_+^{(0)} + b_-^{(0)}c(\nu_k) \le(\frac{\om}{\mu_*}\ri)^{-2i\lam_k}}{a_+^{(0)} + a_-^{(0)}c(\nu_k) \le(\frac{\om}{\mu_*}\ri)^{-2i\lam_k}} \ . \label{oscGr1}
\ee
As discussed in Appendix~\ref{app:master},  for imaginary $\nu_k$, $a_{\pm}^{(0)}, b_{\pm}^{(0)}$  also become complex and in fact 
\be
a_{+}^{(0)} = (a_-^{(0)})^* \qquad b_+^{(0)} = (b_-^{(0)})^* \ .
\ee
As a result, {\it even at $\om = 0$} we find that the spectral weight is non-vanishing for $k < k_o$
\be \label{specG}
\Im G_R(\om = 0, k) = \mu_*^{2\nu_U} \Im \frac{b_+^{(0)}}{a_+^{(0)}} \neq 0   \ .% \= h(k) \sin (\th_b (k) - \th_a (k))
\ee
We stress that this result -- a finite spectral density at precisely zero frequency -- is somewhat unusual, and its interpretation will be discussed further in Section \ref{sec:HFL}. At finite frequency~\eqref{oscGr1} is oscillatory; it is periodic in $\log\om$ with a period given by $\tau_k = \frac{\pi}{\lam_k}$. In particular, it is invariant under a {\it discrete} scale transformation
\be
\om \to e^{n\tau_k} \om, \qquad n \in \mathbb{Z} \  . \label{discSc} 
\ee
We will refer to the region $k < k_o$ as the oscillatory region. 

It can be readily checked that \eqref{oscGr1} has an infinite number of exponentially separated poles in the complex $\om$-half-plane; as \eqref{discSc} shows, there is an accumulation point at $\om = 0$. 
Again using~\eqref{s2} and~\eqref{f2} one finds that for a spinor all the poles lie in the {\it lower} half $\om$-plane while for a scalar they all lie in the {\it upper} half $\om$-plane~\cite{Faulkner09}. % {\bf (see Appendix~\ref{app:poles} for details).}
Thus we immediately conclude that for a scalar, in the parameter region~\eqref{Bcon1}, there is an instability to the condensation of the scalar operator. 
While there appears no obvious pathology for a spinor operator at the current level of discussion, as we will see below and in Sec.~\ref{sec:nfl}, there is nevertheless an instability, and at an exponentially small (in terms of $N$) scale, the system becomes a Fermi liquid.

It is also instructive to understand the above phenomena for both fermions and bosons in terms of 
bulk language. 
 For this purpose it is convenient to slightly rewrite the AdS$_2 \times \RR^{d-1}$ metric~\eqref{ads2M}  by introducing 
\be
\zeta = R_2 e^{y}
\ee
after which~\eqref{ads2M} becomes
\be  \label{nas2m}
ds^2 =  - e^{2y} dt^2 + R_2^2 dy^2 +\mu_*^2 R^2 d\vec{x}^2, \qquad A = {e_d \ov R_2} e^{-y} dt, \qquad
%\mu_L = {e_d \ov R_2}
 \ .
\ee
We note that the spatial part of~\eqref{nas2m} is flat and the chemical potential $\mu_L$ for a local observer with charge $q$  in the bulk is given by 
\be \label{lofc}
\mu_{L} = q \sqrt{g^{tt}} A_t = {|q| e_d \ov R_2} = {g_F |q| \ov \sqrt{2} R}  \ 
\ee
where in the last equality we have used~\eqref{pp2}. Note that $\mu_L$ does not depend 
on the radial direction $y$ of the AdS$_2$. 

For a fermion, in the Thomas-Fermi approximation, we fill all the states up to $\mu_L$ which is the Fermi energy. This gives the local Fermi momentum 
\be \label{ferm1}
k_b^{(F)} = \sqrt{{g_F^2 q^2 \ov  2 R^2} - m^2}  
\ee 
where the local bulk momentum $k_b$ is related to $k$ of the boundary and radial momentum $k_y$ 
conjugate to $y$ as 
\be \label{yu1}
k_b^2 = {k^2 \ov \mu_*^2 R^2} + {k_y^2 \ov R_2^2} \ .
\ee
Note that the bulk Fermi surface defined by~\eqref{ferm1} is one dimensional higher than a momentum 
shell in the boundary theory. Thus projecting it to the boundary we conclude there should be gapless excitations for all $k$ satisfying 
\be 
k < \mu_* \sqrt{{g_F^2 q^2 \ov  2 } - m^2R^2} \ .
\ee
See Fig.~\ref{fig:bal}. Note that the right hand side of the above equation is precisely $k_o$ introduced in~\eqref{Fcon1}. This explains the nonvanishing of spectral weight in~\eqref{specG} as a consequence of the projection of a bulk Fermi surface to the boundary theory. The oscillatory behavior in~\eqref{oscGr1} can similarly be understood in terms of the bulk Fermi surface when taking into account of that the local proper energy is red-shifted by a factor $e^{-y}$ compared with the boundary frequency $\om$ (see~\eqref{nas2m}) and  the behavior of local wave function. 

\begin{figure}[h]
\begin{center}
\includegraphics[scale=0.5]{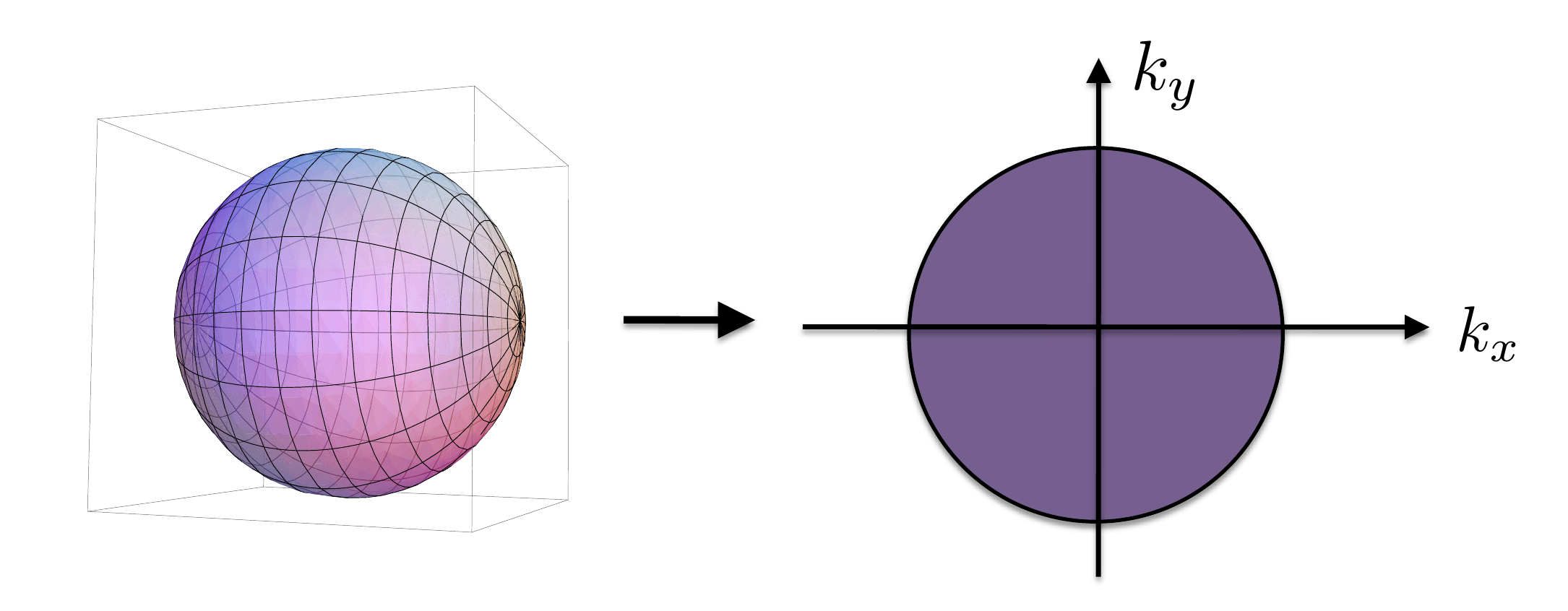}
\end{center}
\caption{In the AdS$_2 \times \RR^{2}$ (take $d=3$ for example) region of the extremal black hole geometry,  at each point in the bulk there is a local three-dimensional Fermi surface with Fermi momentum $k_o$, which upon projection to the boundary theory would result in a Fermi disc, in which there are gapless excitations at each point in the {\it interior} of a disc in the two-dimensional momentum space. }
\label{fig:bal}
\end{figure}

For a scalar, when the local chemical potential~\eqref{lofc} exceeds the mass of a charged particle 
\be \label{nebe}
{g_F |q| \ov \sqrt{2} R} > m 
\ee
the system will simply Bose condense. Equation~\eqref{nebe} agrees with condition~\eqref{Bcon1} up to an additive constant\nicefootnote[1]{Note that for our local approximation to work well, we need the typical local proper wave length to be much smaller than the curvature radius $R$, which translates into $g_F q , m R \gg 1$.} and gives the bulk origin of the scalar instabilities we found earlier from the poles of boundary retarded functions. 

For a charged operator, the above stories can also be understood more heuristically from the black hole geometry as follows. When~\eqref{Bcon1} and~\eqref{Fcon1} are satisfied one can show that the corresponding quanta for the bulk scalar (or spinor) field can be pair produced by the electric field of the black hole~\cite{Pioline:2005pf}.  Suppose that the black hole is positively charged, then 
the negative-charged particle in a pair will fall into the black hole, while the positive-charged one moves to the boundary of the spacetime. However it cannot escape, as the curvature of AdS pulls all matter towards its center, and thus the particle will eventually fall back towards the black hole. It then has some probability of falling into the black hole or being scattered
back toward the boundary. For a scalar it turns out the so-called superradiace is happening, %(see Appendix~\ref{app:poles}), 
i.e. for a wave incident on the black hole, the reflection amplitude becomes greater than 
the incoming amplitude. Thus this pair production process will be magnified and leads to an instability. 
In contrast, for a spinor, there is no superradiance, and this process will eventually reach an equilibrium with a positively charged gas of fermionic quanta hovering outside the horizon.
 The ground state of this fermionic gas is then described by a bulk Fermi surface described earlier. 

For a charged fermionic field, the induced fermionic density scales as $O(N^0)$ in the large $N$ limit as the pair production is a one-loop process in the bulk. Since the charge density carried by the black hole, given in~\eqref{entropy}, scales as $O(N^2)$, naively one might conclude that the fermionic backreaction can be ignored at leading order. However, as we will discuss in more detail in Sec.~\ref{sec:HFL}, the induced local bulk density is independent of radial coordinate $y$, the integration of which over $y$ then gives an infinite answer, signaling an instability~\cite{Hartnoll:2009ns}. In Sec.~\ref{sec:HFL} we show that after taking into account of the fermionic backreaction, \Slql\ becomes a Fermi liquid at 
very low energies.

\subsubsection{Summary}

To summarize, we find two channels for scalar instabilities. The first manifests as the presence of black hole hair, while the second (when the \Slql\ dimension becomes complex) corresponds to the bulk charged scalar condensing near the horizon of a black hole via Bose-Einstein condensation. 

For a spinor, a black hole hair gives rise to an isolated boundary Fermi surface, while 
when the \Slql\ dimension becomes complex, one finds a bulk Fermi surface in the near horizon region and the system settles into a Fermi liquid at very low energies. 

The differences between the scalar and spinor do not depend on any specific details and only on their statistics. 

\subsection{\Slql\ as a universal intermediate-energy phase} \label{sec:inter}

As discussed after~\eqref{bhga2}, the charged black hole~\eqref{RNmetric} provides a ``universal'' geometric description for many different theories at a chemical potential.  Thus the \Slql\ phase appears as a universal IR fixed point among a large class of field theories at finite density, independent of their microscopic details. 
However, from equation~\eqref{entropy}, the extremal charged black hole has a finite zero-temperature entropy density.  This implies that the ground states are highly degenerate. In a system without supersymmetry (as in this case\footnote{Independent of the supersymmetry of the UV conformal field theory, the presence of the chemical potential is expected to break it.}) or other apparent symmetries to protect such degeneracies, we expect this nonzero entropy density has to with the large $N$ limit we are working with. In other words, this nonzero zero-temperature entropy density likely reflects the existence of a large number of closely spaced states which are separated from the  genuine ground state by spacings which go to zero in the $N \to \infty$ limit. One expects that the system should pick a unique ground state, which may or may not be visible in the large $N$ limit. Given that there are many nearly degenerate low energy states, the precise physical nature of the ground state should be sensitive to the specific dynamics of the individual system. In a system with supersymmetry, then ground state degeneracy is indeed possible, and the above argument does not apply. There are many such examples in string theory.

There is also another independent argument~\cite{Jensen:2011su} based on density of states, which appears to indicate that any sort of scaling symmetry in the time direction only cannot persist to arbitrarily low energies. As we discuss in detail in Appendix~\ref{app:dos}, however, there are various subtleties with the argument, which makes it hard to draw a general conclusion from it.

Depending on the spectrum of charged or neutral matter fields  (their charges and masses etc.), it is known that an extremal charged black hole suffers from various bosonic and fermionic instabilities~\cite{Gubser:2008px,Hartnoll:2008vx,Hartnoll:2008kx,Iqbal:2010eh,Hartnoll:2009ns,Nakamura:2009tf}, which will lead to a different lower energy state. Some of the instabilities have already been mentioned earlier and we will discuss them in more details in next two sections.  The Einstein-Maxwell system~\eqref{maxact} can also be generalized to an Einstein-Maxwell-dilaton system in which the charged black hole is no longer a solution. Instead, one finds a solution with zero entropy density and Lifshitz-type scaling in the 
interior (at $T=0$)~\cite{Gubser:2009qt,Goldstein:2009cv,Goldstein:2010aw,Charmousis:2010zz}. In all of these situations, there exist parameter ranges in which a charged black hole provides a good description for some region of the bulk geometry. 

It then appears that among a large class of systems with different microscopics {\it and} different lower energy states, the charged black hole appears as an intermediate energy state. Thus the \Slql\ phase  may be considered a {\it universal intermediate energy phase}~\cite{Iqbal:2011in}  which connects microscopic interactions with  macroscopic, low energy physics, as indicated in Fig.~\ref{fig:phase}.\nicefootnote[1]{The idea that an extremal black hole should be interpreted as an intermediate-energy state has been expressed by many people including the references \cite{Faulkner09,Sachdev:2010um,yamamoto,mac,Faulkner:2011tm}.}  It is characterized by an $SL(2,R)$ symmetry including scaling in the {\it time} direction, but a finite {\it spatial} correlation length, %but 
%an infinite correlation {\it time} and associated nontrivial scaling behavior~\eqref{iiRc} and~\eqref{finiteTch1} in the time direction, 
as well as a nonzero entropy density.

We should emphasize that for a given holographic system there may not always exist an energy range over the \Slql\ manifests itself as an intermediate state. \Slql\ behavior is manifest when there exists a hierarchy between the chemical potential $\mu$ and the energy scale at which a more stable lower energy phase  emerges. On the gravity side there is then an intermediate region of the bulk spacetime which resembles that of AdS$_2 \times \RR^{d-1}$ (or its finite temperature generalization). 
 In a situation where such a hierarchy does not exist, as we will discuss later, the \Slql\ can 
nevertheless provides a useful description for understanding the emergence of the lower energy state.

\begin{figure}[h]
\begin{center}
\includegraphics[scale=0.4]{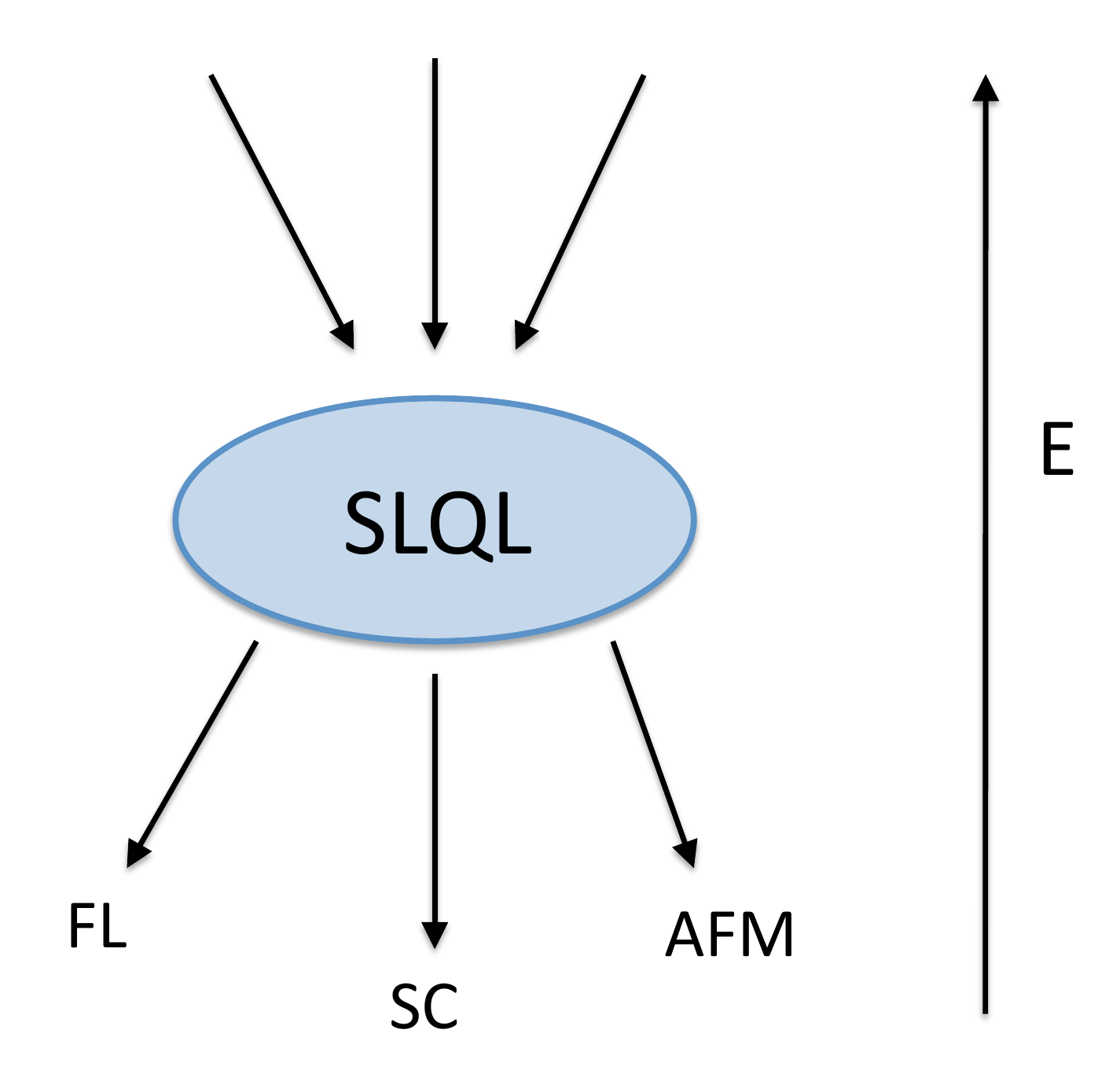}
\end{center}
\caption{The semi-local quantum liquid phase is a useful description at intermediate scales; many different UV theories (i.e. those with gravity duals described by a Maxwell-Einstein sector) flow to it, and at low energies it settles into one of many different ground states, e.g. a Fermi liquid, a superconducting state, or an antiferromagnetic state, as will be discussed in subsequent sections.}
\label{fig:phase}
\end{figure}

As emphasized  by Anderson some time ago~\cite{anderson}, the existence of a ``universal intermediate phase'' appears to be a generic phenomenon in nature; the familiar examples include liquid phases of ordinary matter, through which materials settle into different crystal structures at low temperatures. 
It is thus tempting to speculate that the appearance of a \Slql\ phase in holographic systems at a finite chemical potential may not be specific to these systems and may have wider applications. We will discuss more of these possible applications in the conclusion section. 

% although the large $N$ limit does  magnify the universality by pushing the low energy boundary of the \Slql\ phase to zero temperature.

In the next two sections we will see examples of the evolution of the \Slql\ to various lower energy phases including Fermi liquids,  superconductors, and  AFM-like states. 
We find that the emergence of these lower energy phases can be characterized as a consequence of  bound states formation in the \Slql. Conversely, \Slql\ can be considered as 
a universal deconfined (fractionalized) phase of these lower energy phases.

\section{Holographic (non-)Fermi liquids} \label{sec:nfl}

In this section we consider more explicitly the case when the holographic system has a Fermi surface and study the properties of small excitation around the Fermi surface. We will see that with a choice of parameters, the IR fixed point \Slql\ discussed in Sec.~\ref{sec:slql}  gives rise to non-Fermi liquid behavior which is the same as that of the strange metal phase of cuprate superconductors discussed in Sec.~\ref{fermiliquid}. However, when $N$ is finite, we find that for most of the parameter range, the SLQL does not persist to arbitrarily small energies. Instead, at sufficiently small energies, the \Slql\ instead goes over to a Fermi liquid of heavy fermions through formation of fermionic bound states, where the scale at which this happens is exponentially small in $N$. 

\subsection{Setup} \label{sec:setupF}

Consider a bulk Dirac spinor field $\psi(z,t, \vec x)$ dual to a  boundary theory spinor  operator $\sO(t, \vec x)$.  For definiteness we will now specify to $d=3$. 
We will assume a standard Dirac action in the bulk
\be
S = -\int d^{4}x \sqrt{-g}\;i (\bpsi \Gamma^M \sD_M \psi - m \bpsi \psi), \label{diracac}
\ee
where the covariant derivative $\sD_M$ contains couplings to both the background gauge field and the spin connection of the charged black hole geometry~\eqref{RNmetric},
\be
\sD_M = \p_M + \frac{1}{4} \om_{abM} \Ga^{ab} - i q A_M \ .
\ee
Here $q$ is the charge of $\sO$ under the $U(1)$ current $J^{\mu}$. The asymptotic structure of solutions to \eqref{diracac} is different from the scalar case discussed above, essentially because it is a first order system with more components. In particular, the number of components of the boundary operator $\sO$ is half of that of the bulk spinor $\psi$~\cite{Iqbal:2009fd}, i.e. while $\psi$ has four components, $\sO$ has only two components, as expected for a three-dimensional field theory.  The scaling dimension $\Delta$ of $\sO$ is related to the bulk mass $m$ of $\psi$ by
\be \label{bdimF}
\Delta = \frac{3}{2} + mR  \  .
\ee
We refer readers to~\cite{Iqbal:2009fd,Liu09,Faulkner09} for detailed analysis of the Dirac equation and only mention the main results here.
The boundary theory Green's function can be diagonalized to take the form
\be
G_R(\om, k) = \left(\begin{array}{cc}G_1(\om, k) & 0 \\0 & G_2(\om,k)\end{array}\right),
\ee
where {\it the momentum $k$ can be taken to be along one of the spatial directions} and both $G_{\al = 1,2}$ are the ratio of appropriately defined boundary theory expansion coefficients $A_{\al}, B_{\al}$, as in \eqref{GRdef}
\be
G_{\al} = \frac{B_{\al}}{A_{\al}}.
\ee
The master formula \eqref{roep1} applies to both of the eigenvalues $G_{1,2}$, which will have different values for the prefactor in the AdS$_2$ Green's function and the UV expansion coefficients $a_{\pm}, b_{\pm}$. A useful relation is $G_1(\om, k) = G_2(\om, -k)$; this means that  with no loss of generality we can simply restrict to $G_2$; when we refer to $G_R$ in the rest of this section this is what we mean. For convenience we reproduce 
here~\eqref{roep1}, 
\be
G_R (\om, \vk) = \mu_*^{\nu_U} {b_+(\om, k)+ b_- (\om, k) \sG_k (\om) \mu_*^{-2 \nu_k} \ov
 a_+ (\om, k) + a_- (\om, k) \sG_k (\om) \mu_*^{-2 \nu_k}}, \label{repmast}
 \ee
where $\sG_k (\om)$ is given by~\eqref{iiFc}--\eqref{ffcc} and all of the UV coefficients $a_{\pm}, b_{\pm}$ in \eqref{repmast} have analytic expansion in $\om$:
\be
a_{+}(\om, k) = a_{+}^{(0)}(k) + a_+^{(1)}(k) \om + a_+^{(2)}(k) \om^2 + \dots  \ .
\ee

When $mR \in [0,\ha)$ there is also an alternative quantization for the bulk spinor $\psi$ which leads to the conformal dimension of the boundary operator $\sO$ to be 
\be 
\De = {3 \ov 2} - mR  \ .
\ee
Below we will treat both quantizations together by allowing $mR$ to take a negative value 
in~\eqref{bdimF}.

For simplicity below we will also set $\mu_*$ to unity to avoid cluttering formulas. It can be easily restored on dimensional ground. 

%is now given by that a fermion which differs from the scalar one in the prefactor (see~\eqref{ferM}), and indeed its precise structure {\it is} important for guaranteeing that the poles in the fermionic retarded correlator are always in the lower half plane, as shown in Appendix \ref{app:poles}. 
%We now have all the ingredients we need to search for Fermi surfaces.

 %a Fermi surface if 

 %As discussed in Sec.~\ref{sec:NFL}, a Fermi surface is characterized by 
%low-energy gapless excitations around a whole shell in momentum space and we should see its signature in {\it non-analytic behavior in $G_R(\om, k)$ of certain fermionic operator at a shell in momentum space $k = k_F$}. 
%This motivates the following approach: we will perform ``photoemission'' thought experiments on our black hole;  we will take a fermionic operator $\Psi$ in the field theory and calculate its retarded response function $G_R(\om, k)$ using the techniques developed in last section.

\subsection{Existence of Fermi surfaces} \label{sec:ipp}

As we discussed in Sec.~\ref{sec:hair}, a boundary system has Fermi surfaces 
for a spinor operator $\sO$, if there exists
some $k_F$, $a_+^{(0)} (k_F) =0$, i.e. the charged black hole admits an inhomogeneous fermionic hair.
 To see whether such a $k_F$ indeed exists one needs to solve the static Dirac equation (i.e. at  $\om =0$) for bulk spinor field $\psi$ in the full charged black hole geometry, which can only be done numerically.  Squaring the static Dirac equation and rewriting it as a Schr\"{o}dinger equation one can then reduce the problem of finding a Fermi surface to finding a bound state in the resulting potential as a function of $k$. It can then be checked explicitly that such a $k_F$ does exist for certain range of parameters $(q,m)$ 
and when $q$ is sufficiently large there can also be multiple bound states which corresponds to the system having multiple Fermi surfaces.  In Fig.~\ref{fig:qk} we show the values of $k_F$ found numerically for a few choices of $mR$, taken from~\cite{Faulkner09}.  In the figure, the oscillatory region $k < k_o$ (see~\eqref{Fcon1}), where  $\nu_k$ becomes pure imaginary  and 
$G_R$ becomes oscillatory in $\log \om$ (see~\eqref{oscGr1}), is shaded. As discussed in Sec.~\ref{sec:osc}, 
 the oscillatory region can be attributed to a {\it bulk} Fermi surface in the near horizon AdS$_2$ region, which gives rise to a ``Fermi ball'' in the boundary. In contrast, the Fermi surface at $k_F$ from a fermionic hair is isolated. 
 
For standard quantization, i.e. $m>0$, the parameter region in which a $k_F$ exists always lie inside the 
parameter region~\eqref{Fcon1} for which the oscillatory region exists. But note that 
$k_F$ is always greater than $k_o$, i.e. lying outside the oscillatory region in momentum space. 
For alternative quantization, there exists a parameter range in which we have a single isolated $k_F$ even in the absence of an oscillatory region. 

\begin{figure}[h!]
 \begin{center}
\includegraphics[scale=0.38]{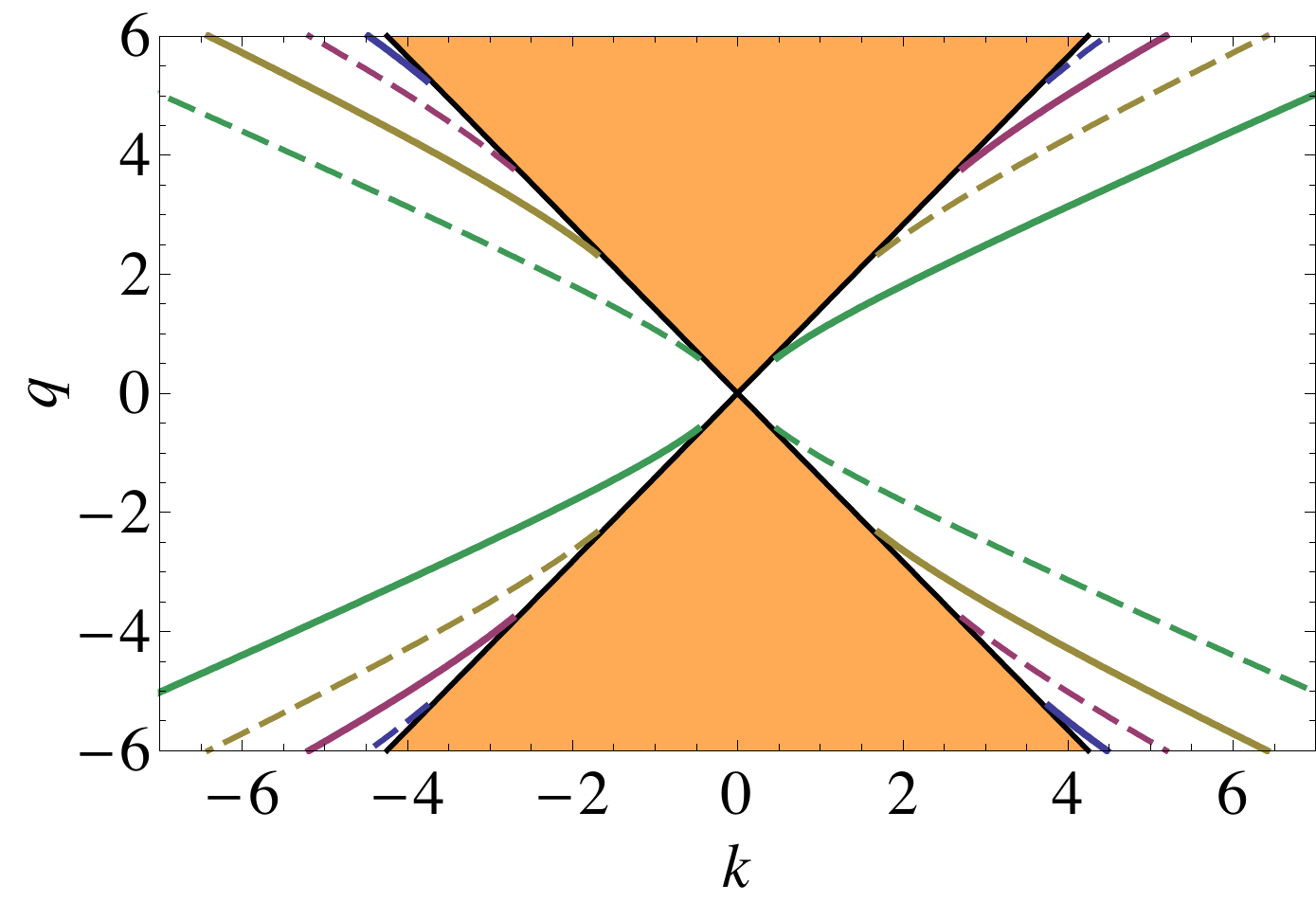} \qquad \quad
\includegraphics[scale=0.38]{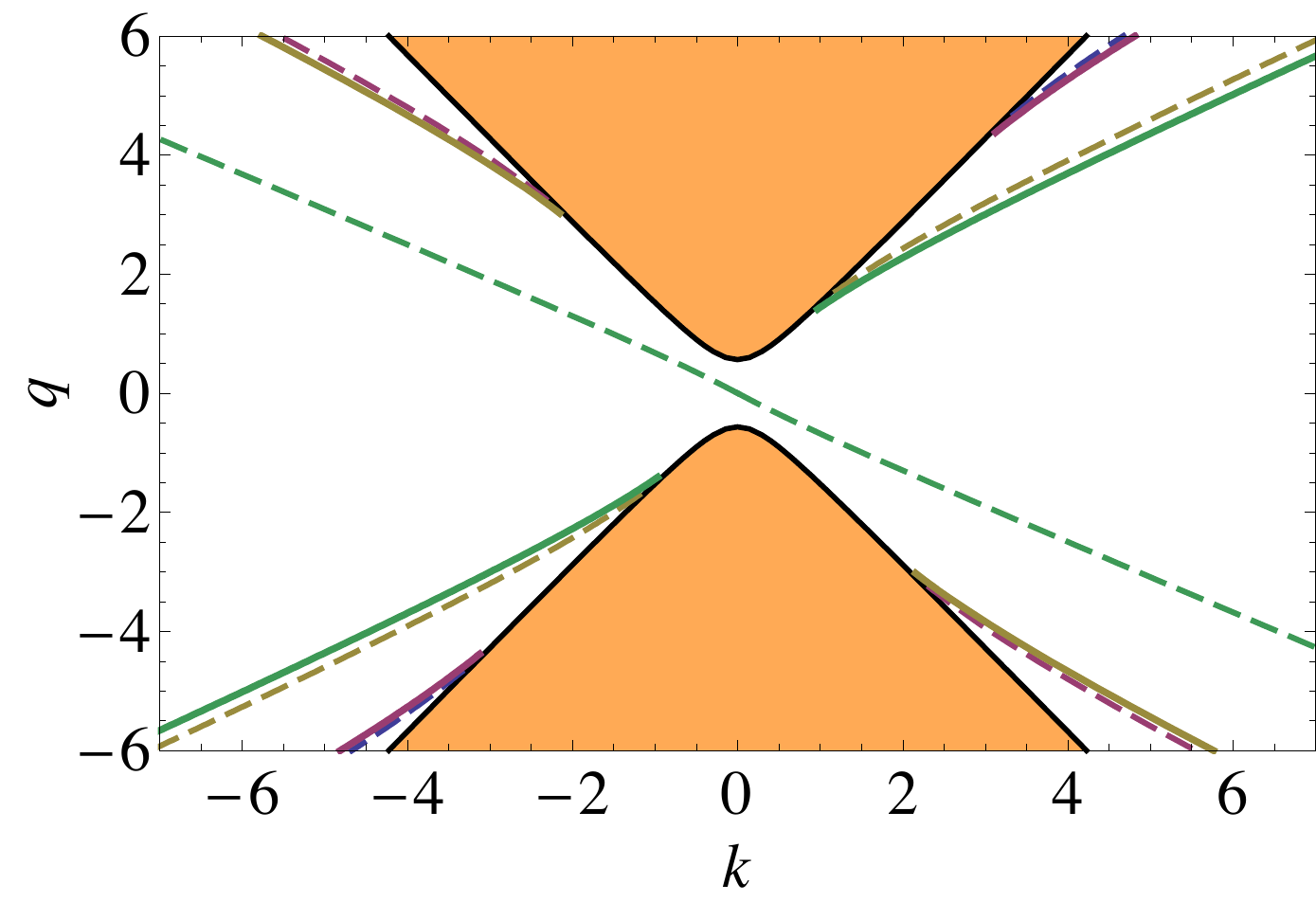}
\caption{\label{fig:qk} The values of $k_F$ as a function of $q$ for the Green function $G_{2}$ are shown by solid lines for $mR =  0, \, 0.4$. In this plot and ones below we use units where $R=1$, $\mu_* =1$, $g_F = 1$ and $d=3$.  The oscillatory region, where $ \nu_k =
{1 \ov \sqrt{6}} \sqrt{k^2 + m^2 -{q^2 \ov 2}}$ is imaginary, is shaded.
Given $G_1 (k) = G_2 (-k)$, so $k_F$ for $G_1$ can be read from these plots by reflection through the vertical $k=0$ axis. 
%The $mR=-0.4$ plot corresponds to alternative quantization for $mR =0.4$. % following from equation~\eqref{altenQ}. 
For convenience we have
included in each plot the values of $k_F$ for the alternative quantization using
the dotted lines. Thus the dotted lines for the second plot ($mR=0.4$) %and the third plot 
correspond to $mR=-0.4$. % in fact contain
%identical information; they are related by taking $k \to -k$ and exchanging dotted and solid lines. 
Also for $m=0$ the alternative quantization is equivalent to the original one. This is reflected in the first plot
in the fact that the dotted lines and solid lines are completely symmetric. Both plots are symmetric with respect to $q , k \to -q , -k$. % as a result of equation~\eqref{jke}.
}
\end{center}
\end{figure}

\begin{figure}[h]
\begin{center}
\includegraphics[scale=0.4]{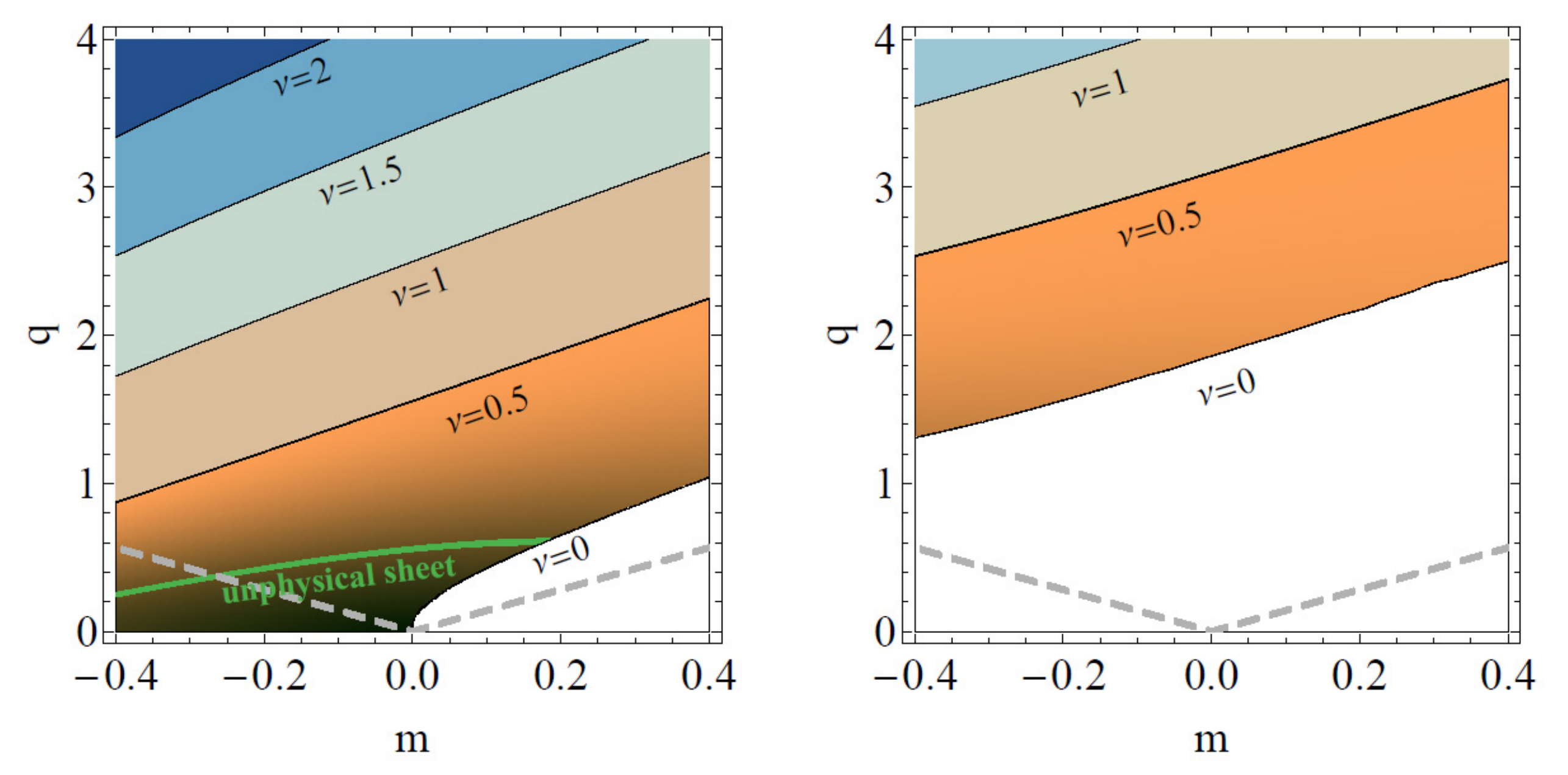}
\end{center}
\vskip -1.0cm
\caption{Distribution of $\nu_{k_F}$ in the $m-q$ plane. When there are multiple Fermi surfaces, $\nu_{k_F}$ is calculated for the one with the largest $k_F$. The two plots correspond to $G_2$ and $G_1$  components. In the white
region, there is no Fermi surface. There is no long-lived quasiparticle in the orange region, $\nu_{k_F} < \ha$. In the
remainder of the parameter space, there are log-lived quasiparticles. Above the dashed lines there is an oscillatory region.}
\label{phase}
\end{figure}

\subsection{Fermi surfaces with or without long-lived quasiparticles} \label{sec:fff}

Near a $k_F$ and at low frequencies, the retarded function $G_R$ is given by~\eqref{FSexp1}--\eqref{selfM1}, which we also reproduce here for convenience
\be
G_R(\om, k) \approx \frac{h_1}{k-k_F - \frac{1}{v_F}\om - \Sigma(\om)} , \qquad
\Sigma(\om) = h \sG_{k_F}(\om) = h c(\nu_{k_F}) \om^{2 \nu_{k_F}}
\label{FSexp}
\ee
with 
 \be
 v_F \equiv -\frac{\p_k a_+^{(0)} (k_F)}{a_+^{(1)} ((k_F)}, \qquad
 h_1 \equiv \frac{b_+^{(0)}(k_F)}{\p_k a_+^{(0)}(k_F)}, \qquad h \equiv -\frac{a_-^{(0)} (k_F)}{\p_k a_+^{(0)}(k_F)} \ .
 \ee
 The self-energy $\Sig$ depends on frequency only and is characterized by a nontrivial scaling exponent $\nu_{k_F}$, which can be computed from~\eqref{ferdim} once $k_F$ determined. The result is presented in fig.~\ref{phase}. 

  Poles of~\eqref{FSexp} in the complex $\om$-plane, which are determined by 
\be
k - k_F - \frac{1}{v_F}\om_c - \Sigma(\om_c) = 0 \label{dispeqn} \qquad \om_c(k) = \om_*(k) - i \Ga(k)  
\ee
give the dispersion of small excitations near the Fermi surface. The real part $\om_* (k)$ of $\om_c (k)$  gives the dispersion relation and the imaginary part $\Ga$ gives the width of an excitation. 
Given that in~\eqref{dispeqn} only $\Sig$ is complex, the width $\Ga$ is controlled by the imaginary part of $\Sig$ which is in turn is proportional to $\Im \sG_{k_F} (\om)$.
The physics turns out to be rather different depending on the value of $\nu_{k_F}$, and we now study the different values in turn. For simplicity of notation in the rest of this section we will drop the subscript $k_F$: $\nu \equiv \nu_{k_F}$ {\it in this section only.}

\subsubsection{$\nu > \ha$}

If $\nu > \ha$, then the analytic linear-in-$\om$ term in \eqref{FSexp} dominates over the self-energy, which goes like $\om^{2\nu}$. Thus the dispersion is essentially controlled by the balancing between the first two terms in \eqref{dispeqn}; note however that the sole contribution to the {\it imaginary} part of the pole comes from the complex self-energy. We find the following expression for the motion of the pole as $k$ is varied near $k_F$:
\be
\om_c(k) = v_F (k-k_F) - v_F h c(\nu)\le(\frac{v_F (k-k_F)}{\mu_*}\ri)^{2\nu} 
\ee
Thus we see that this has a linear dispersion relation $\om_* = v_F (k-k_F)$, with width $\Ga \sim \om^{2\nu}$ that is always much smaller than the energy $\om_*$: as we approach the pole $\frac{\Ga(k)}{\om_*(k)} \sim (k-k_F)^{2\nu -1} \to 0$ as $k-k_F \to 0$. Furthermore the residue at the pole is always finite and given by
\be
Z = -h_1 v_F\ .
\ee 
This is then a long-lived quasiparticle. It is not {\it qualitatively} different from the excitations of a Fermi liquid, except that its width satisfies $\Ga \sim \om^{2\nu}$ rather than $\Ga_{FL} \sim \om^2$.

% {\bf pictures for movement of pole in complex plane.} Finally, note that in this range the IR CFT operator corresponding to the fermion has dimension $\delta = \ha + \nu > 1$ and so the operator is {\it irrelevant} in the AdS$_2$ sense.

\subsubsection{$\nu < \ha$}

If $\nu < \ha$, the situation is different. Now in \eqref{FSexp} the non-analytic contribution from $\Sig(\om)$ always dominates over the analytic linear-in-$\om$ correction, which we may completely neglect. The solution to \eqref{dispeqn} now takes the form
\be
\om_c = \mu_* \le(\frac{k-k_F}{h c(\nu)}\ri)^{\frac{1}{2\nu}}
\ee
As $c(\nu)$ is complex this pole has both a real and imaginary part; however now they are both coming from the same place, i.e. the AdS$_2$ Green's function. As a result the dispersion is now nonlinear, $\om_* \sim (k-k_F)^{\frac{1}{2\nu}}$. Importantly, the width $\Ga$ is now always comparable to the frequency $\om_*$, e.g. for $k-k_F > 0$ we have
\be
\frac{\Ga(k)}{\om_*(k)} = - \tan\le(\frac{\arg(c(\nu))}{2\nu}\ri) = \mbox{const}
\ee
A similar expression exists with a different angle for $k-k_F < 0$. Thus as we vary $k-k_F$ through $0$, the pole moves along a straight line towards the origin, eventually bouncing off and retreating at another angle (but always in the lower-half plane). %for details see Appendix \ref{app:poles}). 
The residue at the pole is
\be
Z = -\frac{\om_c h_1}{2\nu (k-k_F)} \sim (k-k_F)^{\frac{1}{2\nu} - 1},
\ee
and so actually vanishes as we approach the Fermi surface $k-k_F = 0$. The fact that the width of the excitation is always the same  order as its energy and that the residue vanishes at the pole indicate that this can {\it not} considered as a quasiparticle; thus we see that we have an example of a system with a sharp Fermi surface, but with no quasiparticles! 

%Note that in this regime the IR CFT operator corresponding to the fermion has dimension $\delta = \ha + \nu < 1$, and so is {\it relevant} in the AdS$_2$ sense. 

\subsubsection{$\nu = \ha$: Marginal Fermi Liquid}

We turn now to the threshold case between the two studied above. Here we see that the two terms (analytic and AdS$_2$) are both naively proportional to $\om$ and are thus becoming degenerate. As is often the case when there is such a degeneracy, a logarithmic term in $\om$ is generated. Basically in~\eqref{FSexp}, $v_F $ has a simple zero and $c(\nu)$ in the self-energy $\Sig$ (see~\eqref{iiFc})
has a simple pole at $\nu = \ha$; the two divergences cancel and leave behind a finite $\om \log \om$ term with a real coefficient,
\be
G_R(\om, k) \approx \frac{h_1}{k-k_F + \tilde{c}_1 \om \log \om + c_1 \om}
\ee
Here $\tilde{c}_1$ is real and $c_1$ is complex; for more details on the derivation of this formula see Section VI A of \cite{Faulkner09}. Remarkably, this is of precisely the form postulated in \cite{Varma89} for the response function of the ``Marginal Fermi Liquid'' discussed earlier around~\eqref{marsp}. Here this response function emerges from a first-principles holographic calculation (albeit only at the specially tuned value $\nu = \ha$). Note that here the width is suppressed compared to its energy, but only logarithmically $\Gamma \sim \frac{\om_*}{\log (\om_*)}$. Similarly, the residue of the pole scales like $Z \sim 1/\log(k-k_F)$ and so also vanishes logarithmically. The potential physical applications of this $\nu = \ha$ point make it very interesting, although from our gravitational treatment so far we do not really have any way to single it out.

\subsubsection{Perspective from low energy effective theory}

The qualitative behavior discussed above, in particular, the presence (absence) of long-lived quasiparticles for $\nu > \ha$ ($\nu \leq \ha$) can also be seen from the low energy effective theory~\eqref{hybid} based on simple dimensional analysis~\cite{Faulkner:2010tq}. Since in the \Slql, $k$
does not scale~(i.e. has dimension $0$), from~\eqref{freefer} $\Psi$ should have scaling dimension $0$ in the IR. Thus the hybrization term $ \Psi \Phi$ in~\eqref{hybid} has dimension $\ha + \nu$ (which is the dimension of $\Phi$). As a result the hybridization coupling is relevant for $\nu < \ha$ and irrelevant for $\nu > \ha$. 
When the hybridization coupling is relevant, free fermion $\Psi$ is strongly mixed with $\Phi$, resulting a large decay width and a breakdown of the quasiparticle picture. When $\nu = \ha$, the coupling is marginal. Note that while the logarithmic suppression in the decay width was the original motivation for the word ``Marginal'' in the ``Marginal Fermi Liquid'' given in \cite{Varma89}; here we find it is due to a marginal hybridization coupling and the word ``Marginal''  is entirely appropriate. 

%\subsubsection{}

\subsection{Transport of holographic Fermi surfaces} \label{sec:transport}

Having understood the single-particle response functions of our system, we now turn to collective properties, such as transport. A very natural observable to compute is the conductivity; in particular, for Landau Fermi liquids this is essentially always controlled by the single-particle lifetime, and we find for the DC conductivity $\sig_{FL} \sim T^2$. What is $\sig(T)$ in our model?

\begin{figure}[h]
\begin{center}
\includegraphics[scale=0.4]{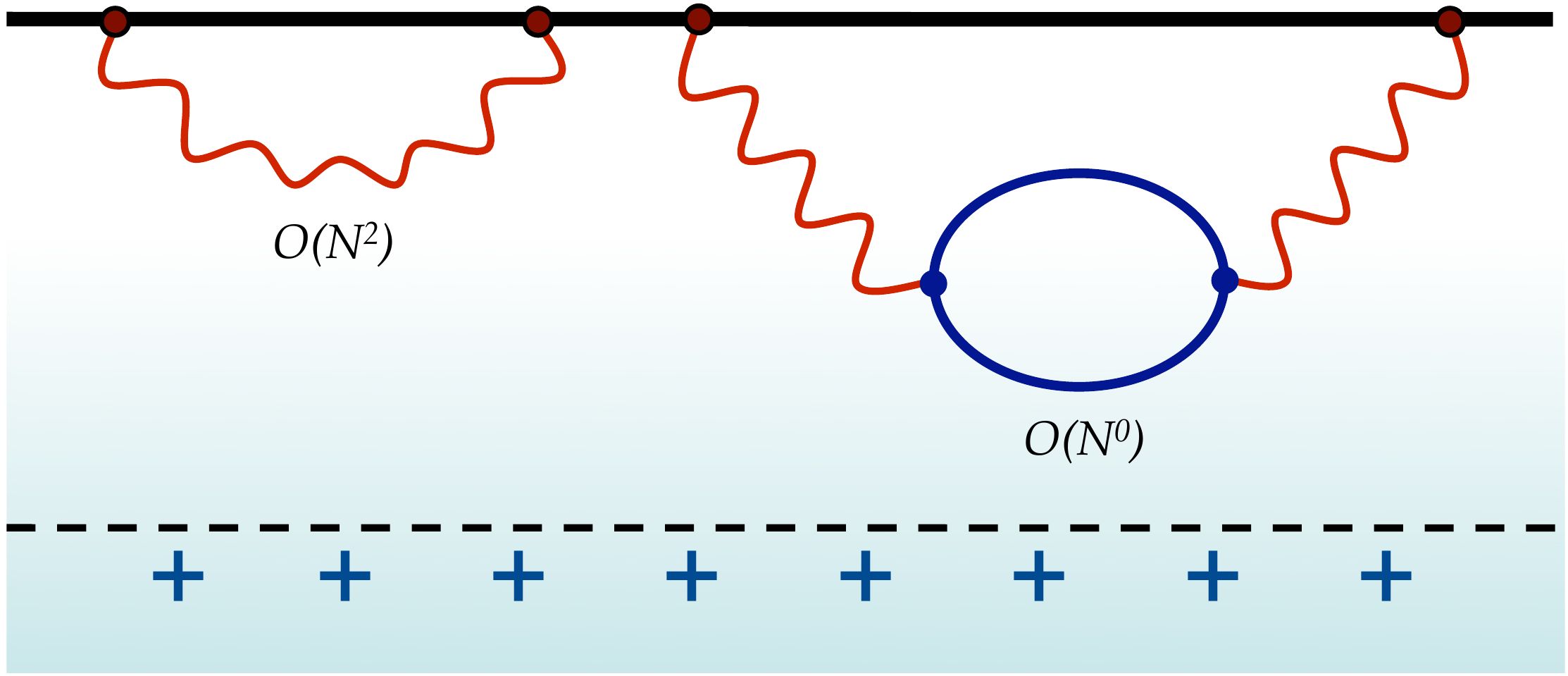}
\end{center}
\vskip -0.3cm
\caption{Conductivity from gravity. The current-current correlator in~\eqref{kubo} can be obtained from the propagator of the gauge field $A_x$ with endpoints on the boundary. Wavy lines correspond to gauge field propagators and the dark line denotes the bulk propagator for
the $\psi$ field. The left-diagram is the tree level $O(N^2)$ contribution while the right diagram 
includes the contribution from a loop of $\psi$ quanta. The contribution from the Fermi surface associated with boundary fermionic operator $\sO$ can be extracted from
the diagram on the right.}
\label{fig:bulkbdy}
\end{figure}

To understand the calculation to follow, it is necessary to understand the answer to the following question: how much of the charge of our system is carried by the Fermi surface that we have just identified? Note from \eqref{chargeD1} that the background charge density of our system scales like $\rho \sim \frac{1}{2\kappa^2}$, where $\frac{1}{\kappa^2}$ is the inverse Newton's constant that multiplies the whole bulk action. In known examples of the duality, this is typically proportional to a positive power of $N$, i.e. $N^2$ or $N^{\frac{3}{2}}$; for simplicity let us just assume the former and say that $\rho \sim O(N^2)$. But $k_F$ identified above has no factors of $\kappa^2$ in its definition; thus we conclude that $k_F \sim O(N^0)$. Since this is the Fermi surface for a singlet operator, it means that the charge density that is associated with our Fermi surface is {\it also} $O(N^0)$. Thus we conclude that {\it our Fermi surface makes up a tiny (i.e. large-$N$-suppressed) portion of the full charge density.} In our gravitational description, the rest of the charge is stored behind the black hole horizon; in field theory terms, this presumably means that it is stored in some other degrees of freedom about which we know very little except that in some sense there are ``$N^2$ of them''. 

This means that none of the leading thermodynamic or transport properties will be sensitive to the Fermi surface. We can directly see this from a bulk calculation. The conductivity can be found from the Kubo formula,
\be \label{kubo}
\sig(\om) = \frac{1}{i\om}\langle J_{x}(\om) J_{y}(-\om) \rangle _{\mathrm{retarded}}.
\ee 
We can find this retarded response function from gravity using the procedures described above; however if we do that it is clear that the leading answer will be of $O(N^2)$ and will depend only on classical fluctuations of the gauge field, which know only about the black hole and appear to know nothing about the Fermi surface. It is clear that to see the effects of the Fermi surfaces identified above, we have to go to higher order in $1/N^2$; this maps to quantum effects in the bulk. Thus we must perform a one-loop calculation on the gravity side, including a fermion loop in the gauge boson propagator as shown in Figure \ref{fig:bulkbdy}.

\begin{figure}[h]
\begin{center}
\includegraphics[scale=0.40]{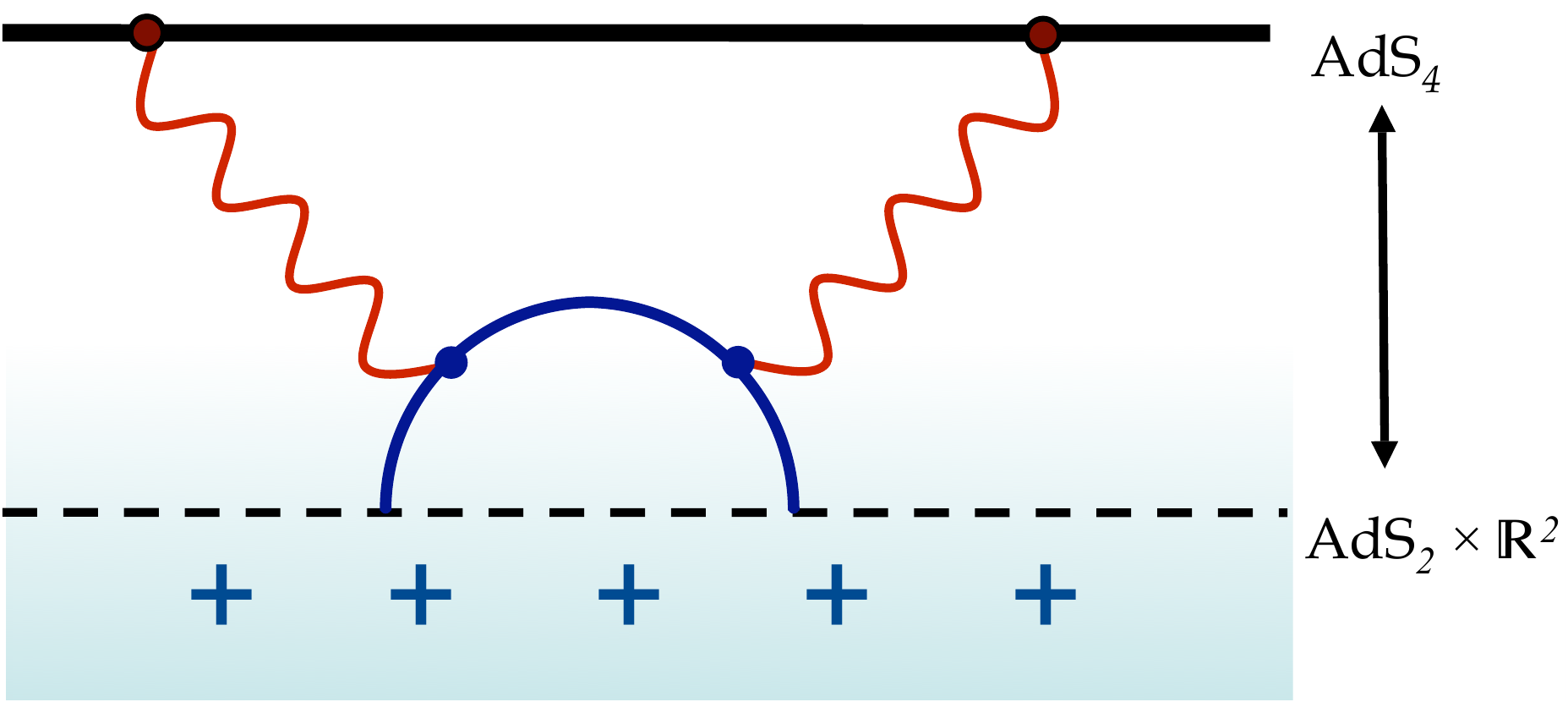}
\end{center}
\vskip -0.3cm
\caption{Possible contribution to conductivity in which fermion loop passes through the horizon. This represents the decay of the current due to current-carrying particles decaying by falling into the black hole, and is the dominant contribution to the DC conductivity.
}
\label{fig:fallin}
\end{figure}

\subsubsection{DC conductivity}

In many ways this calculation parallels that of Fermi liquid theory, except for the extra bulk dimension and complications from the curved geometry. A main difference comes from the fact that we are performing the computation in a black hole spacetime, meaning that processes such as Figure \ref{fig:fallin} where the fermion goes through the horizon can contribute. We do not provide any details of the computation here, only pointing out that the final result can be written in terms of boundary theory quantities as
\be
\sigma(\om) = \frac{C}{i\om} \int d\vec{k} \int \frac{d\om_1}{2\pi} \frac{d\om_2}{2\pi} \frac{f(\om_1) - f(\om_2)}{\om_1 - \om- \om_2 - i\ep} A(\om_1, k) \Lambda(\om_1, \om_2, \om, \vk) \Lambda(\om_2, \om_1, \om, \vk) A(\om_2, \vk) \label{condexp}
\ee

Here $A = \frac{1}{\pi} \Im G_R$ is the single-particle spectral function described in the last section, $f(\om) = \frac{1}{e^\frac{\om}{T} + 1}$ is the Fermi distribution function. All radial integrals have been absorbed into an effective vertex $\Lambda$, which takes the schematic form $\Lambda \sim \int dr\;\bpsi(r) K_{A}(r)\psi(r)$, where $\psi(r)$ is a fermion wavefunction and $K_A(r)$ is a bulk-boundary propagator for the gauge field (which also includes a contribution from the graviton, which mixes with it on this background). This expression can be interpreted in boundary theory terms, as shown in Figure \ref{fig:bubble}; the effect of the radial direction is to provide an exact expression for the renormalized vertex $\Lambda$. 

 \begin{figure}[h]
\begin{center}
\includegraphics[scale=0.35]{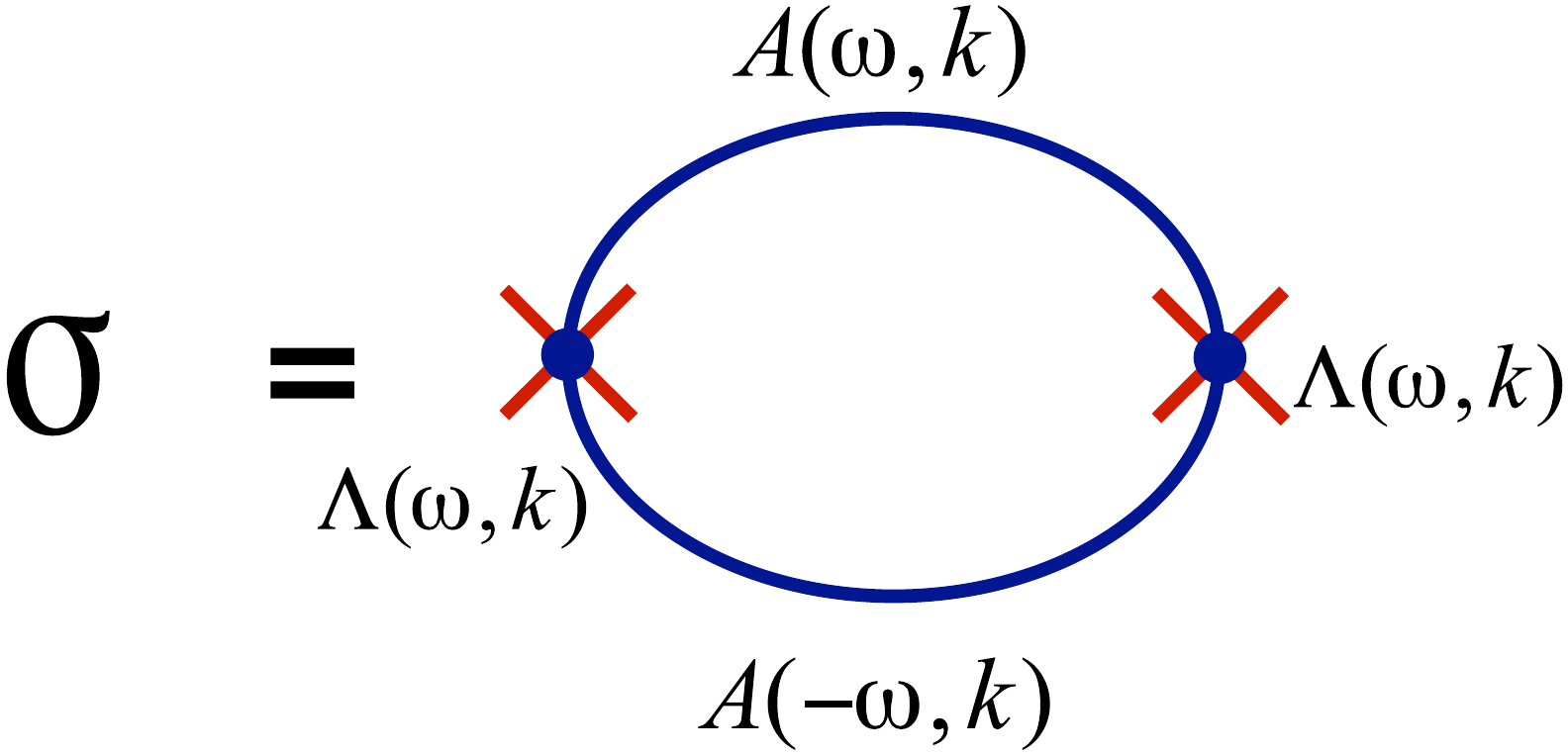}
\end{center}
\caption{Interpretation of \eqref{condexp} in field theory terms; the conductivity can be written as an expression similar to that in Fermi liquid theory, but with exact propagators and renormalized vertexes that can be found from gravity.}
\label{fig:bubble}
\end{figure}

In general $\Lambda$ is a complicated function of all its arguments, but in the low temperature limit and near the Fermi surface it becomes a smooth real function of $|\vec{k}|$, independent of $\om$ and $T$. In this case \eqref{condexp} shows that the conductivity is controlled by the single-particle spectral function near the Fermi surface, and we find the expression

\be
\sigma_{DC} = \al(q,m) T^{-2\nu_{k_F}}. \label{sigdc}
\ee

Here $\al(q,m)$ is a non-universal number that depends on a radial integral over the bulk fermion wavefunctions (and can be found numerically) but the exponent in $T$ is determined by the single-particle decay rate and so by the \Slql. In particular, note that for the Marginal Fermi Liquid case $\nu_{k_F} = \ha$, we find that the contribution to the resistivity from the Fermi surface is {\it linear}: $\rho \sim T$, as observed for the cuprates in the strange metal phase and many other non-Fermi liquid materials \cite{stewart}. 

Note that in general the electrical transport scattering rate -- which emphasizes large momentum transfer -- can be different from the single-particle scattering rate.  On the other hand in our system the exponent in \eqref{sigdc} {\it is} determined by the single-particle scattering rate. On the gravity side this arises from the fact that the dissipative portion of the correlation function is determined from Figure \ref{fig:fallin} essentially by how fast the bulk fermions fall into the black hole, which determines the single-particle decay rate and knows nothing of momentum.\footnote{Note that higher loop diagrams which could potentially spoil this equivalence are suppressed by higher powers of $1/N^2$.}  On the field theory side this is due to the fact that the \Slql\ sees only the time direction and depends very weakly on momentum. 

\subsubsection{Optical conductivity}

In the computation above the formula \eqref{sigdc} for the DC conductivity is rather insensitive to whether or not the Fermi surface has long-lived quasiparticles or not , i.e. the value of $\nu_{k_F}$. The optical conductivity $\sig(\om)$ can also be computed along similar lines; this turns out to depend critically on the value of $\nu_{k_F}$. Again, we present here only results. 

For $\nu_{k_F} > \ha$ there are two regimes. In the first we hold $\om T^{-2\nu_{k_F}}$ fixed in the low temperature limit and find then that we can approximate $\sig(\om)$ by a Drude form,
\be
\sig(\om) = \frac{\om_p^2}{\tau^{-1} - i\om} \qquad \tau^{-1} = 2v_F \Im \Sigma(\om = 0) \sim T^{2\nu_{k_F}} \label{condqp}
\ee
with $\om_p^2$ an overall constant. This is in fact the standard answer for a weakly coupled theory with well-defined quasiparticles, whose lifetime is given by $\tau$. This should not be surprising; in some sense the effect of holography (in this regime) is to provide the nontrivial lifetime $\tau$ by coupling these quasiparticles to a quantum critical bath. In the other regime $\om \gg T$ (but still $\om \ll \mu$) we find
\be
\sig(\om) = i\frac{\om_p^2}{\om} + b(i\om)^{2\nu_{k_F} - 2},
\ee
with $b$ a real constant. The $1/\om$ term gives us a term proportional to $\delta(\om)$ with weight consistent with \eqref{condqp}. 

On the other hand, for $\nu_{k_F} < \ha$, when we have no long-lived quasiparticles, we find a scale-invariant form,
\be
\sigma(\om) = T^{-2\nu_{k_F}}F_1\le(\frac{\om}{T}\ri), \label{scalinvop}
\ee
where $F_1(x)$ is a scaling function that is constant as $x \to 0$. For large $x$ it falls off as $x^{-2\nu_{k_F}}$, giving us an optical conductivity $\sig(\om) \sim a(i\om)^{-2\nu_{k_F}}$ with $a$ a real constant. This behavior is consistent with a system with no scales and no quasiparticles. 

Finally, for the marginal case $\nu_{k_F} = \ha$, we find behavior similar to \eqref{scalinvop} but with logarithmic violations, i.e.
\be
\sig(\om) = T^{-1}F_2\le(\frac{\om}{T}, \log\frac{T}{\mu}\ri)
\ee
where again $F_2$ goes to a constant at small $\om$. For $\om \gg T$ in this case we find
\be
\sig(\om) \sim -C\frac{i}{\om}\le(\frac{1}{\log\frac{\om}{\mu}} + \frac{1}{(\log\frac{\om}{\mu})^2}\frac{1 + i\pi}{2}\ri) + ...
\ee

\subsection{\Slql\ as a fractionalized phase of Fermi liquids} \label{sec:HFL}

In this subsection we further explore the physics of the oscillatory region $k < k_o$ discussed in Sec.~\ref{sec:osc}. Such a region exists  when 
\be \label{newpr}
 {g_F^2 q^2 \ov 2} > m^2 R^2  
 \ee
 and it is shown in the shaded regions in Fig.~\ref{fig:qk}.  
 In the oscillatory region, the IR scaling dimension of a fermionic operator is complex and the spectral function is nonzero at $\om =0$. At a nonzero $\om$, the correlation function~\eqref{oscGr1} is periodic in $\log \om$ and has an infinite number of poles in the lower half $\om$-plane. 
  It can also be readily checked that all the poles  have comparable real and imaginary parts and thus cannot be considered as corresponding to quasiparticles. From the bulk point of view, this region  can be interpreted as corresponding to the projection of a bulk Fermi surface (which is one dimension higher) to the boundary.
 Small excitations at the bulk Fermi surface have a large decay rate since they can fall into the black hole without experiencing any potential barrier.
 
 We will now show that in 
 the parameter range of~\eqref{newpr}, at any finite $N$ the system in fact goes over to a Fermi liquid of ``heavy'' fermions at low energies. We will keep the boundary spacetime dimension $d$ general.

From~\eqref{ferm1}, the local charge density of fermions is given by
\be \label{lode}
\rho_L = D_d (k_b^{(F)})^{d}, \qquad D_d = q s {\Om_d \ov (2 \pi)^d}
\ee
where $\Om_d$ is the volume of a unit sphere in $d$-dimension and $s$ is the number of possible spin degeneracies. $\rho_L$ in~\eqref{lode} is finite and scales as $O(N^0)$ in the large $N$ limit\nicefootnote[1]{In contrast the charge density carried by the black hole, given in~\eqref{entropy} scales as $O(N^2)$.}
which naively implies that the fermionic backreaction can be ignored at leading order. 
However, the contribution to the boundary charge density from $\rho_L$  is obtained from integrating it over the radial direction of the black hole, and 
since in the near horizon region~\eqref{ads2M} the proper radial distance is infinite and the local proper volume of the transverse $\RR^{d-1}$ is constant, the total boundary density coming from integrating~\eqref{lode} over the radial direction is in fact infinite~\cite{Hartnoll:2009ns}. 

\begin{figure}[h]
\begin{center}
\includegraphics[scale=0.5]{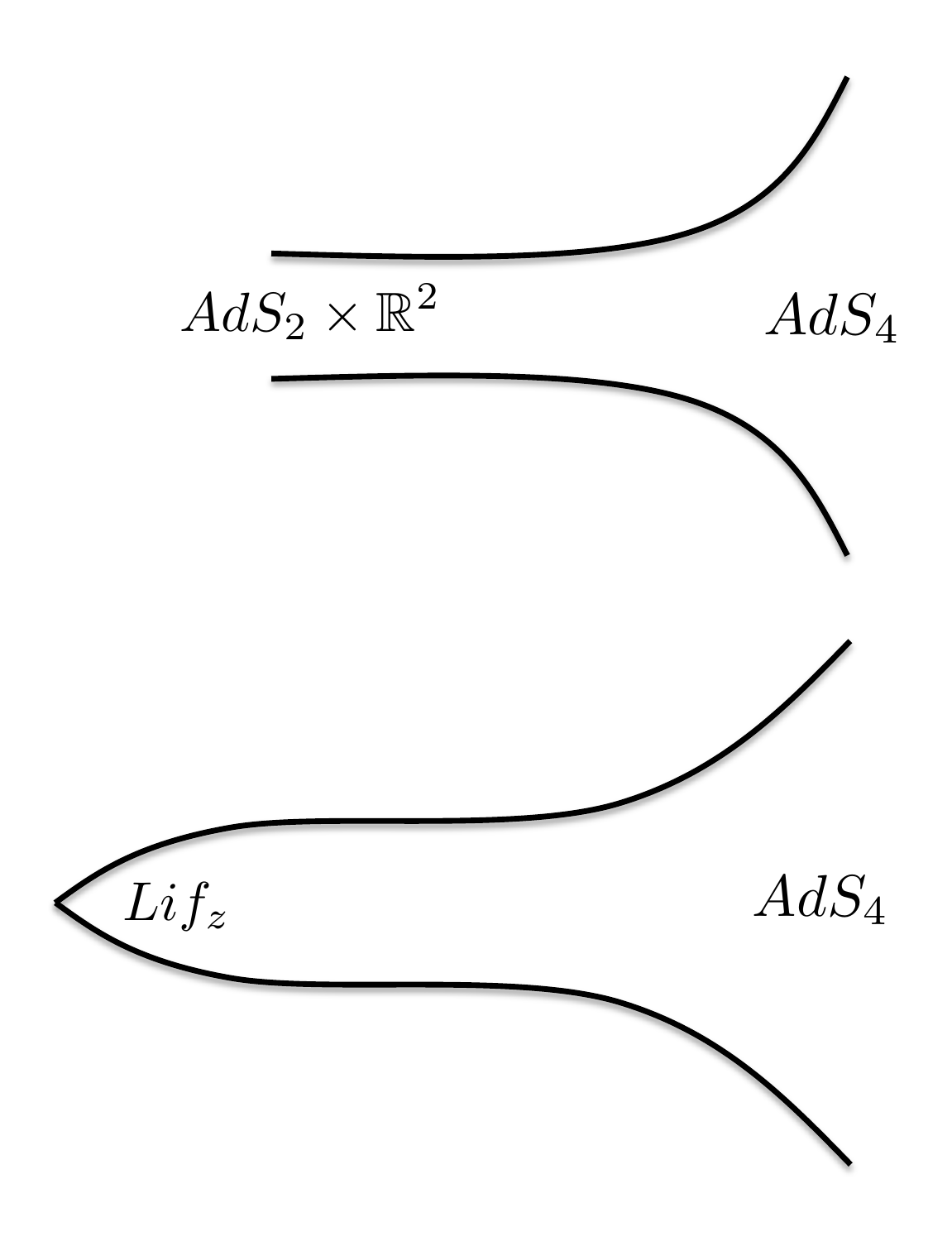}
\end{center}
\caption{Two different geometries (with $d=3$): on the top, the AdS$_2 \times \mathbb{R}^2$ describing the SLQL phase; on the bottom, its resolution into a Lifshitz geometry with a finite $z$ given by \eqref{sizD}. The horizon direction represents the $y$ direction, while the vertical direction represents the transverse $\RR^2$. In the plot for the Lifshitz geometry it should be understood that the tip lies at an infinite proper distance away. When $z$ is large as in~\eqref{sizD}, there is a large range of $y$ for which the Lifshitz geometry resembles that of  AdS$_2 \times \mathbb{R}^2$. Also note that $e^{-y}$ translates into the boundary theory energy scale. In the Lifshitz geometry, the black hole has disappeared; it is replaced by a naked singularity with zero 
horizon size and zero charge. All the charge density is now carried by a bulk fermionic gas and the entropy density vanishes at zero temperature. 
}
\label{fig:comparison}
\end{figure}

The backreaction of the fermionic gas is thus important {\it no matter how large $N$ is}. 
The backreacted geometry can be found by using~\eqref{lode} and its corresponding stress tensor 
as the source for bulk Maxwell and Einstein equations. 
We refer readers to~\cite{Hartnoll:2009ns} for details of the calculation and only state the results here. One finds that after taking into account of the backreaction the near-horizon geometry is modified from~\eqref{nas2m} to a Lifshitz geometry 
\be \label{lifM}
ds^2 = - e^{-2y} dt^2 + e^{-{2 y \ov z}} \mu_*^2 R^2  d \vec x^2  + R_2^2 d y^2 , \qquad A_t = {e_d \ov R_2} e^{-y}
\ee
with 
\be \label{sizD}
z  =  {d \ov \ka^2 \rho_L} {\sqrt{2} R \ov R_2^2}  \sim O(N^2)
\ee
where we have only kept  leading order terms in $1/z$ expansion. It is important to emphasize that even in the large $N$ limit, the exponential prefector before the second term of~\eqref{lifM} cannot be set to unity as 
it becomes important when $y$ is sufficiently large (i.e. $y \sim O(z)$).
Also note that as~\eqref{nas2m},~\eqref{lifM} should be used in the region $y \gtrsim 0$; the geometry  matches to the rest of the rest of black hole (UV region) near $y \sim 0$.

%\be  \label{zexp}
%{1 \ov z} =  {2 q \ov \sqrt{6} \pi} {G_N \ov R^2} \le({k_o \ov \mu} \ri)^3 \sim {1 \ov N^2} \ .
%\ee

In the backreacted geometry~\eqref{lifM}, shown schematically in Figure \ref{fig:comparison}, the local proper volume of the transverse $\RR^{d-1}$ goes to zero as $y \to \infty$, resulting in a {\it finite} boundary fermionic density of order $O(N^2)$, given by
\be \label{boude}
\rho_F = R_2 (\mu_* R)^{d-1} \rho_L \int_{0}^{\infty} d y \, e^{-{d \ov z} y}   
= {R_2 (\mu_* R)^{d-1}  z \rho_L \ov d} %=     {z q k_o^3 \ov 6 \sqrt{2} \pi^2 \mu} 
\ . 
\ee
Plugging the explicit value~\eqref{sizD} of $z$ into~\eqref{boude} we find that~\eqref{boude} becomes identical to the total charge density~\eqref{entropy}, i.e.  all the charge density of the system is now carried by the fermioinic gas. The black hole has disappeared! It is now replaced by a naked singularity\footnote{Note that at this ``singularity'' all scalar curvature invariants are actually finite, but nevertheless an observer freely falling into the singularity will feel divergent tidal forces \cite{Kachru:2008yh}.} of zero horizon size and zero charge. The system now also has zero entropy density. Note that such a singularity appears to be of the ``good'' type allowed by string theory~\cite{Gubser:2000nd,Maldacena:2000mw}.  As we will see below, things could still fall into a singularity, which means in the boundary it should be interpreted as some of ``bath'' which can dissipate things away. While it has zero entropy and charge at order $O(N^2)$, likely such a ``bath'' still carries some charge and entropy at subleading order.

Recall that in obtaining~\eqref{ferm1} and~\eqref{lode} we used the approximation of a local chemical potential (i.e. Thomas-Fermi approximation), which is valid when $q g_F, mR, k_b^{(F)} R $ are taken to be parametrically large. Thus~\eqref{sizD} and~\eqref{boude} are also only valid in this regime.

The Lifshitz geometry~\eqref{lifM} is well approximated by~\eqref{nas2m} until $y \sim z$. From 
the boundary theory perspective, this implies that at an energy scale of order $e^{- z} \sim e^{-O(N^2)}$, physics will start deviating from that described by the \Slql. In other words, at sufficiently low energies, the \Slql\ will order into a new phase which is described by the Lifshitz geometry.  What is this new phase? 
It turns out to be a Fermi liquid of ``heavy'' fermions~\cite{Iqbal:2011in}! Below we outline the basic arguments, following the treatment of ~\cite{Iqbal:2011in}, in which more details can be found. Similar calculations appear in ~\cite{Hartnoll:2011dm,Cubrovic:2011xm}. 

\begin{figure}[h]
\begin{center}
\includegraphics[scale=0.4]{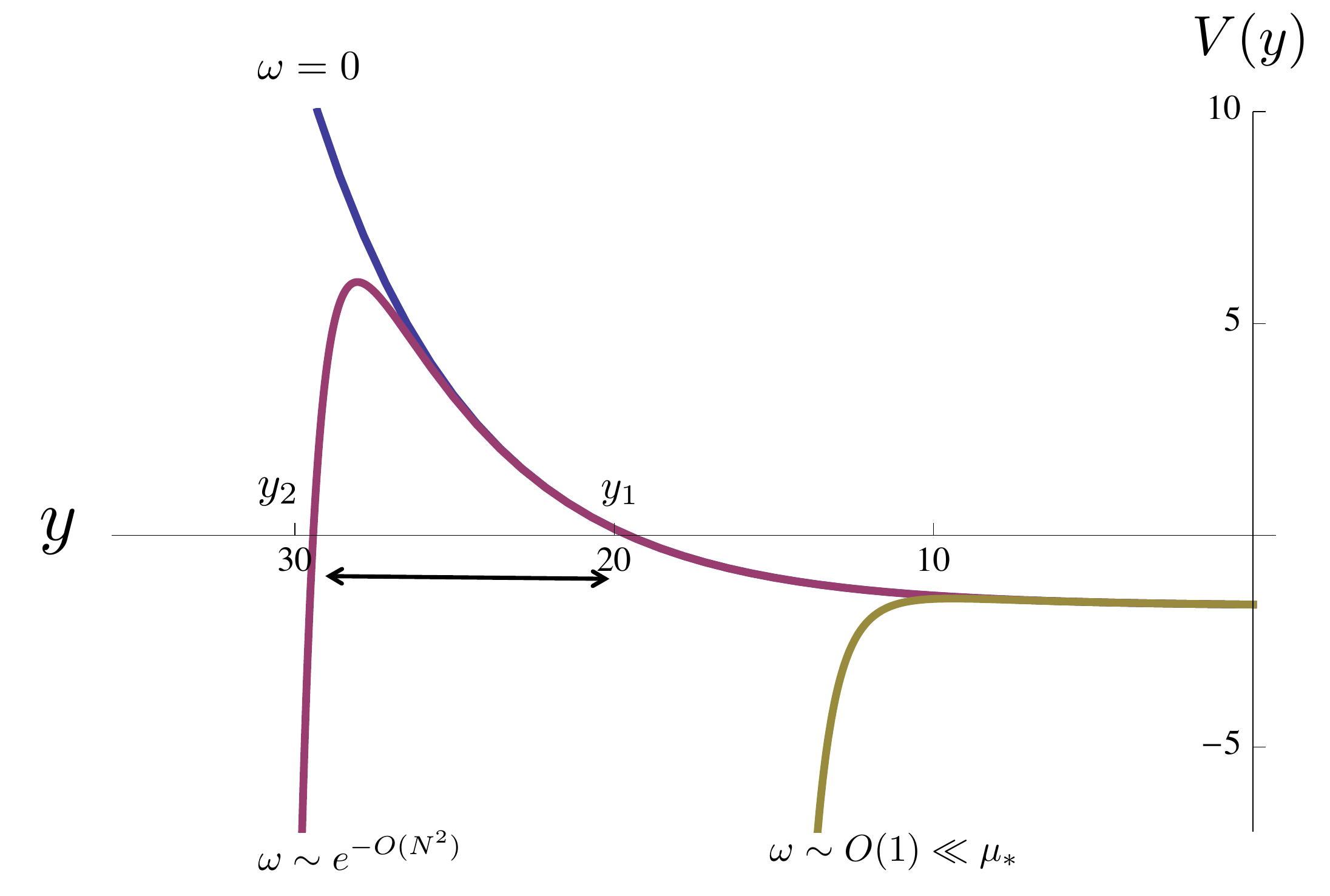}
\end{center}
\caption{Plots of the WKB potential~\eqref{potm} for various values of $\om$ in the Lifshitz region. 
For $\om =0$, the potential is bounded from below and the Bohr-Sommerfeld quantization gives normalizable 
modes in the bulk which correspond to different Fermi surfaces in the boundary theory. For an exponentially small $\om$, although the potential is unbounded from below in the large $y$ region, the excitations have a small imaginary part due to the potential barrier. For small $\om$ but not exponentially small, the potential barrier disappears and the potential becomes the same as that for the AdS$_2 \times \RR^2$. }
\label{fig:wkbP}
\end{figure}

The basic idea is again to examine the retarded Green function for the fermionic operator in which we now expect to find new physics at sufficiently low energies. For this purpose one needs to solve the Dirac equation for the bulk fermions in the backreacted geometry~\eqref{lifM}, which for 
 $q$ and $mR$ parametrically large can be reduced to a Schr\"odinger 
problem describing a particle of {\it zero energy} moving in a potential 
\be \label{potm}
V (y) = {1 \ov 2 \mu^2} \le(k^2 e^{2y \ov z} - k_o^2 - \om^2 e^{2y} - \sqrt{2} \mu_* q \om e^y \ri) \  . 
\ee
For the AdS$_2 \times \RR^{d-1}$ geometry, the first term is simply $k^2$. 

The final results can be readily understood from the qualitative features of the WKB potential~\eqref{potm}, as shown in Fig.~\ref{fig:wkbP} for $k^2 < k_o^2$. First consider $\om =0$, the potential starts being an almost negative constant for $y \ll z$, but eventually rises above zero and approaches $+\infty$ as $y \to + \infty$. The Bohr-Sommerfeld quantization condition then leads a series of normalizable bound states at discrete values of $k$. Each such normalizable mode then gives rise a boundary Fermi surface. 
Alternatively, from the bulk point of view, we now have a bounded potential governing movement in the radial direction. This quantizes the radial modes, and the continuous bulk Fermi surface indicated in Fig.~\ref{fig:bal} becomes discretized. When projected to the boundary this now become a series of Fermi surfaces as indicated in Fig.~\ref{fig:concentric}.
Now notice that in~\eqref{potm}, for any nonzero $\om$ no matter how small, the $\om$-dependent 
terms will eventually dominate for sufficiently large $y$ and drive the potential to $-\infty$ as $ y \to + \infty$. This implies bulk particles will eventually fall into the singularity of the Lifshitz geometry and the corresponding
boundary fermionic excitations should develop a nonzero width. 
For exponentially small $\om$, there is still a potential barrier and the tunneling rate is exponentially small. In this range there are long-lived quasiparticles. More explicitly, one finds that near each $k_F$ the retarded function for $\om$ in the range 
\be \label{2co}
\om  \ll \om_c (k) \equiv {k_o \ov z } \exp \le(- z \log {k_o \ov k_F}\ri)
\ee
can be written as 
\be \label{morGr}
G_R (\om, k) = - {Z  \ov 
 \om - v_F (k-k_F) + \Sig}
\ee
with
\be \label{ferP}
Z, v_F \sim  \exp \le(- z \log {k_o \ov k_F}\ri) ,  \quad
\Sig \sim i  \exp \le(- {\sqrt{2} z k_F \ov \mu} \le({k_F \ov  \om} \ri)^{1 \ov z} - z \log {k_o \ov k_F} \ri) \ .
\ee
Note that since $z \sim O(N^2)$, these are very heavy fermions\nicefootnote[1]{The exponentially small Fermi velocity 
$v_F$ implies that the effective mass $m_*$ is exponentially large, i.e.  
$m_* \sim  \exp \le(+ z \log {k_o \ov k_F}\ri)$.}  and the decay rate is exponentially small 
in $\om$. 

For small but not exponentially small $\om$, the first term in~\eqref{potm} is not important for any value of $y$ and the potential barrier disappears. One thus recovers previous results of \Slql. In particular, the decay rate becomes $O(1)$ and there are no long-lived quasiparticles. 

This procedure of solving the Dirac equation in the backreacted geometry~\eqref{lifM} is self-consistent and extracts the leading non-perturbative behavior in $1/N^2$. 
There are also perturbative loop corrections in the bulk, which give perturbative corrections in $1/N^2$ to the self-energy.  We expect the qualitative features of the above discussion (e.g the family of densely spaced Fermi surfaces etc) to be robust against these corrections, as they have to do with the global structure of the backreacted geometry~\eqref{lifM}.
 Near the Fermi surface perturbative corrections to the self-energy will give rise to a term ${c \ov N^2} \om^2$ with $c$ some complex $O(1)$ coefficient, which will dominate over the imaginary part of $\Sig$ in~\eqref{morGr} for $\om$ in the range~\eqref{2co}. 
Thus we expect that the quasiparticle decay rate should be proportional to $\om^2$ as in a Landau Fermi liquid.\nicefootnote[1]{Nevertheless, one should keep in mind that the much smaller non-perturbative correction does signal some nontrivial underlying physics beyond that of a Landau Fermi liquid.} 

We thus see that at low energies the oscillatory region $k < k_o$ splits into a large number of closely spaced Fermi surfaces whose excitations are Landau quasiparticles.  In this regime the Luttinger theorem should hold, which can be checked explicitly (see~\cite{Iqbal:2011in} for details).

Different Fermi surfaces in the family of densely spaced Fermi surfaces can be interpreted as corresponding to {\it different bound states} generated by the fermionic operator $\sO$, as each of them corresponds to a different radial mode in the bulk. Recall that in our set-up, $\sO$ is a composite operator of fundamental fields. The number of degrees of freedom for the fundamental fields is $O(N^2)$. Thus the Fermi liquid state can be considered a ``confined'' state, in which the low energy degrees of freedom are Fermi surfaces from a discrete set of composite fermionic bound states.  In contrast, the \Slql\ is a ``deconfined'' state in which the composite bound states deconfine and  fractionalize into  more fundamental degrees of freedom. Presumably these more fundamental degrees of freedom account for the entropy density of the \Slql. In the bulk, the emergence of the fractionalized \Slql\ phase is reflected in the 
emergence of a charged black hole description. In the Fermi liquid state, the system is characterized by ``heavy'' fermions. But such coherent quasiparticles disappea in the \Slql. Instead one finds some kind of quantum soup which is characterized by scaling behavior in the time direction for any 
 bosonic or fermionic operators. 
 In the current context the presence of  a large number of Fermi surfaces has to do with a spectrum of densely spaced bound states. The fractionalized picture should be independent of this feature.
We note also that this interpretation of the ~\Slql\ as a fractionalized phase resonates with the lattice model construction of \cite{Sachdev:2010um}, although the details are different.

Note that the WKB potential~\eqref{potm} only includes the near horizon region of the backreacted charged black hole geometry, and  does not include the  
Fermi surfaces discussed in~Sec.\ref{sec:ipp}.  In Fig.~\ref{fig:wkbPF} we show a cartoon of the WKB potential for $\om=0$ for the full spacetime, i.e. including the asymptotic AdS$_{d+1}$ region. 
The isolated Fermi surfaces discussed earlier in~Sec.~\ref{sec:ipp} and~\ref{sec:fff} appear as bound states in a potential well outside the near-horizon region.  Such isolated Fermi surfaces  exist both in the Fermi liquid phase and in the \Slql, i.e. the corresponding fermionic excitations remain confined in the deconfined \Slql\ phase.   At exponentially small energies, the excitations again become Landau quasiparticles, but with an $O(1)$ effective mass.

\begin{figure}[h]
\begin{center}
\includegraphics[scale=0.6]{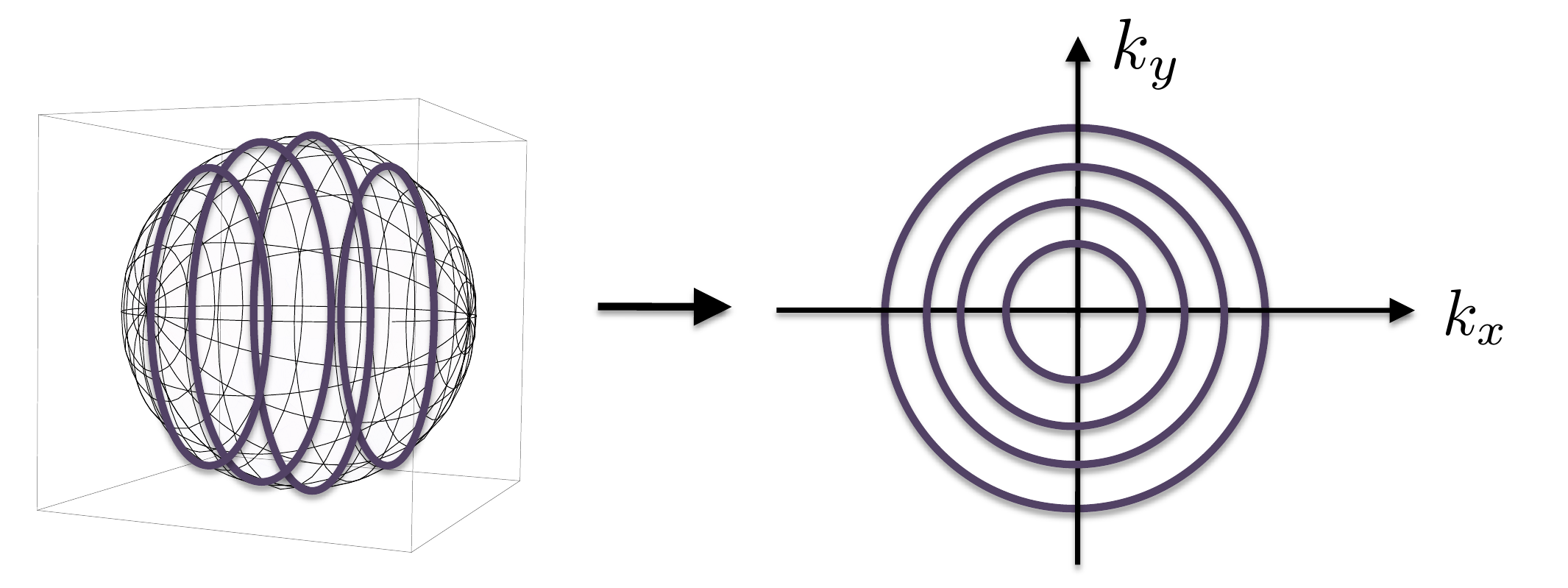}
\end{center}
\caption{In the Lifshitz geometry, we find a set of discrete states in the bulk in which radial motion is quantized. This results in a family of concentric Fermi surfaces in the boundary theory, which resolves the Fermi disk of Fig.~\ref{fig:bal}.}
\label{fig:concentric}
\end{figure}

\begin{figure}[h]
\begin{center}
\includegraphics[scale=0.43]{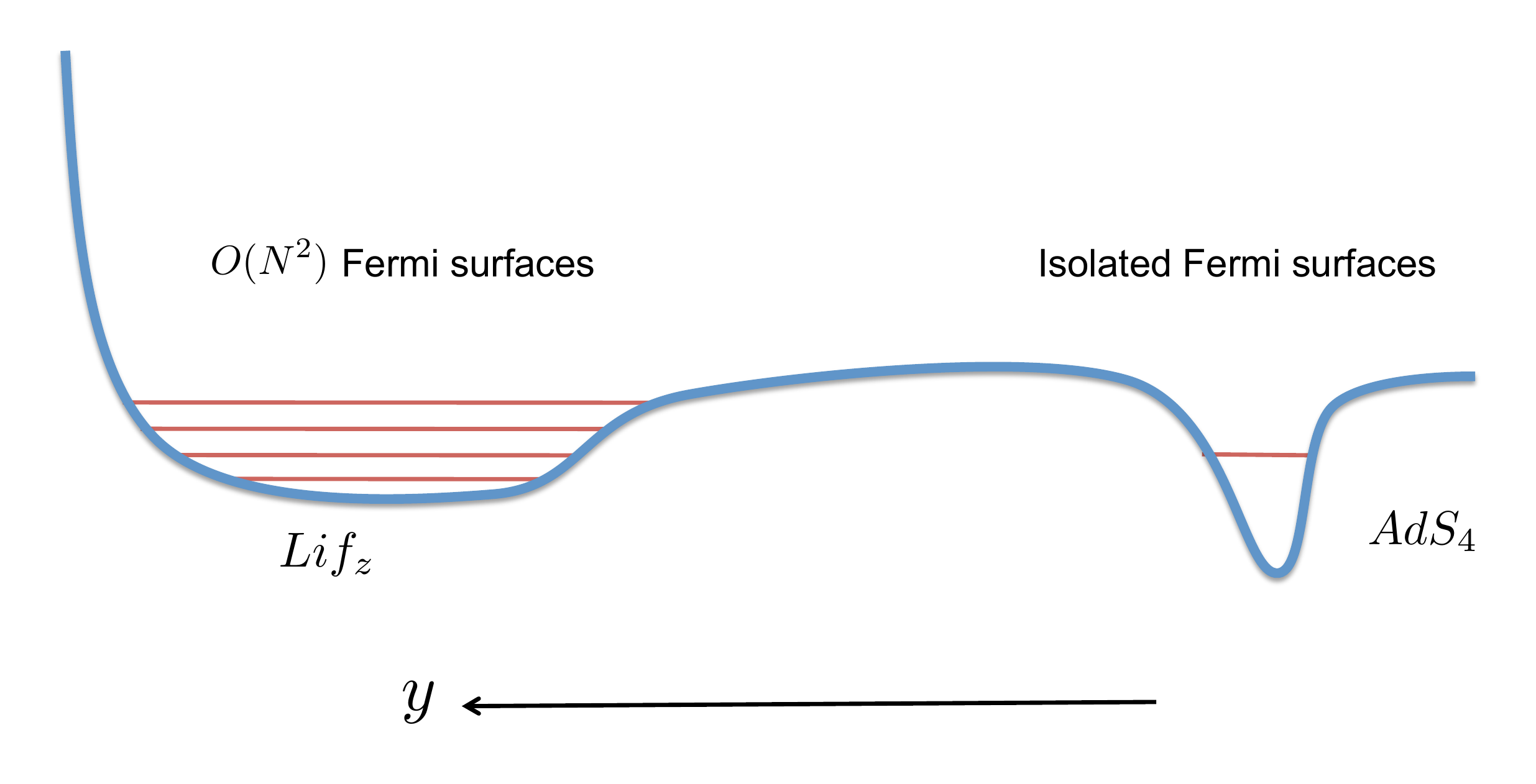}
\end{center}
\caption{A cartoon of the bound-state structure for the full spacetime, i.e. including the asymptotic AdS$_4$ region. The Fermi surfaces in Sec.~\ref{sec:ipp} and~\ref{sec:fff} appear as bound states in a potential well in the UV region. Note that the plot only intends to indicate the locations of bound states in the bulk spacetime and does not reflect the genuine WKB potential.}
\label{fig:wkbPF}
\end{figure}

To summarize, imagine a system with only a single fermionic operator $\sO$ satisfying the condition~\eqref{newpr}, and no other instabilities. Then at very low energies, the system is described by a Fermi liquid with $O(N^2)$ densely spaced Fermi surfaces, each of size $O(N^0)$. 
The quasiparticle excitations have a very large effective mass (proportional to $e^{N^2}$). 
%When $q^2 > 3 m^2 R^2$ 
There could also be some additional isolated Fermi surfaces with an $O(1)$ effective mass. At small but not exponentially small frequencies, there is a wide energy range over which the system is controlled by the \Slql, with scaling behavior in the time direction and various non-Fermi liquid behavior.  In the \Slql\ phase the clue that the system will eventually settles into a Fermi liquid state is the existence of a region $k < k_o$, where the scaling dimensions for fermions become complex.

\subsection{Generalizations}

Before concluding this section, we mention various generalizations which we do not have space to cover. 

In the presence of a magnetic field, one can show that fermionic excitations fall into Landau levels~\cite{Albash:2009wz,Basu:2009qz,Gubankova:2010rc} and that the free energy displays de-Haas van Alphen oscillations in the inverse magnetic field~\cite{Denef:2009kn,Hartnoll:2009kk}. The latter calculation also requires a one-loop calculation in the bulk. In~\cite{Hartnoll:2009kk} it was shown that while the oscillations exist with the expected period, for $\nu_{k_F} < \ha$ the amplitude of the oscillations does not take the textbook Lifshitz-Kosevich form.  
The fermion response in various holographic superconducting phases has been studied in 
\cite{chenkaowen, Faulkner:2009am, fabio, Gubser:2010dm,
Ammon:2010pg,Benini:2010qc,Vegh:2010fc}.
The superconducting condensate opens a gap in the fermion spectrum 
around the chemical potential if a certain bulk coupling between the spinor and scalar is included~\cite{Faulkner:2009am}. In particular, in the condensed phase, one finds stable quasiparticles, even when $\nu_{k_F} \leq \ha$.
The properties of Fermi surfaces and associated excitations with more general fermionic couplings or boundary conditions have also been studied in~\cite{Edalati:2010ww,Edalati:2010ge,Guarrera:2011my,Laia:2011zn} and in more general geometries in
~\cite{Albash:2010dr,Wu:2011bx}. Fermi surfaces in $1+1$-dimensions were studied in~\cite{Wang:2009pp,Balasubramanian:2010sc,Rangamani:2011ae}. 
References~\cite{Hartman:2010fk,Gubankova:2010ny} considered the pairing instability of holographic non-Fermi liquids.

By considering  spinor fields whose mass and charge grows with $N$ (i.e. $m, q \sim N$) and a gauge coupling 
of order $1/N$, the Lifshitz geometry~\eqref{lifM} has been generalized to have a $z \sim O(N^0)$ in~\cite{Hartnoll:2010gu} (see also~\cite{deBoer:2009wk,Arsiwalla:2010bt,vCubrovic:2010bf}). 
Various aspects of  the boundary system dual to such an ``electron star'' were studied in~\cite{Hartnoll:2010xj,Hartnoll:2010ik, Puletti:2010de,Hartnoll:2011dm,Cubrovic:2011xm}. In particular, it was found \cite{Hartnoll:2011dm,Cubrovic:2011xm} that the fermionic spectral function exhibits multiple Fermi surfaces with  
exponentially suppressed (in frequency) decay rates very similar to the behavior discussed earlier around~\eqref{2co}--\eqref{ferP}~(see also~\cite{Sachdev:2011ze} for a hard-wall model for constructing holographic Fermi liquid).

\section{Scalar instabilities and quantum phase transitions} \label{sec:qpt}

The previous section discussed fermions in some detail, describing both non-Fermi liquid physics and the ultimate Fermi-liquid-like ground state of the system. In this section we turn to bosons. Recall from Section \ref{sec:bosvsferm} that while fermionic response functions can have gapless modes that describe propagating degrees of freedom (i.e. are in the lower half complex -$\om$ plane), generically for bosons we instead find {\it instabilities}, i.e. poles in the {\it upper} half complex plane. This indicates that the scalar operator is trying to condense, and if the operator is charged under any symmetries then they will be broken in the condensed phase. 

In this section we examine these bosonic instabilities of the finite charge density 
system. In particular, when there is an instability we describe how to tune external parameters to make this instability {\it vanish}; if this can be done at zero temperature, then this defines a quantum critical point that separates two phases. We will study in detail the critical behavior near such a quantum critical point. 
We will mainly follow the discussion of~\cite{Iqbal:2010eh,Faulkner:2010gj,sdwnew} (related results have also been obtained in~\cite{Jensen:2010ga,Jensen:2010vx,Jensen}, see also~\cite{Edalati:2011yv}). For concreteness we will also specialize to $d=3$. 

We first remind the reader of our setup; consider a CFT$_3$ with a scalar operator $\sO$. We discussed above in Section \ref{lowgreen} how to use holography to compute the retarded Green's function of $\sO$. Before moving on, we will first slightly generalize that discussion to also include double trace deformations in the dual CFT$_3$
\be \label{douv} 
S_{CFT_3} \to S_{CFT_3} + { \ka_+ \ov 2} \int \sO^2 \ .
\ee 
It is straightforward to include the presence of such a deformation in our holographic computation. Details are provided in Appendix~\ref{app:doub}; the main result can be found in \eqref{corrD}, from which we see that \eqref{roep1} becomes 
\be \label{roep23}
G_R (\om, \vk) = \mu_*^{2 \nu_U} {b_+ (k,\om) + b_- (k,\om) \sG_k (\om) \mu_*^{-2 \nu_k}  \ov \tilde a_+ (k,\om) + \tilde a_- (k,\om) \sG_k (\om) \mu_*^{-2 \nu_k}} \  %\qquad {\om \ov \mu} \to 0
\ee
 where 
 \be
 \tilde a_\pm (\om, k) = a_\pm (\om,k) + \ka_+ b_\pm (\om,k)
 = \tilde a_\pm^{(0)} (k)  + O(\om) \ , 
 \qquad \tilde a_\pm^{(0)} (k) = a_\pm^{(0)} (k) + \ka_+ b_\pm^{(0)} (k) \ . 
 \ee
 Note that in general the coefficient $\ka_+$ of the double-trace operator provides a ``knob'' that can be turned to tune the system through a quantum phase transition. 
 
To  follow the standard terminology for discussing phase transitions below we will  start calling~\eqref{roep23} the {\it susceptibility} and use the following notation: 
\be
\chi \equiv G_R (\om=0, \vk=0), \qquad \chi (\vk) \equiv G_R (\om =0, \vk), \qquad 
\chi (\om, \vk) \equiv  G_R (\om, \vk)\ .
\ee
where we distinguish the three only by their arguments. From~\eqref{roep23} we  find the  momentum-dependent and uniform static susceptibilities are given by
\be \label{stsus}
\chi (k) = \mu_*^{2 \nu_U} {b_+^{(0)} (k)  \ov \tilde a_+^{(0)} (k) } , \qquad \chi = \mu_*^{2 \nu_U} {b_+^{(0)} (0)  \ov \tilde a_+^{(0)} (0) } \ 
\ee 
and  the full dynamical susceptibility takes the form
\be \label{roep211}
\chi (\om, \vk) = \mu_*^{2 \nu_U} {b_+ (k,\om) + b_- (k,\om) \sG_k (\om) \mu_*^{-2 \nu_k}  \ov \tilde a_+ (k,\om) + \tilde a_- (k,\om) \sG_k (\om) \mu_*^{-2 \nu_k}} \ . 
\ee
 
Below we first recall the instabilities discussed in Sec.~\ref{sec:bosvsferm} for the system at a finite chemical potential and the corresponding quantum critical points. We will then discuss quantum critical behavior near these quantum critical points.  

\subsection{Finite density instabilities}

Recall that in~\ref{sec:bosvsferm} we discussed two possible channels for scalar instabilities, which manifest  
themselves as presence of poles of~\eqref{roep211} in the {\it upper} half complex-$\om$-plane:

\ben

\item The \Slql\ scaling dimension $\ha + \nu_k$ becomes complex for some $k$, for which there are an infinite number of poles 
in the upper half $\om$-plane.  Writing~\eqref{opep} as  
\be \label{uude}
\nu_k = \sqrt{u + {k^2 \ov 6 \mu_*^2}}, \qquad u \equiv m^2 R_2^2 + {1 \ov 4} - q_*^2  \ 
\ee
$\nu_k$ becomes imaginary for sufficiently small $k$  whenever $u < 0$. For a given $m$, this always occurs for a sufficiently 
large $q$. For a neutral operator $q=0$, $u$ can be negative for $m^2R^2$ lying in the window 
\be \label{winmag}
-{9 \ov 4} < m^2 R^2 < -{3 \ov 2}
\ee
where the lower limit comes from the stability of vacuum theory (i.e. BF bound of AdS$_4$) and the upper limit comes from the condition $u <0$ after using the relation~\eqref{defZ}.

\item $\tilde a_+^{(0)} (k)$ can become zero for some special values of momentum $k_F$.  At $k=k_F$ it is clear from~\eqref{roep23} that since $\tilde a_+^{(0)} =0$, $\chi$ has a singularity at $\om=0$. Furthermore since $a_+^{(0)}$ changes sign near $k=k_F$, as discussed in Sec.~\ref{sec:hair}, % and Appendix~\ref{app:poles}, 
a pole moves from the upper half $\om$-plane (for $k< k_F$) to the lower half $\om$-plane (for $k>k_F)$ through $\om =0$. 
In Fig.~\ref{fig:neu} we show some examples of a  neutral scalar field for which 
$\ta_+^{(0)}$ has a zero at some momentum.

\een

\begin{figure}[h]
\begin{center}
{\includegraphics[scale=0.8]{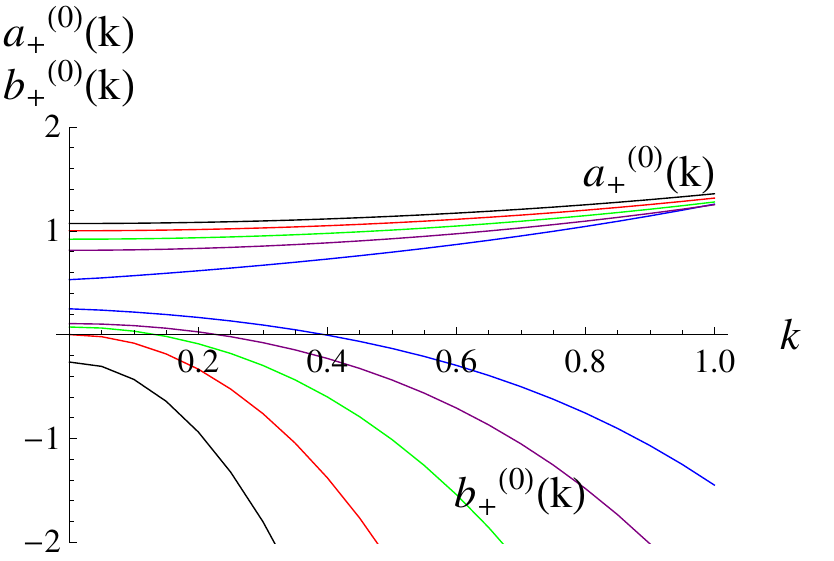}\includegraphics[scale=0.8]{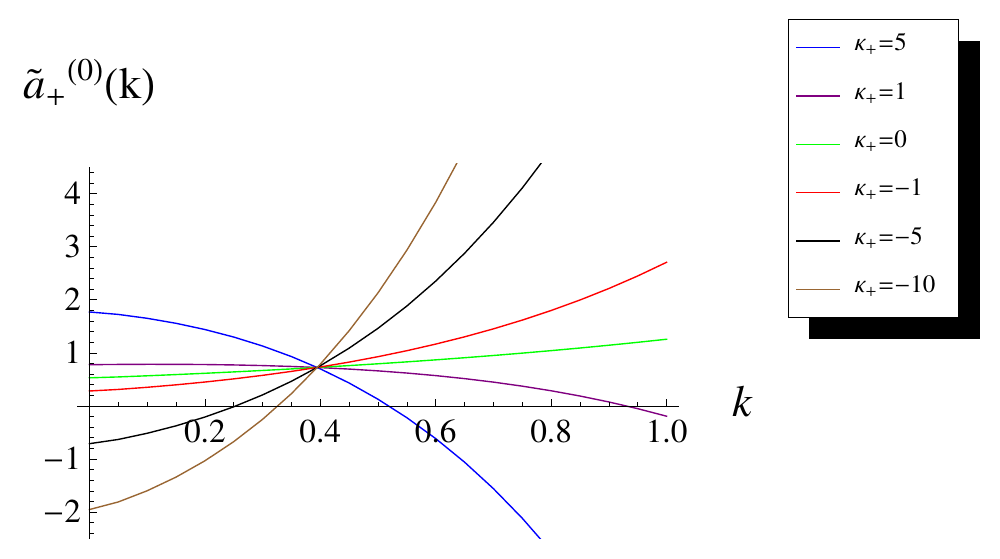}}
\end{center}
\caption{{\it Left}: $a_+^{(0)} (k)$ and $b_+^{(0)} (k)$ plotted for different values of $m^2R^2$ with $q=0$; 
blue is $m^2=-1.4999$, purple $m^2R^2=-1$, green is $m^2R^2=-0.5$, red is  $m^2R^2=0$ and the black line is $m^2R^2=0.5$. $a_+^{(0)} (k)$ is positive and monotonically increases with $k$, while 
 $b_+^{(0)} (k)$ monotonically decreases with $k$. Thus for $\ka_+ < 0$, $\tilde a_+^{(0)} (k)
 = a_+^{(0)} (k) + \ka_+ b_+^{(0)} (k) $ is a monotonically increasing function of $k$. Note that $b_+^{(0)} (k=0) =0$ for $m^2 =0$. This implies that when $m^2 > 0$, $\tilde a_+^{(0)} (k) > 0$ for any $\ka_+ < 0$. 
 {\it Right:} $\ta_+^{(0)} (k)$ for $ m^2 R^2 =-1.4999 $ and $q=0$. $\ta_+^{(0)} (k)$ 
has a zero for some $k$ when $\ka_+ < \ka_c = -2.13$. For $0 > \ka_+ > \ka_c$ there is no instability (see $\ka_+ =-1$ curve). 
When $\ka_+ > 0$, $\ta_+^{(0)} (k)$ can again develop a zero for some $k_F$ (with the value of $k_F$ approaching infinity for $\ka_+ \to 0_+$); this is a {\it UV} instability which is already present in the vacuum. 
}
\label{fig:neu}
\end{figure}

 In the parameter range (say for $m, q, \ka_+$) where either (or both) instability appears, the system is unstable to the condensation of operator $\sO$ (or in bulk language the condensation of $\phi$). For a charged scalar the condensed phase is the well-studied {\it holographic superconductor} \cite{Gubser:2008px,Hartnoll:2008vx}; for reviews see \cite{herzogr,Horowitz:2010gk}. As was first pointed out in \cite{Denef:2009tp}, the first instability essentially drives the original construction of the holographic superconductor. Holographic superconductors due to the second type of instability were first discussed in~\cite{Faulkner:2010gj}. 
 
 For a {\it neutral} scalar, the first type of instability was first pointed out in~\cite{Hartnoll:2008kx}, and as discussed in~\cite{Iqbal:2010eh} the condensed phase can be used as a model for antiferromagnetism when the scalar operator is embedded as part of a triplet transforming under a global $SU(2)$ symmetry corresponding to spin. For a single real scalar field with a $Z_2$ symmetry, the condensed phase can be considered as a model for an Ising-nematic phase. Instabilities of the first type also appear in various other holographic constructions \cite{Jensen:2010ga,Jensen:2010vx}. The potential interaction between these two types of order is discussed in \cite{Basu:2010fa}. 

 Both types of instabilities can be cured by going to sufficiently high temperature; there exists a critical temperature $T_c$, beyond which these instabilities no longer exist and at which the system undergoes a continuous  superconducting (for a charged scalar) or antiferromagnetic (for a neutral scalar) 
phase transition. As have been discussed extensively in the literature \cite{Maeda:2009wv,Iqbal:2010eh,Herzog:2010vz} such finite temperature phase transitions are of mean field type due to that the boundary conditions at a finite-temperature black hole horizon is analytic. 

Alternatively one can continuously dial external parameters of the system at zero temperature
to get rid of the instabilities.\nicefootnote[1]{One might be surprised that in the case of a broken continuous symmetry in $d = (2+1)$ the ordered phase can persist at {\it any} finite temperature, as one generally expects fluctuations of the Goldstone mode to destroy the order. As there is only one Goldstone mode (as compared to $N^2$ other fields) this is a $1/N^2$ effect and so is dual to a quantum effect in the bulk; it is shown in \cite{Anninos:2010sq} that at one-loop order in the bulk the symmetry is indeed restored over exponentially long length scales.}  The critical values of the parameters at which the instabilities disappear then correspond to quantum critical points (QCP) where quantum phase transitions into a superconducting or antiferromagnetic phase occur. We now quickly summarize the nature of each of these quantum critical points. 
\subsection{Quantum critical points}

\subsubsection{Bifurcating quantum critical point}

For the first type of instability a quantum critical point occurs 
when we violate the Breitenlohner-Freedman bound of the AdS$_2$ region at $k=0$~\cite{Iqbal:2010eh,Jensen:2010ga}, i.e. 
from~\eqref{uude}, at 
\be \label{biy}
u = u_c =0 \ .
\ee
For example for a neutral scalar field (with $q=0$) this happens at 
\be \label{crimas}
m^2_c R^2 = -{ 3\ov 2} \ .
\ee
Note that while in AdS/CFT models the mass square $m^2$ for the vacuum theory is typically not an externally tunable parameter, the effective AdS$_2$ mass square can often be tuned. For example, if we place a holographic superconductor in a magnetic field at zero temperature the AdS$_2$ mass can be changed by tuning the magnetic field \cite{Iqbal:2010eh}. In the set-up of~\cite{Jensen:2010ga}, a somewhat different order parameter is dual to a bulk field that has a mass that depends on the external magnetic field. See also~\cite{Iqbal:2010eh} for a phenomenological model. Here we will not worry about the detailed mechanism to realize the $u_c =0$ critical point and will just treat $u$ as a dialable parameter (or just imagine dialing the mass square for the vacuum theory). Our main purpose is to identify and understand the critical behavior around the critical point which is independent of 
the specific mechanism to realize it.  

As will be discussed in subsequent sections, as we approach $u_c=0$ from the uncondensed side ($u > 0$), the static susceptibility remains finite, but develops a cusp at $u=0$ and if we naively continue it to $u < 0$ the susceptibility becomes complex. Below we will refer to this critical point as a {\it bifurcating} QCP. 

\subsubsection{Hybridized quantum critical point}

For the second type of instability, one can readily check numerically that $\ta_+^{(0)}$ is a monotonically increasing function of $k$ for negative $\ka_+$. See fig.~\ref{fig:neu}. Thus the instability goes away 
if $\ta_+^{(0)}  (k=0) > 0$, which corresponds to $\ka_+$ satisfying\nicefootnote[1]{There is also an instability for $\ka_+ > 0$, but one can check that it corresponds to a UV instability of the vacuum. Here we are interested in the IR instability due to finite density effect. See~\cite{sdwnew} for more discussion.} 
\be \label{stare}
0 > \ka_+ \geq \ka_c \equiv -{a_+^{(0)} (k=0) \ov b_+^{(0)} (k=0)}  \ .
\ee 
  At the critical point $\ka_+ = \ka_c$
\be \label{hybq}
\ta_+^{(0)} (k=0, \ka_c)= 0
\ee
and as a result the uniform susceptibility $\chi$ in~\eqref{stsus} diverges.  Such a quantum critical point was first discussed in~\cite{Faulkner:2010gj} and elaborated further in~\cite{sdwnew}. As will be seen below, the presence of the strongly coupled IR sector described by AdS$_2$  gives rise to a variety of new phenomena which cannot be captured by the standard Landau-Ginsburg-Wilson paradigm. For reasons to be clear below, we will refer to such a critical point as a {\it hybridized} QCP.

\subsubsection{A marginal quantum critical point}

We can also tune $\ka_+$ and $u$ together to have a doubly tuned critical point at $u = 0, \ka_+=\ka_+^*$, where the susceptibility both diverges and bifurcates. The value of $\ka_+^*$ can be obtained from the $u \to 0$ limit of the expression for $\kappa_c$ given in \eqref{stare}, leading to 
\be  \label{mcpv1}
\ka_+^* = -{\al \ov \beta}  \  
\ee
where $\al$ and $\beta$ are constants defined in Eq. ~\eqref{req}. For the specific example~\eqref{crimas} of tuning the AdS$_4$ mass of a neutral scalar to reach $u=0$, the values of $\al, \beta$ can be found numerically which give $\ka_* = - 2.10$.

As we will see below, the susceptibility around such a critical point coincide with that of the bosonic fluctuations underlying the ``Marginal Fermi Liquid'' postulated in~\cite{Varma89} for describing the strange metal region of the high $T_c$ cuprates.\footnote{It has also been pointed out by David Vegh~\cite{veghun} that the retarded function for a scalar operator with $\nu=0$ in AdS$_2$ gives bosonic fluctuations of  the ``Marginal Fermi Liquid''.}

We note that  while understanding the end point of the instability, i.e. the stable phase in which $\sO$ is condensed, requires full nonlinear analysis of bulk equations, to diagnose the instability, to identify the onset of instability (i.e. the quantum or finite temperature critical points), and to find the critical behavior near the critical points (from the uncondensed side), the linear analysis which gives rise to~\eqref{roep23} suffices.   

We now proceed to describe the critical behavior near the various critical points identified  above. We will use a {\it neutral} scalar field for illustration. We should also mention that in all critical points identified above, the uncondensed side is described by an extremal charged black hole. Given the interpretation in Sec.~\ref{sec:inter} of an extremal charged black hole as an intermediate state rather than a genuine ground state, the critical behavior discussed below from the uncondensed side should again be interpreted as due to 
intermediate-energy effects; we return to this point in the conclusion.  

\subsection{Quantum Critical behavior: Bifurcating quantum critical point}

\subsubsection{Critical behavior}

Let us first consider the bifurcating critical point $u=0$. Here we will outline the results; for details see the recent discussion in ~\cite{sdwnew}.

To study the critical behavior from the uncondensed side, we take $\nu_k \to 0$ in \eqref{roep211}, i.e. with both $k^2/\mu^2$ and $u$ small (in particular, below we will always consider $k^2/\mu^2$ of order or smaller than $u$ which is the interesting regime). 
This limit is a bit subtle as the factor $\le(\frac{\om}{\mu_*}\ri)^{2\nu_k}$ in the \Slql\ Green function~\eqref{iiRc} behaves differently depending 
on the order in which we take the $\nu_k \to 0$ and $\om \to 0$ limits. For example, the Taylor expansion of such a term in small $\nu_k$  involves terms of the form $\nu_k\log(\om/\mu_*)$, but in the small $\om$ limit, the resulting large logarithms may invalidate the small $\nu_k$ expansion. We thus proceed carefully. 
 Expanding the explicit form of the AdS$_2$ Green's function \eqref{expform} and  $a_\pm ,b_{\pm}$ to leading order in $\nu_k$, but keeping the full dependence on $\om$, we find that 
\be
\chi (\om, \vk) = \chi_0 \frac{\sinh\le(\nu_k \log\le(\frac{-i\om}{\om_b}\ri)\ri)}{\sinh\le(\nu_k \log\le(\frac{-i\om}{\om_a}\ri)\ri)} + \dots  \label{rptfinom}
\ee
where
\be \label{veifp}
\chi_0 \equiv \mu_*^{2 \nu_U}  {\beta \ov \al} , \quad \om_{a} = 2\mu_* \exp \le(\frac{\tilde{\al}}{\al} - \ga_E \ri), \quad 
\om_b = 2\mu_* \exp \le(\frac{\tilde{\beta}}{\beta} - \ga_E \ri),
 \ee 
 $\ga_E$ is the Euler number,  and $\al, \beta, \tilde \al, \tilde \beta$ are some constants introduced in~\eqref{req}.

This expression is exact for all frequencies that are much smaller than the UV scale; however to understand the physics, we should understand that the non-commutativity of the $\om \to 0$ and $\nu_k \to 0$ limits in equation~\eqref{rptfinom} essentially defines a crossover scale 
\be
\Lam_{CO} \sim \mu_* \exp\le(-\frac{\#}{\sqrt{|u|}}\ri), \label{rptco}
\ee 
with $\#$ some $O(1)$ number. 

For
$\om \gg \Lam_{CO}$,  one can expand the arguments of the hyperbolic sine to find
\be 
\chi(\om, \vk ) =\chi_0 \frac{\log\le(\frac{\om}{\om_b}\ri) - i\frac{\pi}{2}  }{\log\le(\frac{\om}{\om_a}\ri) - i\frac{\pi}{2} }  + O(u, k^2)\ \label{quantcritexp}
\ee
with the spectral function given by 
\be \label{crispe}
\Im \chi (\om, k) = {\pi \chi_* \ov (\log \om)^2} + \dots \ .
\ee
However, for $\om \ll \Lam_{CO}$, approaching the critical point from the (stable) $u>0$ side, we find 
\be
\chi (\om \to 0, \vk) = \chi_0 - 2 \nu_k \chis - 4 \nu_k \chis \le({-i \om \ov 2 \mu_*}\ri)^{2 \nu_k} + \dots 
\ .
 \label{rptsusc}
\ee

The static susceptibility is obtained from~\eqref{rptsusc} by taking $\om \to 0$, and interestingly, near the critical point it approaches a finite constant $\chi_0$ for $k=0$. It nevertheless develops a branch point singularity at $u = 0$, 
as $\nu_{k=0} = \sqrt{u}$; the susceptibility is trying to bifurcate into the complex plane as we cross $u = 0$.  Upon Fourier transformation to coordinate space, this branch point singularity leads to a correlation length that diverges at the critical point. Indeed by comparing~\eqref{uude} with~\eqref{defnk}, we find that 
\be \label{zeroXi}
\xi = {1 \ov \sqrt{6} \mu_* \sqrt{u}} \ .
\ee
Thus as $u \to 0$, the correlation length $\xi$ diverges as $u^{-\ha}$ with the same exponent as that in a mean field theory.

When $u < 0$, the \Slql\ scaling dimension of $\sO_\vk$ becomes complex for sufficiently small $k$ as  $\nu_k = \sqrt{u + {k^2 \ov 6 \mu_*^2}} = - i \lam_k$ is now pure imaginary.\nicefootnote[1]{Note that the choice of branch of the square root does not matter as~\eqref{rptfinom} is an even function of $\nu_k$.} For a given nonzero $\om$ and $|u|$ sufficiently small, the corresponding expression for $\chi (k, \om)$ can be obtained from~\eqref{rptfinom} by simply taking $\nu_k = - i \lam_k$, after which we find
 \be 
  \chi (\om, \vk) = \chi_0 \frac{\sin\le(\lam_k \log\le(\frac{-i\om}{\om_b}\ri)\ri)}{\sin\le(\lam_k \log\le(\frac{-i\om}{\om_a}\ri)\ri)} + \dots  \ . \label{usmS}
\ee
 For $\om \gg \Lam_{CO}$ one finds the same critical behavior as~\eqref{quantcritexp}, but 
 for $\om \ll \Lam_{CO}$, we find a 
geometric series of poles in the upper-half complex frequency-plane at 
\be \label{exppoles}
\om_n = i\om_a \exp \le(-\frac{n \pi}{\sqrt{-u}}\ri) \sim i \mu \le({\Lam_{IR} \ov \mu}\ri)^n, \qquad n=1,2, \dots
\ee
with
\be  
\Lam_{IR} \equiv \mu \exp\le(-\frac{\pi}{\sqrt{-u}}\ri), 
\ee
indicating that the disordered state is unstable and the scalar operator condenses in the true vacuum. 
In particular, we expect~\eqref{usmS} to break down for $\om \sim \om_1 \sim \Lam_{IR}$, the largest among~\eqref{exppoles}, and at which scale the physics of condensate sets in. This is indeed consistent with an explicit  analysis of classical gravity solutions in~\cite{Iqbal:2010eh,Jensen:2010ga,sdwnew} where it was found that  $\sO$ develops an expectation value of order 
\be \label{exvev1}
{\vev{\sO} \ov \mu^\De} \sim \le({\Lam_{IR} \ov \mu} \ri)^{\ha}   \ . 
\ee
The exponent $\ha$ in~\eqref{exvev1} is the scaling dimension of $\sO$ in the \Slql\
for $u=0$, while $\De$ is its scaling dimension in the vacuum. Furthermore, one in fact finds~\cite{Iqbal:2010eh,Jensen:2010ga,sdwnew} an infinite tower of  excited condensed states in one to one correspondence with the poles
in~\eqref{exppoles}
\be \label{vevex}
\vev{\sO}_n \sim \mu^\De  \exp \le(-\frac{n \pi}{2 \sqrt{-u}}\ri) 
\sim \mu^\De \le({\Lam_{IR} \ov \mu}\ri)^{n \ov 2}% \le({\Lam_{IR} \ov \mu} \ri)^{n \ov 2} , 
\quad n=1,2, \dots \ .  % , \quad \Lam_n  \sim \mu
\ee
This tower of condensed states with geometrically spaced expectation values is reminiscent of Efimov states~\cite{efimov}.\nicefootnote[1]{In fact the gravity analysis reduces to a quantum mechanical problem which is exactly the same as that for the Efimov states.}
The $n=1$ state is the ground state with the lowest free energy which scales as  (with that of the disordered state being zero)
 \be
 F \sim - \Lam_{IR} \ .
\ee

\begin{figure}[h]
\begin{center}
\includegraphics[scale=0.5]{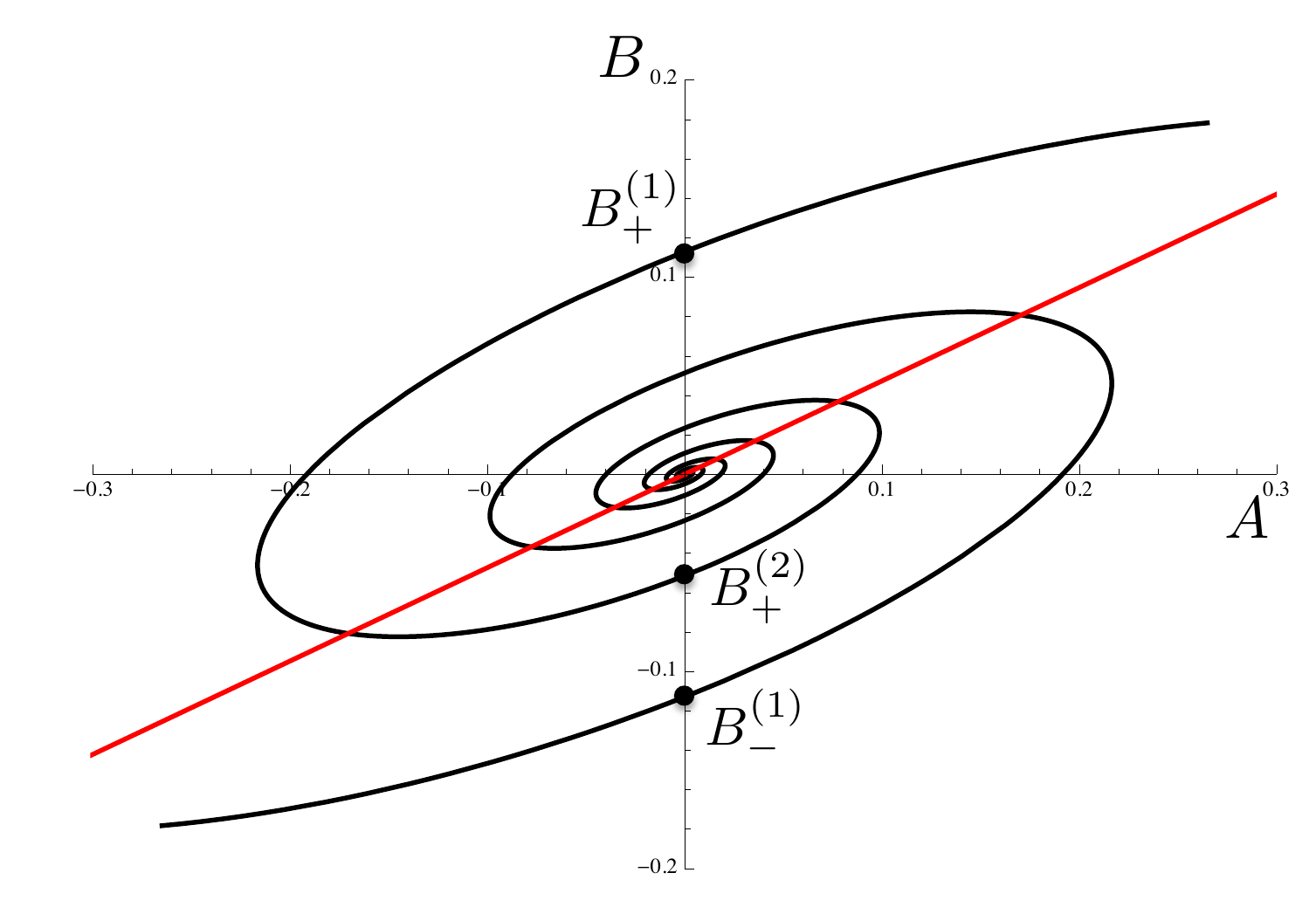} 
\includegraphics[scale=0.5]{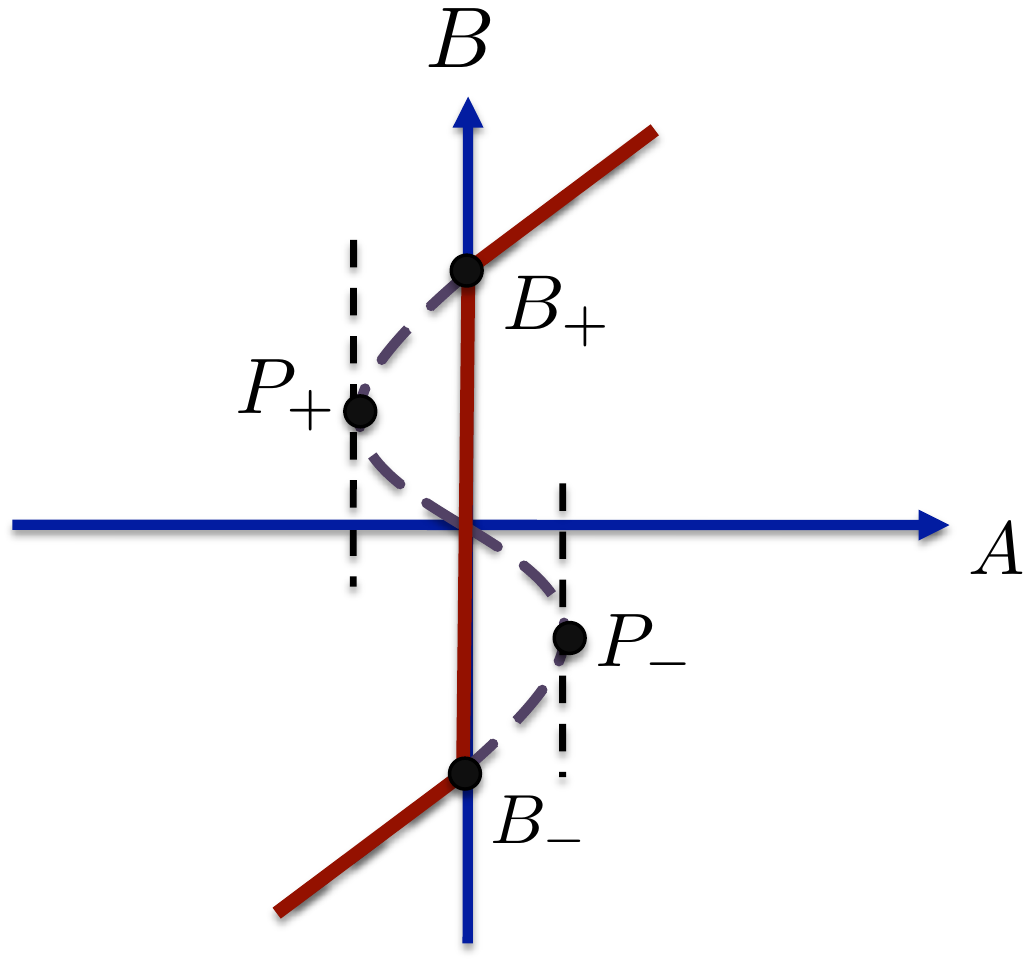}
\end{center}
\caption{{\it Left plot:} the spirals describing responses of the tower of ``Efimov'' states. $A$ and $B$ in the figure denote the source and expectation value for $\sO$ respectively.
The system under consideration has a $Z_2$  $\sO \to - \sO$ reflection symmetry, which gives rise to a pair of infinite spirals related by this symmetry. The condensed states are given by the intersections of the spirals with  the $B$-axis with $B_\pm^{(1)}$ the ground states and $B_\pm^{(2)}$ the first excited states and etc.  
The red straight line has slope given by~\eqref{limitslope}. As $\sqrt{-u} \to 0$ most part
of the spiral becomes parallel to it.  For ease of visulization a nonlinear mapping has been performed along the the major and minor axes of the spiral; while the zeros of $A$ and $B$ are preserved by this mapping the location of divergences and zeros of $\frac{dB}{dA}$ are {\it not}. 
{\it Right plot:}  Response curve in a Landau-Ginsburg mean-field model 
$ \sL_{LG} =  \ha u \psi^2 + \ha (\p \psi)^2 + \lam \psi^4 + \dots$,
where $\psi$ is a $Z_2$ order parameter and the critical point lies at $u=0$. The plot is for a small negative $u$ (condensed side). Note that the part to the right of $B_+$ behaves as $A^{1 \ov 3}$ instead of the linear behavior of the left plot since at the critical point the uniform susceptibility diverges.  
The part between $B_+ $ ($B_-$)  and $P_+$ ($P_-$) describes a metastable region while  the part between $P_+$ and $P_-$ describes an unstable region. In the left plot the region between $P_+$ and $P_-$ is replaced by a pair of infinite spirals which give rise to a tower of infinite excited states.  }
\label{fig:spiral1}
\end{figure}

It is also interesting to compute the response of the system to a  static and uniform external source (i.e $\om = k =0$) in this tower of ``Efimov'' states, which turns out to be described by a pair of continuous spiral curves given in Fig.~\ref{fig:spiral1} (see~\cite{sdwnew} for more details). $A$ and $B$ in the figure denote the source and expectation value for $\sO$ respectively. The set of Efimov states are obtained by setting the source $A=0$, i.e. they can be read from the intersections of the spirals with the $B$-axis. For comparison we also show the response curve for the standard Landau-Ginsburg  story in the right plot of~Fig.~\ref{fig:spiral1}.
 Approaching the critical point $\sqrt{-u} \to 0$, the spirals exhibit the following features:

\bi

\item The spirals are being squeezed into a straight line, with limiting slope
\be
\frac{B}{A}\bigg|_{\sqrt{-u} \to 0} = \chi_0 \  \label{limitslope}
\ee
which agrees with the value found from linear response approaching the critical point from the other side \eqref{rptsusc}.

\item The susceptibility from the condensed side, which is defined by 
\be \label{lsus}
\chi_L = {dB \ov d A}\biggr|_{A=0}, 
\ee
however,  differs from~\eqref{limitslope}, and is given by 
 \be
\begin{split}
\chi_L 
&=  \chi_0 + \chis + O(u) , \qquad \chis \equiv \mu_*^{2 \nu_U} {1 \ov 4\nu_U  \al^2}  \  .
\label{slopechange1}
\end{split}
\ee
Essentially, even though the spiral is squished into a straight line as we approach the transition, each {\it intersection} of the spiral with the $A = 0$ axis has a different slope than the limiting slope of the entire spiral. 
Note that this result is independent of $n$ and in particular applies to $n=1$, the ground state. Since $\chi_0$ is the value at $u= 0$ from the uncondensed side, we thus find a jump in the value of uniform susceptibility  in crossing $u=0$.
% and the difference is precisely the same coefficient as the divergent terms in~\eqref{nonan1}, which has also appeared in other places 
%such as~\eqref{zerwL}. 

\ei

The existence of a tower of  ``Efimov'' states with geometrically spaced expectation values may be considered as a consequence of spontaneous breaking of the discrete scaling symmetry
of the system. With an imaginary scaling exponent,~\eqref{usmS} exhibits a  discrete scaling symmetry with (for $k=0$)
\be 
\om \to e^{ {2 \pi \ov \sqrt{-u}}}  \om \  
\ee
which is, however, broken by the condensate.\nicefootnote[1]{Note that for $n=1$ state, since the physics of the condensate sets in already at $\Lam_1$, the range of validity for~\eqref{usmS} is not wide enough for the discrete scaling symmetry to be manifest.} The tower of ``Efimov'' states may then be considered as the ``Goldstone orbit'' for this broken discrete symmetry. 

At a finite temperature, using the finite $T$ generalization of~\eqref{roep211} (as discussed at the end of Sec.~\ref{lowgreen}), equation~\eqref{rptfinom} generalizes to 
\be
\chi^{(T)}(\om, k) = \chi_0 \frac{\sinh\le(\nu_k\le[ \log\le(\frac{2\pi T}{\om_b}\ri) + \psi\le(\ha - i\frac{\om}{2\pi T}\ri)\ri]\ri)}{\sinh\le(\nu_k \le[\log\le(\frac{2\pi T}{\om_a}\ri)  + \psi\le(\ha - i\frac{\om}{2\pi T}\ri)\ri]\ri)} \ ,\label{finTomsusc}
\ee 
where $\psi$ is the digamma function. Let us first look at the static regime $\om =0$, for which we find 
\be
\chi^{(T)}(\vk) = \chi_0 \frac{\sinh\le(\nu_k \log\le(\frac{T}{T_b}\ri)\ri)}{\sinh\le(\nu_k \log\le(\frac{T}{T_a}\ri)\ri)} , \label{finTsinh}
\ee
where $T_{a,b}$ some constants of order $\mu$ given by 
\be 
{T_a  }=  {4 \mu_* \ov \pi} e^{\tilde \al \ov \al}  , \qquad 
{T_b  }=  {4 \mu_* \ov \pi} e^{\tilde \beta \ov \beta}
\ . \label{defTab}
\ee
Similarly to~\eqref{usmS}, the expression for $u < 0$ is obtained by analytically continuing~\eqref{finTsinh} to  obtain 
\be
\chi^{(T)} (\vk) =  \chi_0 \frac{\sin\le(\lam_k \log \le(\frac{T}{T_b}\ri)\ri)}{\sin\le(\lam_k \log \le(\frac{T}{T_a}\ri)\ri)} \ . \label{imagFinTsusc}
\ee
Both~\eqref{finTsinh} and~\eqref{imagFinTsusc} are analytic at $u=0$ and for $T \gg \Lam_{CO}$  reduce to 
\be
\chi^{(T)} (\vk) =\chi_0  {\log {T \ov T_b} \ov \log {T \ov T_a}} + O(u, k^2)
\ .  \label{finiteTsusc}
\ee
When  $T \ll \Lam_{CO}$, for $u > 0$ equation~\eqref{finTsinh} crosses over to an expression almost identical to~\eqref{rptsusc} with $\om$ replaced by $T$. For $u < 0$, equation~\eqref{imagFinTsusc} has poles at (for $k=0$)
\begin{align}
T_n  = T_a \exp\le(-\frac{n\pi}{\sqrt{-u}}\ri) = {4 \mu_* \ov \pi}  \exp\le(-\frac{n\pi}{\sqrt{-u}}+ {\tilde \al \ov \al}\ri), 
\qquad n \in \mathbb{Z}^+ \ . \label{Tc} 
\end{align} 
Comparing to~\eqref{Tc} and \eqref{exppoles}, we see that $T_n\sim \om_n $. The first of these temperature should be interpreted as the critical temperature
\be 
T_c =  {4 \mu_* \ov \pi}  \exp\le(-\frac{\pi}{\sqrt{-u}}+ {\tilde \al \ov \al}\ri)
\ee
below which the scalar operator condenses and~\eqref{imagFinTsusc} is no longer valid.\footnote{Including frequency dependence, one can check that as $T$ is decreased through each $T_{n}$, a pole of $\chi^{(T)} (\om, \vk)$ moves from the lower half $\om$-plane to the upper half-plane.  Thus we see the interpretation of each of these $T_n$; as the temperature is decreased through each of them, one more pole moves through to the upper half-plane. There exist an infinite number of such temperatures with an accumulation point at $T = 0$; and indeed at strictly zero temperature there is an infinite number of poles in the upper half-plane, as seen earlier in~\eqref{exppoles}.  Of course in practice once the first pole moves through to the upper half-plane at $T_c = T_1$, the uncondensed phase is unstable and we should study the system in its condensed phase.  One can further study the critical behavior near the finite temperature critical point $T_c$.  Here one finds mean field behavior.} Note that equation~\eqref{imagFinTsusc} also has poles $T_n$ with $n \leq 0$. But such $T_n$ are of order or much larger than $\mu$ for which our approximation is no longer valid, and so we ignore them.

For nonzero $\om$, in the quantum critical region $T \gg \Lam_{CO}$, equation~\eqref{finTomsusc}
becomes 
\be
\chi^{(T)}(\om,\vk) = \chi_0 \frac{\log\le(\frac{2\pi T}{\om_b}\ri) + \psi\le(\ha - i\frac{\om}{2\pi T}\ri)}{\log\le(\frac{2\pi T}{\om_a}\ri)  + \psi\le(\ha - i\frac{\om}{2\pi T}\ri)} \  \label{finTexp}
\ee
which can now be applied all the way down to zero frequency.  Equation~\eqref{finTexp} reproduces~\eqref{quantcritexp} for $\om \gg T$ and~\eqref{finiteTsusc} for $\om \ll T$.

In Fig.~\ref{fig:funnel} we summarize the finite temperature phase diagram.
\begin{figure}[!ht]
\begin{center}
\includegraphics[scale=0.4]{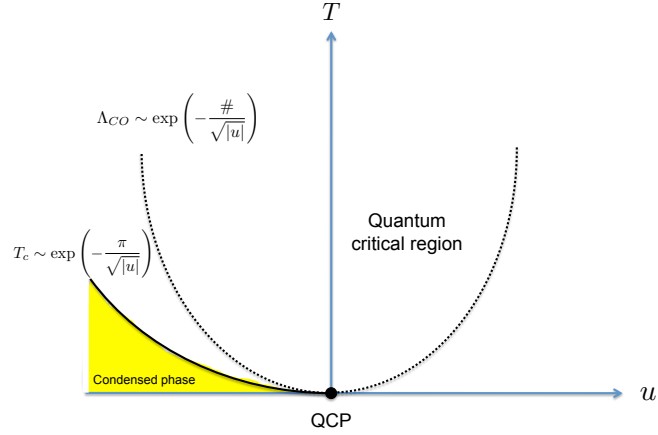}
\end{center}
\caption{Finite temperature phase diagram with the quantum critical region for bifurcating criticality as a function of $u$. The dotted line is given by $\Lam_{\rm CO}$ in~\eqref{rptco}. But note that the discussions there are not enough to determine the $O(1)$ factor in the exponent for $\Lam_{\rm CO}$. 
 The dynamical susceptibility in the bowl-shaped quantum critical region is given by~\eqref{finTexp} with the zero temperature limit given by~\eqref{quantcritexp}.}
\label{fig:funnel}
\end{figure}

\subsubsection{Physical interpretation: \Slql\ as a fractionalized phase of Bose condensates}

Now let us now try to interpret the above results. 
First we emphasize that nowhere on the uncondensed side do we see a coherent and gapless quasiparticle pole in the dynamical susceptibility, which usually appears close to a quantum phase transition and indicates the presence of soft order parameter fluctuations.  The facts that at the critical point the susceptibility~\eqref{rptsusc} does not diverge and the spectral function~\eqref{crispe} is logarithmically suppressed at small frequencies are also  manifestations of the lack of soft order parameter fluctuations. 

It has been argued in~\cite{Iqbal:2011in,sdwnew}  that the quantum phase transition at a bifurcating critical point corresponds to a confinement/deconfinement transition.  Across the critical point, the scalar operator $\sO_{k=0}$  develops a complex dimension in the \Slql\ and conformality is lost. The  loss of conformality is realized through a dynamically generated ``confinement'' scale $\Lam_{IR}$ below which an infinite tower of geometrically separated bound states of operator $\sO$ form and then Bose condense.  The physical picture here is similar to the BEC regime in a strongly interacting ultracold Fermi system where fermions form bound molecules and then Bose condense. The bifurcating QCP is thus characterized by the onset of forming bound states 
rather than by soft order parameter fluctuations; this may help explain why the susceptibility does not diverge at the critical point and the 
spectral density is suppressed.
This should be contrasted with the fermionic case discussed in Sec.~\ref{sec:HFL}, where fermionic bound states each form a Fermi surface.  In both cases, \Slql\ serves as a deconfined high energy state for the more stable lower energy states which arise from \Slql\ via bound state formation.

To summarize, for a bifurcating QCP, while the phase transition can still be characterized by an 
order parameter, the order parameter remains gapped at the critical point and the phase transition is not driven by its fluctuations.  Instead the phase transition appears to be driven by formation of bound states.

\subsection{Quantum critical behavior: a hybridized QCP} \label{sec:hybrid}

\subsubsection{From uncondensed side}
 
We now examine the critical behavior around a hybridized QCP~\eqref{hybq} (again using a neutral scalar) following~\cite{Faulkner:2010gj,sdwnew,Jensen}. 

The full dynamical behavior from the uncondensed side can be obtained by expanding 
$\tilde a_+ (\om, k)$ in~\eqref{roep211} around $\om =0$, $k=0$ and $\ka_+ = \ka_c$, which leads to 
\be 
\chi(\om, \vk) \approx \frac{1}{\kappa_+ - \kappa_c + h_k \vk^2  - h_\om \om^2 +  h \sG_k(\om) } \label{hybridizedG1}
\ee
with 
\be 
h_k = {\p_{k^2} \tilde a^{(0)}_+ (k) \ov b^{(0)}_+ (k)} \biggr|_{k=0, \ka_+ = \ka_c} , \qquad h_\om \equiv {\tilde a_+^{(2)}  (k) \ov b_+^{(0)}  (k) } \biggr|_{k=0, \ka_+ = \ka_c}, \quad h \equiv { \mu_*^{-2 \nu_k} \tilde a_-^{(0)}  (k) \ov b_+^{(0)}  (k) } \biggr|_{k=0, \ka_+ = \ka_c} \ .
\ee
Recall that the \Slql\ retarded function $\sG_k (\om) \sim \om^{2 \nu_k}$. 
From explicit gravity calculation one finds that the various constants in~\eqref{hybridizedG1}  have the following behavior: $h_k> 0,\; h < 0$, and $h_\om > 0$~(for  $\nu_{k=0} > 1$).

Setting $\om =0$ in~\eqref{hybridizedG1},  the static susceptibility $\chi (k)$
 can be written as 
\be 
\chi(\vk) \approx \frac{1}{(\kappa_+ - \kappa_c) + h_k \vk^2  } 
%, \quad  
%h_k = {\p_{k^2} \tilde a^{(0)}_+ (k) \ov b^{(0)}_+ (k)} \biggr|_{k=0, \ka_+ = \ka_c} %\ka_+ \to \ka_c  \ 
 \label{hybridizedG0}
\ee
which has a standard mean field behavior, with the spatial correlation length scaling as 
\be  \label{colen}
\xi \sim (\kappa_+ - \kappa_c)^{- \nu_{crit}}, \qquad \nu_{crit}=\ha \ .
\ee
and  critical exponents (see Appendix~\ref{app:critexp} for a review of definition of critical exponents)
\be
\gamma=1 ,  \qquad \eta=0 \ .
\ee 

The behavior of the full dynamical susceptibility~\eqref{hybridizedG1} depends on the competition between the analytic contribution $h_\om \om^2$ and the 
non-analytic contribution $\sG_k (\om)$ from \Slql. 
When $\nu_{k=0} \in (0,1)$, the non-analytic part dominates at low energies and the analytic contribution can be ignored, leading to the interesting and decidedly non-mean-field behavior
\be
 \chi(\om, \vk) \approx \frac{1}{(\kappa_+ - \kappa_c) + h_k k^2 + h C(\nu) (-i \om)^{2 \nu} } \label{univform}
\ee
where $\nu \equiv \nu_{k=0}$ and constant $C(\nu) $ is real and negative for $\nu \in (0,1)$. We will consider $k=0$ below.  At the critical point $\ka_+ = \ka_c$ we find that
\be 
\chi (\om, k=0) \sim (-i \om)^{-2 \nu} \ . \label{critom}
\ee
Away from the critical point, the relative magnitude of the two terms (with $k=0$) in the denominator of~\eqref{univform} defines a crossover energy scale 
\be 
\Lam_{\rm CO}^{(\om)} \sim |\kappa_c - \kappa_+|^{\frac{1}{2\nu}} \ . \label{Tcexp}
\ee 
For $\om \ll \Lam_{\rm CO}^{(\om)}$ one finds~\eqref{hybridizedG0} 
while for $\om \gg \Lam_{\rm CO}^{(\om)}$~\eqref{critom}.  This crossover scale~\eqref{Tcexp} defines the correlation time $\xi_\tau$ of the system
\be 
\xi_\tau \sim {1 \ov \Lam_{\rm CO}^{(\om)}} \sim  |\kappa_c - \kappa_+|^{-\frac{1}{2\nu}} \ .
\ee
Comparing the above expression with~\eqref{colen} we then find that $\xi_\tau  \sim \xi^z$ with
the dynamical exponent $z$ given by
\be \label{timeco}
z = {1 \ov \nu} \ . 
\ee
Of course this exponent can equivalently be seen by balancing the $k^2$ term and the $\om^{2 \nu}$ term in~\eqref{univform}.  Also note that when $\ka_+ < \ka_c$ equation~\eqref{univform} has a pole  in the 
upper half plane (since $h C(\nu) > 0$),
\be 
\om_{\rm pole} \sim i \Lam_{\rm CO}^{(\om)}
\ee
indicating an instability. 

When $\nu > 1$, in~\eqref{hybridizedG1}, the non-analytic part $\sG_k(\om) \sim \om^{2\nu}$ from \Slql\ becomes less important than the analytic corrections $\sim \om^2$. Since 
the term proportional to $\sG_k(\om)$ is the only complex term, it should nevertheless be kept. 
Now~\eqref{hybridizedG1} describes a long-lived (nearly gapless) relativistic particle with a small width $\Gamma \sim \om^{2\nu}$. The dynamical exponent is now $z=1$. 

At a finite temperature $T \ll \mu$,  equation~\eqref{hybridizedG1} generalizes at leading order in $T/\mu$ to~(see discussion at the end of Sec.~\ref{lowgreen}), 
\be 
\chi(\om, \vk; T) \approx \frac{1}{\kappa_+ - \kappa_c + h_k \vk^2  - h_\om \om^2 + h_T T +  h \sG_k^{(T)} (\om)  } \label{hybridizedGT}
\ee
where $h_T T$ ($h_T$ a constant)\footnote{From explicit gravity calculation one finds  $h_T > 0$ for $\nu_{k=0} > \ha$.}  comes from (analytic) finite temperature corrections to $a_+$ and $b_+$.  The finite temperature \Slql\ retarded function $\sG_k^{(T)} (\om)$ has the form  $\sG_k^{(T)} (\om)= T^{2 \nu_k} g_b({\om \ov T}, \nu_k)$ with $g_b$ a universal scaling function~(see~\eqref{finiteTch1} and~\eqref{univgF}).  Consider first $\om =0$
in~\eqref{hybridizedGT}, which is 
\be \label{finrT}
\chi^{(T)} \sim {1 \ov \ka_+ - \ka_c + h_T T + h C(\nu) T^{2 \nu}} \ .
\ee 
It is interesting that the analytic contribution now dominates for $\nu > \ha$. For $\nu < \ha$ we find that there is a pole at 
\be 
T_c \sim (\ka_c - \ka_+)^{1 \ov 2 \nu} 
\ee
for $\ka_+ < \ka_c$. This should be interpreted as the critical temperature for a thermal phase transition, above which the instability disappears. From the uncondensed side, such a temperature scale coincides with the crossover scale~\eqref{Tcexp}. For $\nu > \ha$, we find instead mean field behavior 
\be 
T_c \sim (\kappa_c - \kappa_+) \  \label{mfT_c}
\ee
and the finite temperature crossover scale becomes 
\be 
\Lam_{\rm CO}^{(T)} \sim  |\kappa_c - \kappa_+|
\ee
which no longer tracks the scale~\eqref{Tcexp} of zero temperature. With nonzero $\om$, when $\nu < \ha$, we 
find $\om/T$ scaling, while for $\nu > \ha$ there is no $\om/T$ scaling. 

\subsubsection{Low energy effective action}

The discussion in Sec.~\ref{sec:semi} leading to the low energy effective action~\eqref{hybid} for the fermionic case can be applied almost identically to a scalar field. In particular, near a hybridized QCP~\eqref{hybq}, $K$ in~\eqref{eff3} can be expanded near $\om =0$ and $k=0$ as (say for a neutral scalar)
\be \label{kcoe}
K = {\tilde a_+ (\om, k) \ov b_+ (\om,k)} = \kappa_+ - \kappa_c + h_k \vk^2 - h_\om \om^2 + \dots \ .
\ee
We thus find that near a hybridized QCP, the low energy effective theory can be written as 
\be \label{thep}
S_{eff}[\psi,\Phi] = \tilde S_{\Slql} [\Phi] + \lam \int  dt \, \Phi \psi  + S_{LG} + \int \psi J
\ee
where $\tilde S_{\Slql}$ is the same as~\eqref{newsl} (except that $\Phi$ is now a scalar operator)
and 
\be \label{LGQ}
S_{LG} =  - \ha \int \, \psi K (k,\om) \psi 
\ee
can be considered as the quadratic part of the Landau--Ginsburg action for a scalar order parameter $\psi$ near criticality. The full low energy action~\eqref{thep} describes 
the {\it hybridization} between a Landau-Ginsburg sector and a strongly coupled sector controlled by \Slql. This mixing between two sectors is the key to understanding the critical physics, and is why we call this a {\it hybridized} critical point. 

The behavior of the dynamical susceptibility at $T=0$ discussed earlier and in particular the crossover to mean field behavior at $\nu = 1$ can be readily seen from the effective action~\eqref{thep}. Recall that $\Phi$ has \Slql\ scaling dimension $\ha + \nu$, and from the second term in~\eqref{thep},  $\psi$ has dimension $\ha - \nu$. 
For $\nu > 1$, the dimension for $\psi$ become smaller than $-\ha$, and as a result the kinetic term $(\p_t \psi)^2$ (coming from $\om^2$ term  in~\eqref{kcoe}) in~\eqref{LGQ}  becomes relevant and  more important than the hybridization term $\Phi \psi$ (which is marginal by definition) in~\eqref{thep}. Alternatively, we can now assign $-\ha$ as dimension of $\psi$  using  $(\p_t \psi)^2$, under which the 
hybridization term $\Phi \psi$ will have dimension $\nu$ which becomes irrelevant for $\nu > 1$. 

The  crossover to mean field behavior at $\nu = \ha$  at a finite temperature discussed in~\eqref{finrT}--\eqref{mfT_c}  can also be readily seen from the effective action~\eqref{thep}. 
Finite temperature can be thought to generate a term $\int dt \, T \psi^2$, which becomes relevant when the dimension of $\psi$ becomes smaller than zero, i.e. for $\nu > \ha$. Alternatively we can now use  $T  \psi^2$ term to assign dimension $0$ to $\psi$, under which the 
hybridization term $\Phi \psi$ then becomes irrelevant for $\nu > \ha$ as now the dimension for $\Phi$ becomes larger than $1$. 

Finally, we mention that for $\nu < \ha$ there is an alternative description in which the SLQL theory is studied in its alternative quantization, i.e. a different IR operator $\Phi_-$ is used that has dimension $\ha- \nu$. We do not go into details of this description here except to point out that in this description the critical point $\ka_+ = \ka_c$ can be understood as tuning the relevant deformation $\Phi_-^2$ in the SLQL to $0$. This is thus consistent with the standard Landau-Ginsburg-Wilson paradigm of a critical point as a quantum field theory with a single relevant direction; however the interaction of this relevant direction with the sector described above by the field $\psi$ is nontrivial, resulting in the physics described above. We refer the reader to \cite{Faulkner:2010gj,sdwnew} for more discussion of this.

\subsubsection{Free energy: condensed side}

Finally let us look at the the scaling of the expectation value and the free energy from the condensed side. This can be done by analyzing the condensed solution on the gravity side~\cite{Faulkner:2010gj}. Alternatively, one could use the low energy effective action~\eqref{thep}~\cite{Jensen,sdwnew} which provides a significant simplification. We will use the latter approach. We first generalize~\eqref{thep} by including the next nonlinear term for $\psi$ in~\eqref{LGQ}, i.e. 
\be \label{LGQ1}
S_{LG} =  - \ha \int \, \psi K (k,\om) \psi %+ h_\om \om^2 ) 
-  u  \int dt \, \psi^4 + \dots
\ee
with $u$ some positive constant.

%We will consider 
%$k=0$ and again denote $\nu_{k=0}$ simply as $\nu$.  Since  $\Phi$ has a scaling dimension $\ha + \nu$ and $\psi$ has dimension $\ha - \nu$, $J$ has the same dimension $\ha + \nu$ as $\Phi$. The spatial momentum $k$ does not scale, thus its IR dimension is zero. 
Now consider that $\psi$ develops some nonzero expectation value. From the relative scaling dimensions between $\Phi$ and $\psi$, we can then write $\Phi$ as 
\be 
\Phi \sim \psi^{\ha + \nu \ov \ha - \nu}
\ee
and the free energy density $F$ corresponding to~\eqref{thep} can then be written as 
\be  \label{nifree}
F \sim C  \psi^{1 \ov \ha - \nu}  + \ha (\ka_+ - \ka_c) \psi^2 + u \psi^4 
\ee
where the first term comes from the $\Phi \psi$ term with $C$ some constant. Equation~\eqref{nifree} can also be derived from a detailed bulk analysis  which also gives that $C > 0$ for $\nu < \ha$. 
 Now notice that 
for $\psi$ small, the first term dominates over the $\psi^4$ term if $\nu < {1 \ov 4}$, while the Landau-Ginsburg $\psi^4$ term dominates for $\nu > {1 \ov 4}$. In other words, since the first term is marginal by assignment, $\psi^4$ term becomes relevant when $\nu > {1 \ov 4}$.\nicefootnote[1]{Some readers might worry that higher powers like $\psi^6$ may also become relevant at some point (for example for $\nu > {1 \ov 3}$). But note that once the last two terms in~\eqref{nifree}
dominate we should reassign the dimension of $\psi$ and the standard Landau-Ginsburg story applies.} 

For $\nu < {1 \ov 4}$ we can ignore the last term in~\eqref{nifree} and for $\ka_+< \ka_c$ find that 
\be \label{ovev}
\vev{\sO} \sim \psi \sim  (\ka_c - \ka_+)^{\ha - \nu \ov 2 \nu}
\ee
and as a result 
\be 
F \sim (\ka_c - \ka_+)^{1 \ov 2 \nu} \ . 
\ee
Including the source $J$, which has dimension $\ha + \nu$, the free energy should then be given by a scaling function 
\be \label{scalFr}
F = (\ka_c - \ka_+)^{1 \ov 2 \nu} f_1 \le(J  (\ka_c - \ka_+)^{-{\ha +\nu \ov 2 \nu}}\ri)
= \xi^{-{1 \ov \nu}} f_2 \le(J \xi^{\ha + \nu \ov \nu}\ri)
\ee
where in the second equality we have expressed the free energy in terms of correlation length using~\eqref{colen}. From~\eqref{scalFr} we can also deduce that at the critical point we should have 
\be \label{onMexp}
\vev{\sO} \sim \psi \sim J^{\frac{\ha - \nu}{\ha + \nu}}  \ .
\ee 
From~\eqref{ovev},~\eqref{onMexp} and~\eqref{scalFr} we can collect the values of various scaling exponents (see Appendix~\ref{app:critexp} for a review of exponents)
\be \label{varexp}
\al=  2-{1\ov 2\nu}, \quad  \beta= {\ha - \nu \ov 2\nu} , \quad 
\delta =  {\ha + \nu \ov \ha - \nu}
\ee 

For $\nu > {1 \ov 4}$, we can ignore the first term in~\eqref{nifree} and the analysis becomes the standard Landau-Ginsburg one. As a result, the behavior near the critical point becomes those of mean field~\cite{Faulkner:2010gj}. 
We thus find that for $\nu > {1 \ov 4}$, 
\be \langle\sO \rangle \sim \psi \sim  (\kappa_c - \kappa_+)^\ha \qquad F \sim -(\kappa_c - \kappa_+)^{2} \qquad \langle\sO\rangle_{\kappa _+= \kappa_c} \sim J^{\frac{1}{3}} \ 
\ee
and various exponents become
\be 
\al=  0, \quad  \beta= \ha , \quad 
\delta =  3 \ 
\ee 
which agree with the values of~\eqref{varexp} for $\nu = {1 \ov 4}$. 

\subsubsection{Summary}

We summarize the somewhat intricate critical behavior near a hybridized QCP for various values of $\nu$ in the following table 
\begin{center}
\begin{tabular}{c||c|c|c|c}
Quantity& $\nu\in\le(0,\frac14\ri)$ & $\nu\in\le(\frac14,\frac12\ri)$ &  $\nu\in\le(\frac12,1\ri)$  &  $\nu>1$ \\
\hline
$\om/T$ scaling & yes & yes & no & no \\ 
$\Lam^{(\om)}_{\rm CO}, z  $ & HYB & HYB & HYB & MFT\\
$T_c $ & HYB & HYB & MFT & MFT\\
$\al, \ \beta,\ \delta$ & HYB & MFT & MFT & MFT\\
$\gamma, \ \eta, \nu_{crit}$ & MFT & MFT & MFT & MFT 
\end{tabular}
\end{center} 
In the above ``MFT'' denotes mean-field behavior and ``HYB'' (which is for hybridized) refers to the following  non-mean-field exponents for zero-T crossover scale $\Lam^{(\om)}_{\rm CO}$, the dynamical exponent $z$ and critical temperature $T_c$, 
\bea	 
\Lam^{(\om)}_{\rm CO} \sim  \le(\kappa_c-\kappa_+\ri)^{1 \ov 2\nu}, \quad
 z = {1\ov \nu} , \quad T_c  \sim  \le(\kappa_c-\kappa_+\ri)^{1 \ov 2\nu}
 \eea
and 
\be \label{varexp1}
\al=  2-{1\ov 2\nu}, \quad  \beta= {\ha - \nu \ov 2\nu} , \quad 
\delta =  {\ha + \nu \ov \ha - \nu} \ .
\ee 
For all values of $\nu$ the static susceptibility from the uncondensed side is always given by the mean field behavior~\eqref{hybridizedG0}.

Note in particular that for $\nu \in (0,\ha)$ one  finds an expression 
\be \label{laos}
\chi(\om, \vk; T) \approx \frac{1}{\kappa_+ - \kappa_c + h_k \vk^2   +  h T^{2\nu} g_b ({\om \ov T})  }
\ee
with $g_b$ given by~\eqref{univgF} (with $q_* =0$). Equation~\eqref{laos} is of the form~\eqref{cechi} 
for the dynamical susceptibility of $CeCu_{6-x} Au_x$ with a choice of $2 \nu = 0.75$~\cite{Faulkner:2010gj}, although the universal function $g_b$ appears to be different from the $f (\om/T)$ there obtained by fitting experimental data.

\subsection{Doubly fine-tuning to a marginal critical point} \label{sec:doucri}

We have just described two different kinds of fixed point; the bifurcating transition involves tuning $u \to 0$, and the hybridized transition involves tuning $\ka_+ \to \ka_c$.
We can obtain a {\it new} kind of critical point by tuning $u \to 0$ and $\ka_+ \to \ka_c$ at the same time, at some critical value $\kappa_c^*$ given by
\be
\kappa_c^* = -\frac{\al}{\beta} \label{kappstar}
\ee 
where we have used the $\nu \to 0$ limit of the expression for $\kappa_c$ given in \eqref{stare} and $\al, \beta$ are defined in~\eqref{11as}--\eqref{req}. %What is the nature of this critical point? 
We will fix $u=0$ and vary $\ka_+$.

The dynamical susceptibility can be obtained from 
 the $\kappa_+ = 0$ result at the bifurcating critical point~\eqref{quantcritexp} and~\eqref{corrD} 
 \be
\chi^{(\kappa_+)}(\om, k) = \frac{\log\le(\frac{\om}{\om_b}\ri) - i\frac{\pi}{2}}{\frac{\al}{\beta}\le(\log\le(\frac{\om}{\om_a}\ri) - i\frac{\pi}{2}\ri) + \kappa_+\le(\log\le(\frac{\om}{\om_b}\ri) - i\frac{\pi}{2}\ri)}.
\ee
Now using the definitions of $\om_{a,b}$ in~\eqref{veifp}, applying the Wronskian relation \eqref{W1}, and expanding near $\kappa_+ \sim \kappa_c^* \equiv -\frac{\al}{\beta}$, we find:
\be
\chi^{(\kappa_+)}(\om, k) =  \frac{\log\le(\frac{\om}{\om_b}\ri) - i\frac{\pi}{2}}{(\kappa_+ - \kappa_c^*)\le(\log\le(\frac{\om}{\om_b}\ri) - i\frac{\pi}{2}\ri) - \frac{1}{2\nu_U \beta^2}} \ . \label{margT0}
\ee
This contains a good deal of interesting physics. First let us keep $\kappa_+ > \kappa_c^*$ and study the very low-frequency behavior: expanding in powers of the inverse logarithm, we find
\be
\chi^{(\kappa_+)}(\om \to 0, k) \sim \frac{1}{\kappa_+ - \kappa_c^*} + \sO\le(\frac{1}{\log{\om}}\ri). 
\ee
Thus the static susceptibility diverges as $\kappa \to \kappa_c^*$, just as for the hybridized critical point discussed above. Away from the critical point the leading contribution to the spectral density appears at order $\log^{-2}(\om)$,  
\be
\Im \chi^{(\kappa_+)}(\om \to 0, k) \sim \frac{\pi}{4 (\kappa_+ - \kappa_c^*) \nu_U^2}\log^{-2}\le(\frac{\om}{\om_b}\ri) + \sO\le(\frac{1}{\log^{3}{\om}}\ri) \label{ImGnearcrit}
\ee
This is the same as the corresponding result close to a bifurcating critical point with zero double-trace deformation. 

Next, we note that~\eqref{margT0} contains a pole in the upper half plane at
\be
\om_* = i\om_b \exp\le(\frac{1}{2\nu_U \beta^2} \frac{1}{\kappa_+ - \kappa_c^*}\ri)\ee
Note however that if $\kappa_+ > \kappa_c^*$, the pole is at an exponentially high energy scale, this low-frequency formula breaks down far before the pole, and the pole is not physical. This is the stable phase. However, if $\kappa_+ < \kappa_c^*$, then close to the critical point the pole is very close to the origin and this formula is valid. The instability pole is now physical, and we see that a new low-energy scale has been exponentially generated. This is reminiscent of the effect of a marginal coupling growing strong in the infrared, as in QCD or the BCS instability. Indeed as discussed in~\eqref{fulef} and~\eqref{eff1}, we expect in the IR a double trace deformation to \Slql\ will be generated.  At $u=0=k$, the double trace deformation 
\be
\delta S =  \frac{\kappa^{(IR)}_{m}}{2} \int dt \, \Phi^2, 
\ee
has dimension $1$ and is thus {\it marginal}. It can be checked explicitly that the value of the effective marginal double-trace parameter $\kappa^{(IR)}_m$ changes sign at $\kappa_+ = \kappa_c^*$ (although one must be somewhat careful; for details see~\cite{sdwnew}). Thus what we have found above is that if $\kappa_+ > \kappa_c^*$ then the coupling is marginally irrelevant and has no effect in the IR, whereas if $\kappa_+ < \kappa_c^*$ it is marginally {\it relevant} and drives the theory to a new fixed point (i.e. the scalar will condense).

Similarly, one can compute the corresponding finite temperature correlator using~\eqref{finTexp}; a precisely analogous calculation gives
\be
\chi^{(\kappa_+)}(\om,k; T) = \frac{\log\le(\frac{2\pi T}{\om_b}\ri) + \psi\le(\ha - i\frac{\om}{2\pi T}\ri)}{(\kappa_+ - \kappa_c^*)\le(\log\le(\frac{2\pi T}{\om_b}\ri) + \psi\le(\ha - i\frac{\om}{2\pi T}\ri)\ri) - \frac{1}{2\nu_U \beta^2}}  \ .\label{margTf}
\ee  
From here we can see that for $\kappa_+ < \kappa_c^*$ the static susceptibility diverges at the critical temperature 
\be
T_c = T_b \exp\le(\frac{1}{2\nu_U \beta^2} \frac{1}{\kappa_+ - \kappa_c^*}\ri) \label{margTc}
\ee
above which the system is stable. The temperature is set by the same dynamically generated scale.

Now we consider the fluctuation spectrum at criticality. Returning to $T = 0$, we now keep $\om \neq 0$ and sit precisely at the critical point $\kappa = \kappa_c^*$. From \eqref{margT0} we now find 
\be
\chi^{(\kappa_+ = \kappa_c^*)}(\om, k) = -2\nu_U\beta^2 \le (\log\le(\frac{\om}{\om_b}\ri) - i \frac{\pi}{2}\ri) \label{zeroTatcrit}
\ee
The appearance of a pure logarithm in the {\it numerator} of this propagator at criticality is interesting. Relatedly, at criticality the spectral density 
\be \label{margc}
\Im \chi^{(\kappa_+ = \kappa_c^*)} = \pi\nu_U\beta^2 \mbox{sgn}(\om)
\ee
is a pure step function \nicefootnote[1]{The logarithm jumps by $i\pi$ as we cross through $\om = 0$, resulting in the step function; note that this was necessary in order to maintain the relation $\om \Im \chi(\om) > 0$, true for any bosonic spectral density.}. This should be contrasted with the situation away from criticality, \eqref{ImGnearcrit}, in which there is a logarithmic suppression at low frequencies; as expected, by sitting at criticality we find significantly more low frequency modes.

Similarly, setting $\kappa \to \kappa_c^*$ in the finite temperature expression \eqref{margTf} we find 
\be 
\chi^{(\kappa_+= \ka_c^*)}(\om,k; T) =- 2\nu_U \beta^2 \le(\log\le(\frac{2\pi T}{\om_b}\ri) + \psi\le(\ha - i\frac{\om}
{2\pi T}\ri) \ri)
 \label{margTf1}
\ee  
Taking the imaginary part we find
\be
\Im \chi^{(\kappa_+ = \kappa_c^*)}(\om, k; T) = \pi \nu_U \beta^2 \tanh\le(\frac{\om}{2\pi T}\ri) \label{finiteTatcrit}
\ee
which is simply a smoothed-out version of the step function that we find at zero temperature. See Fig.~\ref{fig:funnel4} for the finite temperature phase diagram for a marginal critical point. 

This form of the finite-temperature fluctuation spectrum is potentially of great physical interest. Note that \eqref{finiteTatcrit} implies that 
\be
\Im \chi^{(\kappa_+ = \kappa_c^*)}(\om, k;T) \sim \begin{cases} \frac{\om}{T} \qquad\om \ll T \\ \mbox{sgn}(\om) \qquad \om \gg T \end{cases}
\ee
which is {\it precisely} of the form for  spin and charge fluctuations in the phenomenological 
``Marginal Fermi liquid'' \cite{Varma89} description of High-$T_c$ cuprates in the strange metal region (as discussed around equation~\eqref{eqn:fluc} in Sec.~\ref{sec:lqc}). Thus this marginal critical point can be viewed as a concrete realization of the bosonic fluctuation spectrum needed to support a Marginal Fermi liquid.

\begin{figure}[!ht]
\begin{center}
\includegraphics[scale=0.4]{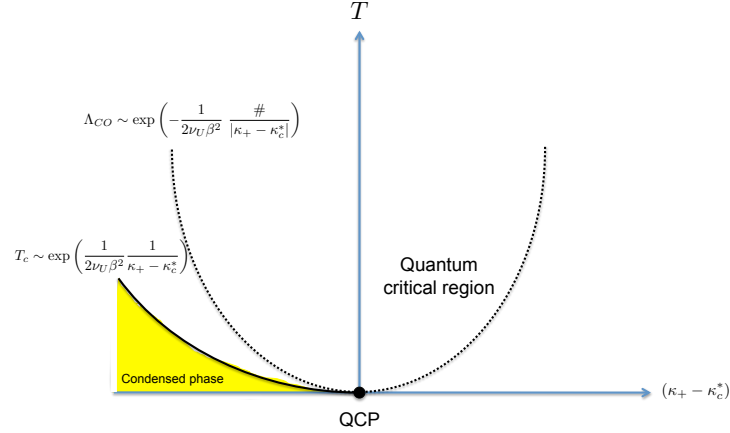}
\end{center}
\caption{Finite temperature phase diagram with the quantum critical region for marginal criticality at $u=0$ and changing $\le(\kappa_+-\kappa_+^*\ri)$. The susceptibility in the bowl-shaped quantum critical region is given by~\eqref{margTf1} with the $\om \gg T$ limit given by~\eqref{zeroTatcrit}}.  
\label{fig:funnel4}
\end{figure}

\subsection{Summary}

In this section we have discussed several types of quantum critical points related to \Slql\ which to different degrees lie  outside the Landau-Ginsburg-Wilson paradigm:

\bi 

\item A hybridized QCP describes an order parameter $\psi$ with a Landau-Ginsburg effective action $S_{LG}$  hybridized with an operator $\Phi$ in \Slql.
Depending  on the scaling dimension of $\Phi$ in the \Slql, the phase transition could exhibit a rich spectrum of critical behavior, including locally quantum critical behavior with nontrivial  $\om/T$ scaling.  At the level of effective theory, this critical point lies mildly outside the standard Landau paradigm, as the phase transition is still driven by soft fluctuations of the order parameter and all the critical behavior is fully captured by~\eqref{thep}. As discussed in Sec.~\ref{sec:bosvsferm}, on the gravity side the Landau-Ginsburg sector is associated with 
the appearance of certain scalar hair in the black hole geometry, which lies outside the AdS$_2$ region.

\item A bifurcating QCP arises from instabilities of the \Slql\ itself to a confined state and is not driven by soft order parameter fluctuations. On the condensed side, a scalar operator develops a complex scaling 
dimension in the \Slql, generating a tower of bound states, which then Bose-Einstein condense (the possible bound-states form a geometric
series of exponentially generated scales).\footnote{ \Slql\ may be considered as a  ``deconfined'' state  in which the composite bound states deconfine and fractionalize into more fundamental degrees of freedom.} 
 In particular, one finds a finite critical susceptibility with a branch point singularity, and the response of condensed states is described by an infinite spiral. As discussed in~\cite{sdwnew}, at a field theoretical level, underlying these features is the annihilation (and moving to the complex plane of a coupling constant) of two conformal fixed points, which is very different from the standard Landau-Ginsburg-Wilson paradigm of phase transitions, which can be characterized as a single critical CFT with some relevant directions. 
On the gravity side, for charged operators the instabilities of the \Slql\ can be understood as the pair production of charged particles which then subsequently backreact on the geometry. For a neutral scalar operator, the instability is related to the violation of the BF bound in the AdS$_2$ region. 

\item A marginal QCP can be obtained by sitting at the critical point of a bifurcating QCP and then dialing  
the external parameter which drives a hybridized QCP. Here given the critical theory describing a bifurcating QCP (which comes from the merger of two fixed points), the phase transition can be described as the appearance of a marginally relevant operator.  
 Interestingly, the fluctuation spectrum that emerges is identical to the bosonic fluctuation spectrum that (when coupled to a Fermi surface) is postulated to underly the ``Marginal Fermi Liquid'' description of the optimally doped cuprates \cite{Varma89}, making this critical point of potential  importance. 

\ei

\begin{figure}[!ht]
\begin{center}
\includegraphics[scale=0.6]{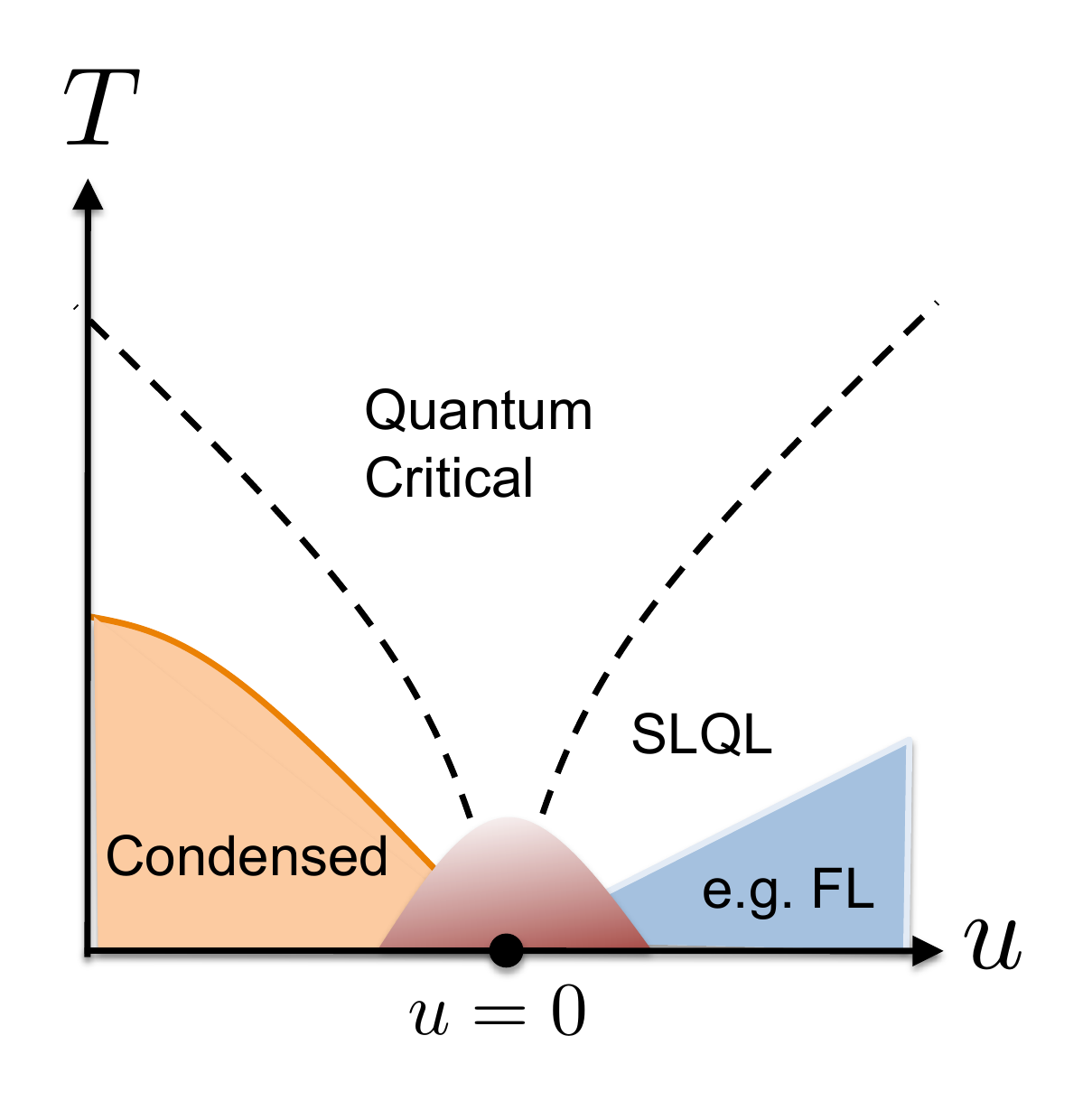}
\end{center}
\caption{How to interpret the quantum critical behavior of this section.  \Slql\ should be interpreted as a universal intermediate phase which orders into some other phases, such as a Fermi liquid, at lower energies. The dome in the figure outlines a region near the critical point to which our current discussion does not have access. 
The discussion of this section only describes the quantum critical behavior outside the 
dome-shaped region. 
} 
\label{fig:inter}
\end{figure}

Finally let us elaborate on an important point, which we have glossed over in our discussion so far. As discussed in Sec.~\ref{sec:inter}, \Slql, which describes the disordered phase in our examples 
above, should be interpreted an intermediate-energy phase, rather than a genuine ground state. 
That is, we expect \Slql\ to order into some other phases at lower energies, which may not be visible 
at the large $N$ limit we are working with. An example discussed in Sec.~\ref{sec:HFL} is that at some exponentially small scale in $N^2$, \Slql\ orders into a Fermi liquid phase.\footnote{See~\cite{Edalati:2011yv} for a recent discussion of nucleation of a neutral order parameter in a Fermi liquid-like phase.} Thus the quantum critical behavior found in this paper should be more correctly interpreted as describing the intermediate-region indicated in Fig.~\ref{fig:inter}.

\section{Discussion and conclusions} \label{sec:conc}

In these lectures, we used gauge/gravity duality to study a class of finite density systems, which are described on the gravity side by a charged black hole. We showed that the duality 
 predicts a universal intermediate-energy
phase, called a {\it semi-local quantum liquid} (\Slql). Such an unstable phase is characterized by a finite spatial correlation length, but an infinite correlation {\it time} and associated nontrivial scaling behavior in the time direction, as well as a nonzero entropy density. We discussed two sets of phenomena related to the \Slql:

\ben 

\item Gapless fermionic or bosonic degrees of freedom (described by mean field)
hybridized with degrees of freedom in \Slql. In either case, the system can be described by the following 
low energy effective  theory 
\be \label{hybid1}
S_{eff} =  \tilde S_{\Slql} [\Phi] + \int \, {\lambda (k, \om) } \Phi_{-\vk} \Psi_\vk  + S_{\rm mean \; field} [\Psi] \ .
\ee
In the fermionic case, $\Psi$ denotes excitations around a Fermi surface with $S_{\rm mean \; field} [\Psi]$ given by that for a free fermion (plus possibly higher order self-interactions).  In the bosonic case, $\Psi$ denotes the order parameter around a quantum critical point with $S_{\rm mean \; field} [\Psi]$ its Landau-Ginsburg effective theory. In both cases, $\Phi$ is some (fermionic or bosonic) operator whose dynamics is controlled by the strongly coupled \Slql. The IR behavior of the theory (i.e. whether one finds a Fermi surface with or without quasiparticles, or a quantum phase transition which obeys locally quantum critical or mean-field behavior) 
depends crucially on the scaling dimension of $\Phi$ in the \Slql. With a suitable choice of the \Slql\ scaling dimension, IR phenomena which strongly resemble those of high-$T_c$ cuprates in the strange metal phase (including a linear resistivity) and that for $Ce Cu_{6-x} Au_x$ near a quantum critical point were found. 

On the gravity side the gapless degrees of freedom are associated with 
the appearance of certain scalar or fermionic hair in the black hole geometry, and lie outside the AdS$_2$ region which describes the \Slql. 

\item Instabilities of the \Slql\ itself. We discussed examples in which \Slql\ orders at lower energies to a Fermi liquid, superconductor, or antiferromagnetic-type state.  In all cases certain operators develop a complex scaling 
dimension in the \Slql\, generating a tower of bound states. In the fermion case, each of these bound states 
forms a Fermi surface. In the scalar case, they form a tower of Bose-Einstein condensed states. In particular, near the bifurcating critical point where the scalar instability sets in one finds a finite susceptibility and the response is described by an infinite spiral.  In all of these examples the lower energy states have no zero-temperature entropy density and  \Slql\ may be considered as a  ``deconfined'' state for them in which the composite bound states deconfine and fractionalize into more fundamental degrees of freedom. This may help explain the nonzero entropy density for the \Slql.  
As discussed in Sec.~\ref{sec:inter}, we expect this picture to be generic;
while the precise nature of the lower energy state depends on the specific dynamics of the individual system, the \Slql\ emerges universally at intermediate energies through {\it fractionalization}. 

Instabilities of \Slql\ in turn provide novel mechanisms for the emergence of Fermi liquids, superconductors and AFM-type of states at low energies.

On the gravity side, for charged operators the instabilities of the \Slql\ manifest themselves as pair production of charged particles which then subsequently backreact on the geometry. For a neutral scalar operator, the instability is related to the violation of the BF bound in the AdS$_2$ region.

\een 

It is natural to ask whether some phase similar to the \Slql\ could underlie some known strongly correlated condensed matter systems, and in particular, whether the observed non-Fermi liquid behavior and novel superconductivities in
various systems could be attributed to a similar intermediate-energy phase.  

 When one encounters scaling behavior in an observable, 
an important immediate question is whether the behavior is due to intermediate-energy or vacuum effects.  
Here we provide an explicit example in which Marginal Fermi Liquid behavior -- such as that proposed for 
high-$T_c$ cuprates in the strange metal phase (including linear resistivity) -- can arise from intermediate energy effects. Local quantum critical behavior similar to that for $CeCu_{6-x}Au_x$ near a quantum critical point also appears here as due to intermediate energy effects. 

In many heavy electron systems, while quantum fluctuations from the quantum critical point corresponding to the onset of magnetism provide a natural starting point for understanding the observed non-Fermi liquid behavior and sometimes novel superconductivity, 
however, such exotic behavior does not always arise in the proximity of a quantum critical point (see e.g. ~\cite{stewart}). There thus appears  plenty of room for them to arise from an intermediate-energy phase.
 
More generally, we expect that candidates for the \Slql\ to occur include  
systems which exhibit frustrated or competing interaction terms in their Hamiltonian. Such systems can have a large number of near-degenerate states near the vacuum, similarly to the holographic systems considered here. Also systems which involve strong competition between tendencies towards itinerancy and localization could exhibit the semi-local behavior found here.
 As mentioned earlier, there are also some tantalizing parallels between properties of the \Slql\ phase and models from dynamical mean field theory (DMFT)~\cite{DMFT}, such as scaling in the time direction with spatial directions as spectators, and a finite entropy. 
 The fact that DMFT techniques have been very successful in treating many materials in some intermediate energy region provides some comfort in thinking 
that an underlying universal intermediate-energy phase like the \Slql\ could be at work.

\vspace{0.2in}   \centerline{\bf{Acknowledgements}} \vspace{0.2in} We thank T.~Faulkner, J.~McGreevy, D.~Vegh, and Q.~Si for collaborations. We also want to thank participants of TASI 2010, KITPC workshop for ``AdS/CM and other approaches,'' and IIP school on ``Holographic View of Condensed Matter Physics''
for many questions and discussions, which helped shape  these lectures.
We also want to thank K.~Balasubramanian, H.~Ebrahim, T. ~Grover, G.~Kotliar, J.~Maldacena, J.~Polchinski, S.~Sachdev for useful discussions. 
  Work supported in part by funds provided by the U.S. Department of Energy (D.O.E.) under cooperative research agreement DE-FG0205ER41360.

\begin{appendix}

%%%%%%%%%%%%%%%%%%%%%%%%%%%%%%%%%%%%%%%%%%%%%%%%
\section{Operator dimensions and retarded functions in AdS$_2$} \label{app:ads2}
%%%%%%%%%%%%%%%%%%%%%%%%%%%%%%%%%%%%%%%%%%%%%%%%

Here we give a derivation of 
the scaling dimension~\eqref{nudef}-\eqref{opep} and retarded Green's function $\sG_k (\omega)$~\eqref{iiRc}
for a charged scalar in AdS$_2 \times \RR^{d-1}$. 
The background metric and gauge field is given by~\eqref{ads2M} which we copy here for convenience
\be 
ds^2 = {R_2^2 \ov \zeta^2} (-dt^2 + d \zeta^2) + \mu_*^2 R^2 d\vec x^2, 
\qquad A_t = {e_d \ov \zeta} \ .
\ee 
We consider the following quadratic scalar action
\be \label{DscaA}
S = -\int  d^{d+1} x \sqrt{-g} \, \left[g^{MN} (\p_M + i q A_M) \phi^* (\p_N - i q A_N ) \phi  + m^2 \phi^* \phi \right] \ .
\ee
Expanding $\phi$ in terms of Fourier modes in spatial directions, 
\be
\phi (t, \vec x , \ze) = \int {d \vk \ov (2 \pi)^d} \, e^{ i \vec k \cdot \vec x} \, \phi_\vk (t,\zeta)
\ee
equation~\eqref{DscaA} reduces to an action for $\phi_\vk$ in AdS$_2$ 
\be \label{2scaA}
S = -\int  d^{2} x \sqrt{-g} \, \left[g^{ab} (\p_a + i q A_a) \phi_\vk^* (\p_b - i q A_b ) \phi_\vk  + m^2_\vk \phi^*_\vk \phi_\vk \right]
\ee
with a mass square 
 \be \label{effM}
 m_k^2 \equiv {k^2  \ov \mu_*^2 R^2} + m^2, \qquad k^2 = |\vk^2| \ .
 \ee
The indices $a,b$ only run over $t, \ze$. Further going to frequency space one finds that equation of motion from~\eqref{2scaA} can be written as (for notational simplicity we will just denote $\phi_\vk (\om, \ze)$ as $\phi$ below)
 \be \label{Aeep2}
 -\p_\zeta^2 \phi + V (\zeta) \phi =0
 \ee
 with
 \be \label{Aeoep}
 V(\zeta) = {m^2_k R_2^2 \ov \zeta^2} - \(\om + {q_*  \ov \zeta} \right)^2 = - \om^2 - {2 \om q_* \ov \zeta} + {\nu^2_k - {1 \ov 4} \ov \zeta^2} 
 \ 
\ee
where 
\be \label{ynuk}
\nu_k \equiv \sqrt{m^2_k R_2^2 - q^2_*  + {1 \ov 4}} = \sqrt{{m^2 R^2_2}- q_*^2 + {1 \ov 4}  +  {k^2 R_2^2 \ov  \mu_*^2 R^2} } 
, \qquad q_* \equiv q e_d \ .
\ee
Approaching the AdS$_2$ boundary $\ze \to 0$, equation~\eqref{Aeep2} has general solution 
\be \label{nuej0}
\phi_\vk (\om, \zeta \to 0) \sim \al \zeta^{\ha - \nu_k} + \beta \zeta^{\ha + \nu_k} \ .
\ee
Compare~\eqref{nuej0} to~\eqref{nearBdyScal}; it implies that the conformal dimension of the operator 
$\Phi_{\vk}$ to $\phi_\vk$ has dimension $\delta_k = \ha + \nu_k$, as claimed in \eqref{nudef} and~\eqref{opep}.

Introducing $x = - 2 i \om \zeta$, equation~\eqref{Aeep2} can be written as 
\be 
\p_x^2 \phi + \le(-{1 \ov 4} + {i q_* \ov x} + {{1 \ov 4} - \nu_k^2   \ov x^2} \ri) \phi =0
\ee
which is precisely the Whittaker's equation and the two linearly independent solutions can be written as 
\be \label{gen2dS}
\phi =  c_{{\rm in}}{\rm W}_{i q_*, \nu_k}\left(x\right)
+c_{{\rm out}}{\rm W}_{-i q_*, \nu_k}\left( -x \right) ,
\ee
where ${\rm W}_{\lambda,\mu}(z)$ is the Whittaker function.
Of these,
the function multiplying $c_{{\rm in}}$,
\be 
{\rm W}_{i q_*, \nu_k}\left(- { 2 i \omega \zeta } \right) = e^{i \om \zeta} (-2 i \omega \zeta )^{i q_*} + \dots, \qquad \zeta \to + \infty
\ee
is ingoing at the horizon. Keeping only the ingoing part, the asymptotic behavior of $\phi$ at $\zeta \to 0$ is given by
\be 
\phi =   {\Ga (2 \nu_k) \ov \Ga (\ha + \nu_k - i q_*)} (-2 i \om \zeta)^{\ha-\nu_k} + {\Ga (-2 \nu_k) \ov \Ga (\ha - \nu_k - i q_*)} (-2 i \om \zeta)^{\ha+\nu_k} \ .
\ee
From~\eqref{GRdef}  the retarded  function is simply the ratio of two coefficients in the above equation, leading to 
 \be \label{AexbG}
 \sG_k (\om) = (- 2 i \om)^{2 \nu_k} \frac{ \Gamma (-2\nu_k ) \Gamma \left(\frac{1}{2}+ \nu_k-i q_*\right)}{\Gamma
(2\nu_k )\Gamma \left(\frac{1}{2}- \nu_k-i q_*\right)   } 
\ee
which is~\eqref{iiRc}.

\section{Derivation of the master formula} \label{app:master}

Here we present the derivation of the master formula \eqref{roep1}, following~\cite{Faulkner09}.  The equation of motion (in momentum space) for $\phi$ with action~\eqref{DscaA} in the charged black hole geometry~\eqref{RNmetric} can be written as
\be \label{eomM}
z^{d+1} \p_z \le({f \ov z^{d-1}} \p_z \phi \ri) + z^2 \le({(\om+ qA_t)^2 \ov f} - k^2 \ri) \phi - m^2 R^2 \phi = 0 \ .
\ee
 We will  consider $T=0$ with $f$ and $A_t$ given by~\eqref{zeroQ}. 
Recall from \eqref{nearBdyScal} that near the AdS$_{d+1}$ boundary $\phi(z)$ has the standard asymptotic behavior $\phi(z \to 0) \sim A z^{d-\Delta_+} + B z^{\Delta_+}$ and that the correlator we seek to compute can be written as\nicefootnote[1]{We will consider the standard quantization in our discussion below. The generalization to the alternative quantization is immediate.}  
 \be
 G_R(\om, \vk) =  \frac{B}{A}  \label{RetD}
\ee 
provided that $\phi$ is an in-falling wave at the horizon. Note that since $A_t \to \mu$ as $z \to 0$, the boundary theory energy corresponds to $\om + q \mu$, i.e. $\om$ should be interpreted as the measured from the effective chemical potential $q \mu$ (which we take to be positive for definiteness).

We are interested in the behavior of $G_R$ in the low frequency limit ($\om \ll \mu$). However, we  
cannot directly perform a perturbative expansion in $\om$ in~\eqref{eomM} since the $\om$-dependent term becomes singular and dominates at the horizon where $f(z_*) \to 0$, no matter how small $\om$ is. This is not just a technicality; physically this reflects the abundance of critical modes coming from the IR fixed point and is of tremendous importance.

To deal with it, it is useful to isolate the near-horizon AdS$_2 \times \mathbb{R}^{d-1}$ region~\eqref{ads2M} (which will be referred to as the IR region) from the rest of the black hole spacetime (which will be referred to as the UV region). In the IR AdS$_2$ region the $\om$ dependence will be treated non-perturbatively.  The crossover between the two regions happens near the boundary of AdS$_2$ with ${1 \ov \mu} \ll \zeta \ll {1 \ov \om}$, where $\mu \zeta \sim {z_* \ov z_*-z}$ remains big so that we are still in the near-horizon region, but the $\om$-dependent term $z_*^2 {\om^2 \ov f} \sim \om^2 \zeta^2 $ in~\eqref{eomM}  becomes small. 
In the UV region we can treat the small $\om$ limit using a standard perturbative expansion with the lowest order equation given by setting $\om =0$ in~\eqref{eomM}.  

So we start with a solution in the AdS$_2   \times \mathbb{R}^{d-1} $ region  and evolve it outwards, matching it to a solution in the UV region to determine the coefficients $A$ and $B$. The scalar wave equation on  the AdS$_2   \times \mathbb{R}^{d-1} $ region~\eqref{ads2M}  is given by~\eqref{Aeep2}. 
As we are computing a retarded correlator we are looking for an {\it in-falling} solution at the black hole horizon (i.e. the term proportional to $c_{in}$ in~\eqref{gen2dS} even though the explicit form is not important for the current discussion).
We now take this infalling solution and evolve it to the boundary of the AdS$_2$ region, where it can be written
as 
\be \label{irsol}
\phi(\zeta) = \zeta^{\frac{1}{2} - \nu_k} + \sG_k(\om) \zeta^{\frac{1}{2}+\nu_k}.
\ee
In~\eqref{irsol} we have normalized the solution so that the first term has coefficient $1$ and then by definition the coefficient of the second term is precisely the retarded Green function in the AdS$_2$ region we computed earlier in~\eqref{AexbG}. 

We now need to continue evolution all the way to the asymptotically AdS$_{d+1}$ region to extract the coefficients $A$ and $B$. We thus need to match the IR solution~\eqref{irsol} to a solution in the UV region. Now that we are out of the dangerous IR region, to leading order we can set $\om =0$ in~\eqref{eomM} in the UV region,
\be \label{leadout}
z^{d+1} \p_z \le({f \ov z^{d-1}} \p_z \phi \ri) + z^2 \le({q^2A_t^2 \ov f} - k^2 \ri) \phi - m^2 R^2 \phi = 0 \ .
\ee 
 The above equation has two independent solutions $\eta_\pm^{(0)}$ which can be specified by their behavior near $z \to z_*$ as
\be
\eta_{\pm}^{(0)} (z) \to  
\le({\zeta \ov z_*}\ri)^{\ha \mp \nu_k} , \quad (z \to z_*)  \ 
\label{IRdefUVsoln}
\ee
with the corresponding asymptotic behavior as $z \to 0$ as 
\be \label{12as}
\eta_{\pm}^{(0)} (z ) \approx  a_{\pm}^{(0)} (k) \le({z\ov z_*}\ri)^{d-\Delta_+ }  + b_{\pm}^{(0)} (k) \le({z\ov z_*}\ri)^{\Delta_+}  \ .
\ee
$a_{\pm}^{(0)} (k)$ and $b_\pm^{(0)} (k)$ thus defined are (dimensionless) functions of $k$ which can be computed numerically.

In the overlapping region both~\eqref{irsol} and~\eqref{IRdefUVsoln} apply, which determines the full UV solution to be 
\be \label{match}
\phi(z) = \eta_{+}^{(0)} (z) + \sG_k(\om) z_*^{2 \nu_k}  \eta_{-}^{(0)} (z) \ . 
\ee
Equation~\eqref{match} can be generalized  to higher orders in $\om$ (for details see~\cite{Faulkner09})
\be \label{match1}
\phi(z) = \eta_{+} (z) + \sG_k(\om) z_*^{2 \nu_k}  \eta_{-} (z) \  
\ee
where 
\be  \label{omex1}
\eta_\pm = \eta^{(0)}_\pm + \om  \eta^{(1)}_\pm + O(\om^2) 
\ee
are the two linearly independent perturbative solutions to the full UV region equation.
Near $z =0$, $\eta_\pm$ have the expansion of the form~\eqref{12as} with various coefficients $a_\pm^{(0)}, b_\pm^{(0)}$ replaced by $a_\pm, b_\pm$ which have an analytic $\om$-expansion such as 
\be \label{omex2}
a_+ (k, \om) = a_+^{(0)} (k) + \om a_+^{(1)} (k) + \dots \  .
\ee  
Note that since both the boundary conditions specifying $\eta_{\pm}$ \eqref{IRdefUVsoln} and the equation~\eqref{eomM} are real, $a_\pm, b_\pm$ are real. From~\eqref{match1} and the expansion of $\eta_\pm$ near $z=0$ we thus find the boundary theory Green's function 
to be 
\be \label{roep11}
\boxed{G_R (\om, \vk) = \mu_*^{2 \nu_U} {b_+(\om, k)+ b_- (\om, k) \sG_k (\om) \mu_*^{-2 \nu_k} \ov
 a_+ (\om, k) + a_- (\om, k) \sG_k (\om) \mu_*^{-2 \nu_k}}} \ . 
\ee

We conclude this discussion with some remarks:

\ben

\item  At $\om =0$, the leading order equation~\eqref{leadout} in the outer region is in fact the full equation of motion; there is no IR region.  $\eta_\pm^{(0)}$ of~\eqref{IRdefUVsoln} now provide a set of basis for the full equation of motion. Note that by construction $\eta_+^{(0)}$ is {\it normalizable} at the horizon. 
 When $a_+^{(0)}$ vanishes, from~\eqref{12as} $\eta_+^{(0)}$ becomes normalizable also at the boundary and thus  is now a genuine normalizable mode in the black hole geometry. 

\item Note that for a neutral scalar with $q=0$, equation~\eqref{eomM} only depends on $\om^2$ and the expansion parameter in~\eqref{omex1} and~\eqref{omex2} should be $\om^2$, i.e.
 \be
a_+ (k, \om) = a_+^{(0)} (k) + \om^2 a_+^{(2)} (k) + O(\om^4) \ 
\ee 
and so on. 

\item The functions $a_\pm (\om, k), \ b_\pm (\om,k)$ are obtained by solving equation~\eqref{eomM}
perturbatively in $\om$ in the UV region. Their $k$-dependence comes from two sources, from dependence on $\nu_k$ via the boundary condition~\eqref{IRdefUVsoln} and from $k^2$ dependence in the equation~\eqref{eomM} itself. Since the geometry is smooth throughout the UV region we expect the dependence on both $\nu_k$ and $k^2$ to be analytic.
In fact we can think of $b_{\pm}$ and $a_{\pm}$ as {\it functions} of $\nu_k$; i.e. there exists a function $b(\nu_k,k^2,\om)$, {\it analytic in all its arguments}, such that $b_{\pm} = b(\pm\nu_k,k^2,\om)$. This is clear from the boundary condition~\eqref{IRdefUVsoln} (and its generalization for higher orders in $\om$) and from the fact that there is no other dependence on $\nu_k$ from the equation of motion itself.  

\item  As $\nu_k  \to 0$, the two solutions in~\eqref{IRdefUVsoln} become degenerate, i.e.
\be
a_+ (k, \om) \to  a_- (k,\om),  \qquad b_+ (k, \om) \to  b_- (k,\om)  \ .
\ee
 At $\nu_k=0$, the basis of functions in~\eqref{IRdefUVsoln} should be replaced by
\be
\eta^{(0)} (z) = \le({\zeta \ov z_*}\ri)^\ha , \qquad \tilde \eta^{(0)} (z) = -\le({\zeta \ov z_*}\ri)^\ha  \log {\zeta \ov z_*} \label{nu0etas}
\ee
where the asymptotic behavior for them at $z \to 0$ is
\be \label{11as}
\eta^{(0)} (z) \approx  \al  \le({z\ov z_*}\ri)^{3-\Delta }  + \beta  \le({z\ov z_*}\ri)^{\Delta }  , \qquad 
\tilde \eta^{(0)} (z ) \approx  \tilde \al  \le({z\ov z_*}\ri)^{3-\Delta } + \tilde \beta   \le({z\ov z_*}\ri)^{\Delta }
\ee
$\al, \beta, \tilde \al, \tilde \beta$ are now dimensionless real numbers which can again be found numerically.  More explicitly, in the limit of $\nu_k \to 0$ the basis of functions~\eqref{IRdefUVsoln} can be expanded as
\be \label{expsi}
\eta_\pm^{(0)} = \eta^{(0)} (z) \pm \nu_k \tilde \eta^{(0)} (z) + O(\nu_k^2) 
\ee
which leads to 
\be \label{req}
b_\pm^{(0)} = \beta \pm {\nu_k} \tilde \beta + \dots , \qquad   
a_\pm^{(0)} = \al \pm {\nu_k} \tilde \al +  \dots \ .
\ee
 
\item Coefficients $a_\pm, b_\pm$ are not independent. For example evaluating the Wronskian of~\eqref{eomM}~(for $\om =0$)\nicefootnote[1]{The Wronskian of equation~\eqref{eomM} is given by 
\be
W [\phi_1, \phi_2]=  {f \ov z^2}  (\phi_1 \p_z \phi_2 - \phi_2 \p_z \phi_1)
\ee
which is independent of $z$.} for $\eta_\pm^{(0)}$ and demanding that it be equal at infinity and at the horizon, we find the elegant relation:
\be \label{pp3}
a_+^{(0)} (k) b_-^{(0)} (k) - a_-^{(0)} (k)  b_+^{(0)} (k) = {\nu_k \ov \nu_U}  \ .
\ee
A similar analysis on $\eta, \tilde\eta$ results in
\be \label{W1}
\al \tilde \beta - \beta \tilde \al = -{1 \ov 2\nu_U}  \ .
\ee

\item When $\nu_k =- i \lam_k$ is pure imaginary (i.e. when~\eqref{Bcon1} is satisfied), and the basis of solutions~\eqref{IRdefUVsoln} now has the form 
\be
\eta_{\pm}^{(0)} (z) \to  %\le(\frac{6 (z_* -z) }{z_*}\ri)^{-\frac{1}{2} \mp i   \lam_k} = 
\le({\zeta \ov z_*}\ri)^{\ha \pm i \lam_k} , \quad z \to z_*  \ 
\label{UVsolnS}
\ee
These boundary conditions are now complex, and thus so are the $\eta_{\pm}$. As the $\eta_{\pm}$ actually obey a real wave equation, the full analytic structure is determined by the boundary conditions in the infrared; thus we find that now $\eta_{+} = \eta_{-}^*$. This also implies that $a_\pm, b_\pm$ are complex and 
\be  \label{comab}
a_+^* = a_-, \qquad b_+^* = b_- \ .
\ee
 
\item The derivation leading to~\eqref{roep11} can be immediately generalized to a finite temperature $T \ll \mu$. In the IR region one replaces the equation~\eqref{Aeep2}  by the corresponding equation for the  AdS$_2$ black hole~\eqref{ads2T}. As a result, $\sG_k (\om)$ in~\eqref{irsol} is now replaced its finite-temperature generalization $\sG_k^{(T)} (\om)$~\eqref{finiteTch1}. The outer region discussion is exactly the same as before except that one should use finite temperature counterparts of $f$ and $A_t$ in~\eqref{leadout}. For $T \ll \mu$, the outer region equation depends on $T$ analytically and the asymptotic behavior in equations~\eqref{IRdefUVsoln} and~\eqref{12as} is unchanged. 
Thus for $T \ll \mu$, one simply replaces $\sG_k (\om)$ in~\eqref{roep11}  by $\sG_k^{(T)} (\om)$,
but keeping in mind that now $a_{\pm}, b_{\pm}$ also have analytic temperature dependence.

\een

%%%%%%%%%%%%%%%%%%%%%%%%%%%%%%%%%%%%%%%%%%%%%%%%
\section{Double trace deformations} \label{app:doub} % and the RG  flow from the alternative to standard quantization} 
%%%%%%%%%%%%%%%%%%%%%%%%%%%%%%%%%%%%%%%%%%%%%%%%

%\subsection{Double trace deformation}

Consider a large $N$ gauge theory in Euclidean signature deformed by a  
 double-trace operator
\be \label{defF}
\de S =  \ha \ka \int d^d x \, \sO^2
\ee
where $\sO$ is a scalar operator whose Euclidean correlation function (in the absence of deformation) is given by $G (x)$. 
With this deformation the two point function for ${\cal O}$ now becomes:
\bea
G_\ka &=&{1 \ov Z_\ka}  \vev{\sO (x) \sO(0) \, e^{-\ha \ka \int d^d y \, \sO(y)^2}}, \cr
Z_\ka &=& \vev{e^{-\ha \ka \int d^d y \, \sO(y)^2}}
\eea 
which leads to
%\bwt
\be
G_\ka (x) = {1 \ov Z_\ka} \sum_{n=0}^\infty {(-\ka)^n \ov 2^n n!} \le(\prod_{m=1}^n \int d^dy_m\ri)
\vev{\sO(x) \sO(0) \sO(y_1)^2 \dots \sO (y_n)^2} 
\ee 
%\ewt
The disconnected diagrams cancel between the numerator and denominator, leaving us with only connected diagrams. To leading order in the large $N$ limit, the $n$-th term in the above equation becomes 
\be
(-\ka)^n \int d^dy_1 \dots d^d y_n \, G(x-y_1) 
\dots G(y_{n-1}- y_n) G(y_n) \ .
\ee
In momentum space we thus find  a simple geometric sum
\be \label{Eudou}
G_\ka (k) = \sum_{n=0}^n (-\ka)^n G^{n+1} (k) = {G (k) \ov 1 + \ka \,G(k)} = {1 \ov G^{-1} (k) +\ka} \ .
\ee
Analytically continuing~\eqref{Eudou} to the Lorentzian signature and using the standard relation between the Euclidean and retarded  functions we conclude that 
\be  \label{corrD}
G_R^{(\ka)} (\om, \vk)=  {1 \ov G^{-1}_R (\om, \vk) +\ka}
\ee
where $G_R (\om, \vk)$ and $G_R^{(\ka)} (\om, \vk)$ are retarded functions for $\sO$ before and 
after the double trace deformation. 

%\subsection{Flow from alternative to standard quantization}

For a CFT,  the Euclidean correlator $G(k)$ in momentum space for an operator $\sO$ can be written as 
\be 
G(k) = C(\De) k^{2 \De - d}
\ee
where where $\De$ is the scaling dimension of $\sO$ and $C(\De)$ is some constant factor. Plugging the above equation into~\eqref{Eudou} we find that for $\De > {d \ov 2}$, 
\be 
G_\ka (k) \to G (k), \qquad k \to 0 \ .
\ee
This is expected, since for $\De > {d \ov 2}$,  the double trace deformation~\eqref{defF} is irrelevant and 
the theory flows back to the original fixed point in the IR. 
When  $\De< {d \ov 2}$, for which~\eqref{defF} is relevant, we find that in the IR
\be 
G_\ka (k)  \to {1 \ov \ka} - {1 \ov \ka^2} G^{-1} (k) + \dots, \qquad k \to 0 \ .
\ee
The first term in the above equation is a constant (which gives rise to a contact term in coordinate space) and can be ignored. The second term corresponds to an operator of dimension $ d - \De$. We thus  conclude that under the double trace deformation the system flows to a new CFT in the IR in which $\sO$ has dimension $d - \De$. Note that two-point functions of other operators (which are orthogonal to $\sO$) are unaffected. 

In the bullet below that of equation~\eqref{GRdef} in Sec.\eqref{sec:ads} we mentioned that for 
 $\nu_U \in (0,1)$, there are two ways to quantize a scalar field $\phi$ in AdS. In the standard quantization (Dirichlet boundary condition) the corresponding boundary operator $\sO_+$ has dimension $\De_+ = {d \ov 2} + \nu_U$, while in the alternative quantization (Neumann boundary condition), the corresponding boundary operator $\sO_-$ has dimension $\De_- = {d \ov 2} - \nu_U$.
From our discussion above, upon turing on a double trace operator $\sO_-^2$ in the alternative quantization, in the large $N$ limit the system should flow in the IR to a new fixed point in which $\sO_-$ has  dimension $d-\De_- = \De_+$ with the dimensions of all other operators unchanged. Clearly this new fixed point should be identified with that corresponding to the standard quantization. 

 \section{Fourier transforms} \label{app:fourier}

Consider an integral of the form
\be
I(x) \equiv \int dk \; F(\nu_k) e^{ikx}
\ee
where $\nu_k = c \sqrt{k^2 + k_0^2}$ with $c$  and $k_0$ some constants.  We will assume that the analytic structure of $F$ is such that the only singularities in the complex $k$-plane are at the beginning of the branch cuts in $\nu$, that is at $k = \pm {i  k_0} $. If that is so, then we can take the two branch cuts to go upwards and downwards in the complex $k$-plane and then (for positive $x$) by deforming the contour upwards we can rewrite the above integrand as
\be
I(x) = i\int_{k_0}^{\infty} d\kappa \le[F\le(-i\sqrt{\kappa^2 - k_0^2}\ri) -  F\le(+i\sqrt{\kappa^2 - k_0^2}\ri)\ri]e^{-\kappa x}
\ee
Now in the limit of large $x$ we expect the integral to be dominated by the lower limit of the integrand $\kappa \sim k_0$. We can expand the sub-exponential part of the integrand in powers of $(\kappa - k_0)$ and perform the integrals directly; thus it is clear that the leading $x$ behavior will be of the form
\be
I(x) \sim \frac{e^{-k_0x}}{x^{\#}}
\ee
where the power appearing in $\#$ depends on the exact form of the integrand. The key point here is the the location of the branch cut sets a scale which determines the correlation length in the integral. 

\section{Density of states of semi-local quantum liquids}  \label{app:dos}

In this Appendix we consider the density of states of a semi-local quantum liquid. The discussion is motivated by a recent 
argument of~\cite{Jensen:2011su}, which appears to indicate that scaling symmetry only in the time direction cannot persist to arbitrarily low energies. 
The argument is very simple and goes as follows. On dimensional grounds, the scaling symmetry in the time direction implies that the density of states of such a theory should have the form 
\be \label{roo}
\rho (E) = A \delta (E) + {B \ov E}
\ee
where $A$ and $B$ are constants.  The term proportional to $\delta (E)$ gives rise to an entropy density, while the term proportional to $1/E$ appears to be needed to have any nontrivial dynamics (as otherwise the Hamiltonian will be proportional to the identity operator in this sector). 
With $B \neq 0$, however, the integral of the $1/E$ term is divergent at low energies, suggesting an infinite number of states at arbitrarily low energies. This peculiar result suggests that the scaling symmetry must always break down at sufficiently low energies, say below some new dynamically generated scale $E_{IR}$.  Note that the argument is independent of the existence of a ground state entropy density, i.e. it should apply even if $A =0$.\footnote{For example, it will rule out the kind of local quantum critical point discussed in the literature (e.g.~\cite{Si.01}) as a genuine low energy description of of certain quantum phase transitions in heavy fermion materials.} 

We note, however, there are some subtleties with the argument.  In order to have a finite density of states we should introduce an IR cutoff, i.e. we should put the system in a finite box with volume $V_{d-1} \equiv L^{d-1}$, and~\eqref{roo} applies to the system in a finite box. Even {\it without} a new dynamically generated scale $E_{IR}$, in a finite box, we expect that the system should develop a finite 
gap $\ep (L)$ with 
\be
\ep (L) \to 0
\ee
in the thermodynamic limit $L \to \infty$. Thus, more precisely, the argument for a new dynamically generated scale $E_{IR}$ should be made in the thermodynamic limit, i.e. $E_{IR}$ should remain finite for $L \to \infty$.  Let us now assume that such an $E_{IR}$ does not exist, and see whether one can find an immediate contradiction from~\eqref{roo} in the thermodynamic 
limit. 

In general it is hard to deal with the total density of states of an interacting system. In the thermodynamic limit, we can instead consider entropy whose exponential can be considered as giving the total density of states. Thus we first translate equation~\eqref{roo} into a statement regarding the entropy. For $E \neq 0$, taking the logarithm of~\eqref{roo} we find 
\be \label{4}
S (E) = \log B - \log {E } \ .
\ee
The $\log E$ term is singular in the $E \to 0$ limit, but note that this singular term is not proportional to the volume, and thus appears to be {\it negligible
in the thermodynamic limit}. This by itself does not imply that~\eqref{roo} is fine with $B \neq 0$. But it does indicate that drawing a conclusion requires going beyond the thermodynamic limit, in which case one will need to worry about the precise boundary conditions, shape of the box, and etc.  In particular, given that the finite volume cutoff breaks the scaling symmetry, it is not clear whether~\eqref{roo} is really valid. To further highlight possible subtleties associated with going outside the thermodynamic limit, note that there is an alternative way to obtain the entropy from~\eqref{roo}, 
\be \label{2}
N(E) = \int_0^{E} dE' \, \rho (E') \equiv  e^{S (E)} \ .
\ee
In the thermodynamic limit this definition is equivalent to simply taking the logarithm of the density of states. But applying~\eqref{2} to~\eqref{roo} we find that 
\be  \label{3}
N(E) = B \log {E \ov \ep (L)}, \quad \to \quad S(E) = \log B + \log\log {E \ov \ep (L)}, \quad E \geq \ep (L)
\ee
where we have put an IR cutoff $\ep (L)$ in the integration of~\eqref{2}.  The $E$-dependence in~\eqref{3} differ from that in~\eqref{4}. In the $L \to \infty$ limit the second term in $S(E)$ is singular, but 
the divergence is much smaller than volume factor in the usual thermodynamic limit. 

The above discussion does motivate a stronger argument based on the thermodynamic limit. Instead of~\eqref{roo} one may argue that the scaling symmetry in the time direction implies that the entropy 
of the system should have the form 
\be \label{eoo}
S (T) %= V_{d-1} f (T) 
= V_{d-1} (a + b \log T)
\ee
where for convenience we have used the canonical ensemble and $a, b$ are $T$-independent constants. Argument similar to that below~\eqref{roo} again seems to say that nontrivial dynamics requires $b \neq 0$, which then implies that the system should have a dynamically generated cutoff in the thermodynamic limit. We note, however, the coefficient $b$ of the $\log T$ term may not be required to be nonzero to have nontrivial dynamics, as the systems which exhibit such a scaling symmetry have a UV completion. The leading contribution could come from an irrelevant operator. We will discuss a potential example below.  

For a boundary theory at finite density described by a charged black hole in the bulk, the entropy has the form 
\be \label{coo}
S (T) = V_{d-1} \le(N^2 a(T) +  b(T)  + O(1/N^2)  \ri) \ .
\ee
The $O(N^2)$ term comes from the black hole geometry. The term proportional to $b(T)$ comes from a one-loop calculation in the bulk. 
It is a well defined question to ask whether $b(T)$ has a singular $ \log T$ piece in the $T \to 0$ limit.
Here we will not attempt a full one-loop calculation. To get an indication of the behavior of $b(T)$, we compute instead the {\it single-particle density of states}\footnote{It should be kept in mind this bulk single-particle density of states does not correspond to any single-particle density of states in the boundary.} $\rho_s (\om)$ for a scalar field in AdS$_2 \times \RR^{d-1}$. Using $\rho_s$ one obtains an ``off-shell'' partition function $\log Z_s (T)$ using the standard formulas
\be 
\log Z_s (T) =  \sum_{n=1}^\infty f_s \le({T \ov n} \ri) , \qquad  f_s (T) = \int_0^\infty d \om \, e^{-\beta \om} \, \rho_s (\om)
\ee
where $f_s (T)$ is the single-particle partition function in the bulk. We would like to see whether there is a singular $\log T$ term in $\log Z_s (T)$ which can be traced to a similar term in $f_s (T)$\footnote{The standard one loop partition function at a finite $T$ which can be used to directly extract $b(T)$ is computed from path integral in the Euclidean black hole geometry with given $T$. There exists argument (see e.g.~\cite{Fursaev:1997th}) that the ``off-shell'' partition function we are considering is equivalent to that computed from Euclidean path integral. Clearly it would be desirable to compute the Euclidean 
partition function directly. }. 

Before going into technical details, let us summarize the results:

\ben

\item[A.] For a scalar in AdS$_2 \times \RR^{d-1}$, introducing a finite volume in the field-theory spatial directions does not  result in a finite density of states in the bulk\footnote{As emphasized earlier the relation of this bulk single-particle density of states with the density of states in the boundary theory is rather indirect. In particular, one should not view the cutoff $\ep$ introduced below as a new emergent dynamical scale. As discussed in~\cite{Fursaev:1997th} the divergence of $\log Z_s (T)$ in the $\ep \to 0$ limit in fact is identical to UV divergences in the Euclidean partition functions.
 }, due to the presence of the horizon. To proceed, we will thus introduce an additional cutoff $\ep \sim \Lam^{-1}$ where $\Lam$ is a radial coordinate cutoff of the spacetime above the horizon. 
%that would remain finite in the thermodynamic limit. Rather, it is a reflection of our ignorance of the true IR spectrum, and should be thought of as going to zero as the size of the box goes to infinity, i.e.
%\be
%\ep (L\to \infty) = 0 \ .
%\ee 
%Such an $L$-dependent IR cutoff should presumably dynamically emerge in a fully stringy treatment of this problem. The fact that this IR cutoff is invisible in the supergravity limit appears to be dual to the fact that in a many-body system, the spacings of the low-lying energy spectrum could be much smaller than the scale $1/L$ from finite volume, even though on general grounds we do still expect them to go to zero in the $L \to \infty$ limit.
\label{iyna} 

\item[B.] Within this regularization scheme, for a scalar operator of real scaling dimension $\nu_k$ in SLQL, the dangerous $\log T$ piece is {\it not} present. Instead the leading cutoff-independent contribution in the density of states comes from an irrelevant operator (a double trace deformation) and is integrable at small $\om$. One finds that $f_s (T) \sim T^{2 \nu_{k=0}}$ and thus the corresponding contribution to $\log Z_s (T)$ goes to zero in the $T \to 0$ limit. 

\item[C.] For an operator with {\it complex} scaling dimension in SLQL, one does find a non-vanishing (cutoff-independent) $1/\om$ piece in the single-particle density of states. Upon integration, this leads to a singular single-particle partition function $f_s \sim \log T$ and thus a singular $\log Z_s (T)$.  
We would conclude that in this case, a new dynamical scale $E_{IR}$ should emerge in the thermodynamic limit. This is in nice correspondence with the discussion of sec.~\ref{sec:osc}, where we show that precisely in this case the system is unstable to the condensate of the scalar operator and thus a new dynamical scale must emerge. We expect similar results should also apply to a fermionic operator with a complex dimension, which will again be in nice correspondence with the fermionic instability discussed in Sec.~\ref{sec:HFL}. 

\een

From the above, it then appears that the argument based on~\eqref{roo} or~\eqref{eoo} alone is not enough to rule out the presence of a scaling symmetry in the time direction persisting to arbitrarily low energies; it seems that dynamics still plays a role in determining whether this is possible. Clearly further study of this is desirable.

\subsection{Single-particle density of states in the bulk}

Here we calculate the density of states of a neutral scalar field $\phi$ in  AdS$_2 \times \RR^{d-1}$, whose equation of motion is given by~\eqref{Aeep2}--\eqref{ynuk} with $q_* =0$. We will consider solutions which are normalizable at the boundary\footnote{For definiteness we will restrict to the standard quantization.}, i.e. $\al =0$ in~\eqref{nuej0}. 
At the horizon $\zeta \to \infty$, the correponding wave function is given by a plane wave 
\be 
\phi (\om, k, \zeta) = e^{i \om \zeta + {i \de (\om, \nu_k)}} + e^{-i  \om \zeta - {i \de (\om,\nu_k) }} 
\ee
with $\de (\om, \nu_k)$ a phase shift.  Note that the normalizable condition and a discrete $k$ does not impose a quantization condition on $\om$; the system has a continuous spectrum (we are dealing with a scattering problem) even with a finite volume cutoff. While we expect that the system should develop a discrete spectrum in a finite volume, the spacings of low-lying spectrum 
could be much smaller than $1/L$, with $L$ is the system size, in the large $\lam$ and large $N$ limit, and thus not immediately visible at our level of approximation. 
Thus in order to obtain a finite density of states, we will impose an IR cutoff by hand 
at $\zeta = \Lam$ near the horizon\footnote{$\Lam$ should be $k$-dependent, but we will not specify the dependence.}. $\Lam$ has units of time and is the inverse of the energy cutoff scale $E_c$ discussed above. This now requires that $\phi$ satisfies 
\be 
\phi (\om, \Lam) = 0 \quad \to \quad \om  \Lam + \de (\om, \de_k) =   \le(n+\ha \ri)\pi , \qquad n=0,1,\cdots
\ee
Taking differential on both sides of the last equation we find that 
\be
d\om \le(\Lam + {d \de \ov d \om} \ri) = \pi dn \quad \to \quad {dn \ov d\om} = {1 \ov \pi} \le(\Lam + {d \de \ov d \om} \ri) 
\ee
We thus conclude that the density of states is given by 
\be \label{doe}
\rho (\om) = {V_{d-1}}   \int {d^{d-1} k \ov (2 \pi)^{d-1}} \, {d n \ov d\om}  
%= {V_{d-1} \ov \pi}   \int {d^{d-1} k \ov  (2 \pi)^{d-1}} \, \le(\Lam + {d \de \ov d \om} \ri)  
= {V_{d-1} \ov \pi}   \int {d^{d-1} k \ov  (2 \pi)^{d-1}} \, \Lam 
+ f (\om) 
\ee
where $f(\om)$ is the cutoff-independent part given by
\be \label{doe}
f (\om) = {V_{d-1} \ov \pi}   \int {d^{d-1} k \ov  (2 \pi)^{d-1}} \,  {d \de \ov d \om} \ .
\ee
Now note that the scaling symmetry of AdS$_2$ implies that the phase shift $\de (\om, \de_k)$ is in fact $\om$ {\it independent} as the only dependence of a wave function on $\om$ should be through $\om \zeta$. Thus~\eqref{doe} is identically zero
and we are left with a piece which depends on the cutoff. 

\subsubsection{Double trace deformation}

The above discussion is, however, not the full story as the AdS$_2$ region appears in the IR part of the full black hole geometry, which provides the UV completion. 
As discussed in Sec.~\ref{lowgreen} (see discussion around equation~\eqref{eff1}), the leading irrelevant deformation of SLQL is provided by the double trace operator, which we now consider. With a double trace deformation, the normalizable solution is now given by  
setting 
\be 
\al = \xi_k \beta \ 
\ee
in~\eqref{nuej0}. Note that the double trace parameter $\xi_k$ has dimension $-2 \nu_k$. Up to an overall scaling the normalizable condition can also be written as 
\be 
\phi = \xi_k \ze^{\ha - {\nu_k}} + \ze^{\ha + {\nu_k}} \qquad \ze \to 0
\ee
The full normalizable solution in AdS$_2$ then has the form 
\be \label{sol1}
\phi = \xi_k \Ga (1-\nu_k) \le({\om \ov 2} \ri)^{\nu_k} \sqrt{\ze} J_{-\nu_k} (\om \ze) +
\Ga (1+\nu_k) \le({\om \ov 2} \ri)^{-\nu_k} \sqrt{\ze} J_{\nu_k} (\om \ze)
\ee
We thus find the phase shift is given by
\be 
e^{i 2 \de (\om, \de_k)} =  e^{-{i } (\nu_k+\ha)\pi} {1 + C e^{i \nu_k \pi} \ov 1 + C e^{- i \nu_k \pi}} , \qquad
C = \xi_k \le({\om \ov 2} \ri)^{2\nu_k} {\Ga (1-\nu_k) \ov \Ga (1+\nu_k)}
\ee
For small $\om$ we thus find that
\be 
\de (\om) = {\rm const} + C \sin \nu_k \pi + \cdots 
\ee
which when plugged into~\eqref{doe} leads to 
\be 
f(\om) = {V_{d-1} }   \int {d^{d-1} k \ov  (2 \pi)^{d-1}} \,  \xi_k  \le({\om \ov 2} \ri)^{2\nu_k-1} {1 \ov (\Ga (\nu_k))^2} + \cdots 
\ee
where $\cdots$ denotes higher order corrections in $\om$. The above density of states is integrable at $\om=0$.\footnote{Recall that for $k \ll \mu$, $\nu_k \approx \nu_{k=0} > 0$.}  
We thus find that that the cutoff-independent part of the density of states is controlled by leading irrelevant perturbation.

\subsection{Imaginary $\nu$} 

Let us now consider an imaginary $\nu = - i \lam$. For notational simplicity we will now drop the momentum subscript. Embedding the AdS$_2$ into the full extremal BH geometry) amounts to requiring the normalizable wave function to have the asymptotic behavior near the AdS$_2$ boundary 
\be 
\phi = e^{-i \th} \ze^{\ha + i \lam} + e^{i \th} \ze^{\ha - i \lam}
\ee
for some choice of $\th$ (which depends on $\lam$), whose precise value will not concern us. The full normalizable solution in AdS$_2$ is then almost identical to~\eqref{sol1} with $\nu$ now imaginary 
\bea
\phi &=& e^{-i \th} \Ga (1-\nu) \le({\om \ov 2} \ri)^\nu \sqrt{z} J_{-\nu} (\om z) +e^{i \th}
\Ga (1+\nu) \le({\om \ov 2} \ri)^{-\nu} \sqrt{z} J_{\nu} (\om z) \cr
& = & e^{- i \Th} \sqrt{z} J_{i \lam} (\om z)+ e^{ i \Th} \sqrt{z} J_{-i \lam} (\om z)
\eea
with 
\be 
\Th =\lam \log \om + {\rm const}
\ee
Note that in contrast to the earlier case, $\th$ is now dimensionless, which indicates that $f(\om)$ now has to be proportional to $\om^{-1}$ on dimensional ground. The phase shift is now given by
\be 
e^{2 i \de (\om)} = - i {e^{i \Th} + e^{- i \Th} e^{\lam \pi} \ov e^{-i \Th} + e^{i \Th} e^{\lam \pi}}
\ee
We thus conclude that 
\be
 {1 \ov \pi} {d \de\ov d \om} = - {\lam \ov \om} {e^{2 \lam \pi} - 1 \ov 1 + e^{2 \lam \pi} + e^{\lam \pi} \cos 2 \Th}
\ee
which has an un-integrable singularity at $\om =0$. Note that including the momentum integrations in~\eqref{doe} will not cure this non-integrability as $\nu_k$ is almost $k$ independent for $k \ll \mu_*$.

\section{Review of critical exponents} \label{app:critexp}

In the vicinity of a critical point we observe scaling behavior of various observable quantities, which is characterized by a set of critical exponents.  We list some of the most commonly used exponents in the following. We will denote the external tuning parameter $g$ with which we tune the system to the critical point $g=g_c$. Near the critical point the spatial correlation length diverges as
\be
\xi \sim \le| g-g_c  \ri|^{-\nu_{crit}} \ .
\ee
The energy gap for elementary excitations scales as
\be
E_{gap} \sim \xi^{-z}\sim \le| g-g_c  \ri|^{-z \nu_{crit}} \ ,
\ee
where $z$ is called the dynamic critical exponent and determines the characteristic time scale of the approach to equilibrium via $\tau_{eq} \sim 1/E_{gap} $. On the condensed side the order parameter $\sO$ also exhibits scaling near the critical point; the corresponding exponent is: 
\be
\le\< \sO \ri\>\sim \le| g-g_c  \ri|^{\beta} \ ,
\ee
and exactly at the critical point it will depend on the source as
\be
\le\< \sO \ri\>\sim J^{1/\delta} \ ,
\ee
where the coupling to the external source is $J \sO$. The correlation function $\chi=\le\< \sO \sO \ri\>$ can also be used to probe the physical properties of the critical point. The next critical exponent we introduce is for $\chi$ at zero momentum:
\be
\chi(k=0,\om=0) \sim \le| g-g_c  \ri|^{-\gamma} \ .
\ee
The decay of $G_R$ at the critical point in the free theory would be $1/x^{d-2}$, the deviation from this is characterized by $\eta$:
\be
\chi (x,\om=0)\vert_{g=g_c} \sim {1 \ov x^{d-2+\eta}} \ .
\ee
To study the scaling of thermodynamic functions we introduce $\al$ as:
\be
f\sim \le| g-g_c  \ri|^{2-\al} \ ,
\ee
where $f$ is the free energy density.

%\section{Pole structure for scalar and spinor operators} \label{app:poles}

%In this appendix we discuss locations of poles of equations~\eqref{FSexp1} and~\eqref{oscGr1}
%for a scalar and spinor operator respectively. Let us first look at~\eqref{FSexp1}. When 
%$\nu_{k_F} > \ha$, the term linear in $\om$ in the denominator dominates over $\Sig (\om)$ the pole can be written as
%\be 
%\om (k) = v_F (k-k_F)

\end{appendix}


\begin{thebibliography}{9}
%
\bibitem{Maldacena:1997re}
  J.~M.~Maldacena,
  ``The large N limit of superconformal field theories and supergravity,''
  Adv.\ Theor.\ Math.\ Phys.\  {\bf 2}, 231 (1998)
  [Int.\ J.\ Theor.\ Phys.\  {\bf 38}, 1113 (1999)]
  [arXiv:hep-th/9711200].
  
\bibitem{Gubser:1998bc}
S.~S.~Gubser, I.~R.~Klebanov and A.~M.~Polyakov, ``Gauge theory correlators from non-critical string theory,''
Phys.\ Lett.\ B {\bf 428}, 105 (1998)
 [arXiv:hep-th/9802109];
%%CITATION = HEP-TH 9802109;%%
%\cite{Witten:1998zw}
% \bibitem{Witten:1998zw}

\bibitem{Witten:1998qj}
  E.~Witten,``Anti-de Sitter space and holography,''
  Adv.\ Theor.\ Math.\ Phys.\  {\bf 2}, 253 (1998)
  [arXiv:hep-th/9802150]

%\cite{Gubser:2009md}
\bibitem{Gubser:2009md}
  S.~S.~Gubser, A.~Karch, ``From gauge-string duality to strong interactions: A Pedestrian's Guide,''
  Ann.\ Rev.\ Nucl.\ Part.\ Sci.\  {\bf 59}, 145-168 (2009).
  [arXiv:0901.0935 [hep-th]].

 \bibitem{CasalderreySolana:2011us}
  J.~Casalderrey-Solana, H.~Liu, D.~Mateos, K.~Rajagopal and U.~A.~Wiedemann,
  ``Gauge/String Duality, Hot QCD and Heavy Ion Collisions,''
  arXiv:1101.0618 [hep-th].
  %%CITATION = ARXIV:1101.0618;%%

\bibitem{coleman} P.~Coleman, ``Introduction to Many-body Physics,'' 
http://www.physics.rutgers.edu/$\sim$coleman.

%\cite{Benfatto:1990zz}
\bibitem{Benfatto:1990zz}
G.~Benfatto and G.~Gallavotti,
``Renormalization-Group Approach to the Theory of the Fermi Surface,''
Phys.\ Rev.\  B {\bf 42} (1990) 9967.
%%CITATION = PHRVA,B42,9967;%%



%\cite{Polchinski:1992ed}
\bibitem{Polchinski:1992ed}
J.~Polchinski,
``Effective Field Theory and the Fermi Surface,''
arXiv:hep-th/9210046.
%%CITATION = HEP-TH/9210046;%%
%\cite{Shankar:1993pf}
\bibitem{Shankar:1993pf}
  R.~Shankar,
  ``Renormalization group approach to interacting fermions,''
  Rev.\ Mod.\ Phys.\  {\bf 66}, 129 (1994).
  %%CITATION = RMPHA,66,129;%%


\bibitem{stewart}
  G.~R.~Stewart,
``Non-Fermi-liquid behavior in d- and f-electron metals,"
  Rev.~Mod.~Phys.~{\bf 73}, 797 (2001).
  
\bibitem{abrahams2000}
E.~Abrahams and C.~M.~Varma, 
``What angle-resolved photoemission experiments tell about the microscopic theory for high-temperature superconductors,''
PNAS {\bf 97}, 5714 (2000).

  

\bibitem{Varma89}
C.~M. Varma, P.~B. Littlewood, S.~Schmitt-Rink, E.~Abrahams,
and A.~E. Ruckenstein,
``Phenomenology of the normal state of Cu-O high-temperature superconductors,''
Phys.~Rev.~Lett. \textbf{63}, 1996 -- 1999 (1989).

\bibitem{Holstein:1973zz}
  T.~Holstein, R.~E.~Norton and P.~Pincus,
  ``de Haas-van Alphen Effect and the Specific Heat of an Electron Gas,''
  Phys.\ Rev.\  B {\bf 8}, 2649 (1973);
  %%CITATION = PHRVA,B8,2649;%%
  P.~A.~Lee and N.~Nagaosa, 
  ``Gauge theory of the normal state of high-Tc superconductors,''
  Phys.\ Rev.\  B {\bf 46}, 5621 (1992);
%\bibitem{reizer}
M. Y. Reizer, Phys. Rev. B {\bf 40}, 11571 (1989);
%\bibitem{Polchinski:1993ii}
  J.~Polchinski,
 ``Low-energy dynamics of the spinon-gauge system,''
  Nucl.\ Phys.\  B {\bf 422}, 617 (1994)   arXiv:cond-mat/9303037;
  %%CITATION = NUPHA,B422,617;%%
%\bibitem{Nayak:1993uh}
  C.~Nayak and F.~Wilczek,
  ``Non-Fermi liquid fixed point in (2+1)-dimensions,''
  Nucl.\ Phys.\  B {\bf 417}, 359 (1994)
  arXiv:cond-mat/9312086,
   ``Renormalization group approach to low temperature properties of a non-Fermi
  liquid metal,''
  Nucl.\ Phys.\  B {\bf 430}, 534 (1994)
  arXiv:cond-mat/9408016;
  %%CITATION = NUPHA,B430,534;%%
  %%CITATION = NUPHA,B417,359;%%
 %\cite{Halperin:1992mh}
%\bibitem{Halperin:1992mh}
  B.~I.~Halperin, P.~A.~Lee and N.~Read,
  ``Theory of the half filled Landau level,''
  Phys.\ Rev.\  B {\bf 47}, 7312 (1993);
  %%CITATION = PHRVA,B47,7312;%%
%  \bibitem{altshuler-1994}
Y.~B. Kim, A.~Furusaki, X.-G. Wen, and P.~A. Lee,
``Gauge-invariant response functions of fermions coupled to a gauge field,''
Phys.~Rev.~B  \textbf{50}, 17917--17932 (1994);
B.~L.~Altshuler, L.~B.~Ioffe and A.~J.~Millis,
``On the low energy properties of fermions with singular interactions,'' Phys.~Rev.~B \textbf{50} 14048 (1994). 
[arXiv:cond-mat/9406024].
%  S.~S.~Lee, arXiv:0905.4532 [cond-mat].

\bibitem{sungsikgaugefield}
S.~S.~Lee,
``Low energy effective theory of Fermi surface coupled with U(1) gauge field in 2+1 dimensions,''
Phys.\ Rev.\ {\bf B 80}, 165102 (2009)
[arXiv:0905.4532 [cond-mat.str-el]].


%\cite{Metlitski:2010pd}
\bibitem{Metlitski:2010pd}
  M.~A.~Metlitski and S.~Sachdev,
  ``Quantum phase transitions of metals in two spatial dimensions: I.
  Ising-nematic order,''
  Phys.\ Rev.\  B {\bf 82}, 075127 (2010)
  [arXiv:1001.1153 [cond-mat.str-el]].
  %%CITATION = PHRVA,B82,075127;%%


%\cite{Mross:2010rd}
\bibitem{Mross:2010rd}
  D.~F.~Mross, J.~McGreevy, H.~Liu and T.~Senthil,
  ``A controlled expansion for certain non-Fermi liquid metals,''
  Phys.\ Rev.\ B {\bf 82}, 045121 (2010) 
  [arXiv:1003.0894 [cond-mat.str-el]].
  %%CITATION = ARXIV:1003.0894;%%
  
\bibitem{Hertz.76}
J.~A.~Hertz,``Quantum critical phenomena,''
Phys.~Rev.~B \textbf{14}, 1165--1184 (1976).


\bibitem{Sachdev_book}
S.~Sachdev, \emph{Quantum Phase Transitions}, Cambridge University Press,
Cambridge (1999).

\bibitem{mucio}
M.~A.~Continentino, \emph{Quantum scaling in many-body systems}, World Scientific, Singapore, (2001).

\bibitem{Sondhi}
S.~L.~Sondhi, S.~M.~Girvin, J.P.~Carini, D.~Shahar, ``Continuous quantum phase transitions,''' Rev. Mod. Phys. {\bf 69}, 315-333 (1997) [	arXiv:cond-mat/9609279]

\bibitem{Vojta}
M.~Vojta, ``Quantum phase transitions,'' Rep. Prog. Phys. {\bf 66}, 2069 (2003), [arXiv:cond-mat/0309604].

%~\cite{Natphys.08,Gegenwart.08,Lohneysen.07}
\bibitem{Natphys.08}
{F}ocus issue: Quantum~phase transitions,
Nature~Phys. \textbf{4}, 167--204 (2008).


\bibitem{Gegenwart.08}
P.~Gegenwart, Q.~Si, and F.~Steglich, ``Quantum criticality in heavy-fermion metals,''
Nat.~Phys. \textbf{4}, 186--197 (2008).


\bibitem{Lohneysen.07}
H.~v.~L\"{o}hneysen, A.~Rosch, M.~Vojta, and P.~W\"{o}lfle, ``Fermi-liquid instabilities at magnetic quantum phase transitions,''
Rev.~Mod.~Phys. \textbf{79}, 1015--1075 (2007).


\bibitem{xiaogang0}
X.-G. Wen,  ``Vacuum degeneracy of chiral spin states in compactified space,''Phys. \ Rev. \ B {\bf40}, 7387 (1989). 


\bibitem{xiaogang} X.-G. Wen and Q. Niu, ``Ground-state degeneracy of the fractional quantum Hall states in the presence of a random potential and on high-genus Riemann surfaces,'' Phys.\ Rev.\ B{\bf 41}, 9377 (1990). 

\bibitem{coleman06}
P.~Coleman,  ``Heavy Fermions: electrons at the edge of magnetism'',
arXiv:cond-mat/0612006.


\bibitem{heavy2} Q, Si and F.~Steglich, ``Heavy Fermions and
Quantum Phase Transitions,'' Science {\bf 329}, 1161 (2010).


\bibitem{Si.01}
Q.~Si, S.~Rabello, K.~Ingersent, and J.~L.~Smith, ``Locally critical quantum phase transitions in strongly correlated metals,''
Nature \textbf{413}, 804--808 (2001).

\bibitem{Senthil.04}
T.~Senthil, A.~Vishwanath, L.~Balents, S.~Sachdev, and M.~P.~A. Fisher, ``Deconfined quantum critical points,''
Science \textbf{303}, 1490--1494 (2004).

\bibitem{schroeder:00}
A. Schr\"oder, G. Aeppli, R. Coldea, M. Adams, O. Stockert, H. v. Lohneysen, E. Bucher, R. Ramazashvili, and P. Coleman, ``Onset of antiferromagnetism in heavy-fermion metals,''
Nature {\bf 407}, 351 (2000). 

\bibitem{coleman99}
P.~Coleman, ``Theories of non-Fermi liquid behavior in heavy fermions,'' Physica {\bf B 259-261},
353 (1999).

\bibitem{DMFT}
A.~Georges, G.~Kotliar, W.~Krauth, M.~J.~Rozenberg, ``Dynamical mean-field theory of strongly correlated fermion systems and the limit of infinite dimensions,'' 
Rev.\ Mod.\ Phys.\ 
{\bf 68}, 13-125 (1996); 

G. Kotliar, S. Y. Savrasov, K. Haule, V. S. Oudovenko, O. Parcollet, and C. A. Marianetti,  ``Electronic structure calculations with dynamical mean-field theory,'' Rev.\ Mod.\ Phys.\ 
{\bf 78}, 865 (2006).





\bibitem{Bagger:2007vi}
  J.~Bagger and N.~Lambert, ``Comments On Multiple M2-branes,''
  JHEP {\bf 0802}, 105 (2008)
  [arXiv:0712.3738 [hep-th]];
  %\cite{Bagger:2007jr}
 % J.~Bagger and N.~Lambert,
 ``Gauge Symmetry and Supersymmetry of Multiple M2-Branes,''
  Phys.\ Rev.\  D {\bf 77}, 065008 (2008)
  [arXiv:0711.0955 [hep-th]];
  %\cite{Bagger:2006sk}
%\bibitem{Bagger:2006sk}
  %J.~Bagger and N.~Lambert,
  ``Modeling multiple M2's,''
  Phys.\ Rev.\  D {\bf 75}, 045020 (2007)
  [arXiv:hep-th/0611108].
  %%CITATION = PHRVA,D75,045020;%%
  %%CITATION = PHRVA,D77,065008;%%
  %%CITATION = JHEPA,0802,105;%%
  
  \bibitem{Gustavsson:2007vu}
  A.~Gustavsson,
  ``Algebraic structures on parallel M2-branes,''
  Nucl.\ Phys.\  B {\bf 811}, 66 (2009)
  [arXiv:0709.1260 [hep-th]].
  %%CITATION = NUPHA,B811,66;%%
  

%\cite{Bagger:2007vi,Aharony:2008ug}
\bibitem{Aharony:2008ug}
  O.~Aharony, O.~Bergman, D.~L.~Jafferis and J.~Maldacena,
  ``N=6 superconformal Chern-Simons-matter theories, M2-branes and their
  gravity duals,''
  JHEP {\bf 0810}, 091 (2008)
  [arXiv:0806.1218 [hep-th]].
  %%CITATION = JHEPA,0810,091;%%

%\cite{Lee09,Liu09,Cubrovic09,Faulkner09,Faulkner:2010zz,Faulkner:2011tm,Faulkner:2010tq,Iqbal:2011in,Iqbal:2010eh,Faulkner:2010gj,sdwnew}
\bibitem{Lee09}
S.-S. Lee, ``A Non-Fermi Liquid from a Charged Black Hole; A Critical Fermi Ball,'' Phys.~Rev.~D \textbf{79}, 086006 (2009) [arXiv:0809.3402 [hep-th]].


\bibitem{Liu09}
H.~Liu, J.~McGreevy, and D.~Vegh, ``Non-Fermi liquids from holography,'' arXiv:0903.2477 [hep-th]. 
 %%CITATION = ARXIV:0903.2477;%%


\bibitem{Cubrovic09}
M.~Cubrovic, J.~Zaanen, and K.~Schalm, ``String Theory, Quantum Phase Transitions and the Emergent Fermi-Liquid,''
Science \textbf{325}, 439--444
  (2009) [arXiv:0904.1993 [hep-th]].


\bibitem{Faulkner09}
T.~Faulkner, H.~Liu, J.~McGreevy, and D.~Vegh, ``Emergent quantum criticality, Fermi surfaces, and AdS2,''
arXiv:0907.2694 [hep-th].
 %%CITATION = ARXIV:0907.2694;%%
  
 \bibitem{Faulkner:2010zz}
  T.~Faulkner, N.~Iqbal, H.~Liu, J.~McGreevy and D.~Vegh,
  ``Strange metal transport realized by gauge/gravity duality,''
  Science {\bf 329}, 1043-1047 (2010); arXiv:1003.1728 [hep-th].




%\cite{Faulkner:2010da}
%\bibitem{Faulkner:2010da}
 % T.~Faulkner, N.~Iqbal, H.~Liu, J.~McGreevy and D.~Vegh,
  %``From black holes to strange metals,''
  %arXiv:1003.1728 [hep-th].
  %%CITATION = ARXIV:1003.1728;%%
  
%\cite{Faulkner:2011tm}
\bibitem{Faulkner:2011tm}
  T.~Faulkner, N.~Iqbal, H.~Liu, J.~McGreevy and D.~Vegh,
  ``Holographic non-Fermi liquid fixed points,''
  arXiv:1101.0597 [hep-th].
  %%CITATION = ARXIV:1101.0597;%%
 
%\cite{Faulkner:2010tq}
\bibitem{Faulkner:2010tq}
  T.~Faulkner and J.~Polchinski,
  ``Semi-Holographic Fermi Liquids,''
  arXiv:1001.5049 [hep-th].
  %%CITATION = ARXIV:1001.5049;%%

\bibitem{Iqbal:2011in}
  N.~Iqbal, H.~Liu and M.~Mezei, ``Semi-local quantum liquids,''
  arXiv:1105.4621 [hep-th].
%%CITATION = ARXIV:1105.4621;%%


\bibitem{Iqbal:2010eh}
  N.~Iqbal, H.~Liu, M.~Mezei and Q.~Si,
  ``Quantum phase transitions in holographic models of magnetism and
  superconductors,''
  arXiv:1003.0010 [hep-th].
  %%CITATION = ARXIV:1003.0010;%%

%\cite{Faulkner:2010gj}
\bibitem{Faulkner:2010gj}
  T.~Faulkner, G.~T.~Horowitz and M.~M.~Roberts,
  ``Holographic quantum criticality from multi-trace deformations,''
  arXiv:1008.1581 [hep-th].
  %%CITATION = ARXIV:1008.1581;%%

 %\cite{Iqbal:2011aj}
%\bibitem{Iqbal:2011aj}
\bibitem{sdwnew}
  N.~Iqbal, H.~Liu and M.~Mezei,
  ``Quantum phase transitions in semi-local quantum liquids,''
  arXiv:1108.0425 [hep-th].
  %%CITATION = ARXIV:1108.0425;%%

%
\bibitem{Herzog:2007ij}
C.~P.~Herzog, P.~Kovtun, S.~Sachdev and D.~T.~Son, 
``Quantum critical transport, duality, and M-theory,''
Phys. Rev.Ê D {\bf 75}, 085020 (2007)
[arXiv:hep-th/0701036].
%%CITATION = PHRVA,D75,085020;%%



\bibitem{quantcritbh}
S.~Sachdev, M.~Mueller,
 ``Quantum criticality and black holes,''
0810.2005 [cond-mat.str-el]

\bibitem{Hartnoll:2009sz}
  S.~A.~Hartnoll,
  ``Lectures on holographic methods for condensed matter physics,''
  Class.\ Quant.\ Grav.\  {\bf 26}, 224002 (2009)
  [arXiv:0903.3246 [hep-th]].
  %%CITATION = CQGRD,26,224002;%%

\bibitem{herzogr}
C.~P.~Herzog,
  ``Lectures on Holographic Superfluidity and Superconductivity,''
 J.\ Phys.\ A  {\bf 42}, 343001 (2009)
  [arXiv:0904.1975 [hep-th]].
  %%CITATION = JPAGB,A42,343001;%%
%\cite{Horowitz:2010gk}

\bibitem{Horowitz:2010gk}
  G.~T.~Horowitz,
  ``Introduction to Holographic Superconductors,''
  arXiv:1002.1722 [hep-th].
  %%CITATION = ARXIV:1002.1722;%%


 \bibitem{soojong}
S.~J.~Rey,
``String Theory on Thin Semiconductors,"
Progress of Theoretical Physics Supplement No. 177 (2009) pp. 128-142;
 arXiv:0911.5295 [hep-th].


\bibitem{Sachdev:2010ch}
S.~Sachdev,
``Condensed matter and AdS/CFT,''
arXiv:1002.2947 [hep-th].
%%CITATION = ARXIV:1002.2947;%%

\bibitem{Hartnoll:2011fn}
S.~A.~Hartnoll,
``Horizons, holography and condensed matter,''
arXiv:1106.4324 [hep-th].

%

%\cite{Sachdev:2011wg}
\bibitem{Sachdev:2011wg}
  S.~Sachdev,
  ``What can gauge-gravity duality teach us about condensed matter physics?,''
  arXiv:1108.1197 [cond-mat.str-el].
  %%CITATION = ARXIV:1108.1197;%%

\bibitem{Gauntlett:2011mf}
  J.~P.~Gauntlett, J.~Sonner and D.~Waldram,
  ``The spectral function of the supersymmetry current (I),''
  arXiv:1106.4694 [hep-th].

\bibitem{Belliard:2011qq}
  R.~Belliard, S.~S.~Gubser and A.~Yarom,
  ``Absence of a Fermi surface in classical minimal four-dimensional gauged
  supergravity,''
  arXiv:1106.6030 [hep-th].

%\cite{Bah:2010cu}
\bibitem{Bah:2010cu}
  I.~Bah, A.~Faraggi, J.~I.~Jottar and R.~G.~Leigh,
  ``Fermions and Type IIB Supergravity On Squashed Sasaki-Einstein Manifolds,''
  JHEP {\bf 1101}, 100 (2011)
  [arXiv:1009.1615 [hep-th]].

\bibitem{Bah:2010yt}
  I.~Bah, A.~Faraggi, J.~I.~Jottar, R.~G.~Leigh and L.~A.~Pando Zayas,
  ``Fermions and $D=11$ Supergravity On Squashed Sasaki-Einstein Manifolds,''
  JHEP {\bf 1102}, 068 (2011)
  [arXiv:1008.1423 [hep-th]].

\bibitem{Gauntlett:2009dn}
  J.~P.~Gauntlett, J.~Sonner and T.~Wiseman,
  ``Holographic superconductivity in M-Theory,''
  Phys.\ Rev.\ Lett.\  {\bf 103}, 151601 (2009)
  [arXiv:0907.3796 [hep-th]].
  
\bibitem{Gauntlett:2009bh}
  J.~P.~Gauntlett, J.~Sonner and T.~Wiseman,
  ``Quantum Criticality and Holographic Superconductors in M-theory,''
  JHEP {\bf 1002}, 060 (2010)
  [arXiv:0912.0512 [hep-th]].
  
\bibitem{Gubser:2009qm}
S.~S.~Gubser, C.~P.~Herzog, S.~S.~Pufu and T.~Tesileanu,
  ``Superconductors from Superstrings,''
  Phys.\ Rev.\ Lett.\  {\bf 103}, 141601 (2009)
  [arXiv:0907.3510 [hep-th]].

%
\bibitem{Aprile:2011uq}
  F.~Aprile, D.~Roest and J.~G.~Russo,
  ``Holographic Superconductors from Gauged Supergravity,''
arXiv:1104.4473 [hep-th].
  %%CITATION = ARXIV:1104.4473;%%

%\cite{Donos:2011ut}
\bibitem{Donos:2011ut}
  A.~Donos and J.~P.~Gauntlett,
  ``Superfluid black branes in AdS$_4\times S^7$,''
  arXiv:1104.4478 [hep-th].
  %%CITATION = ARXIV:1104.4478;%%



%\cite{Hartnoll:2009sz,herzogr,Horowitz:2010gk}
\bibitem{Polchinski:2010hw}
  J.~Polchinski, ``Introduction to Gauge/Gravity Duality,''
  arXiv:1010.6134 [hep-th].

  
\bibitem{McGreevy:2009xe}
  J.~McGreevy,
  ``Holographic duality with a view toward many-body physics,''
  Adv.\ High Energy Phys.\  {\bf 2010}, 723105 (2010)
  [arXiv:0909.0518 [hep-th]].
  %%CITATION = 00642,2010,723105;%%
  
 
  
  \bibitem{Susskind:1998dq}
  L.~Susskind and E.~Witten, ``The Holographic bound in anti-de Sitter space,''
  arXiv:hep-th/9805114.
  %%CITATION = HEP-TH/9805114;%%

\bibitem{Peet:1998wn}
  A.~W.~Peet and J.~Polchinski, ``UV / IR relations in AdS dynamics,''
  Phys.\ Rev.\  D {\bf 59}, 065011 (1999)
  [arXiv:hep-th/9809022].
  %%CITATION = PHRVA,D59,065011;%%
  
  
%\cite{Son:2002sd}
\bibitem{Son:2002sd}
  D.~T.~Son, A.~O.~Starinets,
  ``Minkowski space correlators in AdS / CFT correspondence: Recipe and applications,''
  JHEP {\bf 0209}, 042 (2002).
  [hep-th/0205051].
  
\bibitem{SCole}  
See e.g., S. Coleman, ``$1/N$'', in Aspect of Symmetry, Cambridge Univ. Press, (1985).


\bibitem{Romans:1991nq}
  L.~J.~Romans,
  ``Supersymmetric, cold and lukewarm black holes in cosmological
  Einstein-Maxwell theory,''
  Nucl.\ Phys.\  B {\bf 383}, 395 (1992)
  arXiv:hep-th/9203018.
  %%CITATION = NUPHA,B383,395;%%

\bibitem{Chamblin:1999tk}
  A.~Chamblin, R.~Emparan, C.~V.~Johnson and R.~C.~Myers,
  ``Charged AdS black holes and catastrophic holography,''
  Phys.\ Rev.\  D {\bf 60}, 064018 (1999)
  arXiv:hep-th/9902170.
  %%CITATION = PHRVA,D60,064018;%%

 
 


\bibitem{SY} 
S. Sachdev and J. Ye, ``Gapless spin-fluid ground state in a random quantum Heisenberg magnet,'' Phys. Rev. Lett. {\bf 70}, 3339 (1993).  

\bibitem{zhuetal}
L. Zhu and Q. Si, ``Critical local-moment fluctuations in the Bose-Fermi Kondo model,'' Phys.\ Rev.\ {\bf B 66}, 024426 (2002); L. Zhu,
S. Kirchner, Q. Si, and A. Georges, ``Quantum Critical Properties of the Bose-Fermi Kondo Model in a Large-N Limit,'' Phys. \ Rev. \ Lett. \ {\bf 93},
267201 (2004).
%\bibitem{} 
  
\bibitem{bgg} S. Burdin, D. R. Grempel, and A. Georges, ``Heavy-fermion and spin-liquid behavior in a Kondo lattice with magnetic frustration,'' Phys. Rev. {\bf B 66}, 045111 (2002).

\bibitem{Sachdev:2010um}
  S.~Sachdev, ``Holographic metals and the fractionalized Fermi liquid,''
  Phys.\ Rev.\ Lett.\  {\bf 105}, 151602 (2010)
  [arXiv:1006.3794 [hep-th]].
  %%CITATION = PRLTA,105,151602;%%

\bibitem{yamamoto}
S.~ J. Yamamoto, Q.~Si, ``Global Phase Diagram of the Kondo Lattice: From Heavy Fermion Metals to Kondo Insulators,'' J. \ Low \ Temp. \ Phys.\ {\bf161}, 233 (2010)
[arXiv:1006.4868].

  
\bibitem{Subir2}
  S.~Sachdev,
  ``The landscape of the Hubbard model,''
  [arXiv:1012.0299 [hep-th]].

\bibitem{Subir3}
  L.~Huijse, S.~Sachdev,
  ``Fermi surfaces and gauge-gravity duality,''
  [arXiv:1104.5022 [hep-th]].
  



\bibitem{Faulkner:2010jy}
  T.~Faulkner, H.~Liu and M.~Rangamani,
  ``Integrating out geometry: Holographic Wilsonian RG and the membrane
  paradigm,''
  arXiv:1010.4036 [hep-th].
 %%CITATION = ARXIV:1010.4036;%%
 
 \bibitem{Heemskerk:2010hk}
  I.~Heemskerk and J.~Polchinski, ``Holographic and Wilsonian Renormalization Groups,''
  arXiv:1010.1264 [hep-th].

 \bibitem{Iqbal:2009fd}
  N.~Iqbal and H.~Liu,
  `Real-time response in AdS/CFT with application to spinors,''
  Fortsch.\ Phys.\  {\bf 57}, 367 (2009)
  [arXiv:0903.2596 [hep-th]].
  %%CITATION = FPYKA,57,367;%%


%\cite{Breitenlohner:1982jf}\cite{Gubser:2005ih,Gubser:2008pf,Denef:2009tp}
\bibitem{Breitenlohner:1982jf}
  P.~Breitenlohner and D.~Z.~Freedman,
  ``Stability In Gauged Extended Supergravity,''
  Annals Phys.\  {\bf 144}, 249 (1982).
  %%CITATION = APNYA,144,249;%%


\bibitem{Gubser:2005ih}
  S.~S.~Gubser, ``Phase transitions near black hole horizons,''
  Class.\ Quant.\ Grav.\  {\bf 22}, 5121 (2005)
  [arXiv:hep-th/0505189].
  %%CITATION = CQGRD,22,5121;%%

%
\bibitem{Gubser:2008pf}
  S.~S.~Gubser and A.~Nellore, ``Low-temperature behavior of the Abelian Higgs model in anti-de Sitter space,''
  JHEP {\bf 0904}, 008 (2009)
  [arXiv:0810.4554 [hep-th]].
  %%CITATION = JHEPA,0904,008;%%


  %\cite{Denef:2009tp}
\bibitem{Denef:2009tp}
  F.~Denef and S.~A.~Hartnoll,
  ``Landscape of superconducting membranes,''
  arXiv:0901.1160 [hep-th].
  %%CITATION = ARXIV:0901.1160;%%

 \bibitem{Hartnoll:2008kx}
  S.~A.~Hartnoll, C.~P.~Herzog and G.~T.~Horowitz, ``Holographic Superconductors,''
  JHEP {\bf 0812}, 015 (2008)
  [arXiv:0810.1563 [hep-th]].  
 
 %\cite{Maeda:2009wv}
\bibitem{Maeda:2009wv}
  K.~Maeda, M.~Natsuume and T.~Okamura, ``Universality class of holographic superconductors,''
  Phys.\ Rev.\  D {\bf 79}, 126004 (2009)
  [arXiv:0904.1914 [hep-th]].
  %%CITATION = PHRVA,D79,126004;%%

\bibitem{Herzog:2010vz}
  C.~P.~Herzog, ``An Analytic Holographic Superconductor,''
  Phys.\ Rev.\  {\bf D81}, 126009 (2010).
  [arXiv:1003.3278 [hep-th]].

%\cite{Pioline:2005pf}
\bibitem{Pioline:2005pf}
  B.~Pioline and J.~Troost,
  ``Schwinger pair production in AdS(2),''
  JHEP {\bf 0503}, 043 (2005)
  [arXiv:hep-th/0501169].
  %%CITATION = JHEPA,0503,043;%%

\bibitem{Jensen:2011su}
  K.~Jensen, S.~Kachru, A.~Karch, J.~Polchinski and E.~Silverstein, ``Towards a holographic marginal Fermi liquid,''
  arXiv:1105.1772 [hep-th].

%Gubser:2008px,Hartnoll:2008vx,Hartnoll:2008kx
\bibitem{Gubser:2008px}
  S.~S.~Gubser,  ``Breaking an Abelian gauge symmetry near a black hole horizon,''
  Phys.\ Rev.\  D {\bf 78}, 065034 (2008)
  [arXiv:0801.2977 [hep-th]];
  %%CITATION = PHRVA,D78,065034;%%
    
 \bibitem{Hartnoll:2008vx}
  S.~A.~Hartnoll, C.~P.~Herzog and G.~T.~Horowitz, ``Building a Holographic Superconductor,''
  Phys.\ Rev.\ Lett.\  {\bf 101}, 031601 (2008)
  [arXiv:0803.3295 [hep-th]].
  
  %\cite{Hartnoll:2009ns}
\bibitem{Hartnoll:2009ns}
  S.~A.~Hartnoll, J.~Polchinski, E.~Silverstein and D.~Tong, ``Towards strange metallic holography,''
  arXiv:0912.1061 [hep-th].
  %%CITATION = ARXIV:0912.1061;%%
  
 \bibitem{Nakamura:2009tf}
  S.~Nakamura, H.~Ooguri and C.~S.~Park, ``Gravity Dual of Spatially Modulated Phase,''
  Phys.\ Rev.\  D {\bf 81}, 044018 (2010)
  [arXiv:0911.0679 [hep-th]];
  %%CITATION = PHRVA,D81,044018;%% 
 %\cite{Ooguri:2010kt}
%\bibitem{Ooguri:2010kt}
  H.~Ooguri and C.~S.~Park, ``Holographic End-Point of Spatially Modulated Phase Transition,''
  Phys.\ Rev.\  D {\bf 82}, 126001 (2010)
  [arXiv:1007.3737 [hep-th]].
  %%CITATION = PHRVA,D82,126001;%%
  
  \bibitem{Gubser:2009qt}
  S.~S.~Gubser and F.~D.~Rocha,
  ``Peculiar properties of a charged dilatonic black hole in AdS$_5$,''
  Phys.\ Rev.\  D {\bf 81}, 046001 (2010)
  [arXiv:0911.2898 [hep-th]].
  %%CITATION = PHRVA,D81,046001;%%


%\cite{Goldstein:2009cv}
\bibitem{Goldstein:2009cv}
  K.~Goldstein, S.~Kachru, S.~Prakash and S.~P.~Trivedi,
  ``Holography of Charged Dilaton Black Holes,''
  JHEP {\bf 1008}, 078 (2010)
  [arXiv:0911.3586 [hep-th]].
  %%CITATION = JHEPA,1008,078;%%

%\cite{Goldstein:2010aw}
\bibitem{Goldstein:2010aw}
  K.~Goldstein, N.~Iizuka, S.~Kachru, S.~Prakash, S.~P.~Trivedi and A.~Westphal,
  ``Holography of Dyonic Dilaton Black Branes,''
  JHEP {\bf 1010}, 027 (2010)
  [arXiv:1007.2490 [hep-th]].
  %%CITATION = JHEPA,1010,027;%%


%\cite{Charmousis:2010zz}
\bibitem{Charmousis:2010zz}
  C.~Charmousis, B.~Gouteraux, B.~S.~Kim, E.~Kiritsis and R.~Meyer,
  ``Effective Holographic Theories for low-temperature condensed matter
  systems,''
  JHEP {\bf 1011}, 151 (2010)
  [arXiv:1005.4690 [hep-th]]; \\
  %%CITATION = JHEPA,1011,151;%%
    R.~Meyer, B.~Gouteraux and B.~S.~Kim,
  ``Strange Metallic Behaviour and the Thermodynamics of Charged Dilatonic
  Black Holes,''
  arXiv:1102.4433 [hep-th]; \\
  %%CITATION = ARXIV:1102.4433;%%
B.~Gouteraux, B.~S.~Kim and R.~Meyer,
  ``Charged Dilatonic Black Holes and their Transport Properties,''
  arXiv:1102.4440 [hep-th].
  %%CITATION = ARXIV:1102.4440;%%

\bibitem{mac}
J. McGreevy, ``In pursuit of a nameless metal,'' A Viewpoint on~\cite{Sachdev:2010um}, Physics {\bf 3}, 83 (2010). 
  
    \bibitem{anderson}
  P.~W.~Anderson, ``In Praise of Unstable Fixed Points: The Way Things Actually Work,'' arXiv:cond-mat/0201431.


%\cite{Gubser:2000nd,Maldacena:2000mw}
\bibitem{Gubser:2000nd}
  S.~S.~Gubser,
  ``Curvature singularities: The Good, the bad, and the naked,''
  Adv.\ Theor.\ Math.\ Phys.\  {\bf 4}, 679 (2000)
  [arXiv:hep-th/0002160].
  %%CITATION = 00203,4,679;%%
  
  
  
%\cite{Maldacena:2000mw}
\bibitem{Maldacena:2000mw}
  J.~M.~Maldacena and C.~Nunez,
  ``Supergravity description of field theories on curved manifolds and a no go
  theorem,''
  Int.\ J.\ Mod.\ Phys.\  A {\bf 16}, 822 (2001)
  [arXiv:hep-th/0007018].
  %%CITATION = IMPAE,A16,822;%%  
  
\bibitem{Kachru:2008yh}
  S.~Kachru, X.~Liu, M.~Mulligan,
  ``Gravity Duals of Lifshitz-like Fixed Points,''
  Phys.\ Rev.\  {\bf D78}, 106005 (2008).
  [arXiv:0808.1725 [hep-th]].


  
%\bibitem{Gegenwart.08}
%P.~Gegenwart, Q.~Si, and F.~Steglich, ``Quantum criticality in heavy-fermion metals,''
%Nat.~Phys. \textbf{4}, 186--197 (2008).



%\cite{Albash:2009wz,Basu:2009qz}

\bibitem{Albash:2009wz}
  T.~Albash and C.~V.~Johnson,
  ``Holographic Aspects of Fermi Liquids in a Background Magnetic Field,''
  arXiv:0907.5406 [hep-th]; \\
  %%CITATION = ARXIV:0907.5406;%%
%\cite{Albash:2010yr}
%\bibitem{Albash:2010yr}
  T.~Albash and C.~V.~Johnson,
  ``Landau Levels, Magnetic Fields and Holographic Fermi Liquids,''
  J.\ Phys.\ A  {\bf 43}, 345404 (2010)
  [arXiv:1001.3700 [hep-th]].
  %%CITATION = JPAGB,A43,345404;%%


   \bibitem{Hartnoll:2011dm}
  S.~A.~Hartnoll, D.~M.~Hofman and D.~Vegh, ``Stellar spectroscopy: Fermions and holographic Lifshitz criticality,''
  arXiv:1105.3197 [hep-th].


\bibitem{Cubrovic:2011xm}
  M.~Cubrovic, Y.~Liu, K.~Schalm, Y.~W.~Sun and J.~Zaanen,
  ``Spectral probes of the holographic Fermi groundstate: dialing between the
  electron star and AdS Dirac hair,''
  arXiv:1106.1798 [hep-th].
  %%CITATION = ARXIV:1106.1798;%%


%\cite{Basu:2009qz}
\bibitem{Basu:2009qz}
  P.~Basu, J.~He, A.~Mukherjee and H.~H.~Shieh,
  ``Holographic Non-Fermi Liquid in a Background Magnetic Field,''
  arXiv:0908.1436 [hep-th].
  %%CITATION = ARXIV:0908.1436;%%

 %\cite{Gubankova:2010rc}
\bibitem{Gubankova:2010rc}
  E.~Gubankova, J.~Brill, M.~Cubrovic, K.~Schalm, P.~Schijven and J.~Zaanen,
  ``Holographic fermions in external magnetic fields,''
  arXiv:1011.4051 [hep-th].
  %%CITATION = ARXIV:1011.4051;%% 
  
  \bibitem{Denef:2009kn}
  F.~Denef, S.~A.~Hartnoll and S.~Sachdev,
  ``Black hole determinants and quasinormal modes,''
  Class.\ Quant.\ Grav.\  {\bf 27}, 125001 (2010)
  [arXiv:0908.2657 [hep-th]];
  %%CITATION = CQGRD,27,125001;%%
  
  %\cite{Denef:2009yy}
%\bibitem{Denef:2009yy}
  F.~Denef, S.~A.~Hartnoll and S.~Sachdev,
  ``Quantum oscillations and black hole ringing,''
  Phys.\ Rev.\  D {\bf 80}, 126016 (2009)
  [arXiv:0908.1788 [hep-th]].
  %%CITATION = PHRVA,D80,126016;%%
  
\bibitem{Hartnoll:2009kk}
  S.~A.~Hartnoll and D.~M.~Hofman,
  ``Generalized Lifshitz-Kosevich scaling at quantum criticality from the
  holographic correspondence,''
  Phys.\ Rev.\  B {\bf 81}, 155125 (2010)
  [arXiv:0912.0008 [cond-mat.str-el]].
  %%CITATION = PHRVA,B81,155125;%%
  
\bibitem{chenkaowen}
J.~W.~Chen, Y.~J.~Kao and W.~Y.~Wen,
``Peak-Dip-Hump from Holographic Superconductivity,''
arXiv:0911.2821 [hep-th].
%%CITATION = ARXIV:0911.2821;%%


%\cite{chenkaowen,Faulkner:2009am,fabio}
\bibitem{Faulkner:2009am}
T.~Faulkner, G.~T.~Horowitz, J.~McGreevy, M.~M.~Roberts and D.~Vegh,
``Photoemission `Experiments' on Holographic Superconductors,''
arXiv:0911.3402 [hep-th].
%%CITATION = ARXIV:0911.3402;%%


\bibitem{fabio}
S.~S.~Gubser, F.~D.~Rocha and P.~Talavera,
``Normalizable Fermion Modes in a Holographic Superconductor,''
arXiv:0911.3632 [hep-th].


 %\cite{Gubser:2010dm}
\bibitem{Gubser:2010dm}
  S.~S.~Gubser, F.~D.~Rocha and A.~Yarom,
  ``Fermion correlators in non-abelian holographic superconductors,''
  arXiv:1002.4416 [hep-th].
  %%CITATION = ARXIV:1002.4416;%
  
  %\cite{Ammon:2010pg}
\bibitem{Ammon:2010pg}
  M.~Ammon, J.~Erdmenger, M.~Kaminski and A.~O'Bannon,
  ``Fermionic Operator Mixing in Holographic p-wave Superfluids,''
  JHEP {\bf 1005}, 053 (2010)
  [arXiv:1003.1134 [hep-th]].
  %%CITATION = JHEPA,1005,053;
  
  %\cite{Benini:2010qc}
\bibitem{Benini:2010qc}
  F.~Benini, C.~P.~Herzog and A.~Yarom,
  ``Holographic Fermi arcs and a d-wave gap,''
  arXiv:1006.0731 [hep-th].
  %%CITATION = ARXIV:1006.0731;%%
 
%\cite{Vegh:2010fc}
\bibitem{Vegh:2010fc}
  D.~Vegh,
  ``Fermi arcs from holography,''
  arXiv:1007.0246 [hep-th].
  %%CITATION = ARXIV:1007.0246;%%  
  

  
  %\cite{Edalati:2010ww,Edalati:2010ge,Guarrera:2011my,Albash:2010dr,}

  
%\cite{Edalati:2010ww}
\bibitem{Edalati:2010ww}
  M.~Edalati, R.~G.~Leigh and P.~W.~Phillips,
  ``Dynamically Generated Mott Gap from Holography,''
  Phys.\ Rev.\ Lett.\  {\bf 106}, 091602 (2011)
  [arXiv:1010.3238 [hep-th]].
  %%CITATION = PRLTA,106,091602;%%
    
 %\cite{Edalati:2010ge}
\bibitem{Edalati:2010ge}
  M.~Edalati, R.~G.~Leigh, K.~W.~Lo and P.~W.~Phillips,
  ``Dynamical Gap and Cuprate-like Physics from Holography,''
  Phys.\ Rev.\  D {\bf 83}, 046012 (2011)
  [arXiv:1012.3751 [hep-th]].
  %%CITATION = PHRVA,D83,046012;%%
    
  %\cite{Guarrera:2011my}
\bibitem{Guarrera:2011my}
  D.~Guarrera and J.~McGreevy,
  ``Holographic Fermi surfaces and bulk dipole couplings,''
  arXiv:1102.3908 [hep-th].
  %%CITATION = ARXIV:1102.3908;%%

%\cite{Laia:2011zn}
\bibitem{Laia:2011zn}
  J.~N.~Laia and D.~Tong,
  %``A Holographic Flat Band,''
  arXiv:1108.1381 [hep-th].
  %%CITATION = ARXIV:1108.1381;%%


\bibitem{Albash:2010dr}
  T.~Albash,
  ``Non-Unitary Fermionic Quasinormal Modes at Zero Frequency,''
  arXiv:1002.4431 [hep-th].
  %%CITATION = ARXIV:1002.4431;%%

%\cite{Wu:2011bx}
\bibitem{Wu:2011bx}
  J.~P.~Wu,
  ``Holographic fermions in charged Gauss-Bonnet black hole,''
  arXiv:1103.3982 [hep-th].
  %%CITATION = ARXIV:1103.3982;%%
    
      
  %\cite{Wang:2009pp,Balasubramanian:2010sc,Rangamani:2011ae}%\cite{Hartman:2010fk,Gubankova:2010ny}
\bibitem{Wang:2009pp}
  J.~R.~Wang and G.~Z.~Liu,
  ``Non-Fermi liquid behavior due to U(1) gauge field in two dimensions,''
  Nucl.\ Phys.\  B {\bf 832}, 441 (2010)
  [arXiv:0907.1022 [cond-mat.supr-con]].
  %%CITATION = NUPHA,B832,441;%%

%\cite{Balasubramanian:2010sc}
\bibitem{Balasubramanian:2010sc}
  V.~Balasubramanian, I.~Garcia-Etxebarria, F.~Larsen and J.~Simon,
  ``Helical Luttinger Liquids and Three Dimensional Black Holes,''
  arXiv:1012.4363 [hep-th].
  %%CITATION = ARXIV:1012.4363;%%

  
  %\cite{Rangamani:2011ae}
\bibitem{Rangamani:2011ae}
  M.~Rangamani and B.~Withers,
  ``Fermionic probes of local quantum criticality in one dimension,''
  arXiv:1106.3210 [hep-th].
  %%CITATION = ARXIV:1106.3210;%%

 
\bibitem{Hartman:2010fk}
  T.~Hartman and S.~A.~Hartnoll,
  ``Cooper pairing near charged black holes,''
  JHEP {\bf 1006}, 005 (2010)
  [arXiv:1003.1918 [hep-th]].
  %%CITATION = JHEPA,1006,005;%% 
  
%\cite{Gubankova:2010ny}
\bibitem{Gubankova:2010ny}
  E.~Gubankova,
  ``Particle-hole instability in the $AdS_4$ holography,''
  arXiv:1006.4789 [hep-th].
  %%CITATION = ARXIV:1006.4789;%%

\bibitem{Hartnoll:2010gu}
  S.~A.~Hartnoll and A.~Tavanfar,
  ``Electron stars for holographic metallic criticality,''
  arXiv:1008.2828 [hep-th].
  %%CITATION = ARXIV:1008.2828;%%


\bibitem{deBoer:2009wk}
  J.~de Boer, K.~Papadodimas and E.~Verlinde,
  ``Holographic Neutron Stars,''
  JHEP {\bf 1010}, 020 (2010)
  [arXiv:0907.2695 [hep-th]].
  %%CITATION = JHEPA,1010,020;%%
 
 \bibitem{Arsiwalla:2010bt}
  X.~Arsiwalla, J.~de Boer, K.~Papadodimas and E.~Verlinde, ``Degenerate Stars and Gravitational Collapse in AdS/CFT,''
  JHEP {\bf 1101}, 144 (2011)
  [arXiv:1010.5784 [hep-th]].
  %%CITATION = JHEPA,1101,144;%%  
 
  \bibitem{vCubrovic:2010bf}
  M.~Cubrovic, J.~Zaanen and K.~Schalm,
  ``Constructing the AdS dual of a Fermi liquid: AdS Black holes with Dirac
  %hair,''
  arXiv:1012.5681 [hep-th].
  %%CITATION = ARXIV:1012.5681;%% 
  
\bibitem{Hartnoll:2010xj}
  S.~A.~Hartnoll, D.~M.~Hofman and A.~Tavanfar,
  ``Holographically smeared Fermi surface: Quantum oscillations and Luttinger
  count in electron stars,''
  arXiv:1011.2502 [hep-th].
  
  
  
  \bibitem{Hartnoll:2010ik}
  S.~A.~Hartnoll and P.~Petrov,
  ``Electron star birth: A continuous phase transition at nonzero density,''
  arXiv:1011.6469 [hep-th].
  %%CITATION = ARXIV:1011.6469;%%
  
  \bibitem{Puletti:2010de}
  V.~G.~M.~Puletti, S.~Nowling, L.~Thorlacius and T.~Zingg,
  ``Holographic metals at finite temperature,''
  JHEP {\bf 1101}, 117 (2011)
  [arXiv:1011.6261 [hep-th]].
  %%CITATION = JHEPA,1101,117;%%
  

%\cite{Sachdev:2011ze}
\bibitem{Sachdev:2011ze}
  S.~Sachdev,
  ``A model of a Fermi liquid using gauge-gravity duality,''
  arXiv:1107.5321 [hep-th].
  %%CITATION = ARXIV:1107.5321;%%

%\cite{Jensen:2010ga}
\bibitem{Jensen:2010ga}
  K.~Jensen, A.~Karch, D.~T.~Son and E.~G.~Thompson,
  ``Holographic Berezinskii-Kosterlitz-Thouless Transitions,''
  arXiv:1002.3159 [hep-th].
  %%CITATION = ARXIV:1002.3159;%%

%\cite{Jensen:2010vx}
\bibitem{Jensen:2010vx}
  K.~Jensen,
  ``More Holographic Berezinskii-Kosterlitz-Thouless Transitions,''
  Phys.\ Rev.\  D {\bf 82}, 046005 (2010)
  [arXiv:1006.3066 [hep-th]].
  %%CITATION = PHRVA,D82,046005;%%

\bibitem{Basu:2010fa}
  P.~Basu, J.~He, A.~Mukherjee, M.~Rozali and H.~H.~Shieh,
  %``Competing Holographic Orders,''
  JHEP {\bf 1010}, 092 (2010)
  [arXiv:1007.3480 [hep-th]].
  %%CITATION = JHEPA,1010,092;%%

%\cite{Jensen:2011af}
%\bibitem{Jensen:2011af}
\bibitem{Jensen} 
  K.~Jensen,
  ``Semi-Holographic Quantum Criticality,''
  arXiv:1108.0421 [hep-th].
  %%CITATION = ARXIV:1108.0421;%%


%\cite{Edalati:2011yv}
\bibitem{Edalati:2011yv}
  M.~Edalati, K.~W.~Lo and P.~W.~Phillips,
  ``Neutral Order Parameters in Metallic Criticality in d=2+1 from a Hairy
  Electron Star,''
  arXiv:1106.3139 [hep-th].
  %%CITATION = ARXIV:1106.3139;%%


\bibitem{Anninos:2010sq}
  D.~Anninos, S.~A.~Hartnoll and N.~Iqbal,
  ``Holography and the Coleman-Mermin-Wagner theorem,''
  Phys.\ Rev.\  D {\bf 82}, 066008 (2010)
  [arXiv:1005.1973 [hep-th]].

\bibitem{veghun} D.~Vegh, ``Marginal Fermi liquids at holographic quantum critical points,''
unpublished. 


\bibitem{efimov}
V.~Efimov, ``Energy levels arising from resonant two-body forces in a three-body system,'' Phys. Lett. {\bf B33}, 563 (1970).

%\cite{Fursaev:1997th}
\bibitem{Fursaev:1997th}
  D.~V.~Fursaev,
  %``Euclidean and canonical formulations of statistical mechanics in the
  %presence of killing horizons,''
  Nucl.\ Phys.\  B {\bf 524}, 447 (1998)
  [arXiv:hep-th/9709213].
  %%CITATION = NUPHA,B524,447;%%

%\cite{Albash:2009wz,Basu:2009qz,Denef09,soojong,chenkaowen,Faulkner:2009am,fabio,hofman,Hartnoll:2009ns}





\end{thebibliography}
\end{document}